\documentclass[letterpaper,11pt]{article}

\usepackage{graphicx}
\usepackage{dcolumn}
\usepackage{bm}
\usepackage[bottom]{footmisc}
\usepackage[font=small,labelfont=bf,
   justification=raggedright,
   format=plain]{caption}
\usepackage{subcaption}
\usepackage{comment}
\usepackage{braket}
\usepackage{booktabs}
\usepackage{float}
\usepackage{slashed}
\usepackage{placeins}
\usepackage{bbold}
\usepackage{todonotes}
\usepackage{hyperref}
\usepackage{amsfonts}
\usepackage{amsmath}
\usepackage{amssymb}
\usepackage{hhline}
\usepackage{changepage}

\newcommand{\tb}{\bar \theta}

\newcommand{\dslash}[1]{#1 \llap{/\kern-0.5pt}}
\newcommand{\Dslash}[1]{#1 \llap{/\kern+1.5pt}}
\newcommand{\DDslash}[1]{#1 \llap{/\kern+2.3pt}}
\newcommand{\dslashh}[1]{#1 \llap{/\kern+1pt}}

\newcommand{\CP}{C\hspace{-.5mm}P}

\newcommand{\be}{\begin{equation}}
\newcommand{\ee}{\end{equation}}
\newcommand{\overbar}[1]{\mkern 1.5mu\overline{\mkern-1.5mu#1\mkern-1.5mu}\mkern 1.5mu}

\definecolor{bleu_cite}{RGB}{12,127,172}

\begin{document}

\begin{titlepage}
\vspace{1.5cm}

\begin{center}
{\LARGE  \bf

Confirming the Existence of the strong CP Problem\\
\vspace{2mm}
 in Lattice QCD with the Gradient Flow
}
\vspace{2cm}

{\large \bf  Jack Dragos$^{a}$, Thomas Luu$^{b}$, Andrea Shindler$^{a}$, \\
\vspace{3mm}
Jordy de Vries$^{c,d}$, Ahmed Yousif$^{\,a}$}
\vspace{0.25cm}

\vspace{0.5cm}

{\large
$^a$
{\it Facility for Rare Isotope Beams, Physics Department, Michigan State University, East Lansing, Michigan.}}

\vspace{0.25cm}
{\large
$^b$
{\it
Institute for Advanced Simulation (IAS-4), Institut f\"ur Kernphysik (IKP-3) and JARA-HPC, FZJ, J\"ulich, Germany}}

\vspace{0.25cm}
{\large
$^c$
{\it
Amherst Center for Fundamental Interactions, Department of Physics, University of Massachusetts Amherst, Amherst, MA 01003, USA
}}

\vspace{0.25cm}
{\large
$^d$
{\it
RIKEN BNL Research Center, Brookhaven National Laboratory, Upton, New York 11973-5000, USA}}

\end{center}

\vspace{1cm}

\begin{abstract}
\vspace{0.1cm}
We calculate the electric dipole moment of the nucleon induced by the QCD theta term.
We use the gradient flow to define the topological charge and use $N_f = 2+1$ flavors of dynamical
quarks corresponding to pion masses of $700$, $570$, and $410$ MeV, and perform an extrapolation
to the physical point based on chiral perturbation theory.
We perform calculations at $3$ different lattice spacings in the range of $0.07~{\rm fm} < a < 0.11$ fm
at a single value of the pion mass, to enable control on discretization effects.
We also investigate finite size effects using $2$ different volumes.
A novel technique is applied to improve the signal-to-noise ratio in the form factor calculations.
The very mild discretization effects observed suggest a continuum-like behavior of the
nucleon EDM towards the chiral limit. Under this assumption
our results read \({d_{n}=-0.00152(71)\ \bar\theta\  e\text{~fm}}\)
 and  \({d_{p}=0.0011(10)\ \bar\theta\  e\text{~fm}}\).
 Assuming the theta term is the only source of CP violation,
 the  experimental bound on the neutron electric dipole moment
 limits $\left|\bar\theta\right| < 1.98\times 10^{-10}$ ($90\%$ CL).
A first attempt at calculating the nucleon Schiff moment in the continuum resulted in
\({S_{p} = 0.50(59)\times 10^{-4}\ \bar\theta\ e}\)~fm\textsuperscript{3} and
\({S_{n} = -0.10(43)\times 10^{-4}\ \bar\theta\ e}\)~fm\textsuperscript{3}.
\end{abstract}

\vfill
\end{titlepage}


\section{\label{sect:intro}Introduction}
A nonzero measurement of the electric dipole moment (EDM) of the nucleon in the foreseeable future would be a clear signal of new
physics, since the known CP-violating phase of the CKM matrix leads to EDMs that lie orders of magnitude below
current experimental limits.  The source of a nonzero EDM could then either be the QCD $\bar \theta$-term or
higher-dimension CP-violating quark-gluon operators that originate in beyond-the-Standard Model (BSM) physics, or a combination of these two.
To interpret an EDM signal or lack thereof, and to possibly disentangle the source (e.g. $\bar \theta$-term or BSM),
requires a non-perturbative calculation linking the CP-violating sources to the hadronic observables.

Lattice QCD can calculate the nucleon EDM directly in terms of CP-violating operators at the quark level.  Various attempts have
been made in this regard \cite{Shintani:2005xg,Berruto:2005hg,Guo:2015tla,Alexandrou:2015spa,Shintani:2015vsx,Abramczyk:2017oxr,Yoon:2017tag}.
However, the renormalization of CP-violating operators within a lattice
(discretised) formulation of QCD is very non-trivial, and for several operators presents large difficulties in interpreting lattice results.
Further, the $\bar \theta$ term itself introduces a complex phase in the
determinant of the quark matrix, which produces a sign problem and precludes the use of standard stochastic methods.
Several techniques have been used to address the $\bar \theta$-term contribution to the EDM and attempts have been made
to solve the complicated renormalization patterns of the CP-violating operators \cite{Bhattacharya:2015rsa}. We refer to the recent review~\cite{Constantinou:2016ezf} for
a summary.

We proposed to use the gradient flow to calculate all CP-violating source to the EDM
in refs.~\cite{Shindler:2014oha,Shindler:2015aqa}, and presented preliminary results~\cite{Dragos:2018uzd,Rizik:2018lrz,Kim:2018rce,Reyes:2018ucu}.
In this paper we consider the $\bar \theta$-term contribution to the EDM in a perturbative manner
as discussed in~\cite{Shindler:2015aqa}. This is well justified
considering the stringent constraints on $\bar \theta$ set by EDM experiments. In this way we avoid the problem of a complex fermionic determinant.
To define the QCD $\bar \theta$ term we use the gradient flow. The topological charge defined
in this way has a finite and well defined continuum limit~\cite{Luscher:2010iy,Luscher:2011bx,Ce:2015qha}.
It is much faster to compute than using the Ginsparg-Wilson definition
and it is theoretically more robust than definitions using cooling techniques.
Another problem that hinders lattice calculations of the nucleon EDM is the very poor signal-to-noise ratio.
In this respect we explore a novel technique to determine  the space-time region where the signal
in the relevant correlation functions is maximized. A first account of this technique has already been presented \cite{Dragos:2018uzd}.

Additional insights into the EDM of the nucleon~\cite{Crewther:1979pi,Ottnad:2009jw,deVries:2010ah,Mereghetti:2010kp,Guo:2012vf}
and nuclei~\cite{deVries:2011an,Bsaisou:2014zwa} can be provided by chiral effective field theory.
The CP-violating quark operators are translated to effective CP-violating hadronic operators and EDMs
depend on the unknown low-energy constants (LECs) of the theory.
The LECs can be estimated from dimensional analysis or, preferably,
be determined from experiments and/or by lattice-QCD calculations. We use these insights from chiral calculations to understand the pion mass dependence of our results, and to connect our nucleon EDM calculations to nuclear EDMs.

The remainder of the paper is organized as follows: \autoref{sect:phenom} gives a cursory
discussion of the phenomenology of the nucleon EDM, followed by an overview of the
lattice details and parameters in \autoref{sec:LatParams}, where we discuss the general lattice strategy used
and we define the basic observables, including the gradient flow.
Sanity checks are given in \autoref{sec:gf}, 
where we compute and display the topological charge using the gradient flow.
The nucleon two-point correlation function is explored in
\autoref{sec:twopt}, then the setup of the 
computation of the EDM is derived and results shown in \autoref{sec:F3_res}.
A discussion follows the results section in \autoref{sect:discussion},
where we discuss the ramifications of our results and
compare our results to the literature.
Finally, we conclude in \autoref{sect:conclusion}.


\section{Phenomenology of the QCD theta term. \label{sect:phenom}}
The discrete space-time symmetries parity $(P)$ and time-reversal $(T)$, and hence via the $CPT$ theorem also $\CP$
symmetry, are broken in QCD by the  $\bar \theta$ term. In the case of two quark flavors the QCD Lagrangian in Minkowski space is given by 
\begin{equation}\label{QCD1}
\mathcal L_{\mathrm{QCD}} = -\frac{1}{4}G_{\mu\nu}^aG^{a,\mu\nu} + \bar q(i\Dslash{D} - M)q - \tb \frac{g^2}{64\pi^2}\epsilon^{\mu\nu\alpha\beta} G^a_{\mu \nu}G^a_{\alpha \beta}\,\,\,,
\end{equation}
 where  $q = (u\, ,  d)^T$  denotes the quark doublet containing up and down quarks, $G^a_{\mu\nu}$ is the gluon field
 strength tensor, $\epsilon^{\mu\nu\alpha\beta}$ ($\epsilon^{0123}=+1$)
 is the completely antisymmetric tensor, $D_\mu$ the gauge-covariant derivative, $M$ the
real $2\times 2$ quark-mass matrix, and $\tb$ the coupling
of the $\CP$-odd interaction. In eq.~\eqref{QCD1}  the complex phase of the quark-mass matrix has been absorbed in the
physical parameter $\tb = \theta + \mathrm{arg}\,\mathrm{det}(M)$. For the application of chiral perturbation theory
($\chi$PT) it is useful to perform an anomalous axial $U(1)$ transformation to replace the $\CP$-odd gluonic term in
favor of a complex mass term  \cite{Crewther:1979pi, Baluni:1978rf}. Under the assumption that $\tb\ll1$ the QCD Lagrangian can then be written as
\begin{equation}\label{QCD2}
\mathcal L_{QCD} = -\frac{1}{4}G_{\mu\nu}^aG^{a,\mu\nu} + \bar q i\Dslash{D}  q -\bar m \bar q q - \varepsilon \bar m \bar q \tau_3 q + m_\ast \tb \bar q i\gamma^5 q\,\,\,,
\end{equation}
where we have defined the average quark mass  $\bar m = (m_u+m_d)/2$,
the quark-mass difference
$\varepsilon = (m_u-m_d)/(m_u + m_d)$, and the reduced quark mass
$m_\ast = m_u m_d/(m_u + m_d) = \bar m (1-\varepsilon^2)/2$.

The QCD $\bar \theta$ term induces an EDM in hadrons and nuclei.  The first EDM search occured with the neutron in 1957 \cite{Purcell:1950zz,Smith:1957ht}. Till this day, no signal has been found for the EDM, despite measurement sensivities having improved by six orders of magnitude. The current bound
${d_n<3.0\times 10^{-26}}$~e~cm \cite{Baker:2006ts,Afach:2015sja} sets strong limits on the size of $\tb$ and sources of $\CP$
violation from physics beyond the SM \cite{Chupp:2017rkp}.

In order to set a bound on the $\tb$ term, it is necessary to calculate the dependence of the neutron EDM on
$\bar \theta$ \cite{Crewther:1979pi}. One way to do this is by using $\chi$PT. In the first step one derives interactions between the
low-energy degrees of freedom, pion and nucleons (and heavier hadrons), that violate $\CP$ and transform the same
way under chiral symmetry as the complex mass term in eq.~\eqref{QCD2}. In the second step, one combines the chiral
$\CP$-odd interactions with the standard $CP$-even chiral Lagrangian to calculate the nucleon EDM. This calculation
has been done up to next-to-leading order (NLO) in both $SU(2)$ and $SU(3)$ $\chi$PT \cite{Ottnad:2009jw, Mereghetti:2010kp} and gives in the
two-flavored theory for the neutron ($d_n$) and proton ($d_p$) EDM:
\begin{eqnarray}
 \label{neutronEDM}
  d_n(\theta)& = & {\bar d}_n-\frac{e g_A\bar g_0}{8\pi^2 F_\pi} \left(  \ln
\frac{m_\pi^2}{m_N^2} -\frac{\pi m_\pi}{2 m_N} \right)\,\,\,,\nonumber\\
d_p(\theta) & = & {\bar d}_p+\frac{e g_A \bar g_0}{8\pi^2 F_\pi}  \left(  \ln
\frac{m_\pi^2}{m_N^2} -\frac{2\pi m_\pi}{m_N} \right) \,\,\,,
\end{eqnarray}
in terms of $g_A \simeq 1.27$ the strong pion-nucleon coupling constant, $F_\pi \simeq 92.4$ MeV the pion decay constant,
$m_\pi$ and $m_N$ the pion and nucleon mass respectively, $e>0$ the proton charge, and three low-energy constants
(LECs) of $\CP$-odd chiral interactions $\bar g_0$ and $\bar d_{p/n}$. The first term in brackets in
eq.~\eqref{neutronEDM} arises from the leading-order (LO) one-loop diagram involving the $\CP$-odd vertex
\begin{equation}
\mathcal L_{\pi N}(\theta) = \bar g_0\, \bar N \vec \pi \cdot \vec \tau N\,\,\,
\end{equation}
in terms of the nucleon doublet $N = (p\, n)^T$ and the pion triplet $\vec \pi$. The LO loop is divergent and the
divergence and associated scale dependence have been absorbed into the counter terms $\bar d_{p/n}$ which signify
contributions to the nucleon EDMs from short-range dynamics and appear at the same order as the LO loop diagrams. The second term in brackets in eq.~\eqref{neutronEDM} arises from finite next-to-leading-order (NLO) diagrams.

Because the $\tb$ term breaks chiral symmetry as a complex quark mass, the LEC $\bar g_0$ can be related to known
$CP$-even LECs using chiral symmetry arguments \cite{Crewther:1979pi, Mereghetti:2010tp, deVries:2015una}
\begin{equation}\label{eq:g0theta_eq}
\bar g_0 = \frac{(m_n -m _p)^{\mathrm{strong}} (1-\varepsilon^2)}{4F_\pi \varepsilon} \tb = -14.7(2.3)\times 10^{-3}\,\bar \theta\,,
\end{equation}
where $(m_n -m _p)^{\mathrm{strong}}$ is the quark-mass induced part of the proton-neutron mass splitting for which we
used the recent lattice results \cite{Borsanyi:2014jba, Brantley:2016our}. Inserting eq.~\eqref{g0theta} in eq.~\eqref{neutronEDM} we obtain
\begin{eqnarray}
 \label{neutronEDMnum}
  d_n(\theta)& = & {\bar d}_n-2.1(3)\times 10^{-3}\,\tb\,e\,\mathrm{fm}\, ,\nonumber\\
d_p(\theta) & = & {\bar d}_p+2.5(3)\times 10^{-3}\,\tb\,e\,\mathrm{fm}\,.
\end{eqnarray}
Under the assumption that the terms $\bar d_{p/n}$ do not cancel against
the calculable loop contributions, a comparison with the experimental
bound gives the strong constraint  $\tb \leq 10^{-10}$.
Clearly, a more reliable constraint on $\tb$ requires a direct nonperturbative calculation of the full nucleon EDMs. This is the main goal of
this work.

In the isoscalar combination $d_n+d_p$ the loop contribution
cancels out to a large extent. For observables sensitive to this combination,
such as the deuteron EDM whose measurement is the goal of the
JEDI collaboration \cite{JEDI}, a first-principle calculation
of the total nucleon EDM is important. EDMs of light nuclei have been calculated as a function
of $\tb$ in ref.~\cite{Bsaisou:2014zwa}. Nuclear EDMs get contributions from the single-nucleon EDMs and from the CP-violating
nucleon-nucleon potential which is dominated by one-pion-exchange terms. The latter depend mainly on $\bar g_0$ and
are therefore relatively well under control. The dominant remaining uncertainty is the size of the nucleon EDMs as a function of $\tb$. With nonperturbative calculations of nucleon EDMs induced by the $\tb$ term, we immediately obtain predictions for EDMs of light nuclei. With future improvements of nuclear theory even EDMs of diamagnetic atoms such
as ${}^{199}$Hg and ${}^{225}$Ra could be directly given as a function of $\tb$.


\section{Lattice QCD action and numerical details\label{sec:LatParams}}

We discretize the QCD action on an hypercubic lattice with spacing $a$ and
volume $L^3 \times T$. The fermionic part of our QCD lattice action
is the non-perturbatively O($a$)-improved Wilson action
with $N_f=2+1$ dynamical quarks. The gauge part is the Iwasaki gauge action.
For our calculation we have always used valence quarks
with the same lattice action and the same bare parameters as the sea quark action,
that is to say our framework is fully unitary.

We performed calculations using the publicly available PACS-CS gauge fields
available through the ILDG~\cite{Beckett:2009cb}. We used $6$ different ensembles
that allow us to study discretization effects, finite-size effects and pion mass dependence.
We studied the pion-mass dependence with $3$ ensembles at $3$ different bare quark masses,
at $L/a=32$ and $T/a=64$ and a lattice spacing $a = 0.0907(13)$ fm. The lattice spacing and the
physical point were determined with the experimental input of $m_\pi, m_K$, and $m_\Omega$.
More details on these ensembles are available in ref.~\cite{Aoki:2008sm}
and are summarized in the first three M rows of tabs.~\ref{tab:latpar},~\ref{tab:latpar2}.

To study discretization effects we used $3$ ensembles with $3$ different lattice spacings 
but with the same volume, $L \simeq 1.8$ fm. The ratios of masses in the pseudoscalar and vector channels
differ, between the $3$ ensembles, at most by $1\%$ in the light sector and at most of $3\%$ in the strange quark sector.
These very small mismatches are irrelevant
for all purposes for our scaling violation study. The lattice spacings and quark masses in these ensembles are also determined using
$m_\pi$, $m_K$, and $m_\phi$. Details for these ensembles can be found
in ref.~\cite{Ishikawa:2007nn} and summarized
in the last three A rows of tabs.~\ref{tab:latpar},~\ref{tab:latpar2}.

Among the $6$ ensembles described above there are $2$ ensembles, M$_1$ and A$_2$, with the same bare parameters, $\beta=1.9$,
$\kappa_l = 0.13700$, $\kappa_s = 0.1364$ and different lattice volumes with $L/a=20$, $L/a=32$ and $T=2L$.
These $2$ ensembles allow us to investigate finite-size effects.

\begin{table}
  \begin{center}
  \begin{tabular}[h]{|c|c|c|c|c|c|c|c|c|}
    \hline
    & $\beta$ & $\kappa_l$ & $\kappa_s$ & L/a & T/a & c$_{\rm sw}$ &  $N_G$ & $N_{\rm corr}$   \\
    \hhline{|=|=|=|=|=|=|=|=|=|}
    M$_1$ & 1.90 & 0.13700 & 0.1364 & 32 & 64 & 1.715 &      322 & 30094 \\
    \hline
    M$_2$ & 1.90 & 0.13727 & 0.1364 & 32 & 64 & 1.715 &    400 & 20000 \\
    \hline
    M$_3$ & 1.90 & 0.13754 & 0.1364 & 32 & 64 & 1.715 &    444 & 17834 \\
    \hhline{|=|=|=|=|=|=|=|=|=|}
    A$_1$ & 1.83 & 0.13825 & 0.1371 & 16 & 32 & 1.761 &  800 & 15220 \\
    \hline
    A$_2$ & 1.90 & 0.13700 & 0.1364 & 20 & 40 & 1.715 &      789 & 15407 \\
    \hline
    A$_3$ & 2.05 & 0.13560 & 0.1351 & 28 & 56 & 1.628 &    650 & 12867  \\
    \hline
  \end{tabular}
    \caption{Summary of the lattice bare parameters for the ensembles used.
    $N_G$ is the number of gauge configurations and $N_{\rm corr}$ is the number of correlation functions
    calculated using many stochoastically located sources for the same gauge configuration.}
  \label{tab:latpar}
  \vspace*{0.5cm}
  \begin{tabular}[h]{|c|c|c|c|c|}
    \hline
    & $a$ [fm] & $m_\pi$ [MeV] & $m_{N}$ [GeV] & $Z_V$  \\
    \hhline{|=|=|=|=|=|}
    M$_1$ &    0.0907(13) & 699.0(3) & 1.585(2)  & 0.7354(37) \\
    \hline
    M$_2$ &    0.0907(13) & 567.6(3) & 1.415(3)  & 0.7354(37) \\
    \hline
    M$_3$ &    0.0907(13) & 409.7(7) & 1.219(4)  & 0.7354(37) \\
    \hhline{|=|=|=|=|=|}
    A$_1$ &     0.1095(25) & 710(1) &  1.65(1)  & 0.7013(14) \\
    \hline
    A$_2$ &    0.0936(33) & 676.3(7) & 1.549(6)  & 0.7354(37) \\
    \hline
    A$_3$ &     0.0684(41) & 660.4(7) & 1.492(5) & 0.77314(82) \\
    \hline
  \end{tabular}
  \end{center}
  \caption{Summary of some basic lattice quantities computed on the ensembles used.}
  \label{tab:latpar2}
\end{table}

To improve the overlap with the ground state of the relevant matrix elements in the two- and three-point functions,
we applied a Gaussian gauge-invariant smearing~\cite{Gusken:1989qx,Alexandrou:1990dq} at the source and at the sink
of our quark propagators. Using the notation of refs.~\cite{Gusken:1989qx,Alexandrou:1990dq} we use $64$ iterations
of the smearing algorithm with a smearing fraction of $\alpha=0.39$ using the definition in ref.~\cite{Gusken:1989qx}.
These parameters corresponds to a spatial root-mean-square radius for the nucleon
interpolating operator of around $0.4$ fm. The quality of our projection into the ground state can be evaluated from
figs.~\ref{fig:Emass_mpi},~\ref{fig:Emass_latspace}.

For the vector form factors studied here,
we used the renormalization factor $Z_V$ determined using vector Ward identities
in ref.~\cite{Aoki:2010wm} and summarized in tabs.~\ref{tab:latpar},~\ref{tab:latpar2}.

The strategy we use in this paper is a perturbative expansion in powers of $\bar{\theta}$ (the expansion is fully performed in Euclidean space).
This is justified by the small value of $\bar{\theta}$ estimated from experimental constraints.
With this strategy every correlation function $\braket{O}_{\bar{\theta}}$,
evaluated in a $\bar{\theta}$ vacuum, is determined from a small-$\bar{\theta}$ expansion
\be
  \braket{O}_{\bar{\theta}} =
  \braket{O} + i\bar{\theta} \braket{OQ} + \mathcal{O}(\bar{\theta}^{2})\,,
  \label{eq:theta_exp}
\ee
where $O$ is some multi-local operator, $\bar{\theta}$ is the coefficient for the CP-violating
term, and $Q$ is the topological charge. The expectation values on the r.h.s of eq.~\eqref{eq:theta_exp}
are computed on a standard QCD background. This allows us to use lattice QCD gauge configurations
without generating new gauges for this specific calculation.
We define the correlation functions used to determine the nucleon EDM induced by the $\bar{\theta}$ term
in the next section.

The topological charge CP-violating operator that enters the correlation functions 
must in principle be normalized.
We use the gradient flow~\cite{Luscher:2010iy} to define the topological charge which in this way has a finite
continuum limit and does not need any additional normalization~\cite{Luscher:2010iy,Luscher:2011bx,Ce:2015qha}.
The reason is that the flowed fields are free from ultraviolet
divergences~\cite{Luscher:2010iy,Luscher:2011bx} for all positive flow times, $t_{f}>0$.
Additionally it can be shown that  the topological charge defined with the gradient flow
is flow-time independent~\cite{Ce:2015qha} for all positive flow times, $t_f>0$, in the continuum limit.
Details on how we numerically perform the flowing
of the fields have been described in ref.~\cite{Shindler:2015aqa}.

\section{Topological charge and the gradient flow\label{sec:gf}}

\begin{figure}
\begin{adjustwidth}{-0.1\textwidth}{-0.1\textwidth}
\centering
\begin{subfigure}{.5\textwidth}
 \centering
 \includegraphics[trim={11mm 0cm 11mm 0cm},clip,width=\linewidth]{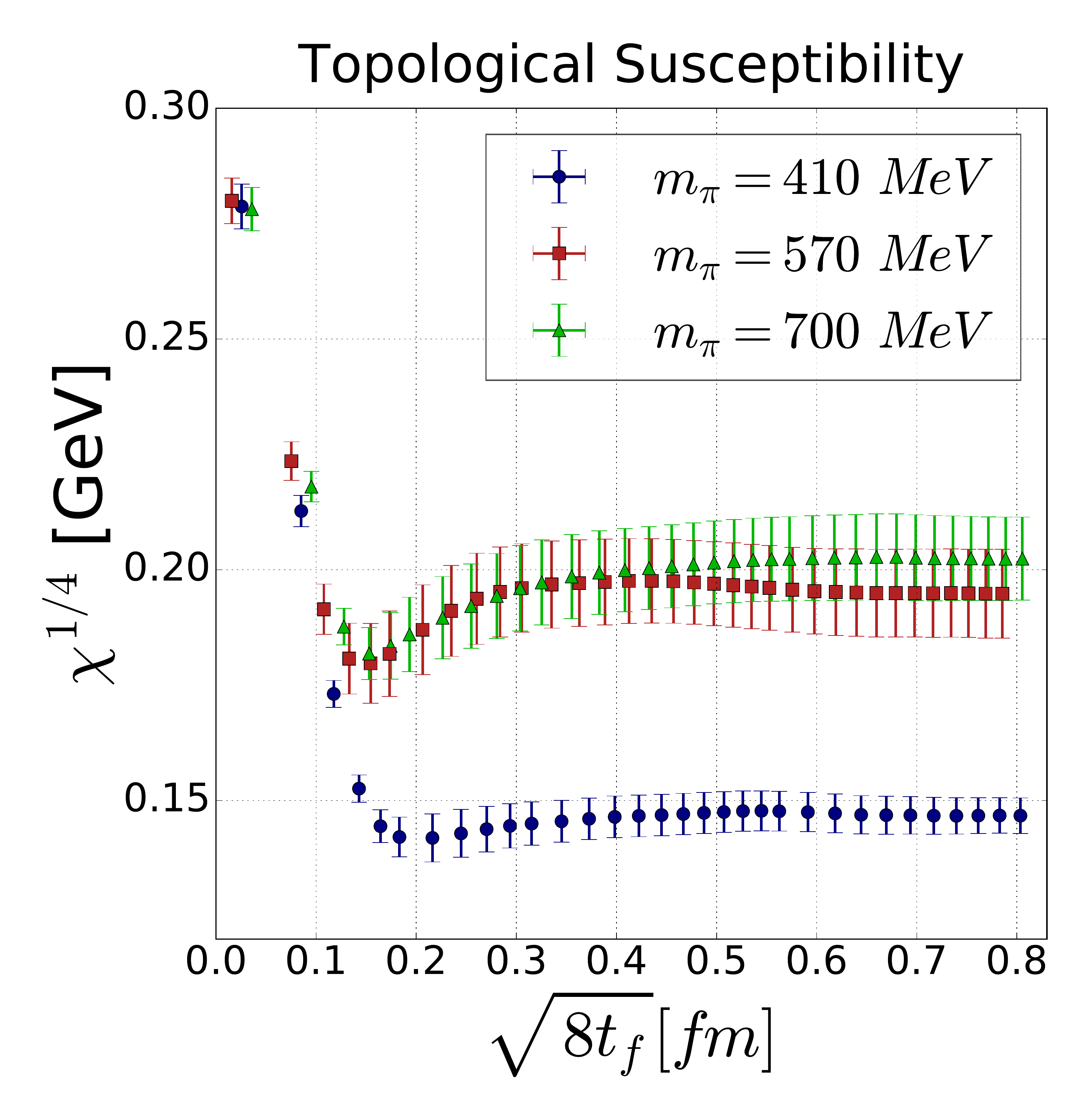}
 \caption{\label{fig:Q2_mpi}}
\end{subfigure}
\quad
\begin{subfigure}{.5\textwidth}
 \centering
 \includegraphics[trim={11mm 0cm 11mm 0cm},clip,width=\linewidth]{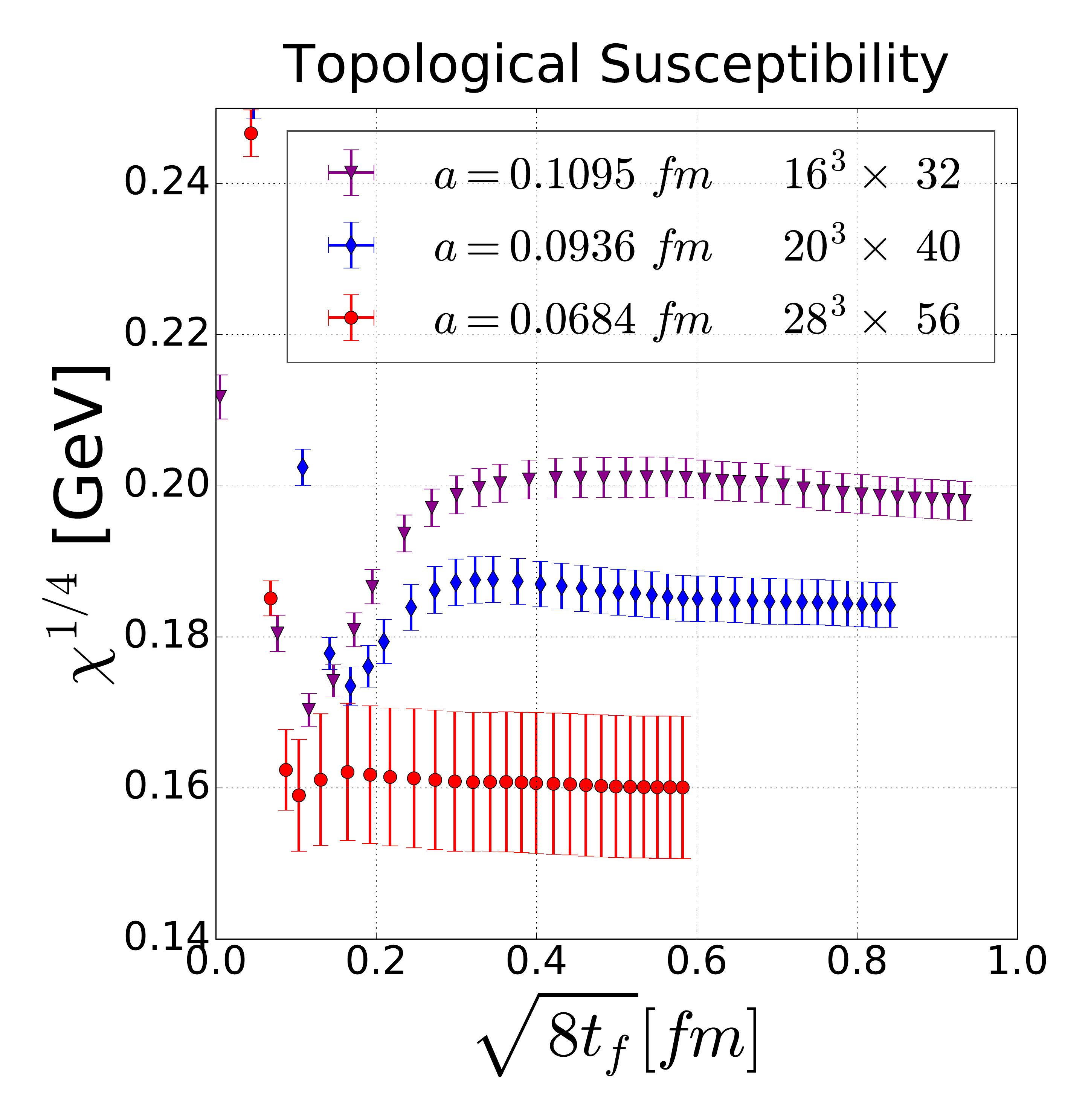}
 \caption{\label{fig:Q2_latspace}}
\end{subfigure}
\end{adjustwidth}
\caption{The topological susceptibility in GeV computed for the M- (left) and A-ensembles (right),
 plotted against the flow-time radius $\sqrt{8t_f}$ in fm.
 \label{fig:Q2_mpi_and_latspace}}
\end{figure}

We define the topological charge at finite lattice spacing as
\begin{equation}\label{eq:top_int}
Q(t_{f}) = a^{4}\sum_{x}q(x,t_{f})\,,
\end{equation}
where the topological charge density reads
\begin{equation}
q(x,t_{f}) = \frac{1}{64\pi^2} \epsilon_{\mu\nu\rho\sigma} G_{\mu\nu}^a(x,t_{f})G_{\rho\sigma}^a(x,t_{f})\ ,
\label{eq:topo_density_t}
\end{equation}
and $G_{\mu\nu}^a(x,t_{f})$ is  a lattice discretization of the continuum field tensor defined with flowed gauge fields.
As a lattice definition for the field tensor, we use the discretization suggested in ref.~\cite{BilsonThompson:2002jk}.
This definition suffers from small discretization effects and, in fact, the corresponding topological susceptibility
\be
\chi(t_f) = \frac{a^{8}}{V} \sum_{x,y} \left\langle q(x,t_f)  q(y,t_f) \right\rangle
\ee
is flow-time independent starting from a flow-time radius,
$\sqrt{8t_f}$, of about $1$~fm for all lattice spacings we have investigated.
This can be seen in fig.~\ref{fig:Q2_mpi_and_latspace},
where we show the flow-time dependence of the topological susceptibility
computed for all M- (left) and A-ensembles (right). 
As expected, the region where the susceptibility is independent of the flow time
extends towards smaller flow-time values for smaller lattice spacings.

We used the topological charge to perform various checks on the quality of the ensembles. 
An important check for EDM calculations is to make sure that the ensembles sample the field space in such a way that no spurious CP-violation is induced.   In other words we must check the expectation value
\mbox{$\langle Q(t_{f}) \rangle=0$} within statistical errors.
\begin{figure}
\vspace*{-2cm}
\begin{adjustwidth}{-0.1\textwidth}{-0.1\textwidth}
\centering
\begin{subfigure}{.55\textwidth}
  \centering
  \includegraphics[trim={11mm 0cm 12mm 0cm},clip,width=\linewidth]{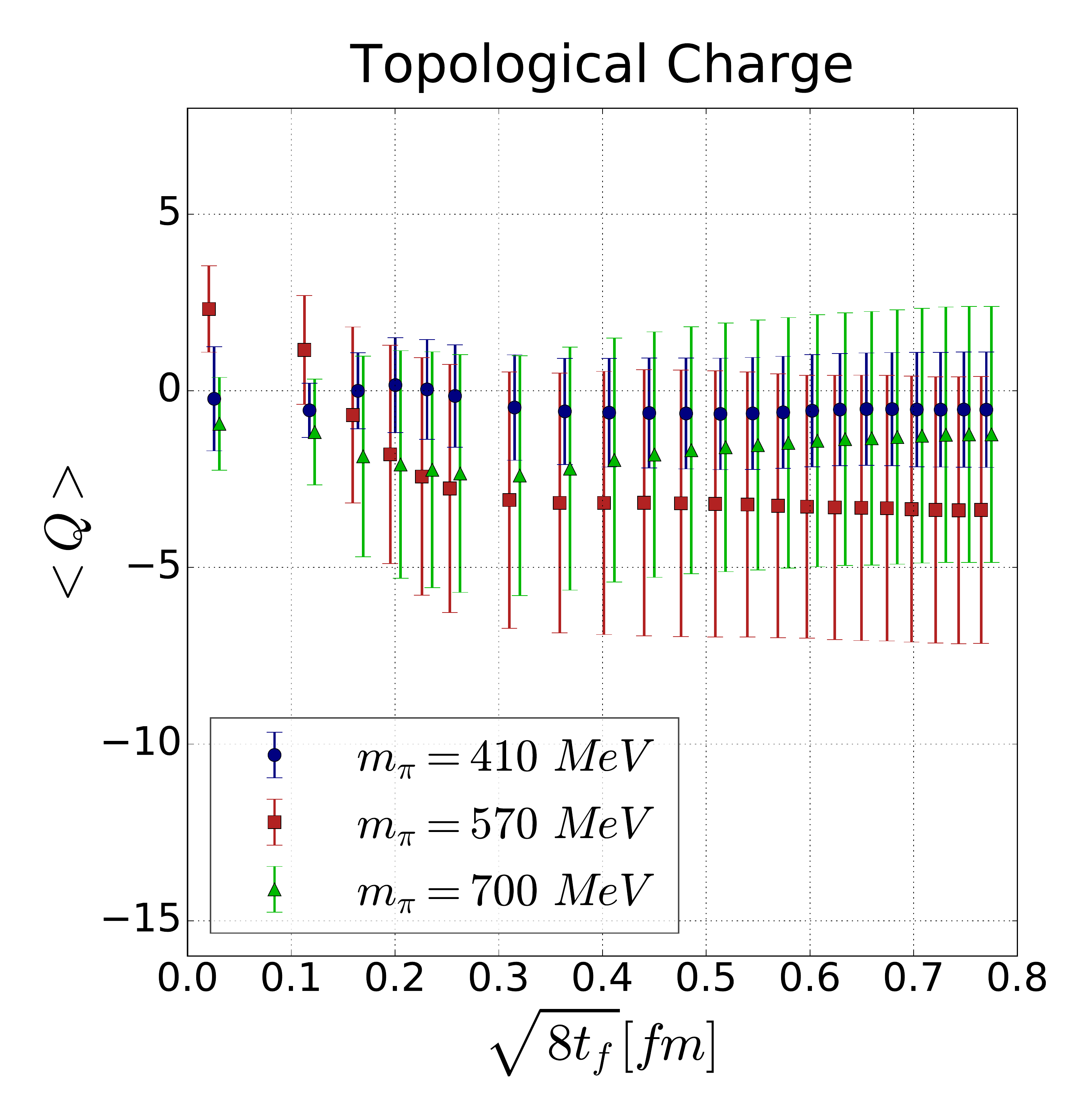}
  \caption{\label{fig:Q_mpi}}
\end{subfigure}
\quad
\begin{subfigure}{.55\textwidth}
  \centering
  \includegraphics[trim={11mm 0cm 0mm 0cm},clip,width=\linewidth]{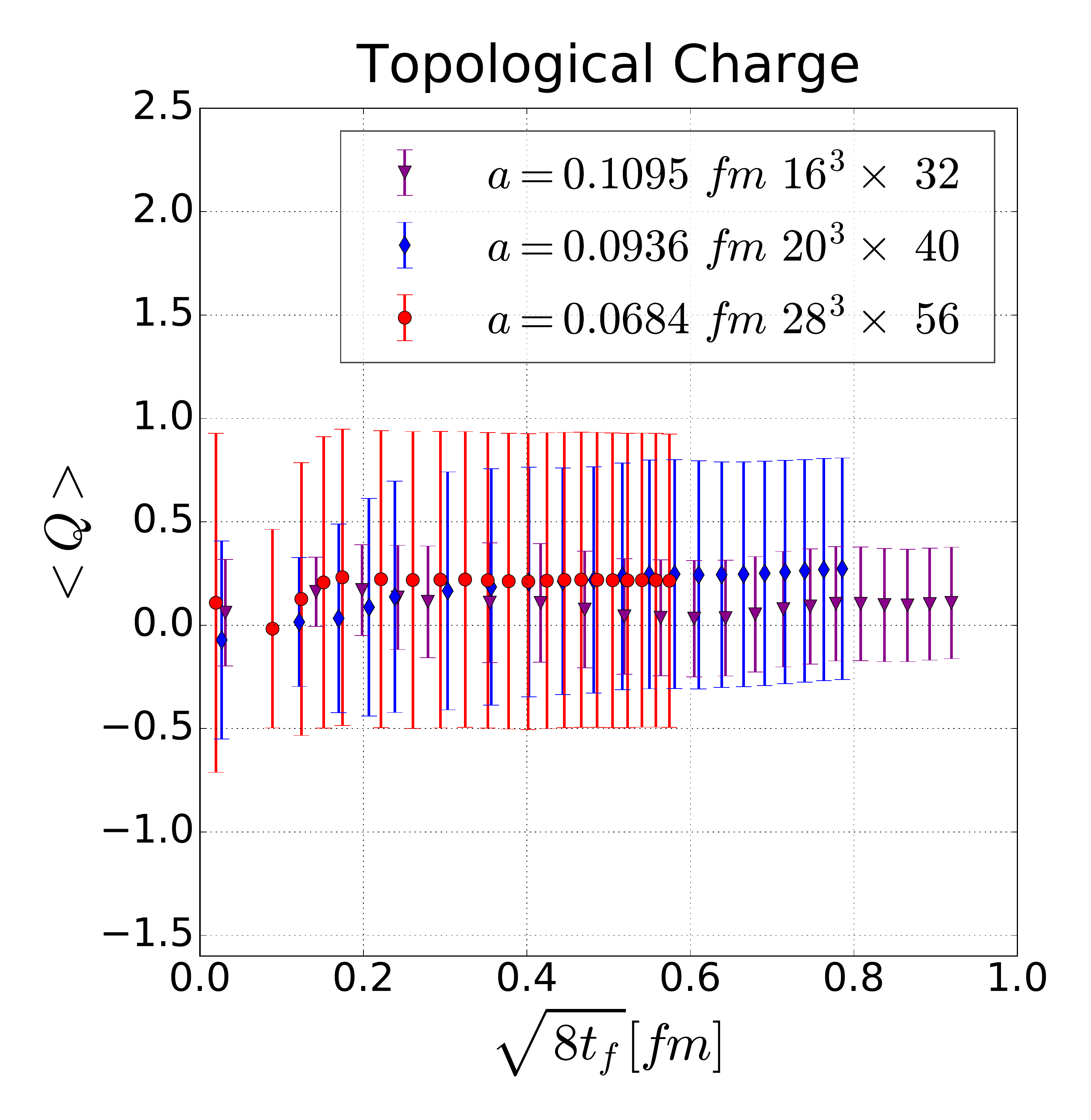}
  \caption{\label{fig:Q_lat}}
\end{subfigure}
\end{adjustwidth}
\caption{
  Flow-time radius $\sqrt{8t_f}$ dependence of the topological charge $\braket{Q}$
  for the M- (left) and A-ensembles (right).
  The errors are computed using an autocorrelation analysis as described
  in ref.~\cite{Wolff:2003sm}.
\label{fig:Q}}
\vspace*{-0.2cm}
\begin{adjustwidth}{-0.1\textwidth}{-0.1\textwidth}
\centering
\begin{subfigure}{.55\textwidth}
  \centering
  \includegraphics[trim={11mm 0cm 12mm 0cm},clip,width=\linewidth]{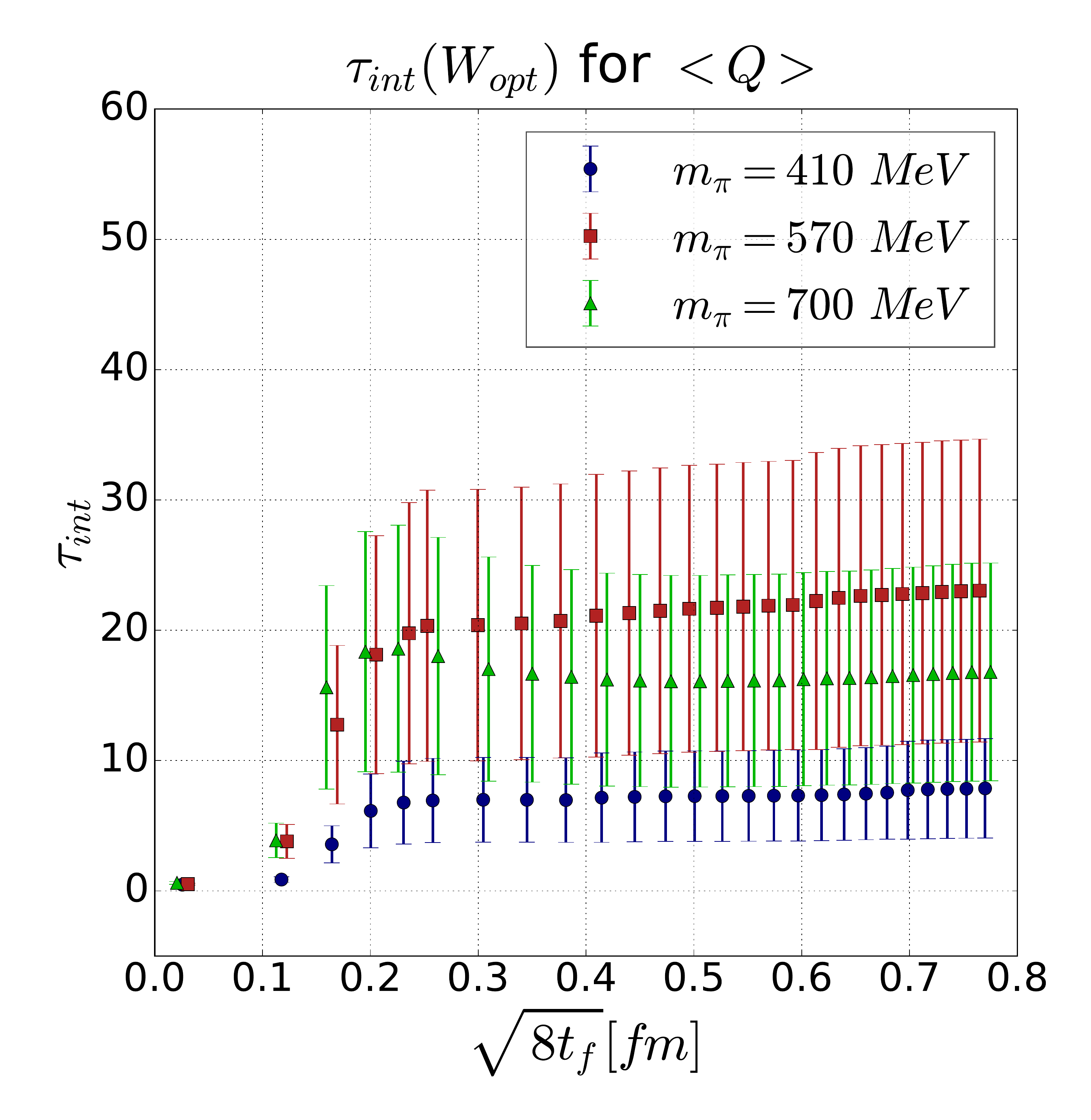}
  \caption{\label{fig:tau_int_mpi}}
\end{subfigure}
\quad
\begin{subfigure}{.55\textwidth}
  \centering
  \includegraphics[trim={11mm 0cm 0mm 0cm},clip,width=\linewidth]{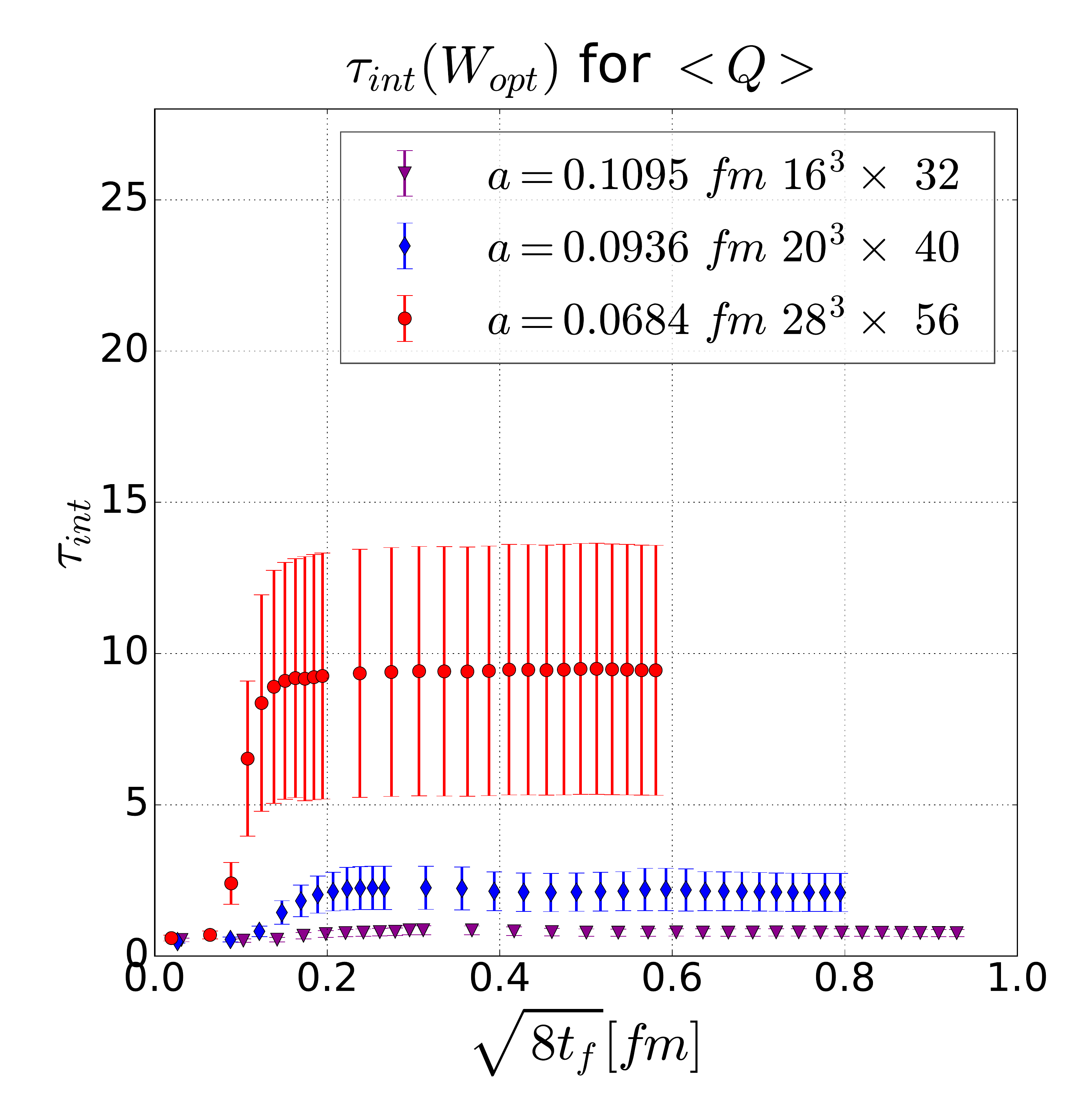}
  \caption{\label{fig:tau_int_lat}}
\end{subfigure}
\end{adjustwidth}
\caption{
  Flow-time radius $\sqrt{8t_f}$ dependence of the optimal integrated correlation time
  \(\tau_{int}\) of the
  topological charge for the M-(left) and A-(right) ensembles.
  The error calculation, as well as the optimal autocorrelation length
  \(W_{opt}\), are computed as described in ref.~\cite{Wolff:2003sm}.
\label{fig:tau_intQ}}
\end{figure}
In fig.~\ref{fig:Q} we show $\langle Q(t_{f}) \rangle$ evaluated
on all our ensembles for various pion masses and lattice spacings.
To properly estimate the statistical uncertainties we evaluate the autocorrelation function
and the corresponding integrated autocorrelation time, $\tau_{\rm int}$, as defined in ref.~\cite{Wolff:2003sm}.
For all our ensembles the average topological charge vanishes within statistical errors.
In fig.~\ref{fig:tau_intQ} we show the flow-time dependence of the integrated autocorrelation time for the topological charge.
As expected the gradient flow, by smoothing out some of the short-distance fluctuations, allows a better determination
of $\tau_{\rm int}$ that reaches a plateau for $\sqrt{8 t_f} \simeq 0.2$ fm for all our ensembles \cite{Luscher:2011kk,Bruno:2014ova}.

The integrated autocorrelation time $\tau_{\rm int}$ we obtain
falls within the range $7 < \tau_{\rm int} < 35$ for the M$_1$ and M$_2$ ensembles
and slightly smaller, $3<\tau_{\rm int}<10$, for our M$_3$ ensemble.
We attribute this behavior with the rather short Markov Chain for the M$_3$ ensemble which most likely does not allow
a more accurate determination of its $\tau_{\rm int}$.
We also observe from fig.~\ref{fig:tau_intQ} that $\tau_{\rm int}$ increases as we decrease the lattice spacing.
This is an expected result~\cite{Schaefer:2010hu,Luscher:2011kk}  since the tunnelling between different topological sectors becomes increasingly difficult with decreasing lattice spacing, meaning that the sampling of different sectors, which would decrease $\tau_{\rm int}$, is lessened.  

\begin{figure}
\begin{adjustwidth}{-0.1\textwidth}{-0.1\textwidth}
\centering
\begin{subfigure}{.55\textwidth}
  \centering
  \includegraphics[trim={11mm 0cm 9mm 0cm},clip,width=\linewidth]{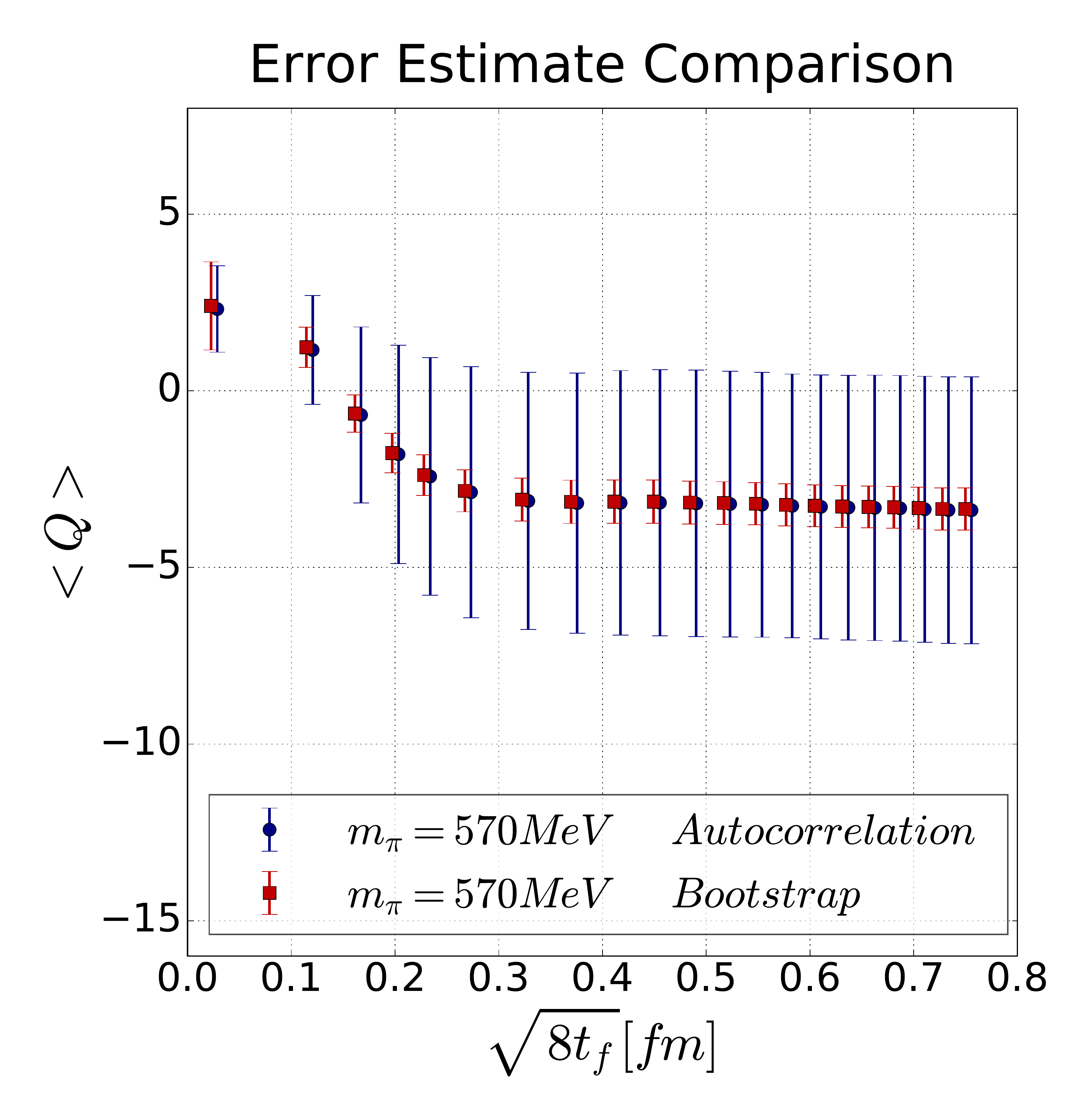}
  \caption{\label{fig:Q_auto_mpi570}}
\end{subfigure}
\quad
\begin{subfigure}{.55\textwidth}
  \centering
  \includegraphics[trim={11mm 0cm 9mm 0cm},clip,width=\linewidth]{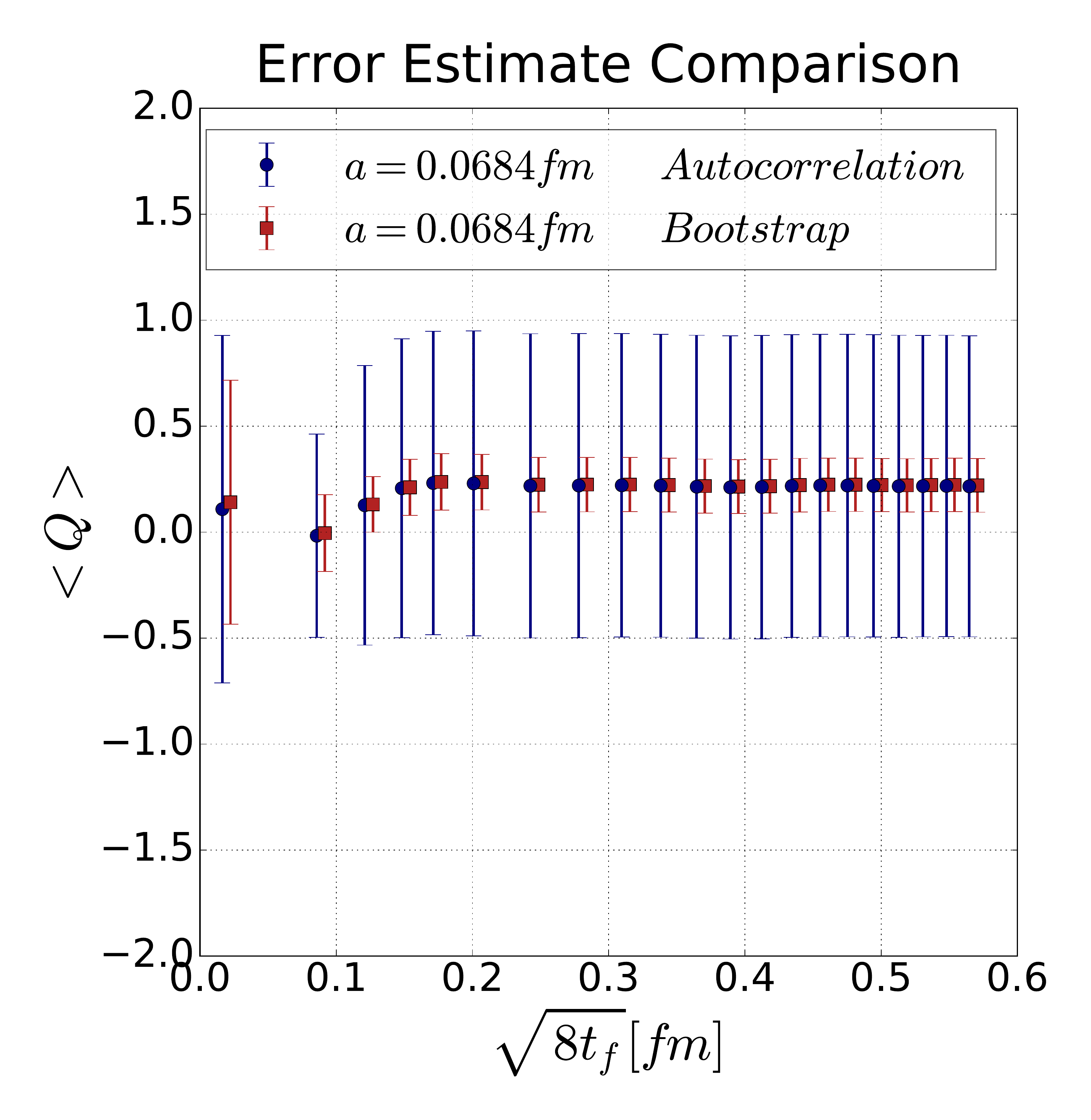}
  \caption{\label{fig:Q_auto_L28}}
\end{subfigure}
\end{adjustwidth}
\caption{Flow-time-radius dependence of the topological charge for the ensemble M$_2$ and A$_3$
  with statistical errors computed using the autocorrelation function (blue data points) and
  a standard bootstrap error estimate (red data points).
\label{fig:Q_auto}}
\end{figure}

For completeness in fig.~\ref{fig:Q_auto} we show the difference of error determination if we were
to use a standard resampling technique, such as bootstrap, instead of the error determination using
the autocorrelation function. In this case, $\langle Q(t_{f}) \rangle \neq 0$ within uncertainties.  This demonstrates how a robust uncertainty determination for the topological charge requires both the estimate of the autocorrelation function and its corresponding integrated autocorrelation time.

\section{Two-point correlation functions and the nucleon mixing angle\label{sec:twopt}}
In this section we analyze the nucleon two-point correlation function, to extract
the effective mass, as well as the nucleon mixing angle.
The standard two-point correlation function with sink momentum ${\bm p}^{\, \prime}$ has the form
\begin{eqnarray}\label{eq:C2}
G_{2}(\bm{p}^{\, \prime},t,\Pi) =
a^{3}\sum_{\bm{x}}
e^{-i\bm{p}^{\, \prime}\cdot\bm{x}}
\,\mathrm{Tr}\left\lbrace \Pi
\braket{ \mathcal{N}(\bm{x},t)
\overbar{\mathcal{N}}(\bm{0},0)}
\right\rbrace ,
\end{eqnarray}
where \(\Pi\) is some spin projector, and \(\mathcal{N}\) is an interpolating field
with the quantum numbers of a nucleon, inserted with a source-sink time separation of $t$.  The spectral decomposition for this
equation in the limit where \(T \gg t \gg 0\), keeping implicit a sum over the polarizations, is
\begin{eqnarray}
  \label{eq:C2_decomp}
    G_{2}(\bm{p}^{\, \prime},t,\Pi)  =
    \frac{
    e^{-E_{\beta_{0}}t}
    }{2E_{\beta_{0}}}
    \,\mathrm{Tr} \left\lbrace \Pi
    \bra{\Omega} \mathcal{N} \ket{\beta_{0}}
    \bra{\beta_{0}}\overbar{\mathcal{N}}\ket{\Omega}
    \right\rbrace ,
\end{eqnarray}
where the lowest energy state \(\beta_{0}\) for which \(\bra{\Omega} \mathcal{N} \ket{\beta_{0}} \ne 0\) and
\(\bra{\beta_{0}}\overbar{\mathcal{N}}\ket{\Omega} \ne 0\) arises from the approximation\footnote{It is clear from the context when we consider operators as in eq.~\eqref{eq:C2_decomp} or interpolating fields as in eq.~\eqref{eq:C2}.} \(T \gg t \gg 0\).

\begin{figure}
\begin{adjustwidth}{-0.1\textwidth}{-0.1\textwidth}
\centering
\begin{subfigure}{.55\textwidth}
  \centering
  \includegraphics[trim={11mm 0cm 11mm 0cm},clip,width=\linewidth]{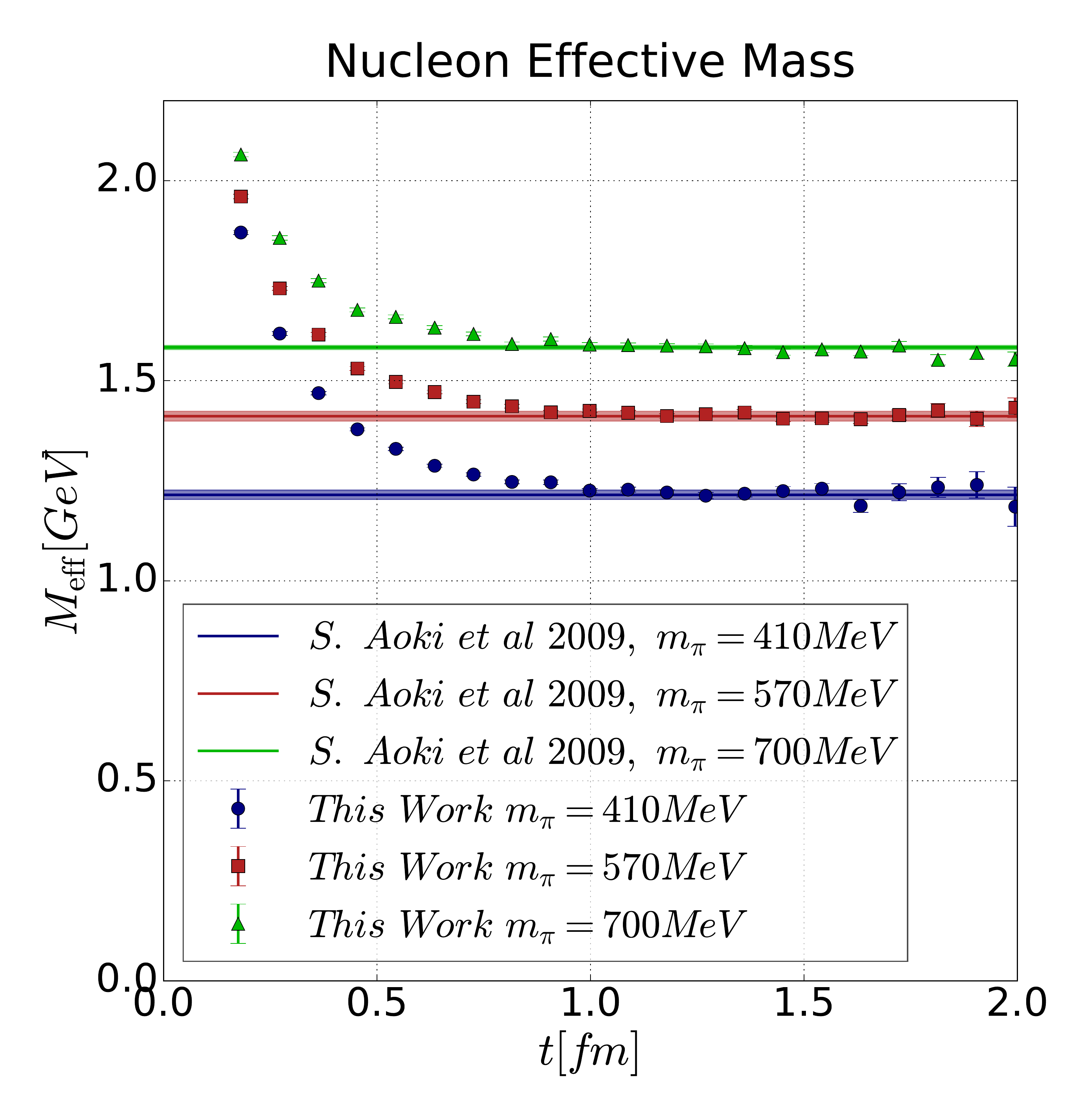}
  \caption{\label{fig:Emass_mpi}}
\end{subfigure}
\quad
\begin{subfigure}{.55\textwidth}
  \centering
  \includegraphics[trim={11mm 0cm 11mm 0cm},clip,width=\linewidth]{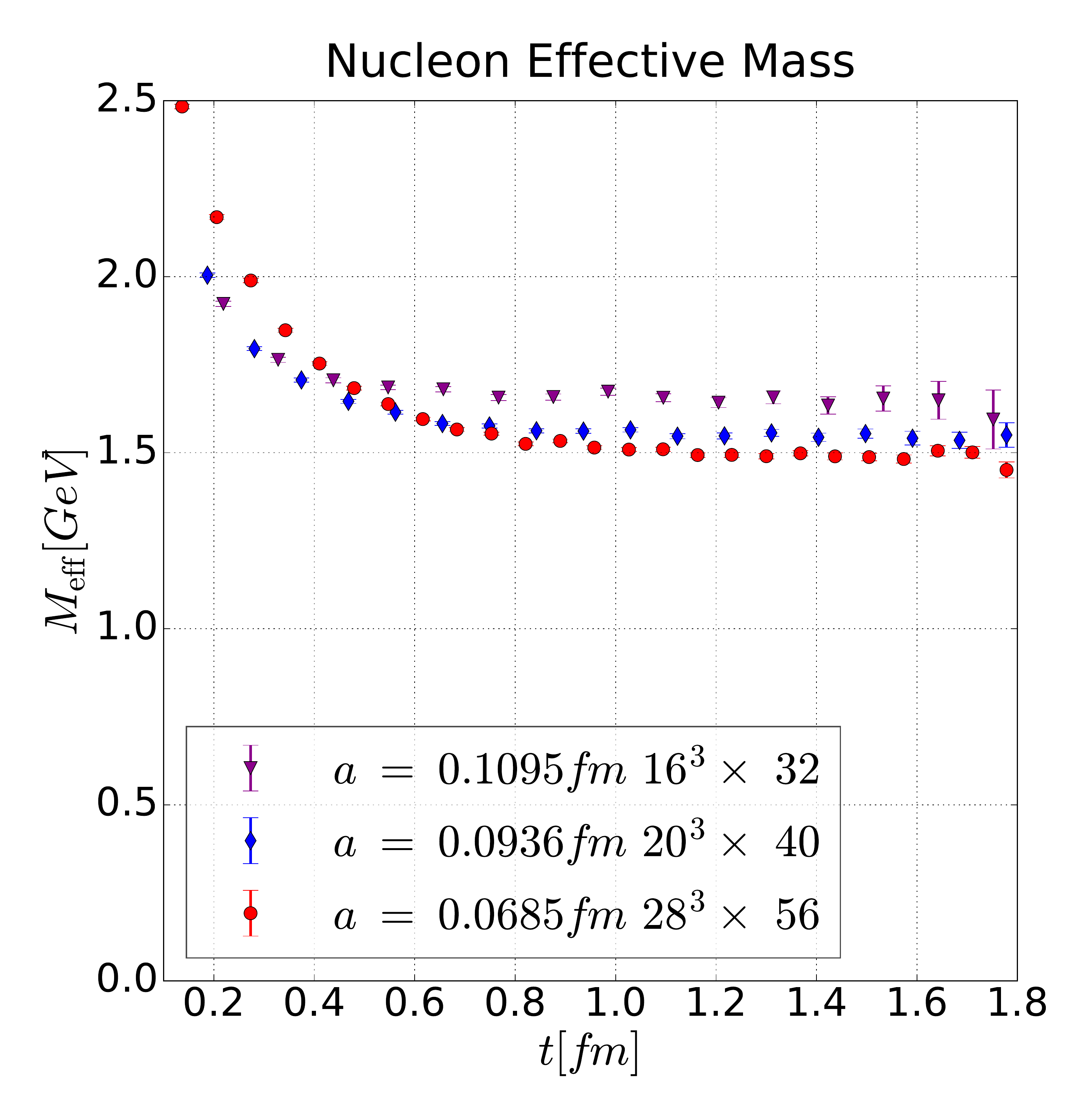}
  \caption{\label{fig:Emass_latspace}}
\end{subfigure}
\end{adjustwidth}
  \caption{
  \label{fig:Emass}
  Left: $m_{\pi}$ dependence of the effective mass (in GeV)
  defined in eq.~\eqref{eq:effmass} plotted against source-sink separation time \(t\). Bands
  correspond to values quoted in ref.~\cite{Aoki:2008sm}.
  Right: lattice-spacing dependence of the effective mass (in GeV)
   plotted against source-sink separation time \(t\).
  }
\end{figure}

The effective mass, which is shown in fig.~\ref{fig:Emass}, is given by the simple
log ratio
\begin{eqnarray}\label{eq:effmass}
  M_{\mathrm{eff}}(\bm{p}^{\, \prime}=\bm{0},t,\Pi_{+}) =
  \log\left[\frac{G_{2}(\bm{p}^{\, \prime}=\bm{0},t,\Pi_{+})}{G_{2}(\bm{p}^{\, \prime}=\bm{0},t+1,\Pi_{+})}\right]
  = m_{\beta^{+}_{0}},
\end{eqnarray}
where \(\Pi_{+} = (I + \gamma_{4})/2\) is the positive parity projector, \(\beta^{+}_{0}\) is the lowest energy positive parity nucleon state, and again, \(T \gg t \gg 0\).  In fig.~\ref{fig:Emass_mpi}, we compare our effective mass determinations
for the M-ensembles to those computed in ref.~\cite{Aoki:2008sm} and find agreement within statistical errors.  As shown in  fig.~\ref{fig:Emass_latspace}, we observe lattice-spacing dependence of the order of \(10\%\) between the finest and coarsest lattices.

The nucleon mixing angle \cite{Shintani:2005xg}, $\alpha_{N}$,
can be extracted by utilizing the two-point correlator from eq.~\eqref{eq:C2},
and the $\bar \theta$-modified two-point correlator
\begin{eqnarray}\label{eq:C2Q}
G^{(Q)}_{2}(\bm{p}^{\, \prime},t,\Pi,t_{f}) =
a^{3}\sum_{\bm{x}}
e^{-i\bm{p}^{\, \prime}\cdot\bm{x}}
\,\mathrm{Tr}\left\lbrace \Pi
\braket{\mathcal{N}(\bm{x},t)
\overbar{\mathcal{N}}(\bm{0},0)
Q(t_{f})}
\right\rbrace\,.
\end{eqnarray}
The mixing angle is defined using the small-\(\bar{\theta}\) expansion as
\begin{eqnarray} \label{eq:alpha_rat}
  \alpha_{N} = \frac{G^{(Q)}_{2}(\bm{p}^{\, \prime}=\bm{0},t,\gamma_{5}\Pi_{+},t_{f})}{G_{2}(\bm{p}^{\, \prime}=\bm{0},t,\Pi_{+})},
\end{eqnarray}
in the region where \(t\gg 0\) and \(t_{f} \gg 0\).

\begin{figure}
\vspace*{-1.5cm}
\begin{adjustwidth}{-0.1\textwidth}{-0.1\textwidth}
\centering
\begin{subfigure}{.55\textwidth}
  \centering
  \includegraphics[trim={11mm 0cm 11mm 0cm},clip,width=\linewidth]{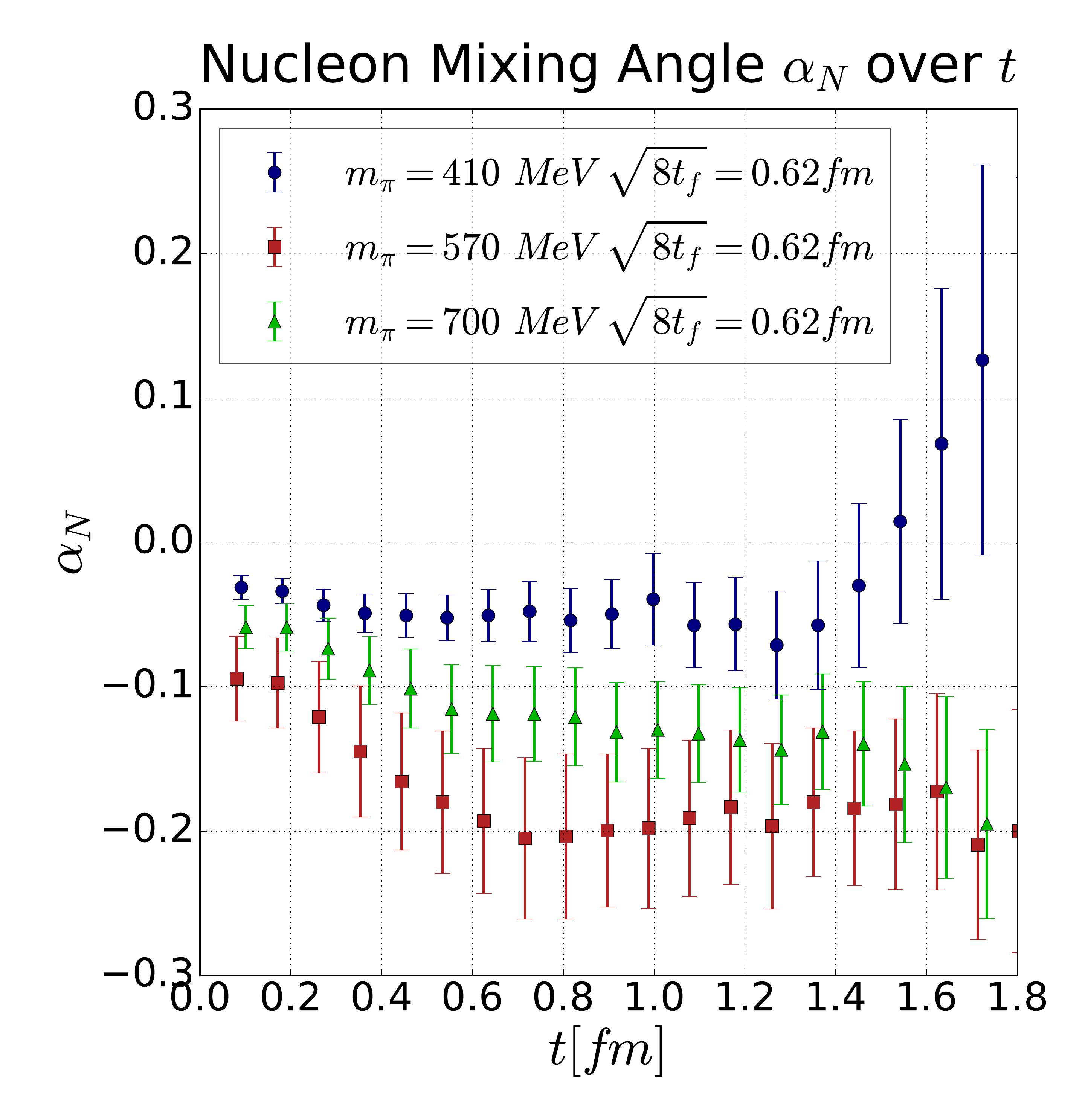}
  \caption{\label{fig:aQ_mpi_vs_t}}
\end{subfigure}
\quad
\begin{subfigure}{.55\textwidth}
  \centering
  \includegraphics[trim={11mm 0cm 11mm 0cm},clip,width=\linewidth]{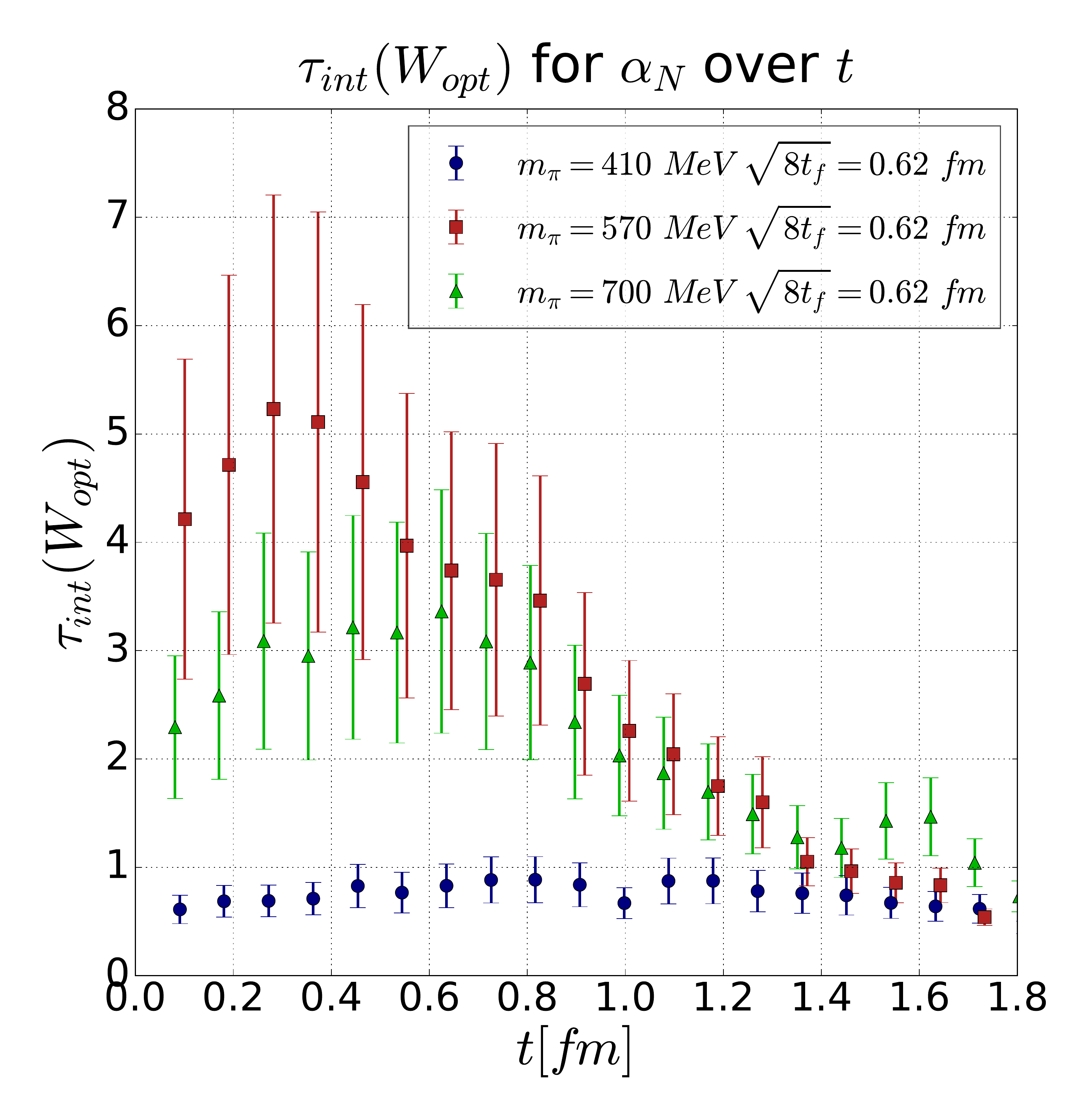}
  \caption{\label{fig:aQ_mpi_vs_t_tauint}}
\end{subfigure}
\end{adjustwidth}
\caption{Left: The nucleon mixing angle as  function of  the source-sink separation \(t\) at fixed
  flow time \(\sqrt{8t_{f}}=0.62 \) fm for different pion masses.
  Right: Integrated autocorrelation of left plot.}
\vspace*{\floatsep}
\begin{adjustwidth}{-0.1\textwidth}{-0.1\textwidth}
  \centering
\begin{subfigure}{.55\textwidth}
  \centering
  \includegraphics[trim={11mm 0cm 11mm 0cm},clip,width=\linewidth]{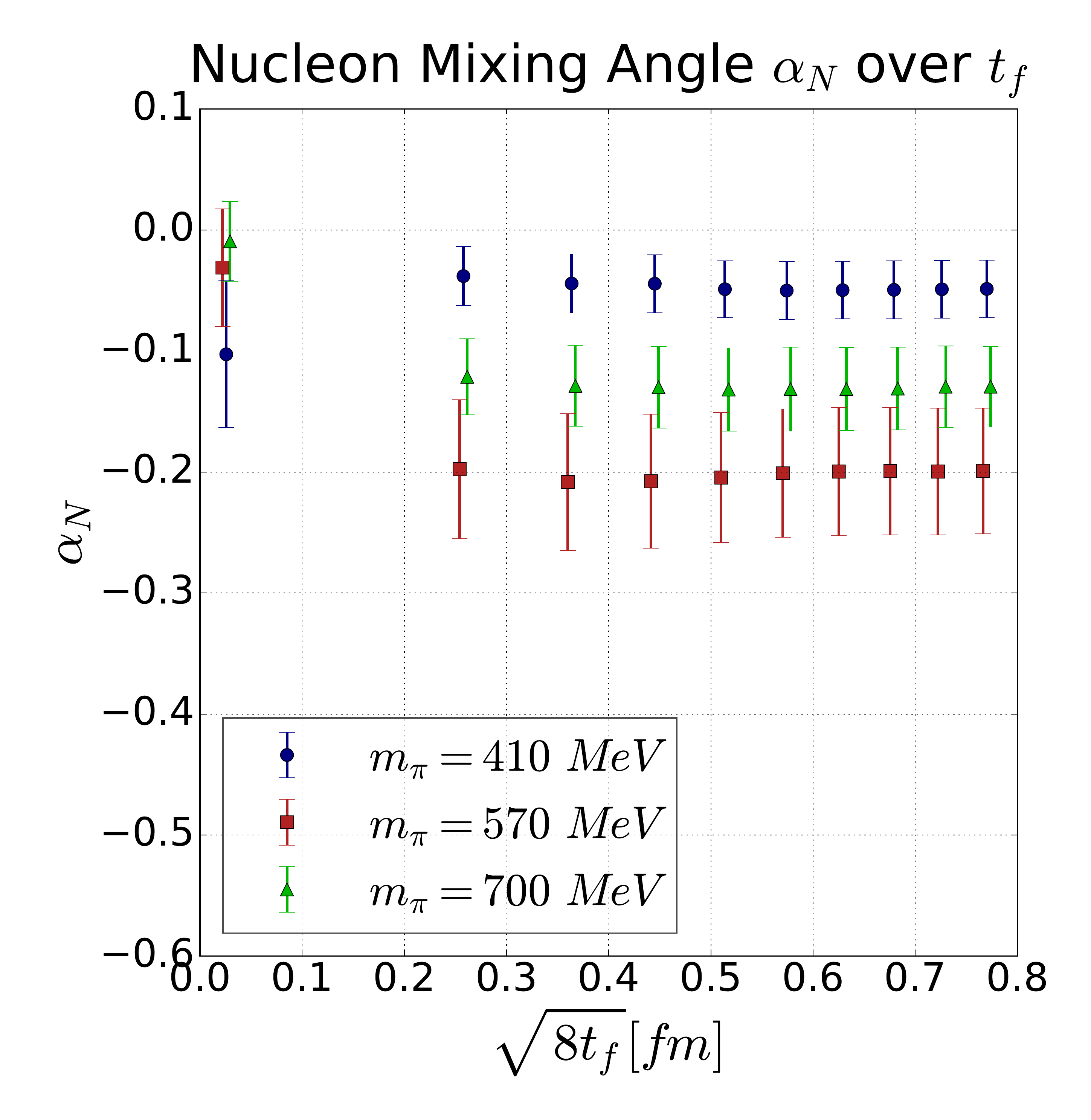}
  \caption{\label{fig:aQ_mpi_vs_tflow}}
\end{subfigure}
\quad
\begin{subfigure}{.55\textwidth}
  \centering
  \includegraphics[trim={11mm 0cm 11mm 0cm},clip,width=\linewidth]{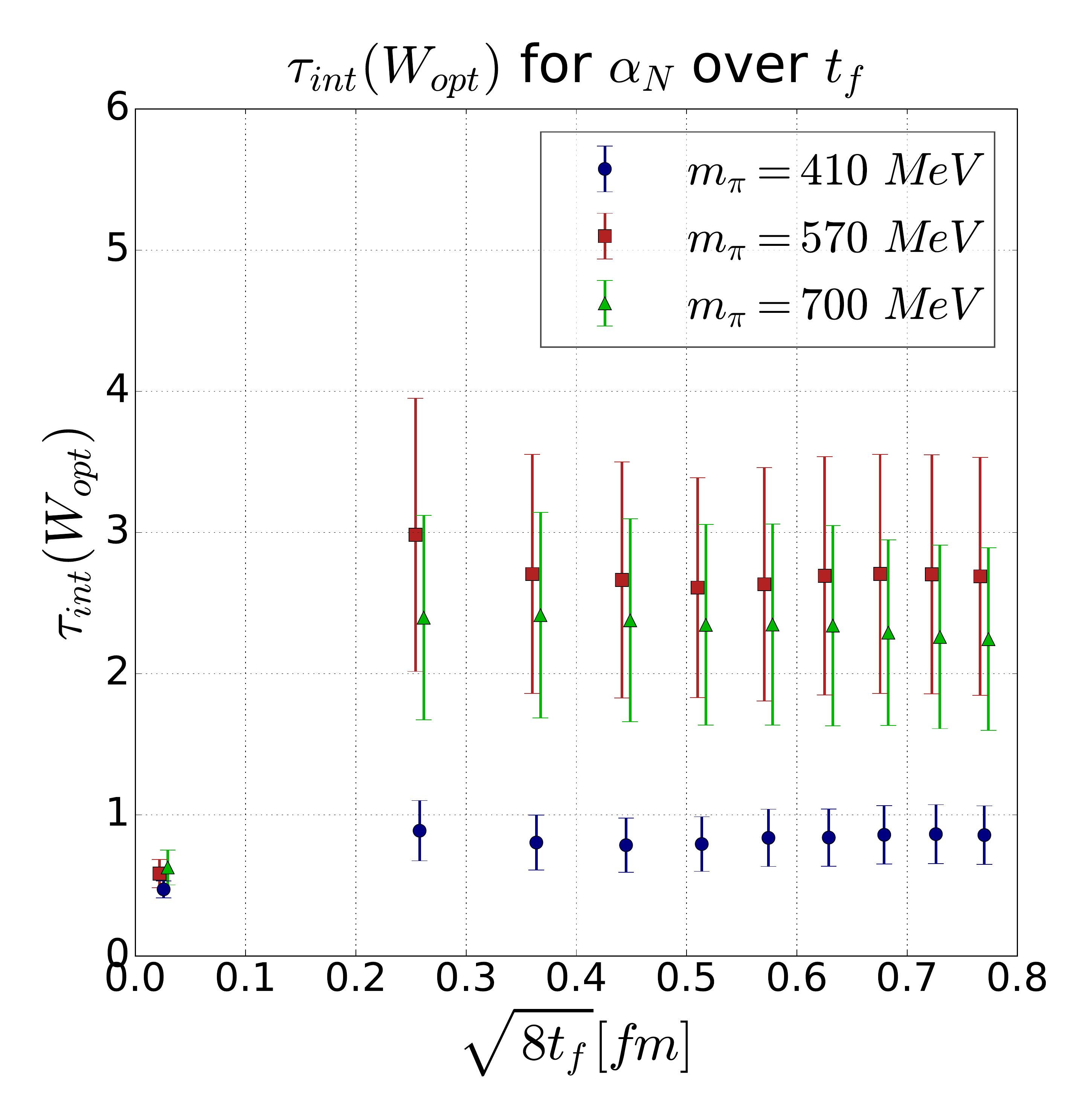}
  \caption{\label{fig:aQ_mpi_vs_tflow_tauint}}
\end{subfigure}
\end{adjustwidth}
\caption{
  Left: The nucleon mixing angle as function of the flow-time radius \(\sqrt{8t_{f}}\),
  at fixed source-sink separation \(t=0.91\) fm for different pion masses.
  Right: Integrated autocorrelation of left plot.}
\end{figure}

\begin{figure}
\vspace*{-1.5cm}
\begin{adjustwidth}{-0.1\textwidth}{-0.1\textwidth}
\centering
\begin{subfigure}{.55\textwidth}
  \centering
  \includegraphics[trim={11mm 0cm 11mm 0cm},clip,width=\linewidth]{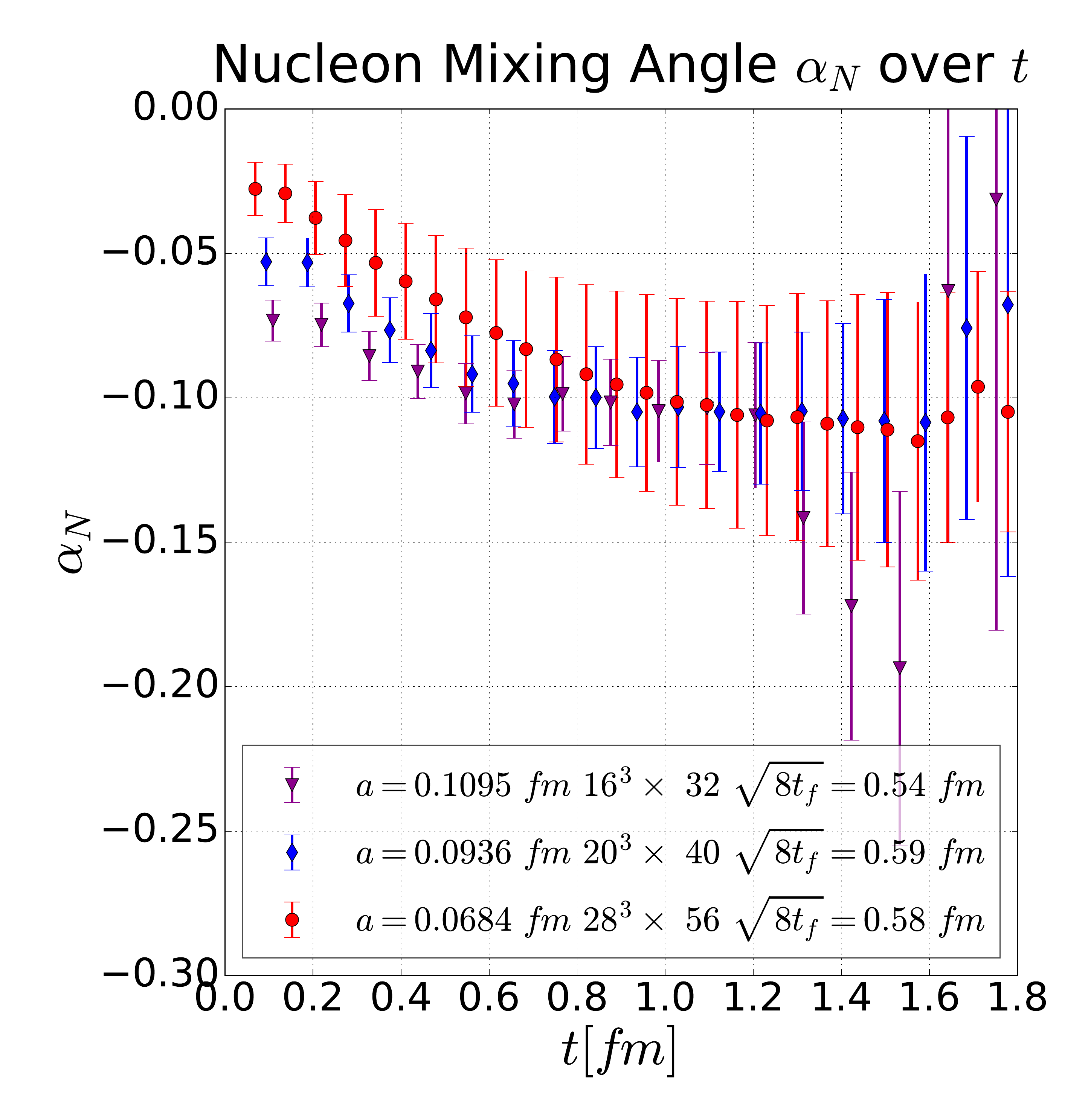}
  \caption{\label{fig:aQ_latspace_vs_t}}
\end{subfigure}
\quad
\begin{subfigure}{.55\textwidth}
  \centering
  \includegraphics[trim={11mm 0cm 11mm 0cm},clip,width=\linewidth]{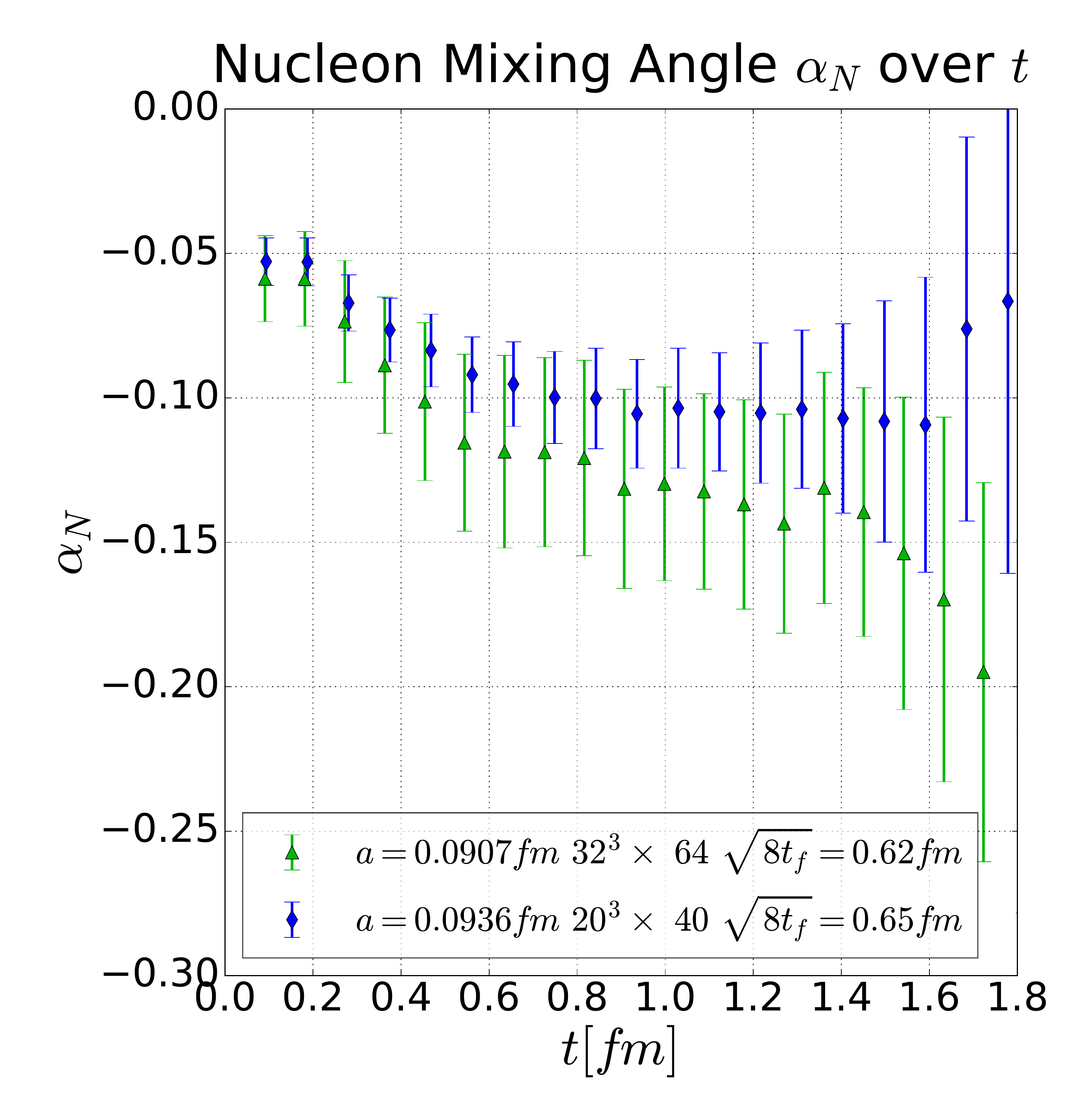}
  \caption{\label{fig:aQ_box_vs_t}}
\end{subfigure}
\end{adjustwidth}
\caption{
  Nucleon mixing angle as function of the source-sink separation \(t\) at fixed flow time for the A-ensembles (left) and box size ensembles (right).
  }
\vspace*{\floatsep}
\begin{adjustwidth}{-0.1\textwidth}{-0.1\textwidth}
\centering
\begin{subfigure}{.55\textwidth}
  \centering
  \includegraphics[trim={11mm 0cm 11mm 0cm},clip,width=\linewidth]{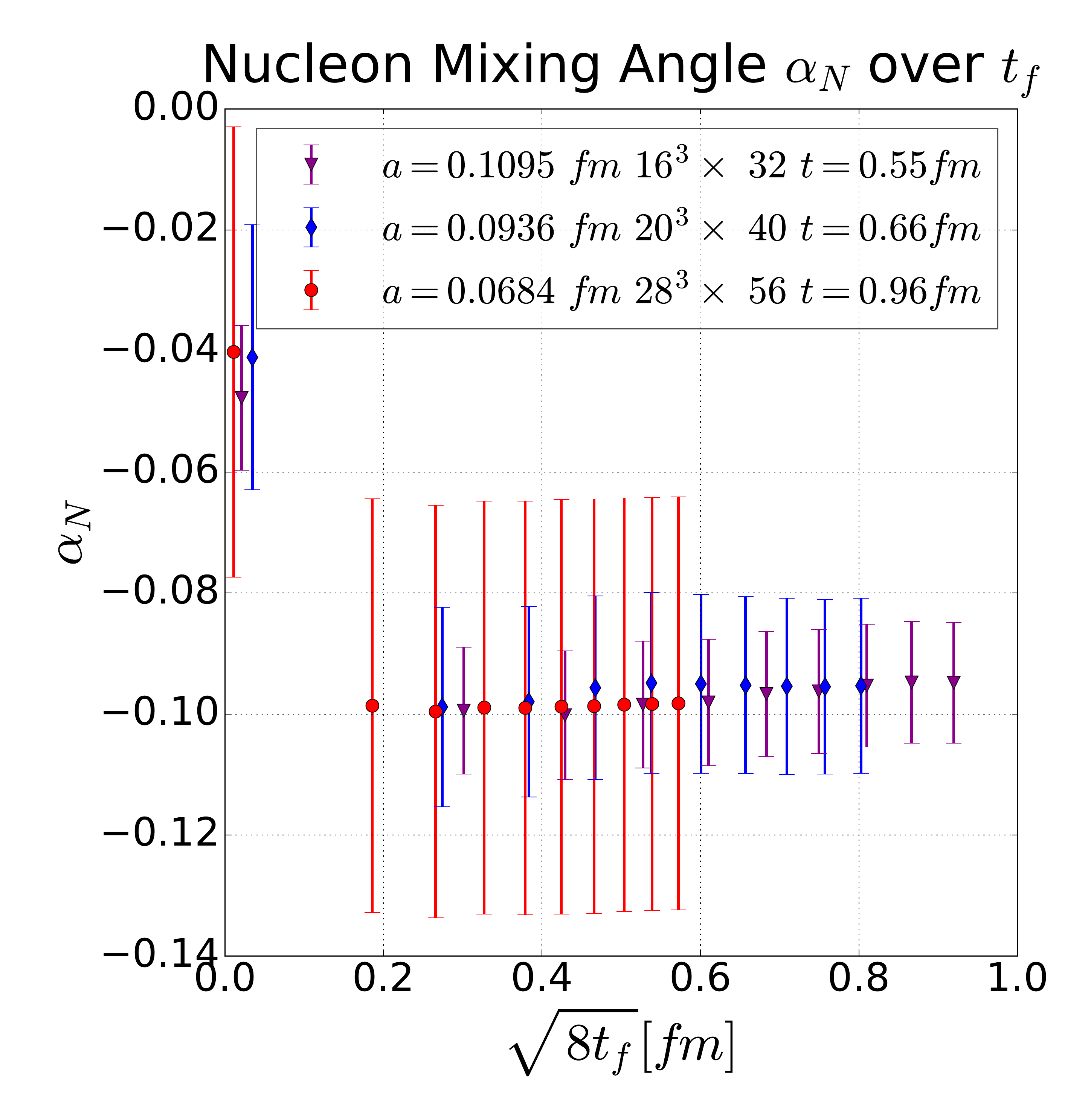}
  \caption{\label{fig:aQ_latspace_vs_tflow}}
\end{subfigure}
\quad
\begin{subfigure}{.55\textwidth}
  \centering
  \includegraphics[trim={11mm 0cm 11mm 0cm},clip,width=\linewidth]{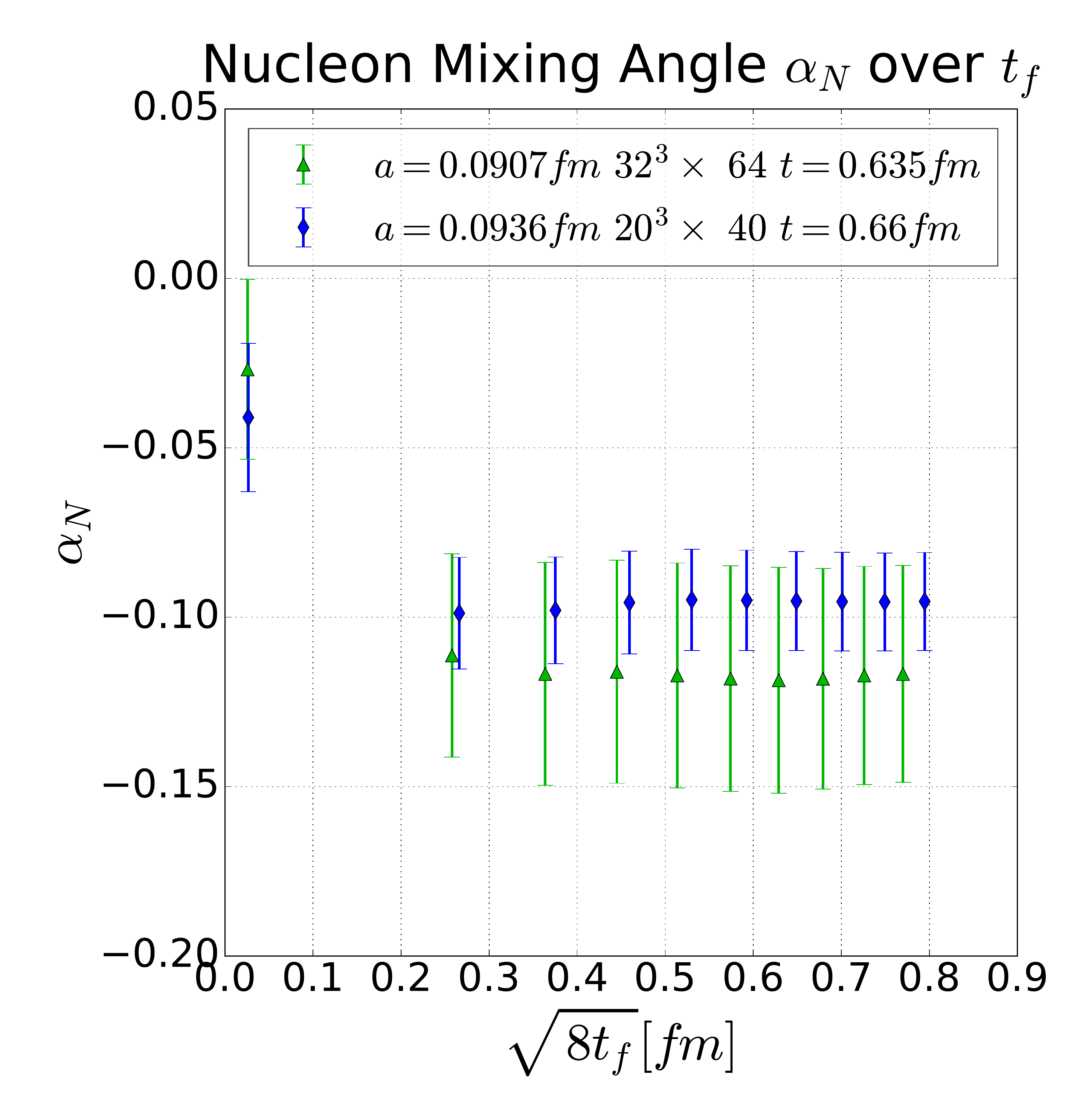}
  \caption{\label{fig:aQ_box_vs_tflow}}
\end{subfigure}
\end{adjustwidth}
\caption{
  Nucleon mixing angle as function of the flow time radius \(\sqrt{8t_{f}}\) at fixed source-sink separation \(t\) for various A-ensembles (left) and box size ensembles (right).
  }
\end{figure}

In figs.~\ref{fig:aQ_mpi_vs_t} and \ref{fig:aQ_latspace_vs_t}, we show the dependence of the nuclear mixing angle
on the source-sink separation $t$ (in fm) for the M- and A-ensembles, respectively.
For the M-ensembles in figs.~\ref{fig:aQ_mpi_vs_t}, there is little to no
excited-state contamination effects for \({t>0.6}\)~fm.
The results suggest a non-trivial chiral behavior for \(\alpha_{N}\).
We will discuss this in detail in sec.~\ref{sec:F3_res} when we describe our EDM determination.
For the lattice-spacing ensembles in fig.~\ref{fig:aQ_latspace_vs_t}, we require a minimum source-sink separation of
$t\approx a\,\{5,7,14\}$ fm (for $a=\{0.1095,0.0936,0.0684\}$ fm ensembles) to plateau and achieve ground state saturation.
The final plateaued quantity for the different A-ensembles all lead to the same angle $\alpha_{N}$ and
all results are consistent within statistical uncertainties.
Similarly, fig.~\ref{fig:aQ_box_vs_t} demonstrates that computing \(\alpha_{N}\) at \(m_{\pi}=700\) MeV
with different box sizes lead to consistent results. In tab.~\ref{tab:alpha_fitr} we summarize the fit ranges and resulting values for \(\alpha_{N}\).

The value of the flow-time radius, \(\sqrt{8t_{f}}\), for all these analyses is fixed around
\(0.5 - 0.6\) fm where the topological charge is least affected by lattice artifacts.
Figs.~\ref{fig:aQ_mpi_vs_tflow},~\ref{fig:aQ_latspace_vs_tflow} show the nucleon mixing angle plotted against the
flow time $\sqrt{8t_f}$ at a fixed source-sink separation $t$ for the M- and A-ensembles.
For the M-ensembles, we see no flow-time dependence after \(\sqrt{8t_{f}} > 0.2 - 0.3\) fm,
confirming that the results obtained in this region are free from gradient-flow discretization effects.
A similar conclusion is reached for the A-ensembles.

\begin{table}
\centering
\caption{Fit ranges $[t^{min},t^{max}]$ over euclidean source-sink separation $t$ used to extract the
nucleon mixing angle \(\alpha_{N}\), along with the resulting value. \label{tab:alpha_fitr}}
  \begin{tabular}{ r|c|c|c|c|c|c }
  ensemble &  M\(_{3}\)& M\(_{2}\)& M\(_{1}\)  & A\(_{1}\)& A\(_{2}\)& A\(_{3}\)  \\
  \hline
  fit range & [10,20] & [10,20] & [10,20]  & [5,11] & [7,17] & [14,21]  \\
  \hline
  fitr [fm] & [0.9,1.8] & [0.9,1.8] & [0.9,1.8]  & [0.6,1.3] & [0.7,1.7] & [0.96,1.43]  \\
  \hline
  $\alpha_N$ & -0.040(21) & -0.190(27) & -0.142(24) & -0.099(11) & -0.103(10) &  -0.105(11) \\
  \end{tabular}
\end{table}

In figs.~\ref{fig:aQ_mpi_vs_t_tauint} we show the integrated
autocorrelation time of $\alpha_N$ for the M-ensemble results shown in fig.~\ref{fig:aQ_mpi_vs_t}.
For the  M$_{1}$ and M$_{2}$ ensembles a factor of \(2-4\) increase in autocorrelation as
the source-sink separation approaches 0. Fortunately, a minimum source-sink separation of \({t\approx1}\)~fm
 greatly decreases the autocorrelation correction that we apply in the determination of the
nucleon mixing angle \(\alpha_{N}\).
Most importantly, in comparison to $\braket{Q}$ from fig.~\ref{fig:tau_intQ} (i.e. not in the presence of a nucleon),
the autocorrelation effect is dramatically decreased by a factor of at least~\(\simeq4\).
We attribute this effect to the presence of a fermionic part, \(\mathcal{N}\overbar{\mathcal{N}}\), in the correlation function.
Numerical evidence suggests that the observables considered in this work containing fermion lines, such
as \(\alpha_{N}\) and the EDM are less coupled to the slow modes contributing to the spectral
decomposition of the autocorrelation function \cite{Luscher:2011kk}.
As this effect will be greater when analyzing the EDM (from three-point correlation functions), we resort to our standard bootstrap error propagation technique for the final EDM computation. We checked explicitly that error estimates from a bootstrap and an autocorrelation analysis give consistent results.

\subsection{Improving the Nucleon Mixing Angle\label{sec:alpha_imp}}
In this section we describe a method previously explored in \cite{Liu:2017man}, that aims to reduce the statistical uncertainty of the determination of the nucleon mixing angle $\alpha_{N}$. The strategy can be described as an attempt to understand the space-time region where the overlap between the topological charge density and the fermionic part of the correlation function is maximal.
To perform this investigation we define a spatially-summed topological charge density
\begin{eqnarray}\label{eq:Q_temp_def}
  \overbar{Q}(\tau_{Q},t_{f}) = a^{3}\sum_{\bm{x}} q(\bm{x},\tau_{Q};t_{f}), \qquad
  Q(t_{f}) = a\sum_{\tau_{Q}}\overbar{Q}(\tau_{Q},t_{f})\,.
\end{eqnarray}
We then  numerically study the dependence on \(\tau_{Q}\) of \(\alpha_{N}\) and corroborate our numerical findings with a spectral decomposition of the relevant correlators.

The ratio \(\alpha_{N}\) and the modified two-point correlator \(G_{2}^{(Q)}\) have the same \(\tau_{Q}\) dependence and we therefore focus on the latter. Setting \(\bm{p}^{\, \prime}=\bm{0}\) and omitting it in our expressions, we define
\begin{eqnarray}\label{eq:C2QImprov}
\Delta^{(\overbar{Q})}_{2}(t,\Pi,t_{f},\tau_{Q}) =
a^{3}\sum_{\bm{x}}
\mathrm{Tr}\left\lbrace \Pi
\braket{
\mathcal{N}(\bm{x},t)
\overbar{Q}(\tau_{Q},t_{f})
\overbar{\mathcal{N}}(\bm{0},0)
}
\right\rbrace
\end{eqnarray}
where the correlator in eq.~\eqref{eq:C2Q} can be obtained by summing \(\tau_{Q}\) from \(0\) to the time extent of the lattice \(T\)
\begin{eqnarray}
  G_{2}^{(Q)}(t,\Pi,t_{f}) = a\sum_{\frac{\tau_{Q}}{a}=0}^{T/a} \Delta^{(\overbar{Q})}_{2}(t,\Pi,t_{f},\tau_{Q})\,.
\end{eqnarray}
To focus on the region where the signal resides, we sum the spatially-summed topological charge density, \(\overbar{Q}(\tau_{Q},t_{f})\), symmetrically starting from the source location. That is we sum \(\tau_{Q}\) starting
from \(0\) (and \(T\)) up to a value \(t_{s}\) (and $T-t_s$). The goal is to find a summation window \(t_{s}\) small enough such that we capture all the signal and avoid the summation of unnecessary ``noise''. We define the partial summed correlator
\begin{eqnarray}\label{eq:C2QImprov_2}
\overbar{G}^{(\overbar{Q})}_{2}(t,\Pi,t_{f},t_{s}) &=&
a\sum_{\frac{\tau_{Q}}{a} = 0}^{t_{s}/a} \left[\Delta^{(\overbar{Q})}_{2}(t,\Pi,t_{f},\tau_{Q})
+ \Delta^{(\overbar{Q})}_{2}(t,\Pi,t_{f},T-\tau_{Q})\right]\,,
\end{eqnarray}
from which, using the periodicity of our lattice, the original correlator in eq.~\eqref{eq:C2Q} is obtained as
\begin{eqnarray}
  G_{2}^{(Q)}(t,\Pi,t_{f}) = \overbar{G}^{(\overbar{Q})}_{2}(t,\Pi,t_{f},t_{s}=T/2)\,.
\end{eqnarray}
Although there are other choices for the starting point of our summation in \(\tau_{Q}\), we only
consider starting from \(\tau_{Q}=0\). In app.~\ref{app:alpha_imp} we derive a spectral decomposition for the correlator in
eq.~\eqref{eq:C2QImprov}. We argue that in the limit \(t_{s} \gg t \gg 0\), the partially-summed correlator
\( \overbar{G}^{(\overbar{Q})}_{2}(t,\Pi,t_{f},t_{s})\) is independent of \(t_{s}\) and \(t\),
up to exponentially suppressed corrections. These corrections seem to be rather small, and in fact our numerical
experiments indicate that we can safely stop the summation over \(\tau_{Q}\) at \(t_{s}\simeq t\).
In this way we avoid to sum in the region between \(t\) and \(T/2\) where numerically the
correlators seem to vanish up to statistical fluctuations.

\begin{figure}
\begin{adjustwidth}{-0.11\textwidth}{-0.11\textwidth}
\centering
\begin{subfigure}{.35\textwidth}
  \centering
  \includegraphics[trim={11mm 0cm 11mm 0cm},clip,width=\linewidth]{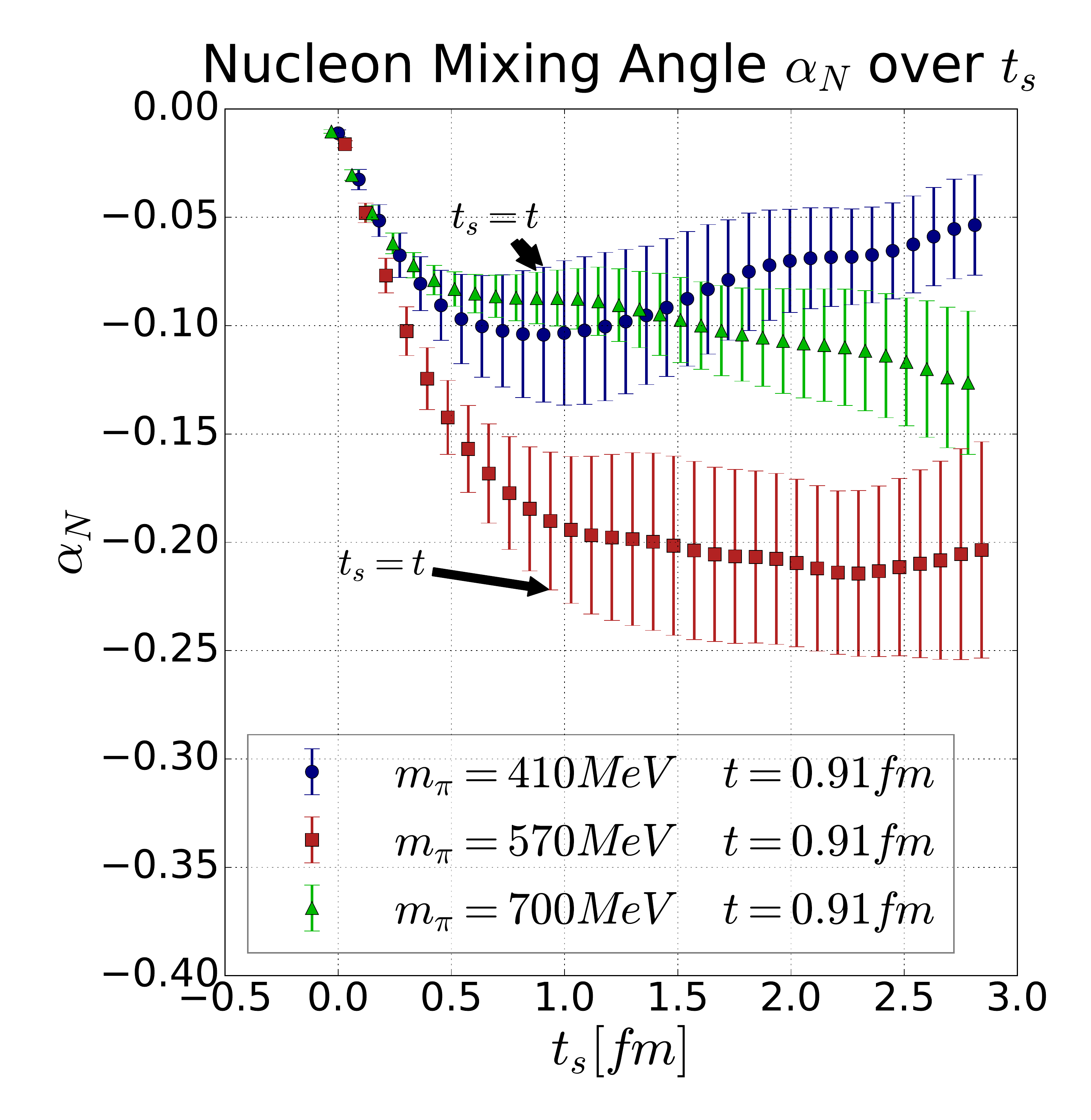}
  \caption{\label{fig:a_imp_mpi}}
\end{subfigure}
\quad
\begin{subfigure}{.35\textwidth}
  \centering
  \includegraphics[trim={11mm 0cm 11mm 0cm},clip,width=\linewidth]{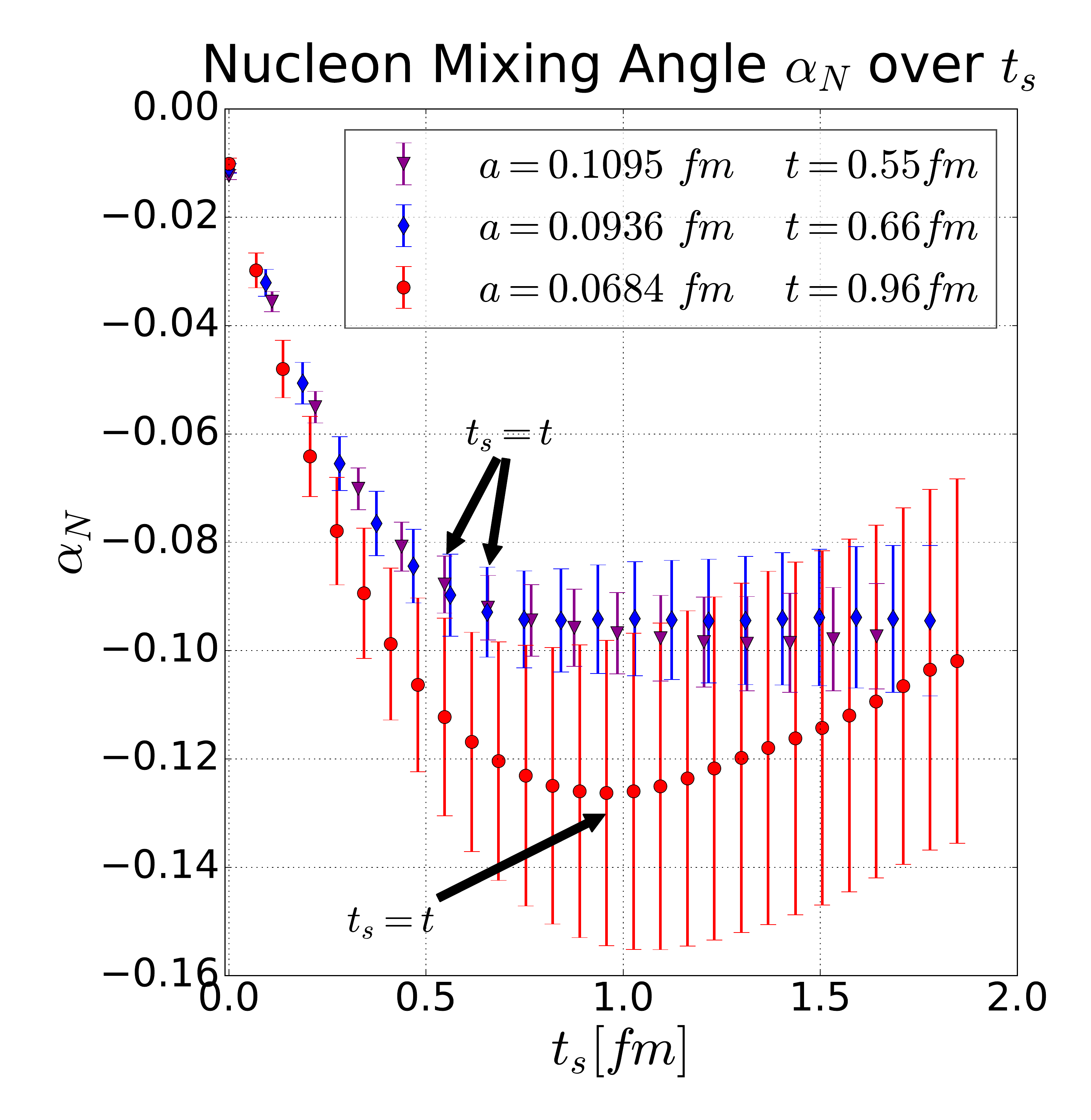}
  \caption{\label{fig:a_imp_latspace}}
\end{subfigure}
\quad
\begin{subfigure}{.35\textwidth}
  \centering
  \includegraphics[trim={11mm 0cm 11mm 0cm},clip,width=\linewidth]{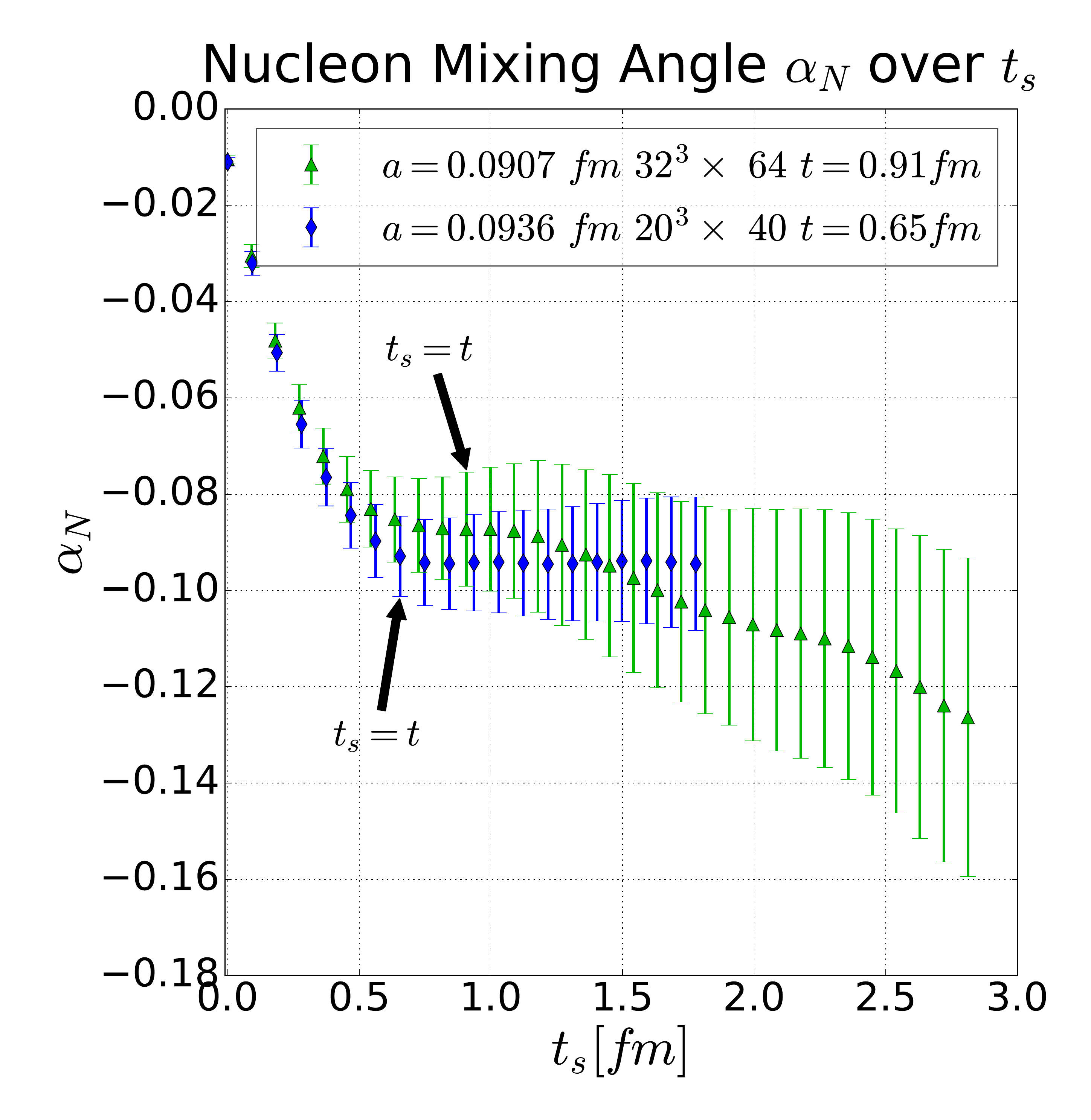}
  \caption{\label{fig:a_imp_box}}
\end{subfigure}
\end{adjustwidth}
\caption{\label{fig:a_imp}
  M- (left), A- (middle), and box-size-ensembles (right) of the improved
  nucleon mixing angle \(\alpha_{N}\) plotted against the sum parameter \(t_{s}\).
  The final point coincides with the regular nucleon mixing angle from sec.~\ref{sec:twopt}.
  }
\end{figure}

We first fix the source-sink separation \(t\) to a large enough value such that effects from excited states are suppressed.
We then study the dependence of \(\alpha_{N}\) on the summation
window \(t_{s}\).
In fig.~\ref{fig:a_imp} we show the \(t_{s}\) dependence of \(\alpha_{N}\), for the M-ensembles (left), and
A-ensembles (middle) and two different physical volumes corresponding to M\(_{1}\) and A\(_{2}\) ensembles (right).
In all ensembles we observe that \(\alpha_{N}\) reaches a plateau when \(t_{s} \simeq  t\), consistent
with the expectation that contributions for \(t_{s} > t\) are exponentially suppressed and below our statistical accuracy.
We do observe a very small drift of  \(\alpha_{N}\) for larger values of $t_s$  for the ensembles M\(_{1}\), and
a smaller drift for the ensemble M\(_{3}\), for small values of \(t\). We attribute this to statistical fluctuations
that could arise from small local parity-violating effects induced by non-vanishing matrix elements of
 \(\overbar{Q}(\tau_{Q},t_{f})\) between two states of the same parity,
 \(\bra{\beta}\overbar{Q}(\tau_{Q},t_{f})\ket{\beta}\ne 0\). These local fluctuations are
 averaged out when the charge density is summed over the whole space-time volume as shown
 in fig.~\ref{fig:Q}. Nevertheless, all values of \(\alpha_{N}\) determined with the improved
 method are statistically compatible with the results obtained with the standard analysis.

\begin{figure}
\vspace*{-1.5cm}
\begin{adjustwidth}{-0.11\textwidth}{-0.11\textwidth}
\centering
\begin{subfigure}{.35\textwidth}
  \centering
  \includegraphics[trim={11mm 0cm 11mm 0cm},clip,width=\linewidth]{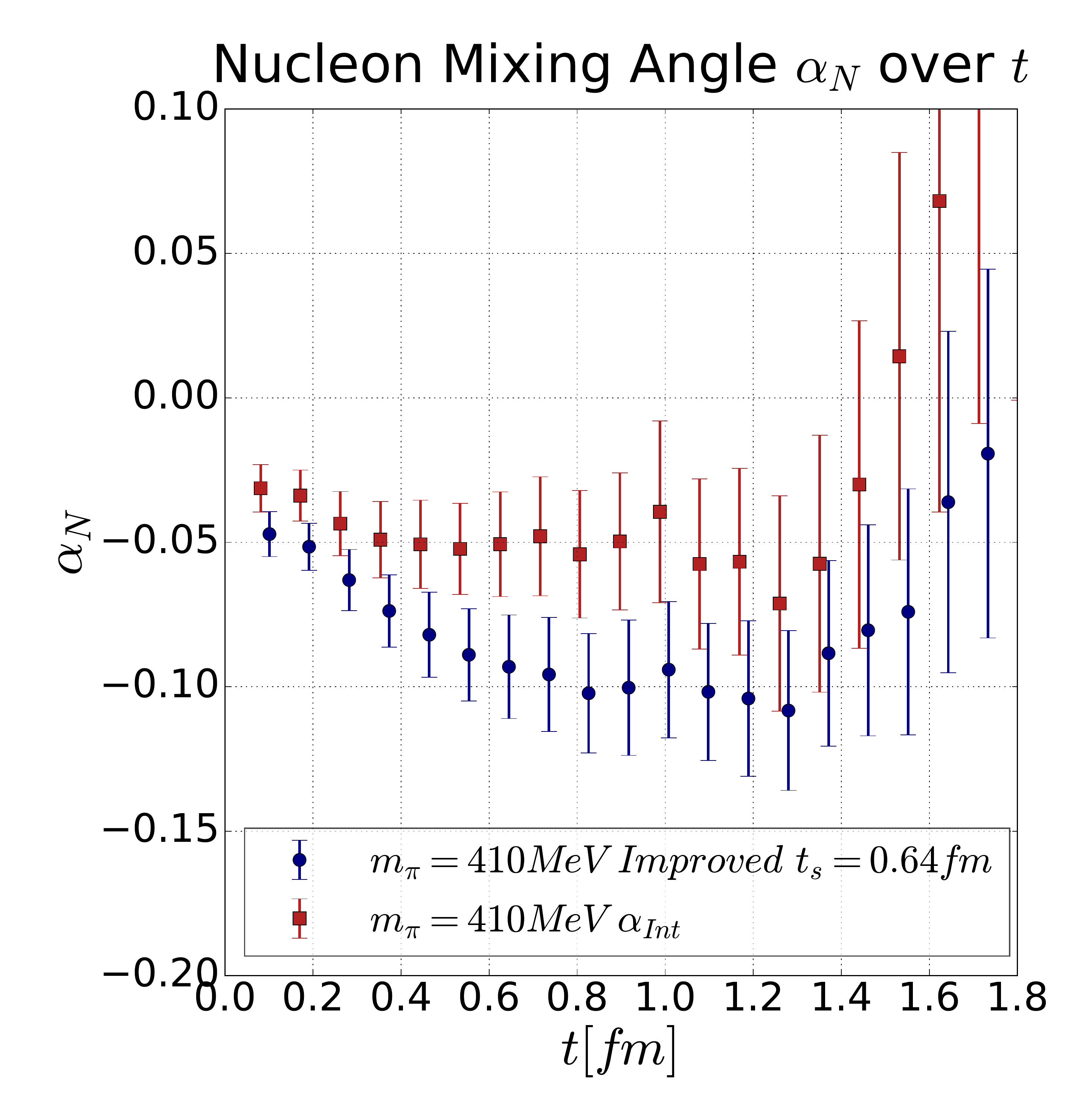}
  \caption{\label{fig:a_imp_mpi410_tsep}}
\end{subfigure}
\quad
\begin{subfigure}{.35\textwidth}
  \centering
  \includegraphics[trim={11mm 0cm 11mm 0cm},clip,width=\linewidth]{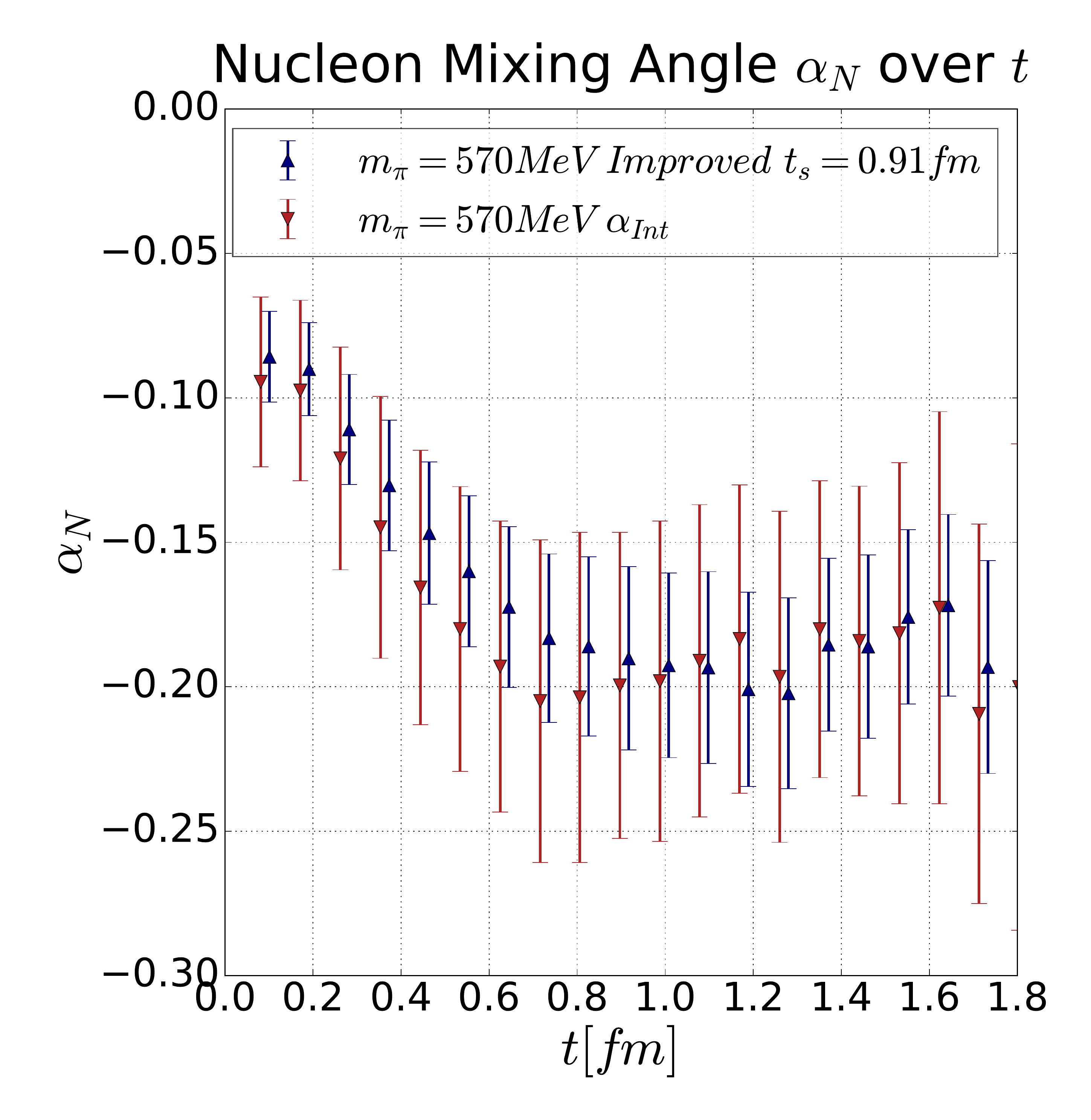}
  \caption{\label{fig:a_imp_mpi570_tsep}}
\end{subfigure}
\quad
\begin{subfigure}{.35\textwidth}
  \centering
  \includegraphics[trim={11mm 0cm 11mm 0cm},clip,width=\linewidth]{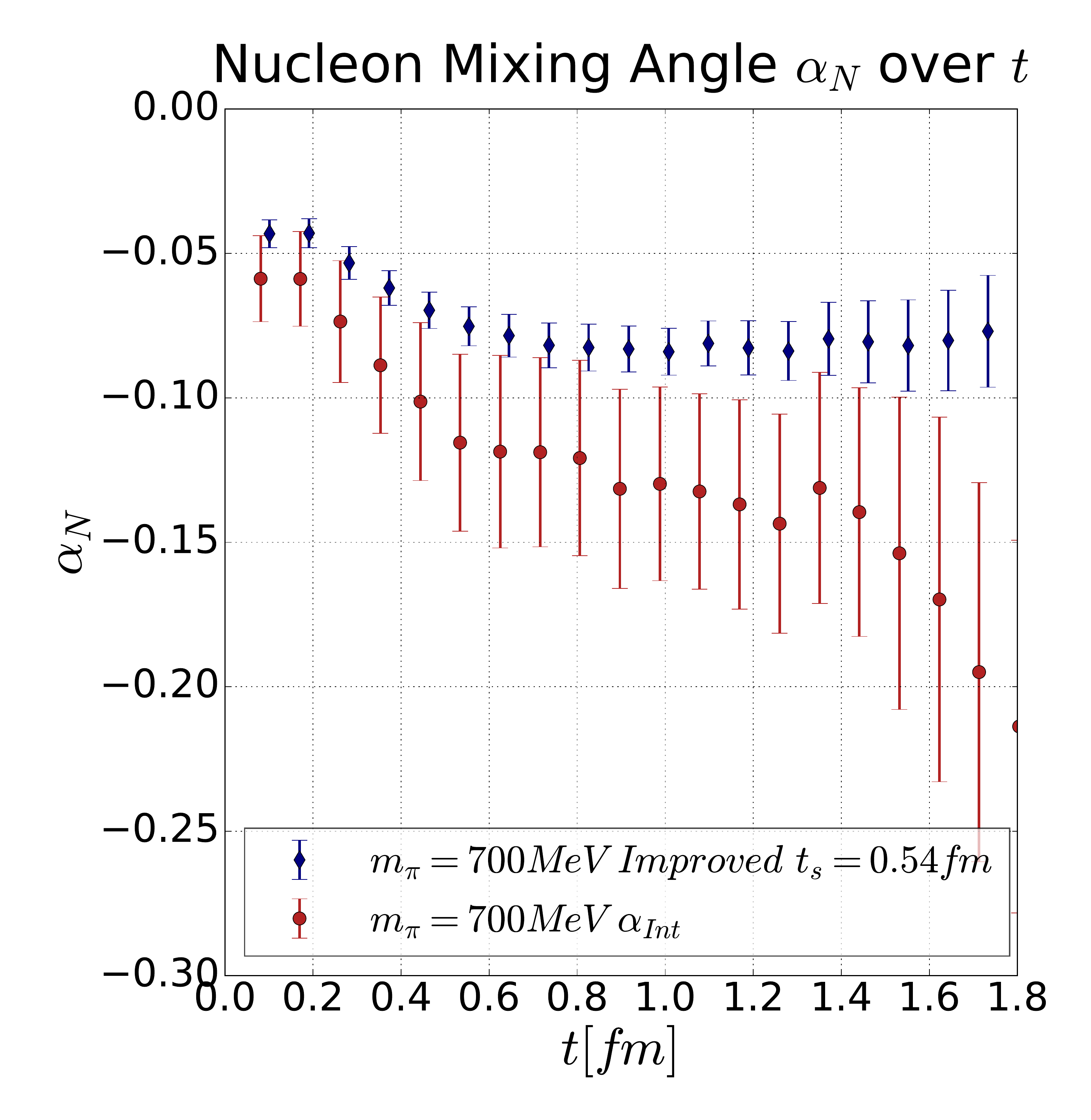}
  \caption{\label{fig:a_imp_mpi700_tsep}}
\end{subfigure}
\end{adjustwidth}
\caption{\label{fig:a_imp_tsep_mpi}
  \(\alpha_{N}\) against \(t\) plots for M-ensembles,
  comparing the improved method (blue) to the regular determination described in sec.~\ref{sec:twopt} (red).}
\vspace*{\floatsep}
\begin{adjustwidth}{-0.11\textwidth}{-0.11\textwidth}
\centering
\begin{subfigure}{.35\textwidth}
  \centering
  \includegraphics[trim={11mm 0cm 11mm 0cm},clip,width=\linewidth]{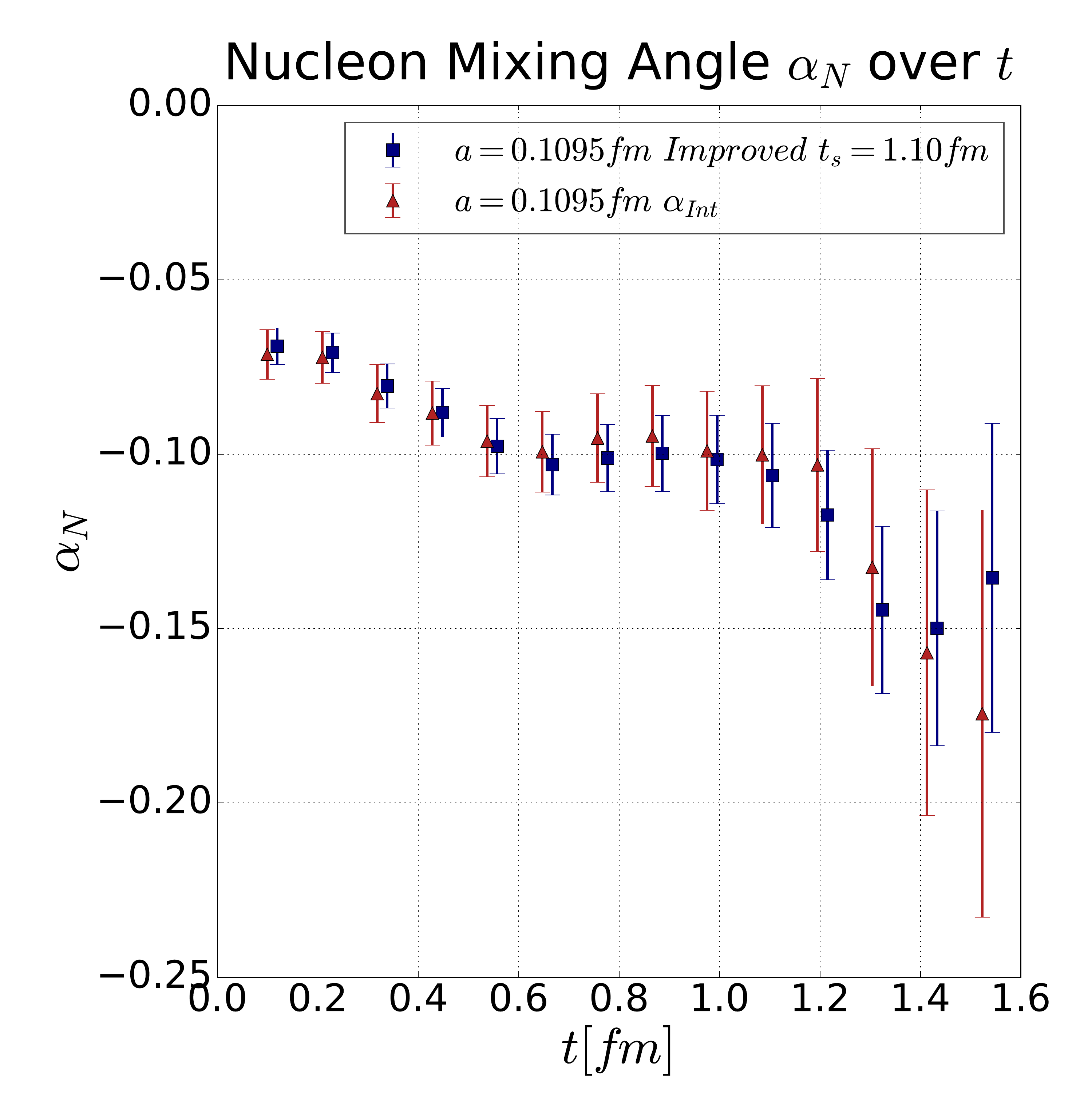}
  \caption{\label{fig:a_imp_L16_tsep}}
\end{subfigure}
\quad
\begin{subfigure}{.35\textwidth}
  \centering
  \includegraphics[trim={11mm 0cm 11mm 0cm},clip,width=\linewidth]{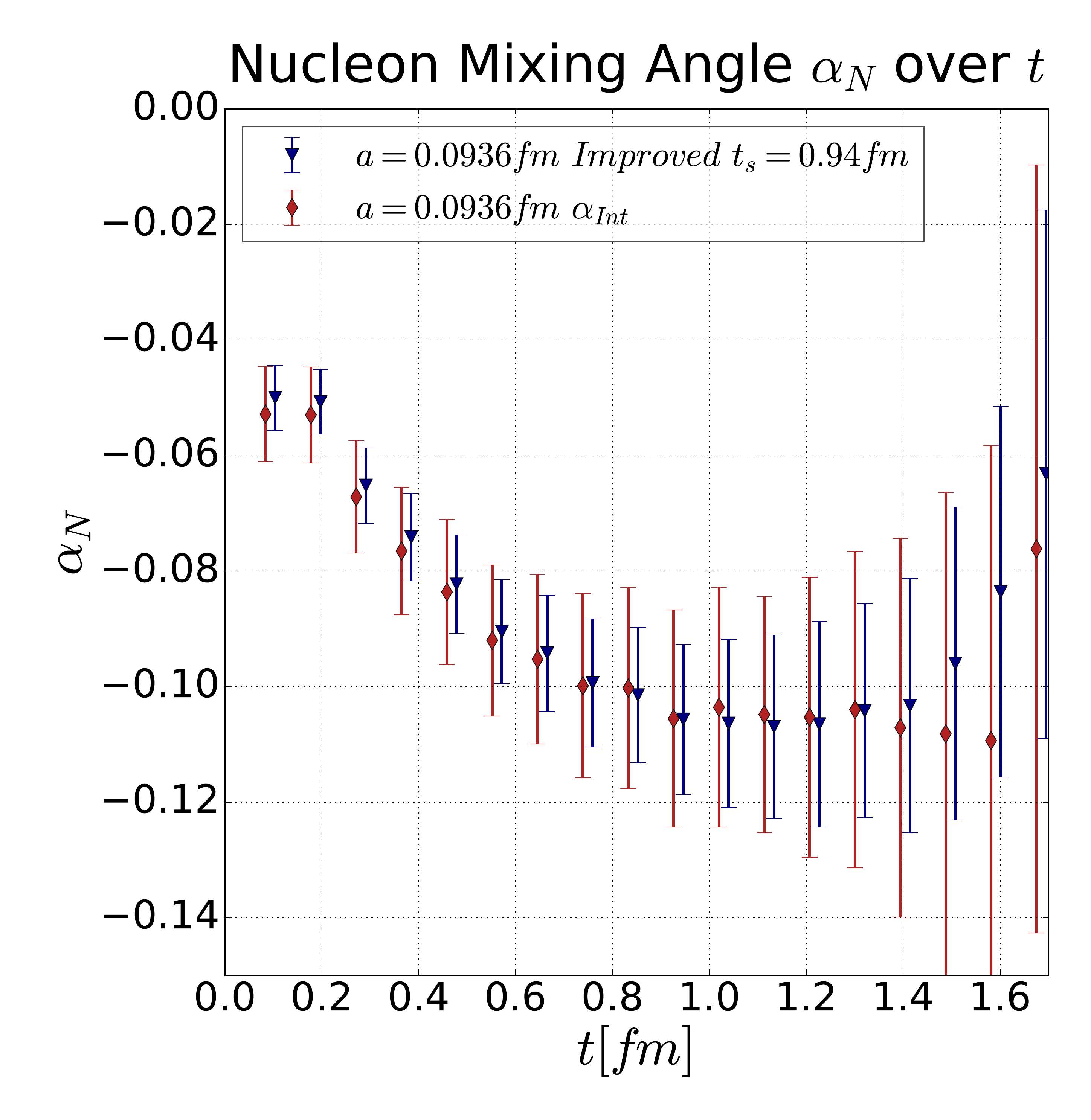}
  \caption{\label{fig:a_imp_L20_tsep}}
\end{subfigure}
\quad
\begin{subfigure}{.35\textwidth}
  \centering
  \includegraphics[trim={11mm 0cm 11mm 0cm},clip,width=\linewidth]{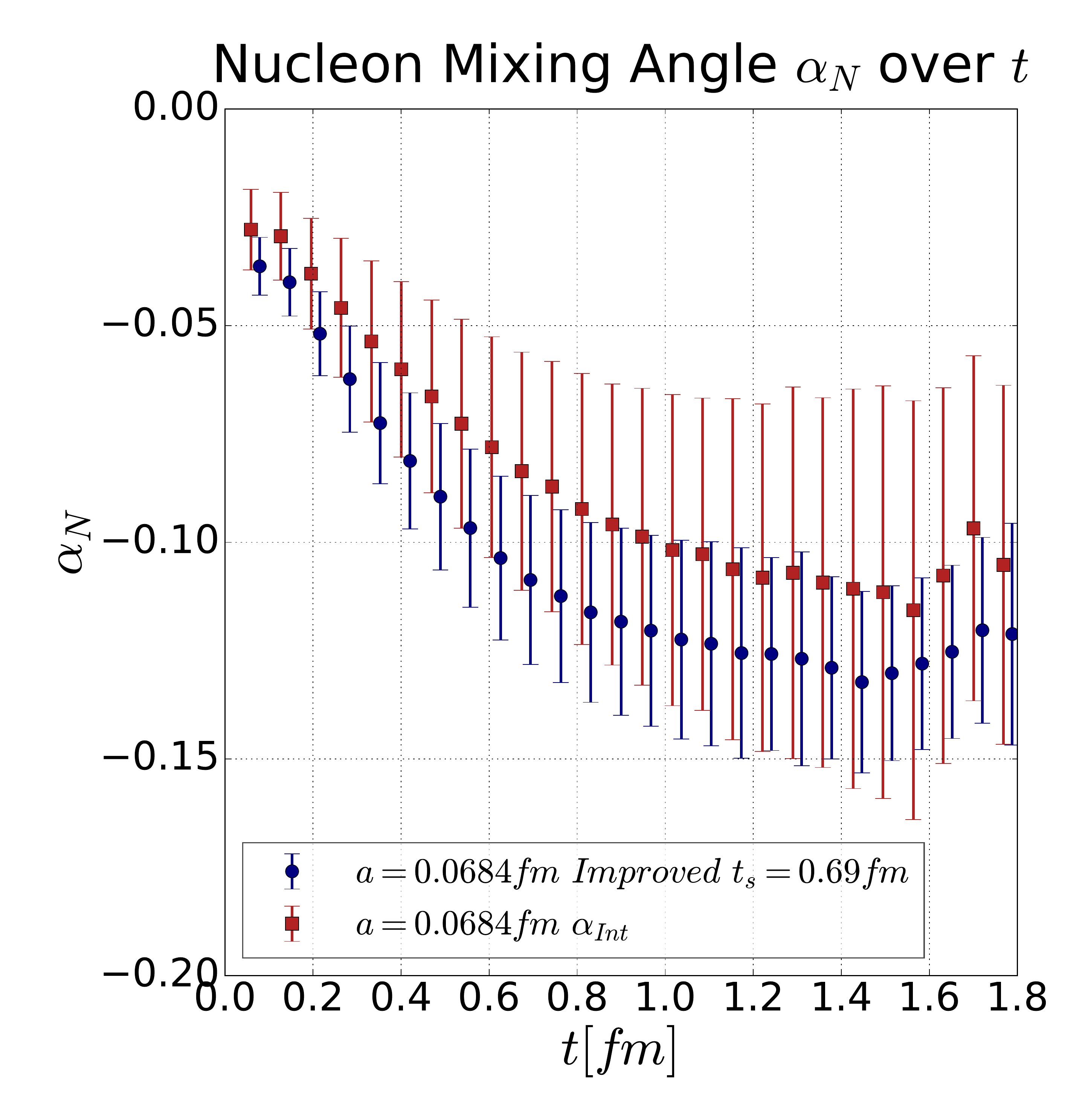}
  \caption{\label{fig:a_imp_L28_tsep}}
\end{subfigure}
\end{adjustwidth}
\caption{\label{fig:a_imp_tsep_latspace}
  \(\alpha_{N}\) against \(t\) plots for A-ensembles,
  comparing the improved method (blue) to the regular determination described in sec.~\ref{sec:twopt} (red).}
\end{figure}

To compare the improved extraction of \(\alpha_{N}\) to the standard determination described in
 sec.~\ref{sec:twopt},
we show in figs.~\ref{fig:a_imp_tsep_mpi},~\ref{fig:a_imp_tsep_latspace} the standard and improved
determination of \(\alpha_{N}\) as a function of the Euclidean source-sink separation \(t\).
The values of \(t_{s}\) considered are summarized in tab.~\ref{tab:alpha_imp_fitr}. We observe
a signal-to-noise improvement in all our ensembles, up to a factor $2$,
with the most significant observed in the ensembles M\(_{1}\), M\(_{2}\) and A\(_{3}\).
We observe the largest discrepancy between the improved and unimproved methods,
of the order of $2.3~\sigma$, in the M\(_{1}\) and M\(_{3}\) ensembles.
We attribute this discrepancy to standard statistical fluctuations of the gauge fields.
A summary of the fit ranges and results for the improved nucleon mixing angle is given in
tab.~\ref{tab:alpha_imp_fitr}, where for companions we added the values of \(\alpha_{N}\) determined in the standard way.

\begin{table}
\caption[Caption for LOF]{The selected starting value $t_{s}^{min}$ for the fit ranges
$[t_{s}^{min},T/2]$
for the summed \(\overbar{Q}\) , the Euclidean
source-sink separation fit range
$[t^{min},t^{max}]$ and the resulting nucleon mixing angle \(\alpha_{N}\)
from the selected parameters.
A comparison between computing $\alpha_{N}$ from $t_{s}^{min}$ and
fitting from $t_{s}^{min}$ onwards showed a negligible difference on all
ensembles.
We add the values determined in a standard way for comparison.
\label{tab:alpha_imp_fitr}}
  \centering
  \begin{tabular}{ r|c|c|c}
  ensemble &  M\(_{3}\) &M\(_{2}\)&M\(_{1}\)  \\
  \hline
  \(t_{s}^{min}/a\) & 7 & 10 & 6 \\
  \hline
  \(t_{s}^{min}\) [fm] & 0.63 & 0.9 & 0.54 \\
  \hline
  \(t/a\) fit range & [10,20] & [10,20] & [10,20]  \\
  \hline
  \(t\) [fm] fit range & [0.9,1.8] & [0.9,1.8] & [0.9,1.8]  \\
  \hline
  \(\alpha_{N}\) improved & -0.098(13) & -0.201(17) & -0.0822(97)  \\
  \hline
  \(\alpha_{N}\) standard & -0.040(21) & -0.190(27) & -0.142(24)  \\
  \\
  ensemble  & A\(_{3}\) &A\(_{2}\)&A\(_{1}\)  \\
  \hline
  \(t_{s}^{min}/a\) &  10 & 10 & 10 \\
  \hline
  \(t_{s}^{min}\) [fm] &  1.21 & 0.98 & 0.69  \\
  \hline
  \(t/a\) fit range &  [5,11] & [7,17] & [14,21]  \\
  \hline
  \(t\) [fm] fit range &  [0.61,1.34] & [0.69,1.67] & [0.96,1.44]  \\
  \hline
  \(\alpha_{N}\) improved &  -0.1016(89) & -0.1012(75) &  -0.1212(67) \\
  \hline
  \(\alpha_{N}\) standard & -0.099(11) & -0.103(10) & -0.105(11)  \\
  \end{tabular}
\end{table}

\section{Electric Dipole Moment Results \label{sec:F3_res}}

The neutron (n) and proton (p) EDMs, \(d_{p/n}\), can be extracted from the CP-odd electric dipole form factor\footnote{The general form not requiring \(Q^{2}\ll m_\pi^2\),
  is given in eq.~\eqref{eq:EDFF_fit} and discussed in detail in secs.~\ref{sec:FF_Rat_imp}
  and~\ref{sec:Schiff_res}. We performed the same analysis with the fit function in eq.~\eqref{eq:EDFF_fit}
  and found insignificant changes to the EDM results (see secs.~\ref{sec:FF_Rat_imp} and~\ref{sec:Schiff_res}).}
\begin{equation}\label{eq:F3TOEDM}
\frac{F^{p/n}_{3}(Q^{2})}{2M_{N}} \xrightarrow{Q^{2}\ll m_\pi^2} d_{p/n} - S_{p/n} Q^{2} + O(Q^{4})\,,
\end{equation}
which requires a lattice QCD computation of \(F_{3}(Q^2)\).
The variable $Q^2$ in this case refers to the momentum transfer and should not be confused with the topological charge.
The small \(\bar \theta\) expansion provides us a way of accessing \(F_{3}\) from three-point correlation functions without
the need for generating new gauge configurations at finite \(\bar{\theta}\) and without relying on a
problematic analytical continuation to imaginary \(\bar{\theta}\). To access \(F_{3}(Q^{2})\), we calculate the
following three-point correlation functions with and without the insertion of the topological charge, respectively,
\begin{eqnarray}\label{eq:C3Q}
  G_{3}^{(Q)}(\bm{p}^{\, \prime},t,\bm{q},\tau,\Pi,\gamma_{\mu},t_{f}) & =&
  a^{6}\sum_{\bm{x},\bm{y}}
  e^{-i ( \bm{p}^{\, \prime}  \cdot \bm{x}-\bm{q}\cdot \bm{y})}
  \mathrm{Tr}\left\lbrace
  \Pi
  \braket{
  \mathcal{N}(\bm{x},t)
  \mathcal{J}_{\mu}(\bm{y},\tau)
  \overbar{\mathcal{N}}(\bm{0},0)
  Q(t_{f})
  }
  \right\rbrace,\nonumber\\
  G_{3}(\bm{p}^{\, \prime},t,\bm{q},\tau,\Pi,\gamma_{\mu})& =&
  a^{6}\sum_{\bm{x},\bm{y}}
  e^{-i ( \bm{p}^{\, \prime}  \cdot \bm{x}-\bm{q}\cdot \bm{y})}
  \mathrm{Tr}\left\lbrace
  \Pi
  \braket{
  \mathcal{N}(\bm{x},t)
  \mathcal{J}_{\mu}(\bm{y},\tau)
  \overbar{\mathcal{N}}(\bm{0},0)
  }
  \right\rbrace\,,
\end{eqnarray}
where the electromagnetic current in terms of the quark currents is given by
\be
  \mathcal{J}_{\mu}(\bm{y},\tau) =
  \frac{4}{3} \overbar{u}(\bm{y},\tau)\gamma_{\mu}u(\bm{y},\tau) -
  \frac{1}{3} \overbar{d}(\bm{y},\tau)\gamma_{\mu}d(\bm{y},\tau)\,, 
\ee
and \(\mathcal{N}\) denotes standard proton or neutron interpolating fields.

Once the three-point correlation functions are computed,
we remove the leading Euclidean time dependence and
nucleon-to-vacuum amplitude contributions via the ratios:
\begin{eqnarray}\label{eq:RatFun}
  R(\bm{p}^{\, \prime},t,\bm{q},\tau,\Pi,\gamma_{\mu}) &=
  \frac{
  G_{3}(\bm{p}^{\, \prime},t,\bm{q},\tau,\Pi,\gamma_{\mu})
  }{
  G_{2}(\bm{p}^{\, \prime},t,\Pi_{+})
  }
  K(\bm{p}^{\, \prime},t,\bm{q},\tau)\,, \nonumber \\
  R^{(Q)}(\bm{p}^{\, \prime},t,\bm{q},\tau,\Pi,\gamma_{\mu},t_{f}) &=
  \frac{
  G^{(Q)}_{3}(\bm{p}^{\, \prime},t,\bm{q},\tau,\Pi,\gamma_{\mu},t_{f})
  }{
  G_{2}(\bm{p}^{\, \prime},t,\Pi_{+})
  }
  K(\bm{p}^{\, \prime},t,\bm{q},\tau)\,,
\end{eqnarray}
where we have implicitly defined ratios for the proton and the neutron.
We define the square-root factor as
\begin{eqnarray}
  K(\bm{p}^{\, \prime},t,\bm{p},\tau) \equiv
  \sqrt{
  \frac{
  G_{2}(\bm{p}^{\, \prime},\tau,\Pi_{+})
  G_{2}(\bm{p}^{\, \prime},t,\Pi_{+})
  G_{2}(\bm{p},t-\tau,\Pi_{+})
  }{
  G_{2}(\bm{p},\tau,\Pi_{+})
  G_{2}(\bm{p},t,\Pi_{+})
  G_{2}(\bm{p}^{\, \prime},t-\tau,\Pi_{+})
  }
  }.
\end{eqnarray}
The spectral decomposition of the ratio function $R$
in eq.~\eqref{eq:RatFun}, in the limit \(T \gg t \gg 0\), reads
\begin{eqnarray}\label{eq:RatFun_specdecomp}
  R(\bm{p}^{\, \prime},t,\bm{q},\tau,\Pi,\gamma_{\mu}) =
  A(E_{\bm{p}^{\, \prime}},E_{\bm{p}})
  \text{Tr}\left\lbrace
  \Pi
  (-i\slashed{p}^{\, \prime} + m)
  \Gamma_{\mu}(Q^{2})
  (-i\slashed{p} + m)
  \right\rbrace,
\end{eqnarray}
where the vector form factor contains all terms allowed by the symmetries of the theory
\begin{eqnarray}
  \Gamma_{\mu}(Q^{2}) = \gamma_{\mu}F_{1}(Q^{2}) +
  \frac{\sigma_{\mu \nu}q_{\nu}}{2m}F_{2}(Q^{2})\,.
\end{eqnarray}
For completeness the expression of \(A(E_{\bm{p}^{\, \prime}},E_{\bm{p}})\) reads
\begin{eqnarray}
  A(E_{\bm{p}^{\, \prime}},E_{\bm{p}}) = \frac{1}{4\sqrt{E_{\bm{p}^{\, \prime}}E_{\bm{p}}(E_{\bm{p}^{\, \prime}}+m)(E_{\bm{p}}+m)}}.
\end{eqnarray}

The data that we computed coming from the fixed-sink method
is such that \(\bm{p}^{\, \prime}=\bm{0}\) (which implies
\(\bm{q} = -\bm{p}\)), simplifying eq.~\eqref{eq:RatFun_specdecomp} to
\begin{eqnarray}
  R(\bm{0},t,\bm{q},\tau,\Pi,\gamma_{\mu}) =
  2mA(m,E_{\bm{p}})
  \text{Tr}\left\lbrace
  \Pi \Pi_{+}
  \Gamma_{\mu}(Q^{2})
  (-i\slashed{p} + m)
  \right\rbrace.
\end{eqnarray}
The analogous modified ratio function with the insertion of the topological charge \(R^{(Q)}\) in
eq.~\eqref{eq:RatFun} has, retaining only the ground state contribution,
the following spectral decomposition at leading order in \(\bar{\theta}\)
\begin{eqnarray}
  &R^{(Q)}(\bm{0},t,\bm{q},\tau,\Pi,\gamma_{\mu},t_{f}) =
  2mA(m,E_{\bm{p}})
  \bigg[&
    \alpha_{N}2m\text{Tr}\left\lbrace \Pi \Pi_{+} \widetilde{\Gamma}_{\mu}(Q^{2})\gamma_{5}\right\rbrace +
    \nonumber \\
    &&\alpha_{N}\text{Tr}\left\lbrace \Pi \gamma_{5} \widetilde{\Gamma}_{\mu}(Q^{2})(-i\slashed{p} + m)
    \right\rbrace + \nonumber\\
    &&\text{Tr}\left\lbrace \Pi \Pi_{+} \frac{\sigma_{\mu \nu}\gamma_{5}q_{\nu}}{2m}
    \widetilde{F}_{3}(Q^{3})(-i\slashed{p} + m)\right\rbrace
  \bigg],
\end{eqnarray}
where
\begin{eqnarray}
  \widetilde{\Gamma}_{\mu}(Q^{2}) = \gamma_{\mu}F_{1}(Q^{2}) +
  \frac{\sigma_{\mu \nu}q_{\nu}}{2m}\widetilde{F}_{2}(Q^{2})\,.
\end{eqnarray}
Due to subtleties between lattice quantities and physical,
the form factor decomposition in presence of a CP-violating operator insertion,
is written in terms of modified form factors, $\widetilde{F}_{2}(Q^{2})$ and $\widetilde{F}_{3}(Q^{2})$,
related to the physical form factors by~\cite{Abramczyk:2017oxr}
\begin{eqnarray} \label{eq:F3_rot}
  F_{3}(Q^{2}) =& \cos(2\alpha_{N})\widetilde{F}_{3}(Q^{2}) + \sin(2\alpha_{N})\widetilde{F}_{2}(Q^{2})\,, \\
  F_{2}(Q^{2}) =& -\sin(2\alpha_{N})\widetilde{F}_{3}(Q^{2}) + \cos(2\alpha_{N})\widetilde{F}_{2}(Q^{2})\,.
\end{eqnarray}
The rotated form factor $ F_{3}(Q^{2}) $ corresponds to the actual electric dipole
form factor as measured in experiments. From now on, we will focus on this quantity.
 
The ratio functions \(R\) and \(R^{(Q)}\) become constant, as long as the large-time approximation \(T\gg t \gg \tau \gg 0\)
is satisfied to ensure ground-state dominance. As the fixed-sink method is employed to compute the three-point
correlation functions, a region in which this large time approximation is satisfied for \(\tau\) can be found and we
denote the results of the fits as
\begin{eqnarray}\label{eq:Rat_fit}
  R(\bm{p}^{\, \prime},t,\bm{q},\tau,\Pi,\gamma_{\mu}) &\rightarrow R_{fit}(\bm{p}^{\, \prime},t,\bm{q},\Pi,\gamma_{\mu})\,, \nonumber \\
  R^{(Q)}(\bm{p}^{\, \prime},t,\bm{q},\tau,\Pi,\gamma_{\mu},t_{f}) &\rightarrow R^{(Q)}_{fit}(\bm{p}^{\, \prime},t,\bm{q},\Pi,\gamma_{\mu},t_{f})\,.
\end{eqnarray}
The technique for fitting these ratio functions over \(\tau\) is described in app.~\ref{app:fitr_select}.
With this construction, a system of equations can be solved for form factors \(F_{i}(Q^{2})\) , \(i=1,2,3\) of the form:
\begin{equation}
  \sum_{i=1}^{3}
  \mathcal{A}(Q^{2})_{Ai} F_{i}(Q^{2}) =
  \left\{
	\begin{array}{ll}
    &R_{fit}(\bm{0},t,\bm{q}_{j},\Pi_{k},\gamma_{l}) \\
    &R^{(Q)}_{fit}(\bm{0},t,\bm{q}_{j},\Pi_{k},\gamma_{l},t_{f}) \\
	\end{array}
  \right.,
  \label{eq:FF_system}
\end{equation}
where the collective index \(A\)  denotes any combination of the indices \(A=\lbrace j,k,l\rbrace\).
In other words, we run over all possible combinations of projectors \(\Pi\), all current momentum \(\bm{q}\) within
a given \(Q^{2}\), and operator gamma matrix \(\gamma_{\mu}\). The index \(A\) of the matrix \(\mathcal{A}_{Ai}(Q^{2})\)
corresponds to the coefficients for each form factor \(F_{i}\) for the corresponding ratio function \(R\) or \(R^{(Q)}\).
These coefficients are found by analyzing the spectral decomposition of \(R\) or \(R^{(Q)}\), which needs to
be done for every evaluated index \(A\).

Using eq.~\eqref{eq:F3TOEDM}, we extrapolate to
\(F^{p/n}_{3}(Q^{2}\rightarrow 0)/(2M_{N})=d_{p/n}\).
We use a linear plus constant fit function, giving
the extrapolated value \(d_{p/n}\) at \(Q^{2}\rightarrow 0\)
(as well as slope in \(Q^{2}\) providing \(S_{p/n}\)).

\begin{figure}
\vspace*{-2cm}
\begin{adjustwidth}{-0.1\textwidth}{-0.1\textwidth}
\centering
\begin{subfigure}{.55\textwidth}
  \centering
  \includegraphics[trim={11mm 0cm 11mm 0cm},clip,width=\linewidth]{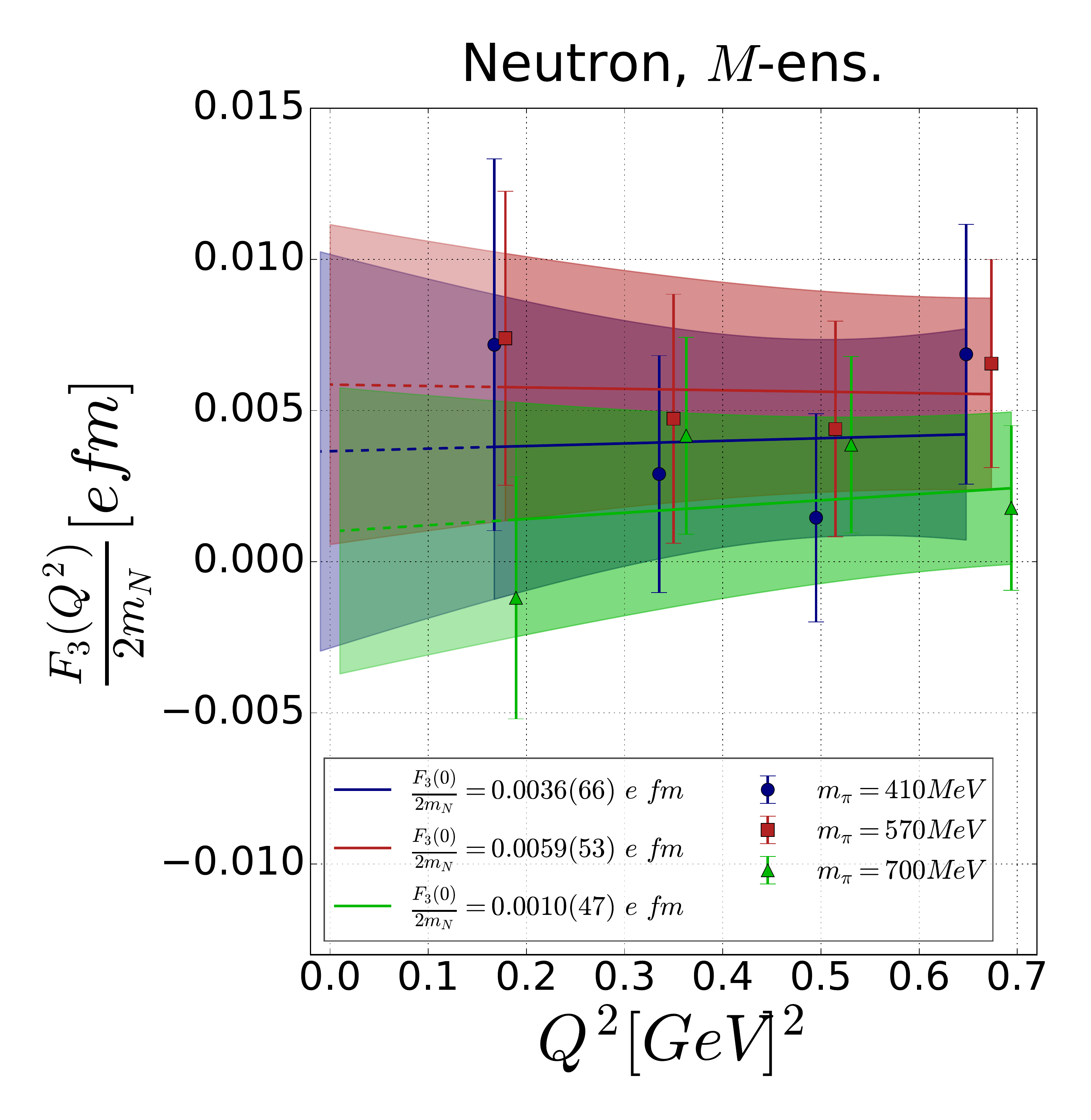}
  \caption{\label{fig:F3_mpi_neutron}}
\end{subfigure}
\quad
\begin{subfigure}{.55\textwidth}
  \centering
  \includegraphics[trim={11mm 0cm 11mm 0cm},clip,width=\linewidth]{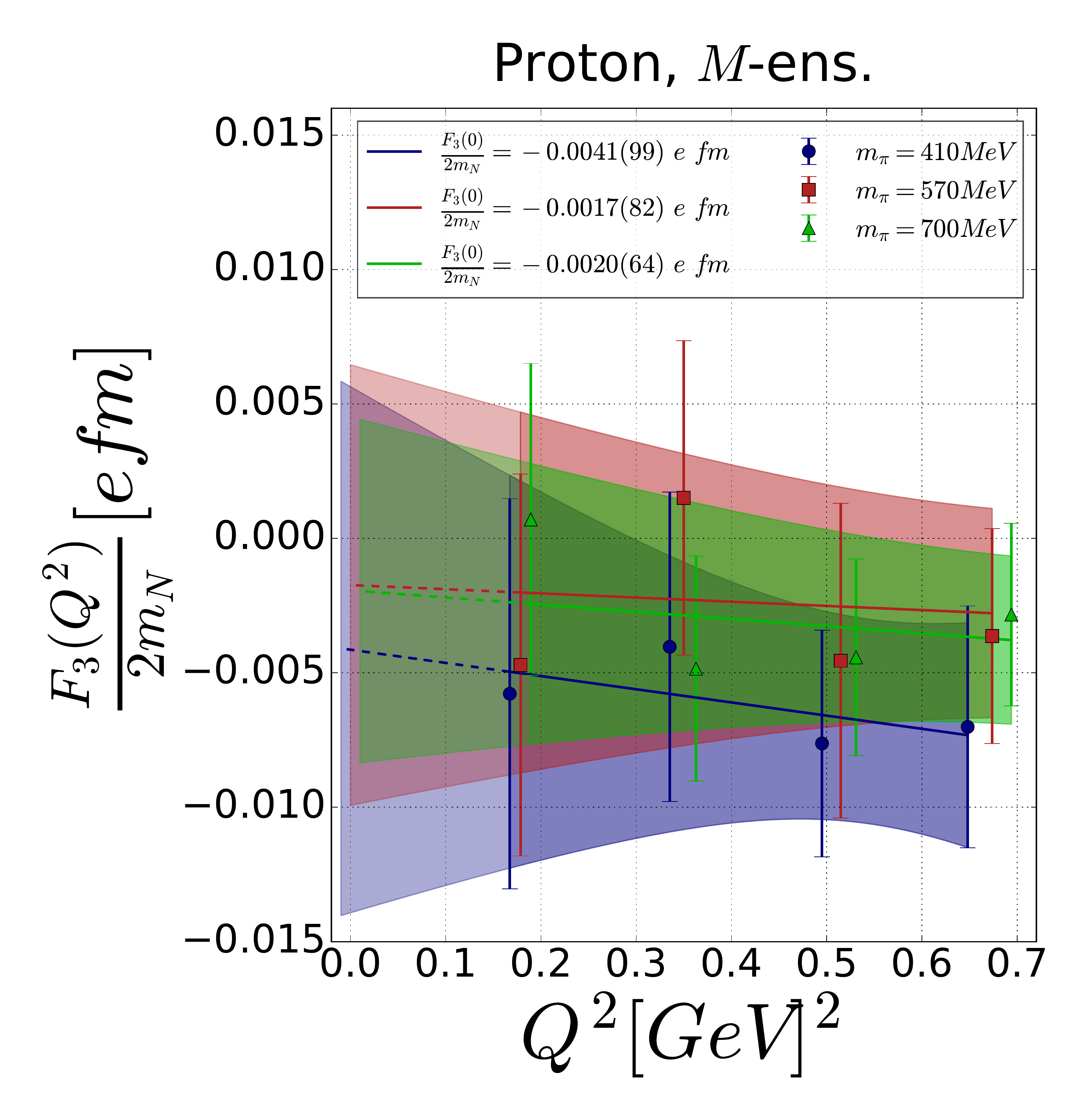}
  \caption{\label{fig:F3_mpi_proton}}
\end{subfigure}
\end{adjustwidth}
\caption{
  M-ensemble results for the neutron (left) and proton (right)
  CP-odd form factor \(\frac{F_{3}(Q^{2})}{2M_{N}}\),
  plotted against the transfer momentum \(Q^{2}\).
  The extrapolation to \(Q^{2}\rightarrow 0\) gives the final EDMs which
  are displayed in tab.~\ref{tab:F3_Q_ext_mpi}.
  \label{fig:F3_mpi}}
\begin{adjustwidth}{-0.1\textwidth}{-0.1\textwidth}
\centering
\begin{subfigure}{.55\textwidth}
  \centering
  \includegraphics[trim={11mm 0cm 11mm 0cm},clip,width=\linewidth]{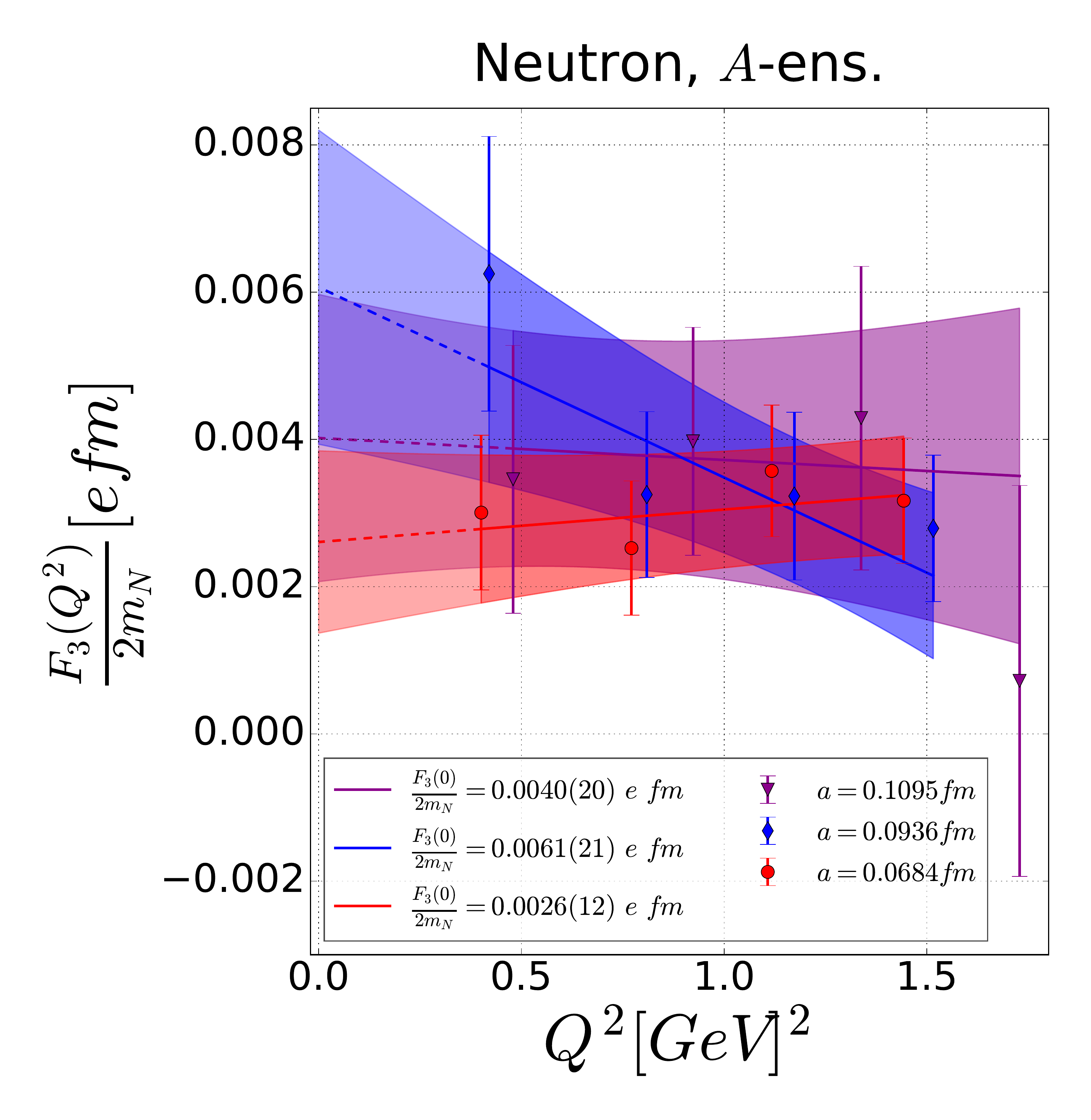}
  \caption{\label{fig:F3_latspace_neutron}}
\end{subfigure}
\quad
\begin{subfigure}{.55\textwidth}
  \centering
  \includegraphics[trim={11mm 0cm 11mm 0cm},clip,width=\linewidth]{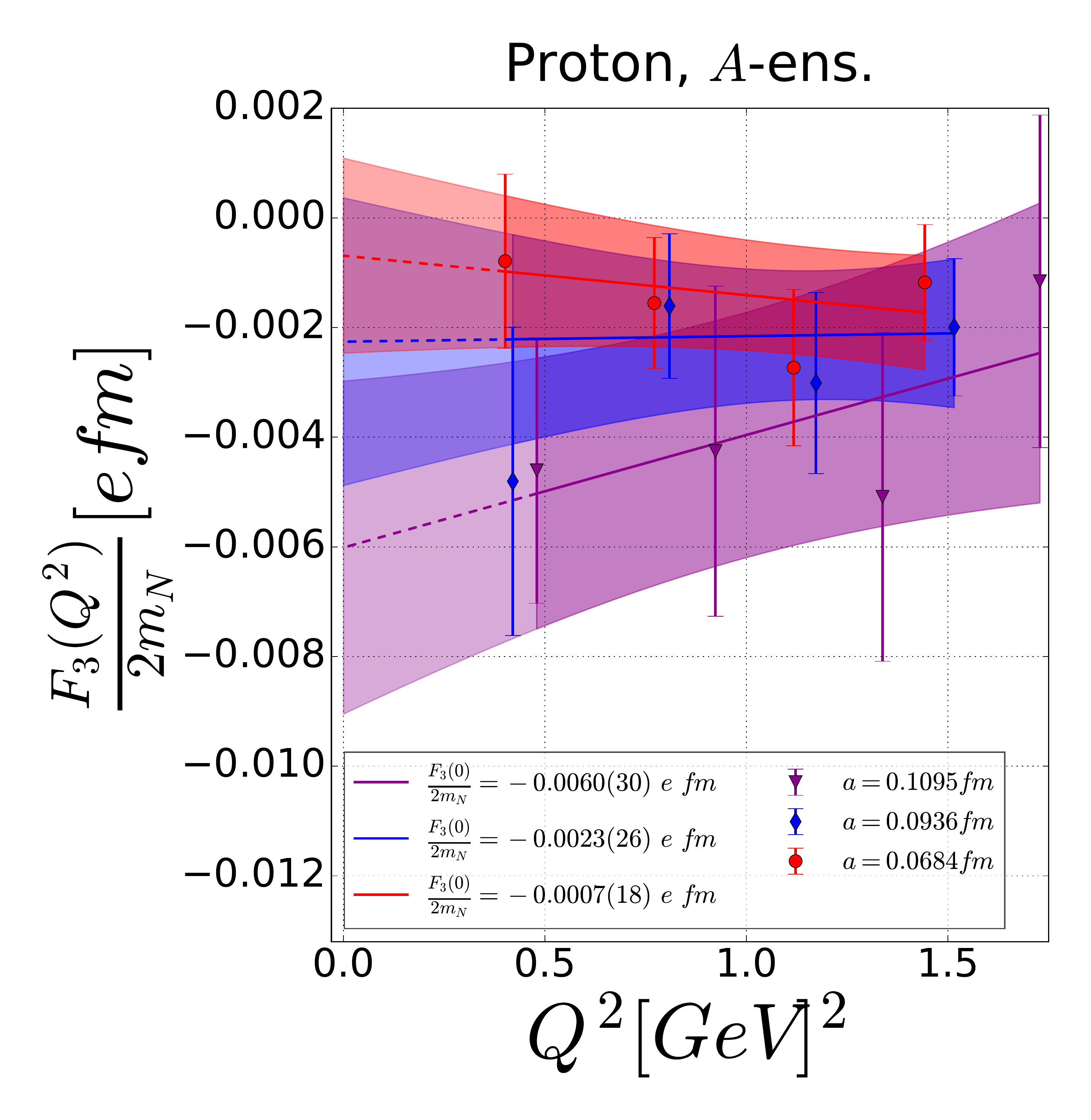}
  \caption{\label{fig:F3_latspace_proton}}
\end{subfigure}
\end{adjustwidth}
\caption{
  A-ensemble results for the neutron (left) and proton (right)
  CP-odd form factor \(\frac{F_{3}(Q^{2})}{2M_{N}}\),
  plotted against the transfer momentum \(Q^{2}\).
  The extrapolation to \(Q^{2}\rightarrow 0\) gives the final EDMs which are
  displayed in tab.~\ref{tab:F3_Q_ext_latspace}.
  \label{fig:F3_latspace}}
\end{figure}

The final extraction of the neutron (left) and proton (right) CP-odd form factor \(\frac{F_{3}(Q^{2})}{2M_{N}}\)
is shown for the M-ensembles in fig.~\ref{fig:F3_mpi} and for the A-ensembles
in fig.~\ref{fig:F3_latspace}.
Fig.~\ref{fig:F3_mpi} shows that all M-ensembles are statistically consistent evaluated, and with zero. Fig.~\ref{fig:F3_latspace} shows that there are no major discretization effects,
as all the extrapolated \(Q^{2}\rightarrow 0\) results are  consistent.

The following figs.~\ref{fig:F3_tflow},~\ref{fig:F3_alpha_fitr},~\ref{fig:F3_alphaimp}
are all displayed to understand
the systematic effects resulting from varying the flow time \(t_{f}\), and different methods of determining
the nucleon mixing angle \(\alpha_{N}\) used in the form factor decomposition
(\(\mathcal{A}_{Aj}(Q^{2})\) in eq.~\eqref{eq:FF_system}).
In fig.~\ref{fig:F3_tflow}, for example, we show how the form factors $F_3$, determined at
different flow-time radii \({\sqrt{8t_{f}}=0.60,0.65,0.70}\)~fm (green, red and blue),
are statistically consistent for all three M-ensembles (left to right).
From both fig.~\ref{fig:F3_alphaimp}, where the improved method (see sec.~\ref{sec:alpha_imp})
of determining the nucleon mixing angle $\alpha_{N}$ (in red)
is compared to the standard method for $\alpha_{N}$ (in blue),
and fig.~\ref{fig:F3_alpha_fitr}, where we vary the
fit range for extracting $\alpha_{N}$,
it is clear that a more precise determination of $\alpha_{N}$ has
a negligible impact on improving the precision of the results
for the CP-odd form factor \(F_{3}\).
A summary of the \(Q^{2}\rightarrow 0\) extrapolations for different ensembles is  given in
tabs.~\ref{tab:F3_Q_ext_mpi},~\ref{tab:F3_Q_ext_latspace}.

\begin{table}
\caption{\(\frac{F^{p/n}_{3}(Q^{2}\rightarrow 0)}{2M_{N}}=d_{p/n}\) fit results over M-ensembles,
taken from fig.~\ref{fig:F3_mpi}.
\label{tab:F3_Q_ext_mpi}}
\centering
  \begin{tabular}{ r|c|c|c }
  ensemble &  \(m_{\pi}=410\) MeV &\(m_{\pi}=570\) MeV&\(m_{\pi}=700\) MeV \\
  \hline
  $d_{p}$ [$e$~fm] & 0.0043(99) & 0.0017(83) & 0.0016(64) \\
  \hline
  $d_{n}$ [$e$~fm] & -0.0035(66) & -0.0060(53) &-0.0009(47)\\
  \end{tabular}
\end{table}
\begin{table}
\caption{\(\frac{F^{p/n}_{3}(Q^{2}\rightarrow 0)}{2M_{N}}=d_{p/n}\) fit results over A-ensembles, taken from fig.~\ref{fig:F3_latspace}.
\label{tab:F3_Q_ext_latspace}}
\centering
  \begin{tabular}{ r|c|c|c }
  ensemble &  \(a=0.1095\) fm  &\(a=0.0936\) fm &\(a=0.0684\) fm  \\
  \hline
  $d_{p}$ [$e$~fm] & 0.0060(30) & 0.0026(25) & 0.0008(18) \\
  \hline
  $d_{n}$ [$e$~fm] & -0.0043(20) & -0.0063(20) & -0.0023(13) \\
  \end{tabular}
\end{table}

\begin{figure}
\begin{adjustwidth}{-0.11\textwidth}{-0.11\textwidth}
\centering
\begin{subfigure}{.35\textwidth}
  \centering
  \includegraphics[trim={11mm 0cm 11mm 0cm},clip,width=\linewidth]{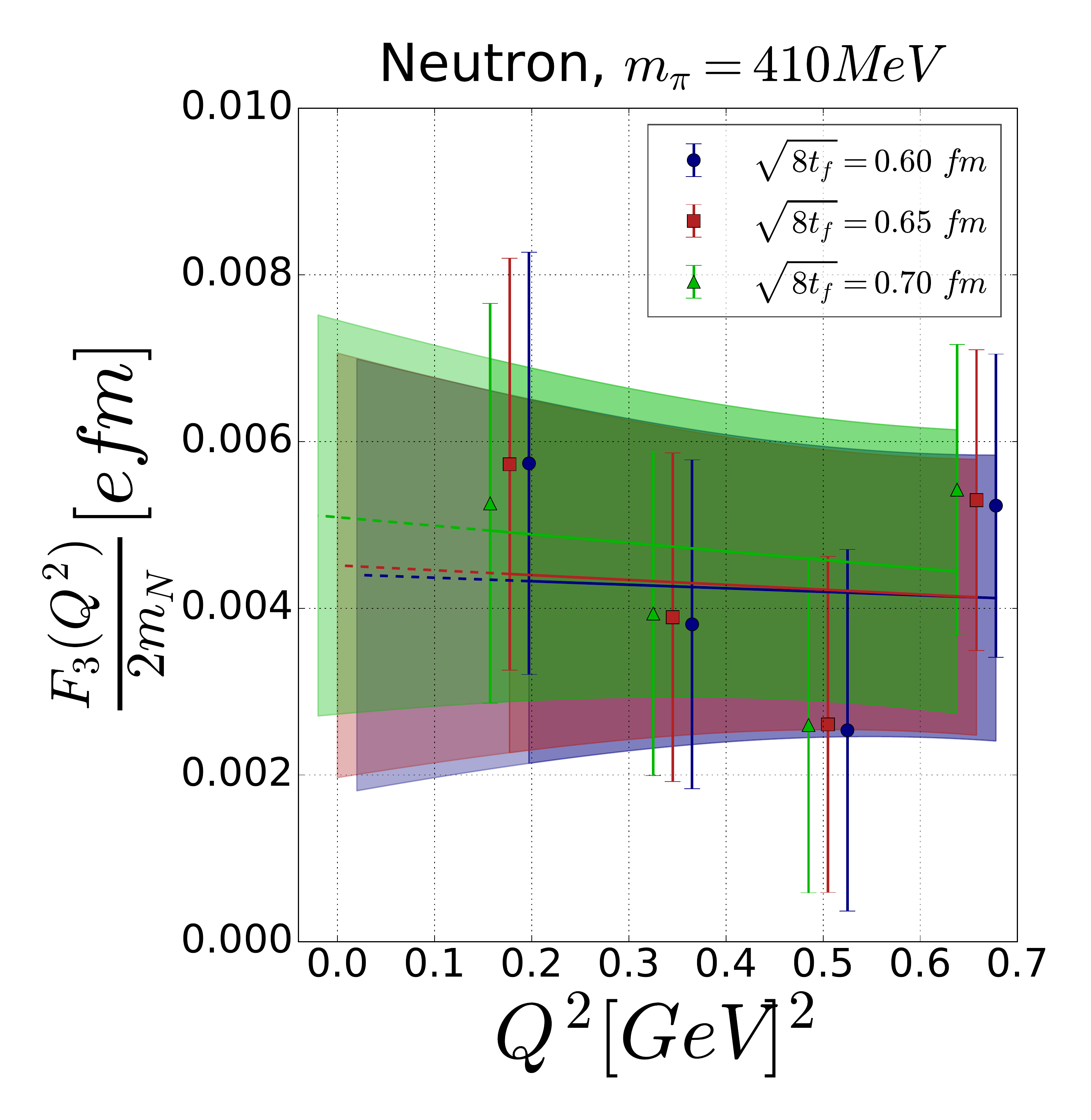}
  \caption{\label{fig:F3_tflow_mpi410}}
\end{subfigure}
\begin{subfigure}{.35\textwidth}
  \centering
  \includegraphics[trim={11mm 0cm 11mm 0cm},clip,width=\linewidth]{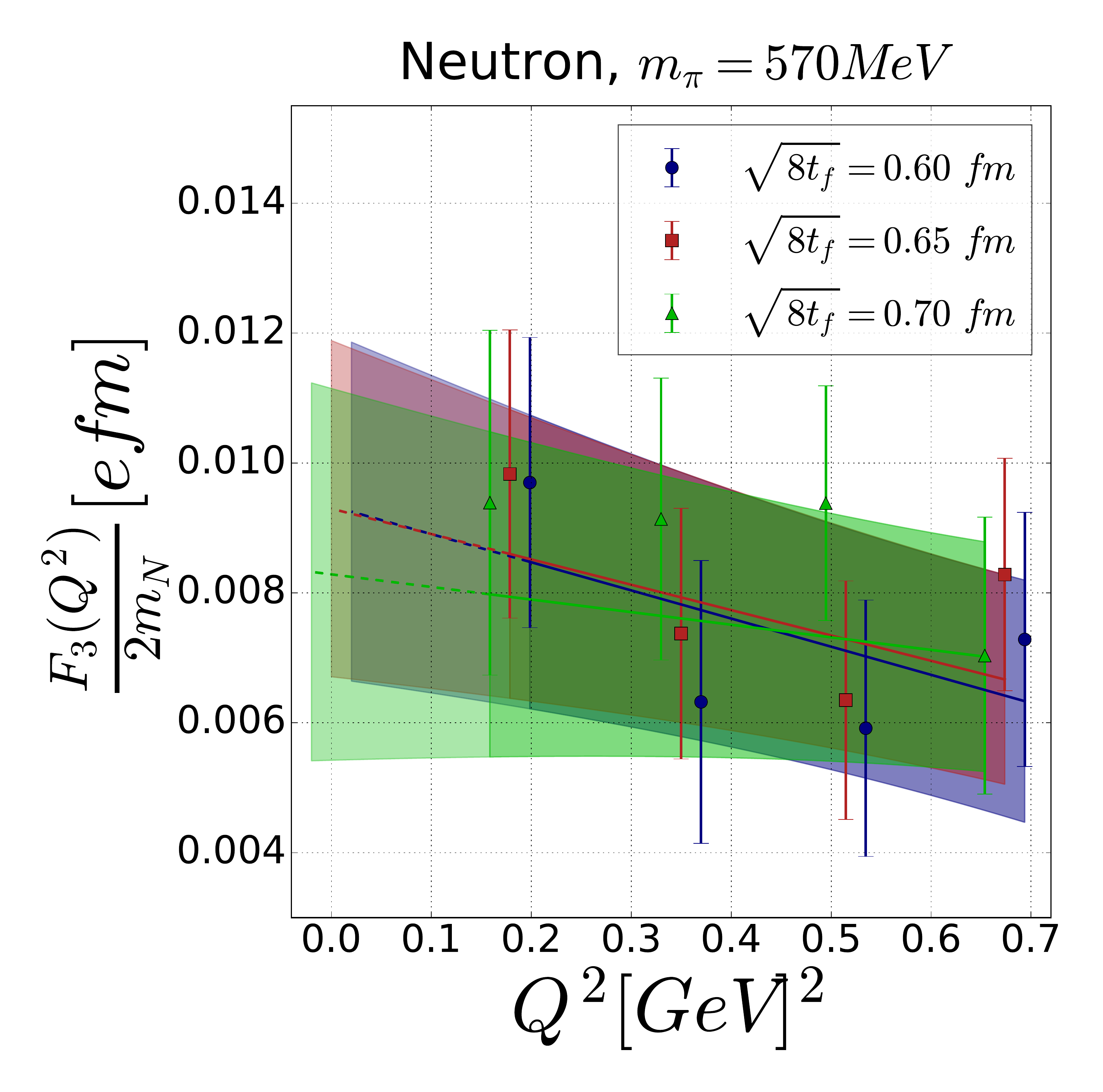}
  \caption{\label{fig:F3_tflow_mpi570}}
\end{subfigure}
\begin{subfigure}{.35\textwidth}
  \centering
  \includegraphics[trim={11mm 0cm 11mm 0cm},clip,width=\linewidth]{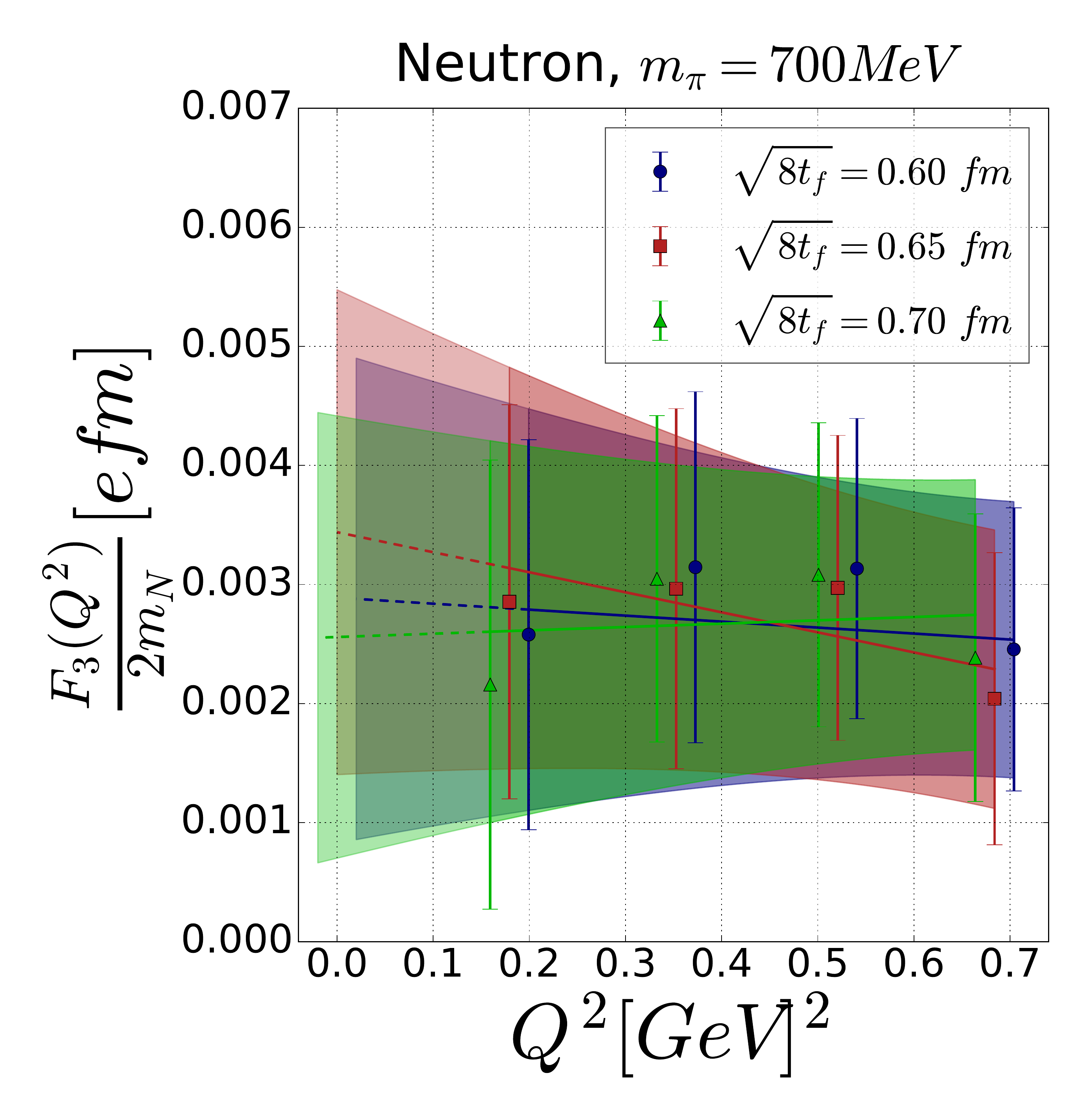}
  \caption{\label{fig:F3_tflow_mpi700}}
\end{subfigure}
\end{adjustwidth}
\caption{
  Flow time radii \(\sqrt{8t_{f}}=0.60,0.65,0.70\) fm (green, red, blue respectively) comparison for
  the neutron CP-odd form factor \(\frac{F_{3}(Q^{2})}{2M_{N}}\)
  using the \(m_{\pi}=\lbrace 410,570,700\rbrace\) MeV (left, middle and right) ensembles.
  The extrapolation to \(Q^{2}\rightarrow 0\) gives the final EDM.
  \label{fig:F3_tflow}}
\end{figure}

\begin{figure}
\begin{adjustwidth}{-0.11\textwidth}{-0.11\textwidth}
\centering
\begin{subfigure}{.35\textwidth}
  \centering
  \includegraphics[trim={11mm 0cm 11mm 0cm},clip,width=\linewidth]{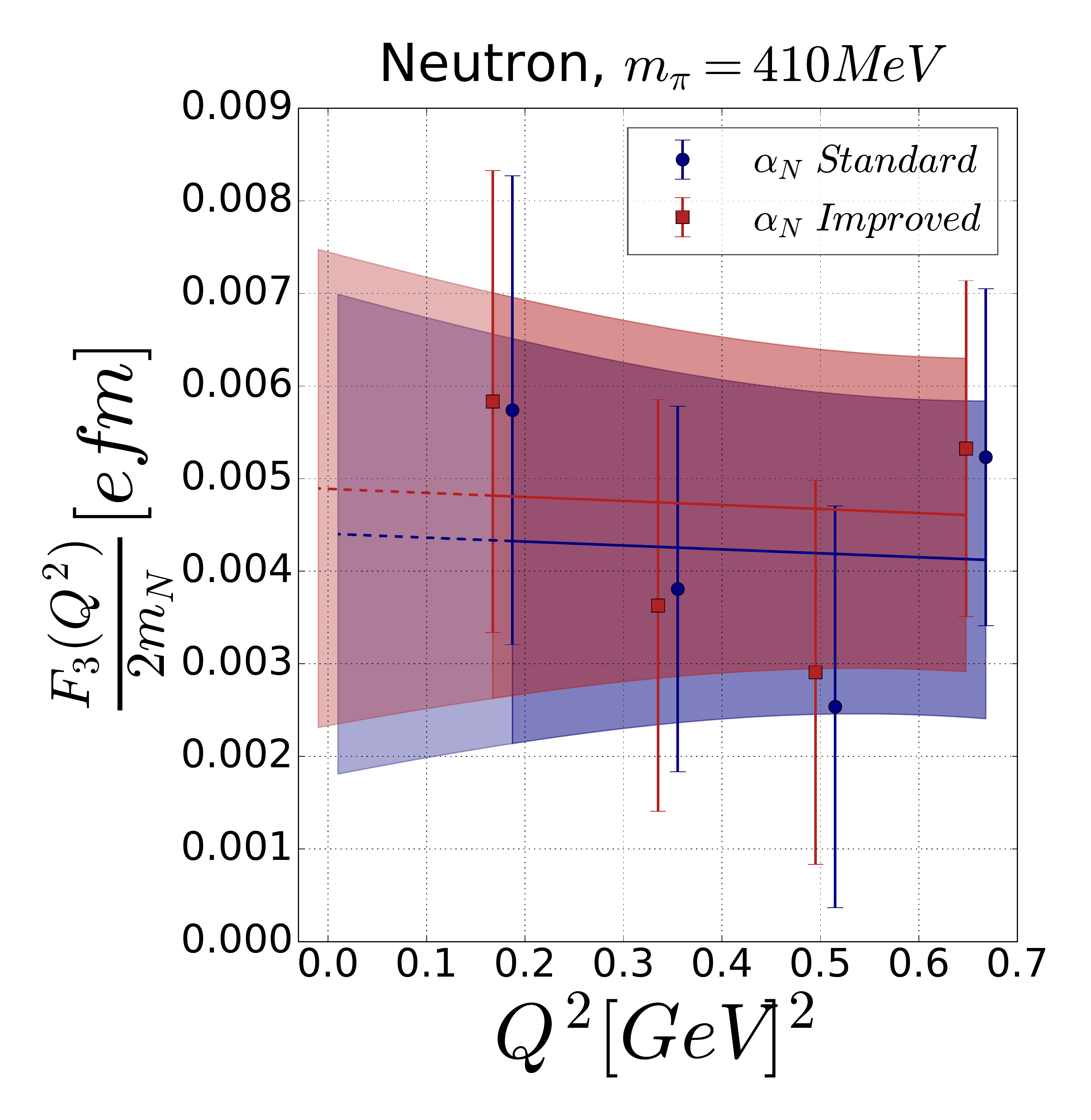}
  \caption{\label{fig:F3_alphaimp_mpi410}}
\end{subfigure}
\begin{subfigure}{.35\textwidth}
  \centering
  \includegraphics[trim={11mm 0cm 11mm 0cm},clip,width=\linewidth]{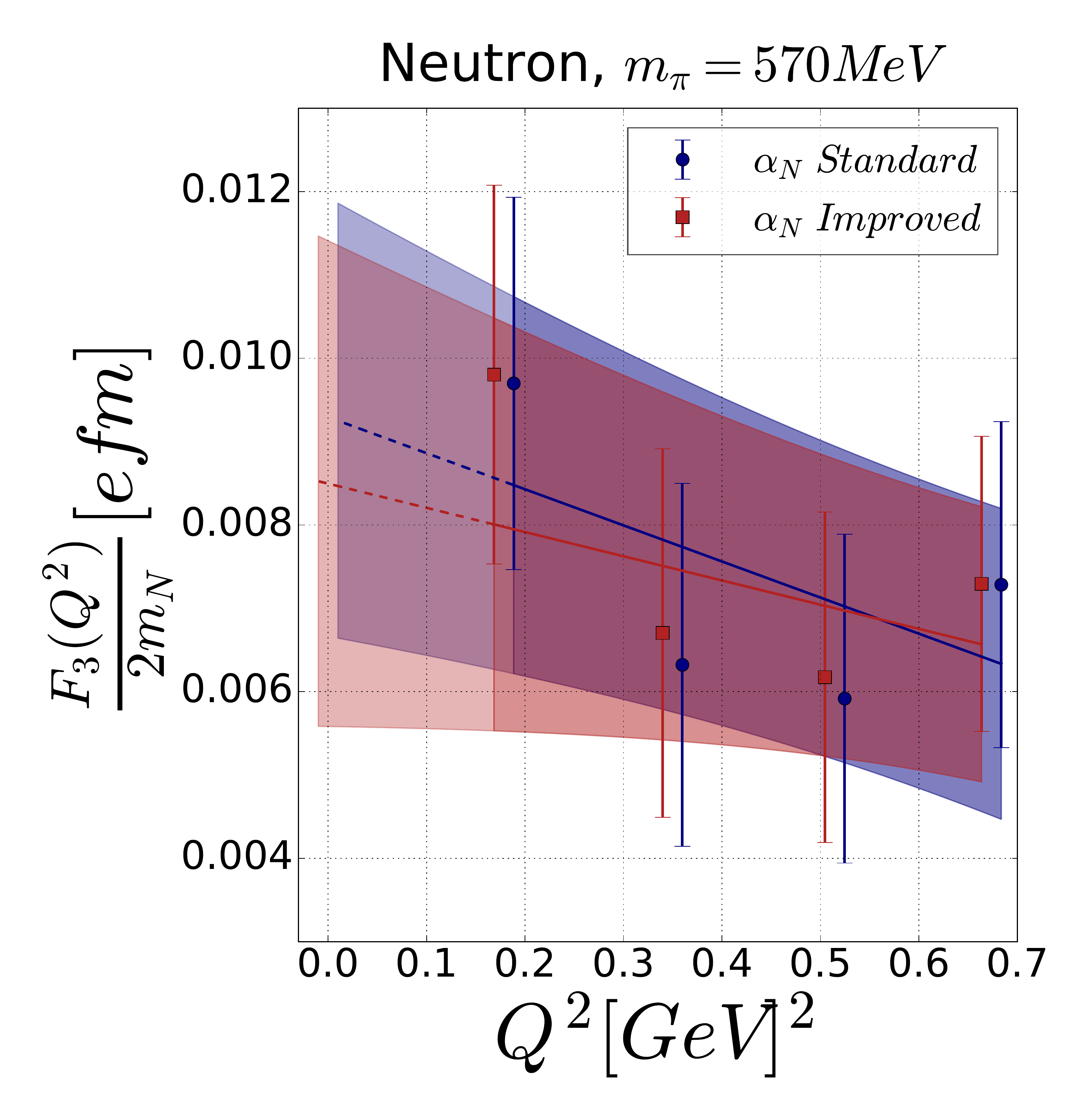}
  \caption{\label{fig:F3_alphaimp_mpi570}}
\end{subfigure}
\begin{subfigure}{.35\textwidth}
  \centering
  \includegraphics[trim={11mm 0cm 11mm 0cm},clip,width=\linewidth]{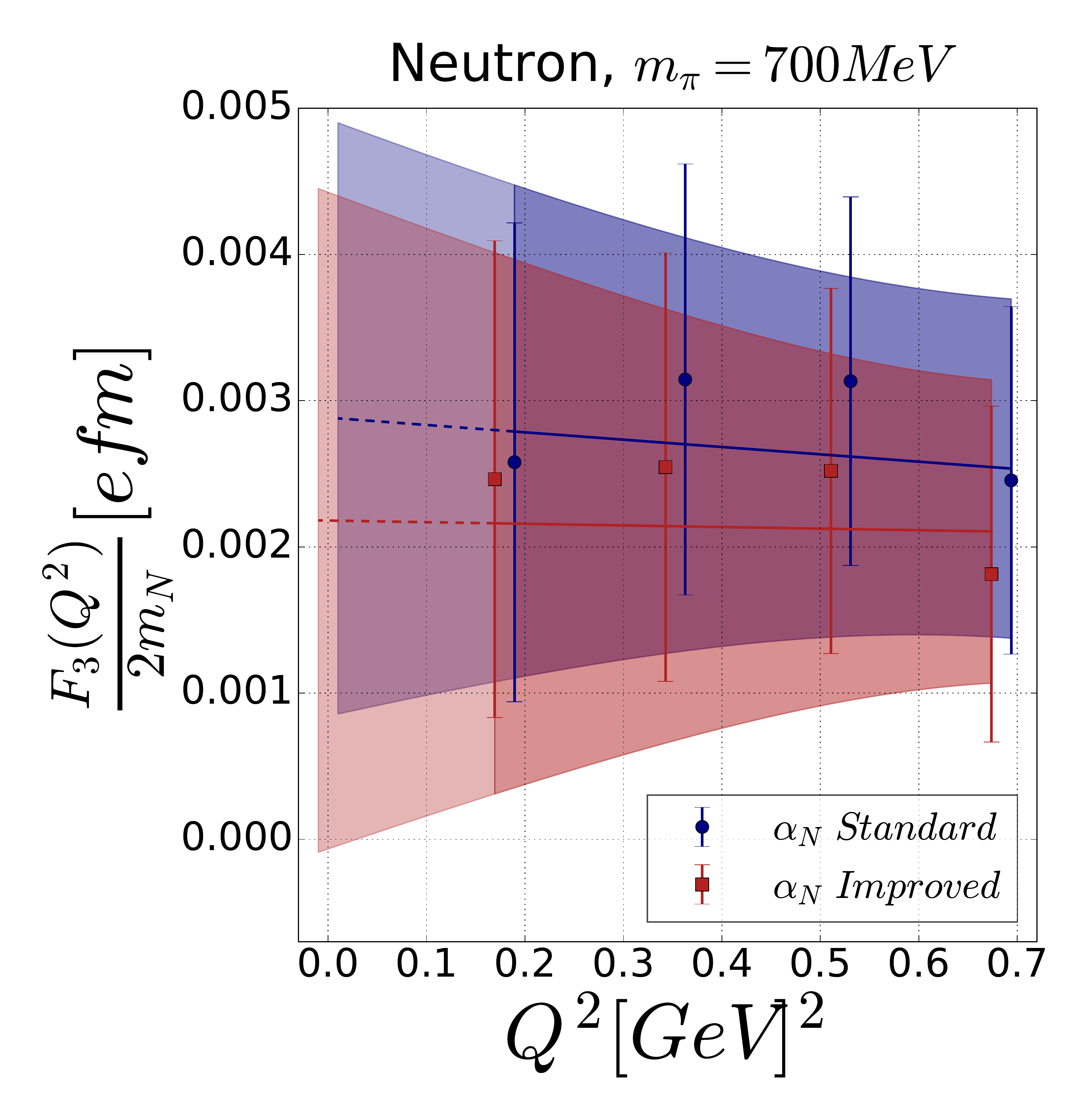}
  \caption{\label{fig:F3_alphaimp_mpi700}}
\end{subfigure}
\end{adjustwidth}
\caption{
  Comparison of
  the neutron CP-odd form factor \(\frac{F_{3}(Q^{2})}{2M_{N}}\) determined
  using improved (red) and unimproved (blue) results form the mixing angle $\alpha_N$/ 
  Shown are the \(m_{\pi}=\lbrace 410,570,700\rbrace\) MeV (left, middle and right) ensembles.
  The extrapolation to \(Q^{2}\rightarrow 0\) gives the final EDM.
  \label{fig:F3_alphaimp}}
\vspace*{\floatsep}
  \begin{adjustwidth}{-0.11\textwidth}{-0.11\textwidth}
\centering
\begin{subfigure}{.35\textwidth}
  \centering
  \includegraphics[trim={11mm 0cm 11mm 0cm},clip,width=\linewidth]{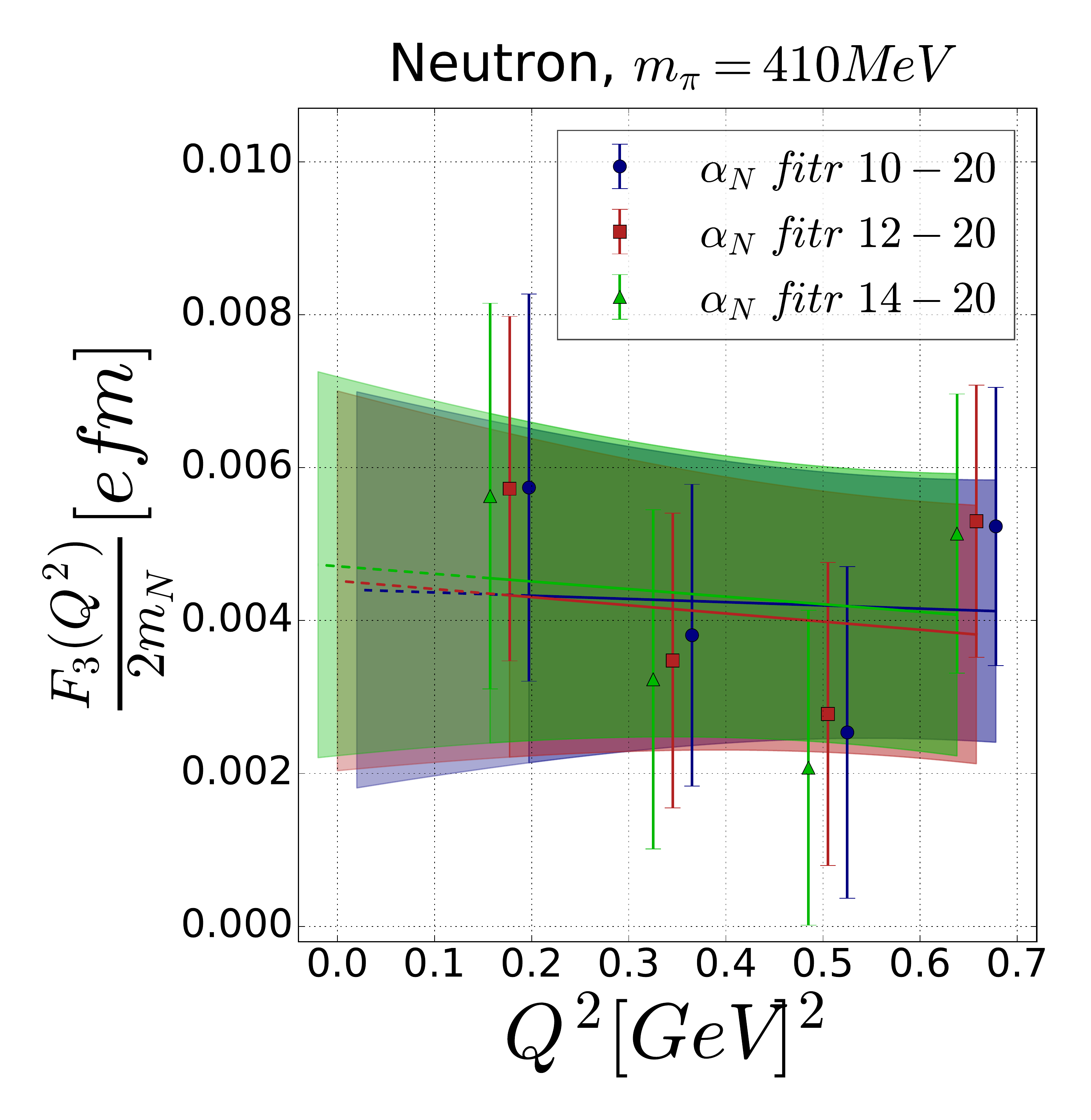}
  \caption{\label{fig:F3_alpha_mpi410}}
\end{subfigure}
\quad
\begin{subfigure}{.35\textwidth}
  \centering
  \includegraphics[trim={11mm 0cm 11mm 0cm},clip,width=\linewidth]{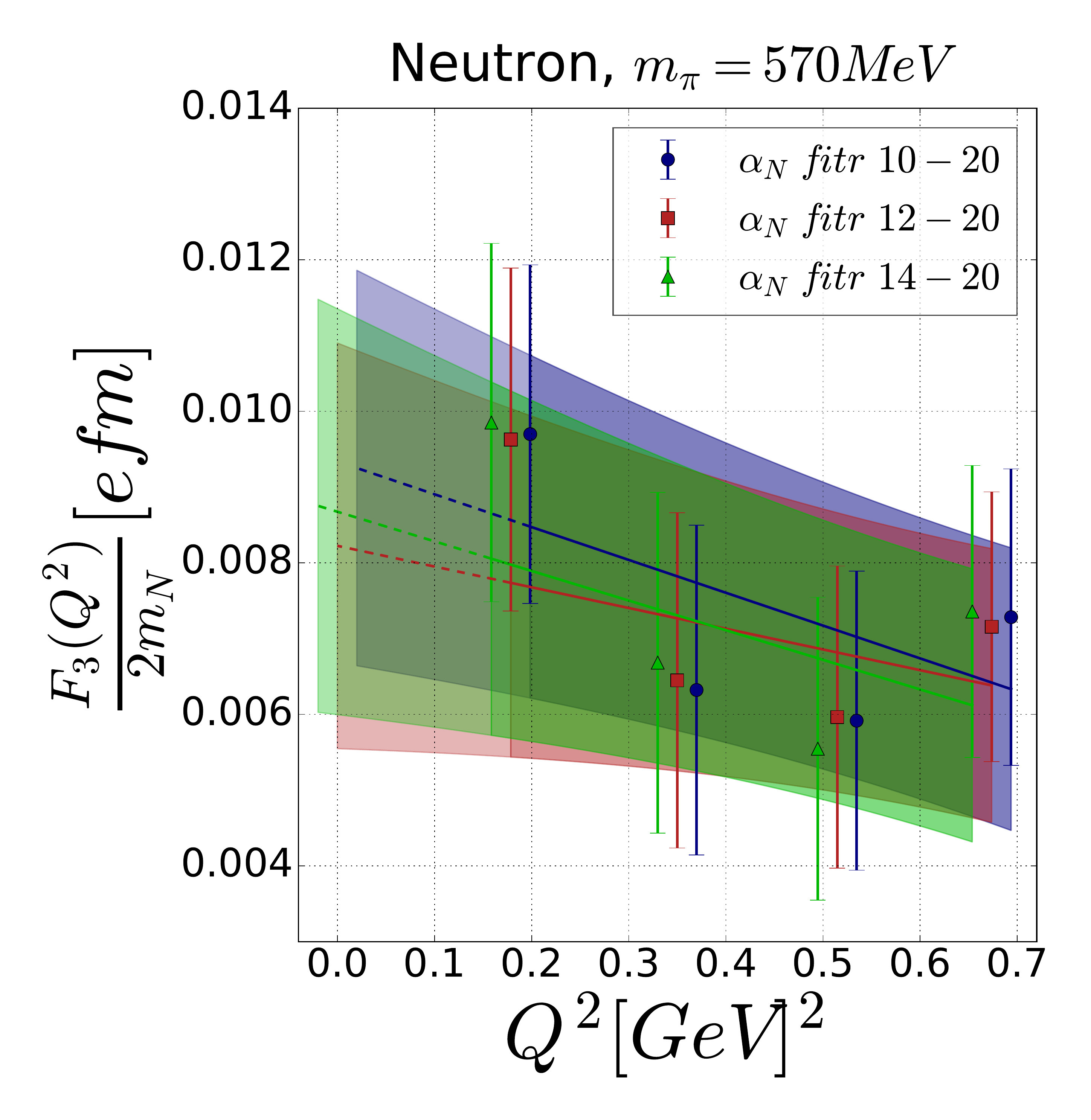}
  \caption{\label{fig:F3_alpha_mpi570}}
\end{subfigure}\quad
\begin{subfigure}{.35\textwidth}
  \centering
  \includegraphics[trim={11mm 0cm 11mm 0cm},clip,width=\linewidth]{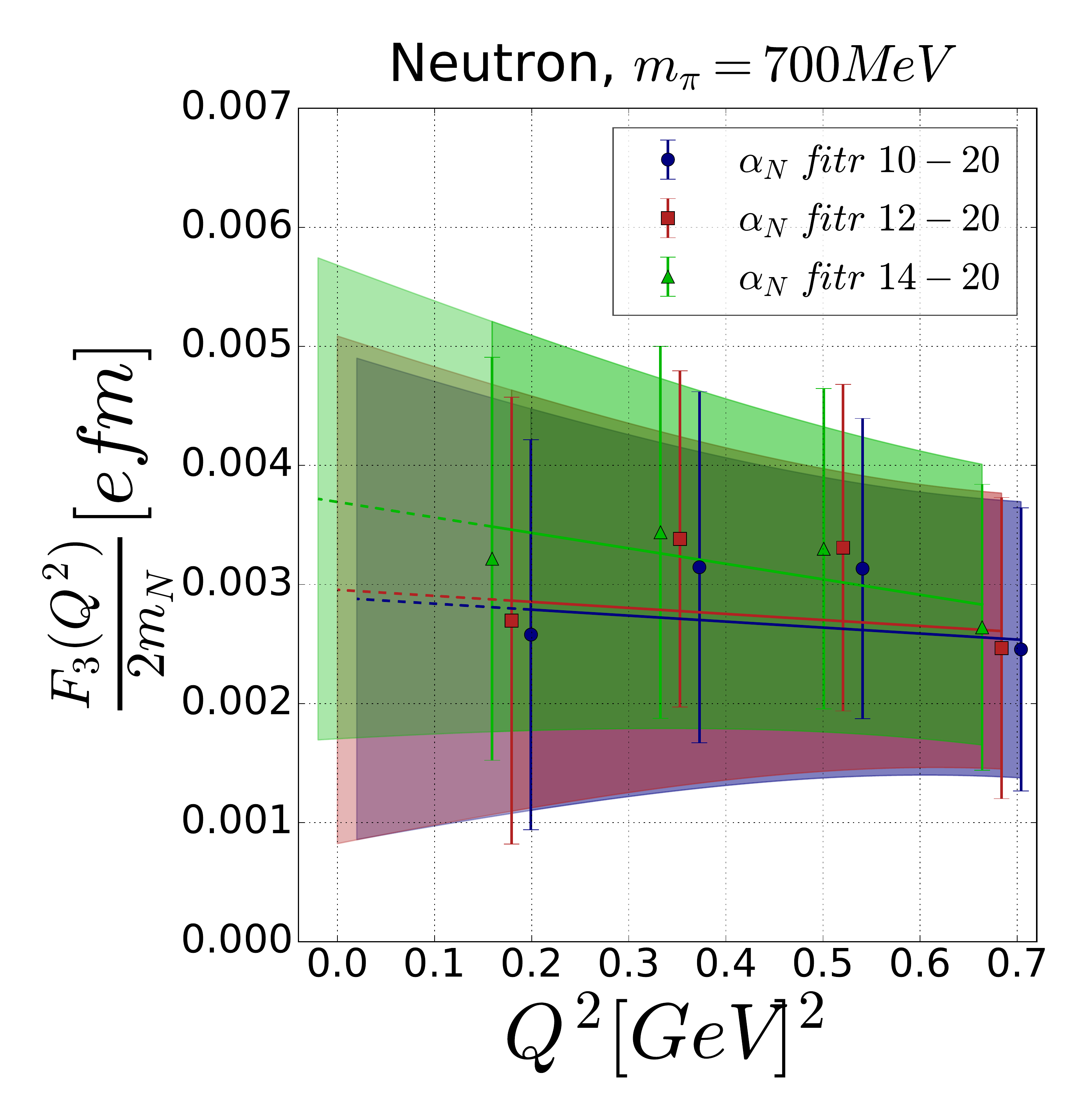}
  \caption{\label{fig:F3_alpha_mpi700}}
\end{subfigure}
\end{adjustwidth}
\caption{
    Comparison of
  the neutron CP-odd form factor \(\frac{F_{3}(Q^{2})}{2M_{N}}\) determined
  using different fit ranges for the determination of the mixing angle $\alpha_N$/ 
  Shown are the \(m_{\pi}=\lbrace 410,570,700\rbrace\) MeV (left, middle and right) ensembles.
  The extrapolation to \(Q^{2}\rightarrow 0\) gives the final EDM.
  \label{fig:F3_alpha_fitr}}
\end{figure}

\subsection{Improving the Modified Three-Point Correlation Function \label{sec:R_imp}}

\begin{figure}
\begin{adjustwidth}{-0.12\textwidth}{-0.12\textwidth}
\centering
\begin{subfigure}{.35\textwidth}
  \centering
  \includegraphics[trim={11mm 0cm 11mm 0cm},clip,width=\linewidth]{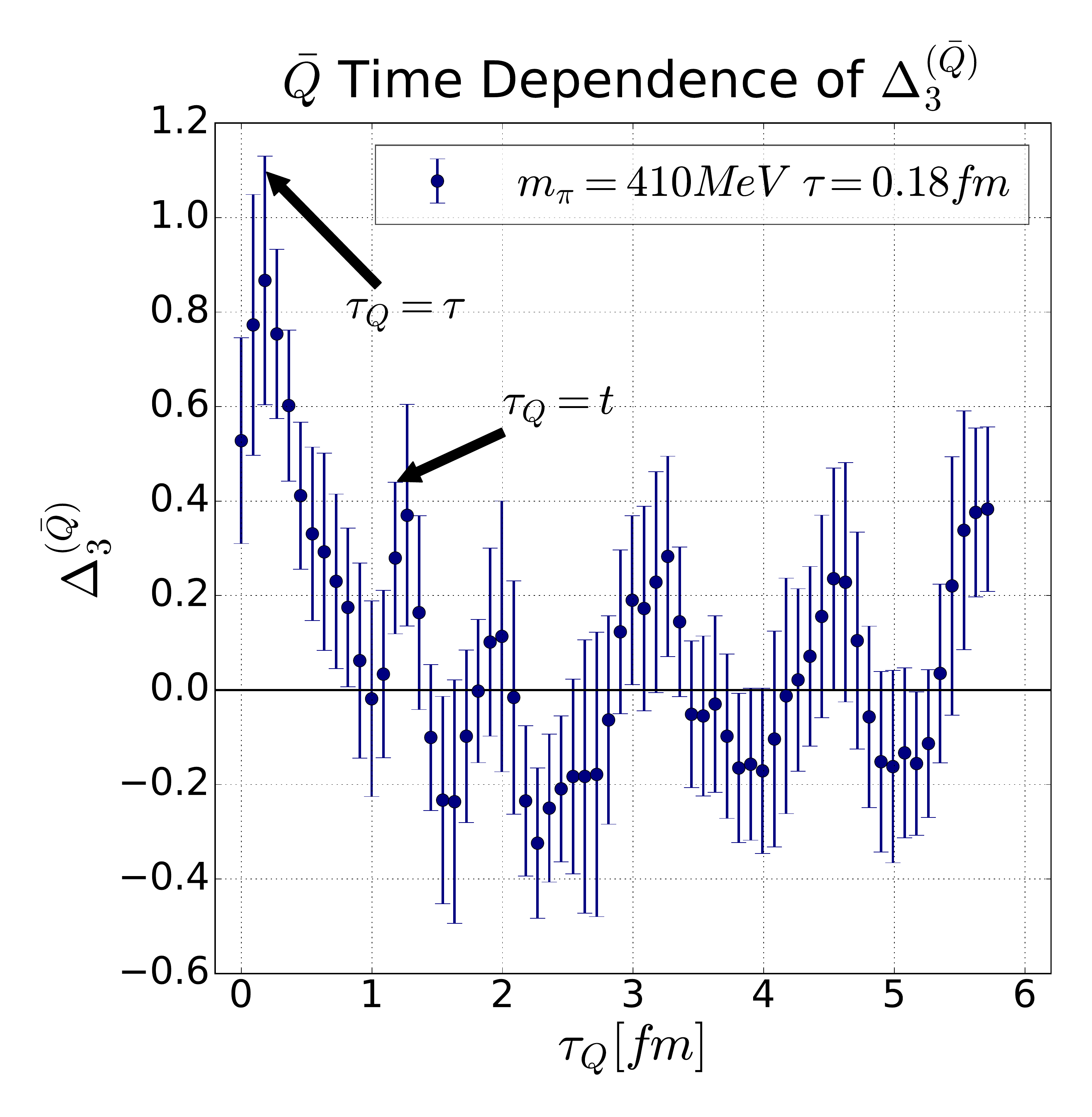}
  \caption{\label{fig:G3_imp_mpi410}}
\end{subfigure}
\quad
\begin{subfigure}{.35\textwidth}
  \centering
  \includegraphics[trim={11mm 0cm 11mm 0cm},clip,width=\linewidth]{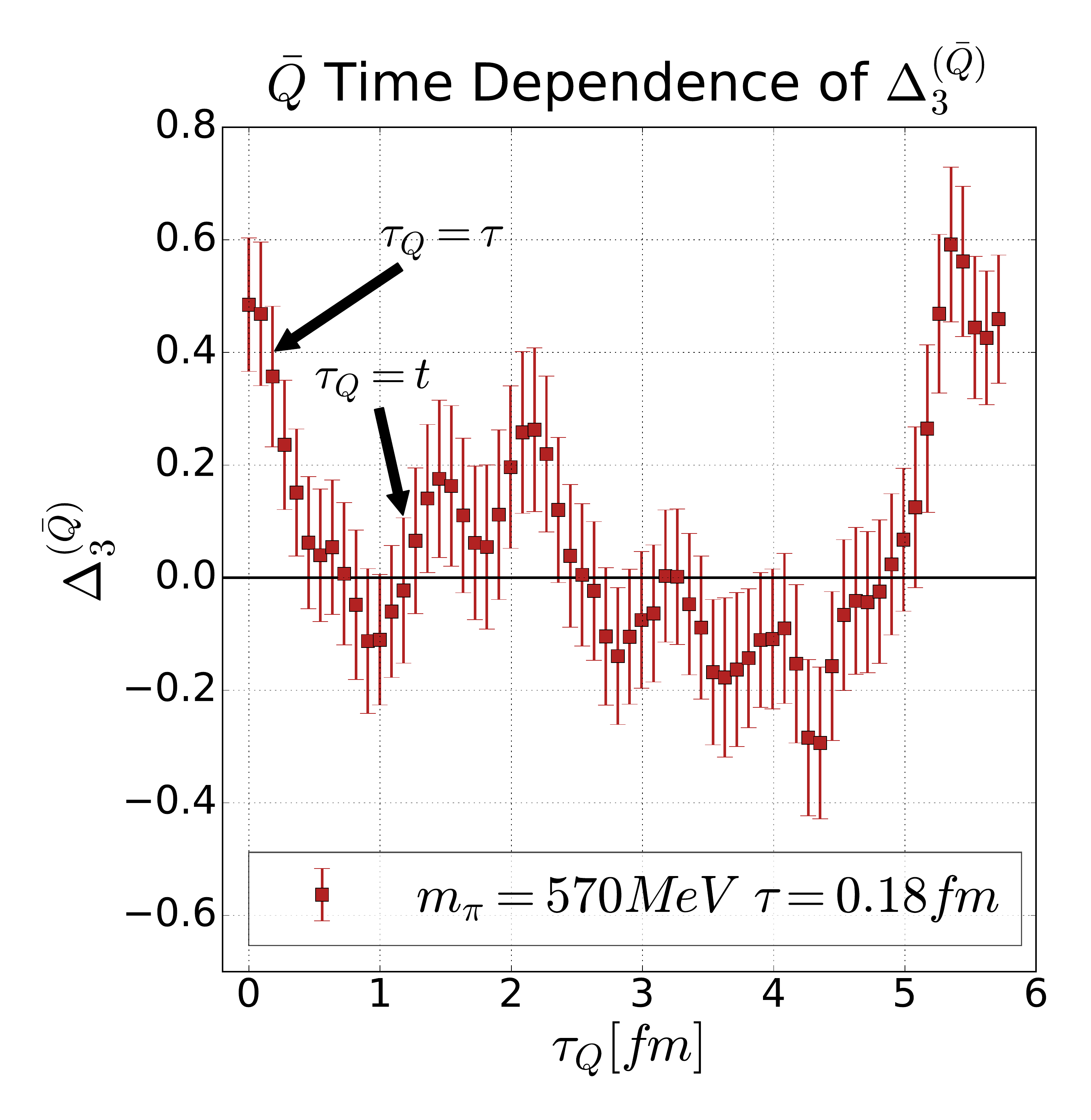}
  \caption{\label{fig:G3_imp_mpi570}}
\end{subfigure}
\quad
\begin{subfigure}{.35\textwidth}
  \centering
  \includegraphics[trim={11mm 0cm 11mm 0cm},clip,width=\linewidth]{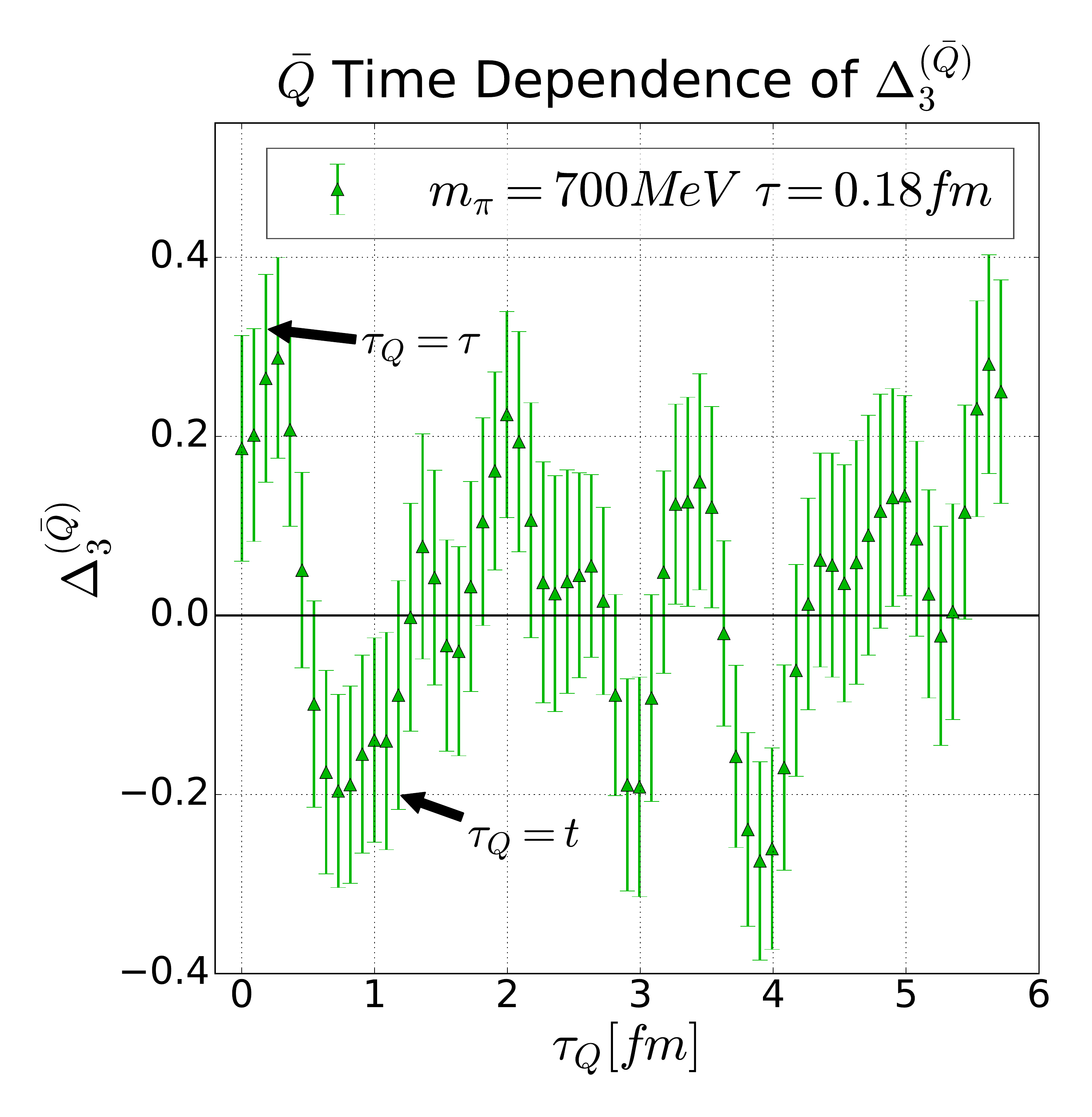}
  \caption{\label{fig:G3_imp_mpi700}}
\end{subfigure}
\vspace*{\floatsep}
\begin{subfigure}{.35\textwidth}
  \centering
  \includegraphics[trim={11mm 0cm 11mm 0cm},clip,width=\linewidth]{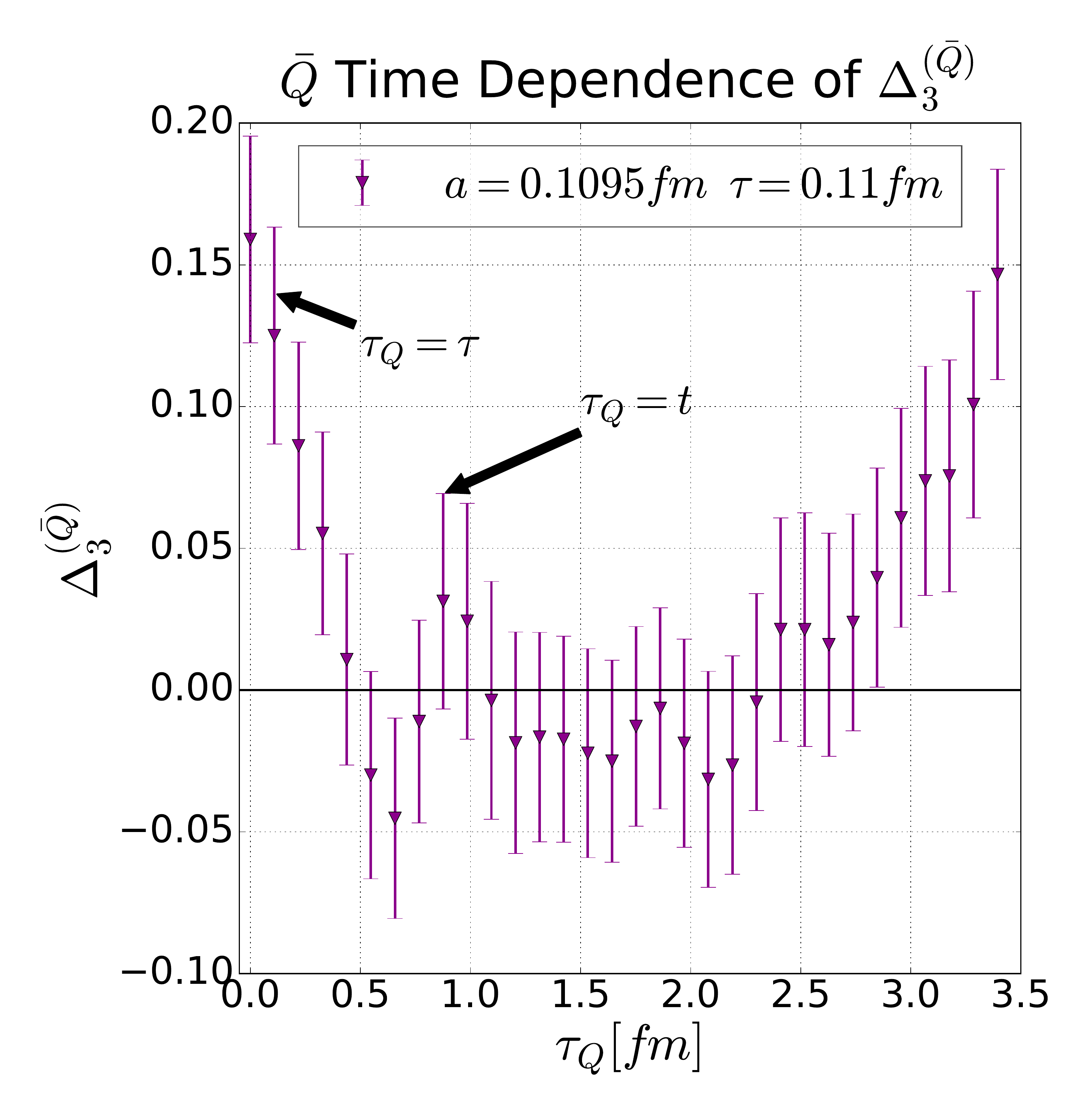}
  \caption{\label{fig:G3_imp_L16}}
\end{subfigure}
\quad
\begin{subfigure}{.35\textwidth}
  \centering
  \includegraphics[trim={11mm 0cm 11mm 0cm},clip,width=\linewidth]{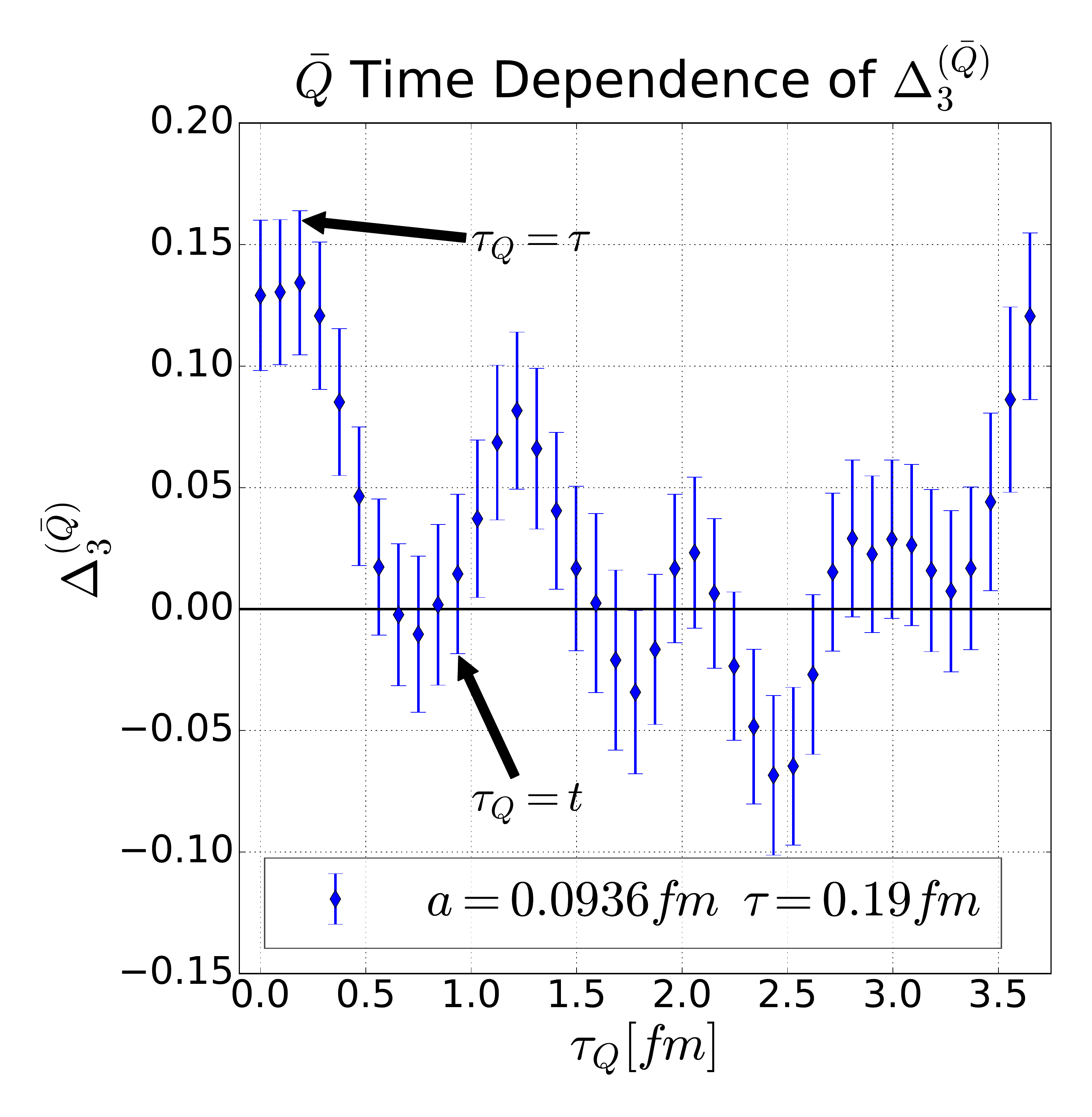}
  \caption{\label{fig:G3_imp_L20}}
\end{subfigure}
\quad
\begin{subfigure}{.35\textwidth}
  \centering
  \includegraphics[trim={11mm 0cm 11mm 0cm},clip,width=\linewidth]{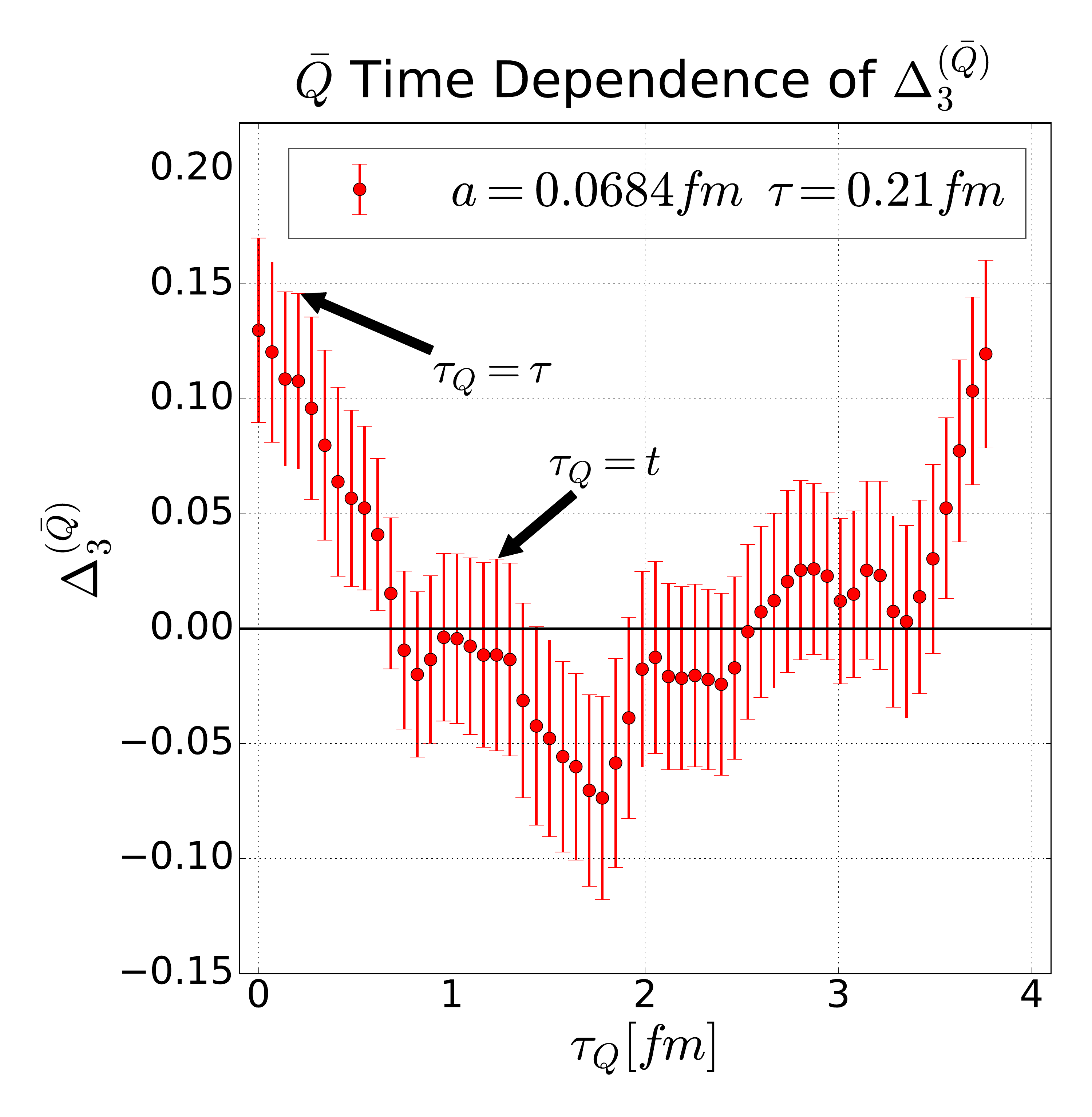}
  \caption{\label{fig:G3_imp_L28}}
\end{subfigure}
\end{adjustwidth}
\caption{
  Plot of the ratio \(\Delta_{3}^{(\overbar{Q})}\) as a function of \(\tau_{Q}\),
  the insertion time of the topological charge \(\overbar{Q}(t_{f},\tau_{Q})\)
  (see eq.~\eqref{eq:Q_temp_def}). We show the result for momentum
   \(\bm{q}=\frac{2\pi}{L}(0,0,2)\),
  \(\gamma_{\mu} = \gamma_{4}\), \(\Pi=\Pi_{+}i\gamma_{5}\gamma_{3}\) and
  the current insertion time \(\tau\) indicated in legend.
  The upper left, middle and right plots are the \(m_{\pi}=\{410,570,700\}\) MeV M-ensembles and
  the lower left, middle and right plots are the \(a=\{0.1095,0.0936,0.0684\}\) fm A-ensembles.
  \label{fig:G3_imp}}
\end{figure}

In this section, we utilize a similar improvement technique used for \(\alpha_{N}\),
but now applied to the modified three-point correlation function \(G^{(Q)}_{3}\).
The improvement starts by analyzing the time dependence of the spatially integrated
 topological charge density
\begin{equation}
  \Delta_{3}^{(\overbar{Q})}(\bm{p}^{\, \prime},t,\bm{q},\tau,\tau_{Q},\Pi,\gamma_{\mu},t_{f}) =
  a^{6}\sum_{\bm{x},\bm{y}}
  e^{-i  \bm{p}^{\, \prime}  \cdot \bm{x}}
  e^{i\bm{q}\cdot \bm{y}}
  \text{Tr}\left\lbrace
  \Pi \left\langle
  \mathcal{N}(\bm{x},t)
  \mathcal{J}_{\mu}(\bm{y},\tau)
  \overbar{Q}(\tau_{Q},t_{f})
  \overbar{\mathcal{N}}(\bm{0},0)\right\rangle
  \right\rbrace,
\end{equation}
where \(\tau_{Q}\) signifies the temporal location of the topological charge \(\overbar{Q}\)
defined in eq.~\eqref{eq:Q_temp_def}.
The spectral decomposition for this correlator has the form:
\begin{equation}
\Delta_{3}^{(\overbar{Q})} =
\left\{
	\begin{array}{ll}
    \sum_{\gamma,\delta}
    \frac{
    e^{-E_{\alpha_{0}}(T-t)}
    e^{-E_{\beta_{0}}(t-\tau)}
    e^{-E_{\gamma}(\tau - \tau_{Q})}
    e^{-E_{\delta}\tau_{Q}}
    }{16E_{\alpha_{0}}E_{\beta_{0}}E_{\gamma}E_{\delta}}& \\
    \text{Tr}\lbrace \Pi
      \bra{\alpha_{0}} \mathcal{N} \ket{\beta_{0}}
      \bra{\beta_{0}}\mathcal{J}_{\mu}\ket{\gamma}
      \bra{\gamma}Q\ket{\delta}
      \bra{\delta}\overbar{\mathcal{N}}\ket{\alpha_{0}}
    \rbrace, & \tau_{Q}< \tau < t  \\
    \vspace{1mm} \\
    \sum_{\beta,\gamma}
    \frac{
    e^{-E_{\alpha_{0}}(T-t)}
    e^{-E_{\beta}(t-\tau_{Q})}
    e^{-E_{\gamma}(\tau_{Q} - \tau)}
    e^{-E_{\delta_{0}}\tau}
    }{16E_{\alpha_{0}}E_{\beta}E_{\gamma}E_{\delta_{0}}}& \\
    \text{Tr}\lbrace \Pi
      \bra{\alpha_{0}} \mathcal{N} \ket{\beta}
      \bra{\beta}Q\ket{\gamma}
      \bra{\gamma}\mathcal{J}_{\mu}\ket{\delta_{0}}
      \bra{\delta_{0}}\overbar{\mathcal{N}}\ket{\alpha_{0}}
    \rbrace, &  \tau < \tau_{Q} < t \\
    \vspace{1mm} \\
    \sum_{\alpha,\beta}
    \frac{
    e^{-E_{\alpha}(T-\tau_{Q})}
    e^{-E_{\beta}(\tau_{Q}-t)}
    e^{-E_{\gamma_{0}}(t-\tau)}
    e^{-E_{\delta_{0}}\tau}
    }{16E_{\alpha}E_{\beta}E_{\gamma_{0}}E_{\delta_{0}}} & \\
    \text{Tr}\lbrace \Pi
      \bra{\alpha} Q \ket{\beta}
      \bra{\beta}\mathcal{N}\ket{\gamma_{0}}
      \bra{\gamma_{0}} \mathcal{J}_{\mu} \ket{\delta_{0}}
      \bra{\delta_{0}}\overbar{\mathcal{N}}\ket{\alpha}
    \rbrace, & \tau < t < \tau_{Q}
	\end{array}
\right.
\end{equation}
where \(\alpha , \beta, \gamma \) and \(\delta\) are labels for the states propagating, and the \(_{0}\) subscript
indicates the lowest energy state propagating with the appropriate quantum numbers.
We stress that \(t_{f} \neq 0\) implies the absence of any contact terms.
From fig.~\ref{fig:G3_imp}, a clear signal is observed at \(\tau_{Q}=0\) on all ensembles. This motivates summing
\(\tau_{Q}\) symmetrically around \(\tau_{Q}=0\) to obtain the summed three-point correlator:
\begin{eqnarray}\label{eq:C2QImprov_3}
\overbar{G}^{(\overbar{Q})}_{3}(\bm{p}^{\, \prime},t,\bm{q},\tau,\Pi,\gamma_{\mu},t_{f},t_{s}) =
a\sum_{\frac{\tau_{Q}}{a} = 0}^{t_{s}/a}
&\Big[&\Delta^{(\overbar{Q})}_{3}(\bm{p}^{\, \prime},t,\bm{q},\tau,\tau_{Q},\Pi,\gamma_{\mu},t_{f}) +\nonumber \\
&&\Delta^{(\overbar{Q})}_{3}(\bm{p}^{\, \prime},t,\bm{q},\tau,T-\tau_{Q},\Pi,\gamma_{\mu},t_{f})\Big]\,.
\end{eqnarray}
The resulting fit function to the sum range \(t_{s}\), for the range
\(t_{s} > t\), is:
\begin{equation}  \label{eq:G3_imp_fit_full}
    \overbar{G}_{3}^{(\overbar{Q})}(t_{s}) =
    \left\{
	   \begin{array}{ll}
        A_{0}+
        \sum_{\gamma_{\pm}\ne 0_{\pm}}
        A_{\gamma_{\pm}0_{\mp}}
        e^{-E_{\gamma_{\pm}}(\tau-t_{s})}
        e^{-E_{0_{\mp}}t_{s}}+
        A_{0_{\pm}0_{\mp}}
        e^{-E_{0_{\pm}}t_{s}}
        e^{-E_{0_{\mp}}[(T-t)-t_{s}]} & 0 < t_{s} < \tau \\
        \vspace{1mm} \\
        A_{0}+
        \sum_{\beta_{\pm},\gamma_{\mp}}
        A_{\beta_{\pm}\gamma_{\mp}}
        e^{-E_{\beta_{\pm}}(t-t_{s})}
        e^{-E_{\gamma_{\mp}}(t_{s}-\tau)} +
        A_{0_{\pm}0_{\mp}}
        e^{-E_{0_{\pm}}t_{s}}
        e^{-E_{0_{\mp}}[(T-t)-t_{s}]} & \tau < t_{s} < t \\
        \vspace{.5mm} \\
        A_{0}+
        \sum_{\beta_{\pm}\ne 0_{\pm}}
        A_{\beta_{\pm}0_{\mp}}
        e^{-E_{0_{\mp}}(T-t_{s})}
        e^{-E_{\beta_{\pm}}(t_{s}-t)} +
        A_{0_{\pm}0_{\mp}}
        e^{-E_{0_{\pm}}t_{s}}
        e^{-E_{0_{\mp}}[(T-t)-t_{s}]} & t < t_{s} < T/2
    	\end{array}
    \right.
\end{equation}
Where \(\gamma_{\pm}\) and \(\beta_{\pm}\)
represent the positive and negative (\(\pm\)) parity nucleon states.
\(A_{0}\), \(A_{\gamma_{\pm}}, A_{\beta_{\pm},\gamma_{\mp}}\), and \(A_{\beta_{\pm}}\) are combinations
of nucleon matrix elements, \(E_{\gamma_{\pm}}\) is the energy of the propagating state
\(\gamma_{\pm}\), and \(E_{0_{\pm}}\) is the lowest energy of the positive and
negative parity nucleon states \(0_{\pm}\).
We construct the improved ratio function \(\overbar{R}^{(\overbar{Q})}\) in the same way as in
eq.~\eqref{eq:RatFun}, but with the replacement of \(G^{(Q)}_{3}\rightarrow \overbar{G}^{(\overbar{Q})}_{3}\).
The value of the correlation function $G^{(Q)}_{3}$, used to extract the CP-odd form factor,
is obtained in the limit $t_s \rightarrow T$.
If the summation over $\tau_Q$ is performed up to a value $t_s <T$ the neglected terms
will be exponentially small as one can deduce from eq.~\eqref{eq:G3_imp_fit_full}.
Our numerical results seem to indicate that indeed
the neglected contributions for intermediate values of $t_s$
are well below the statistical accuracy of our calculation.

\begin{figure}
\begin{adjustwidth}{-0.12\textwidth}{-0.12\textwidth}
\centering
\begin{subfigure}{.35\textwidth}
  \centering
  \includegraphics[trim={8mm 0cm 11mm 0cm},clip,width=\linewidth]{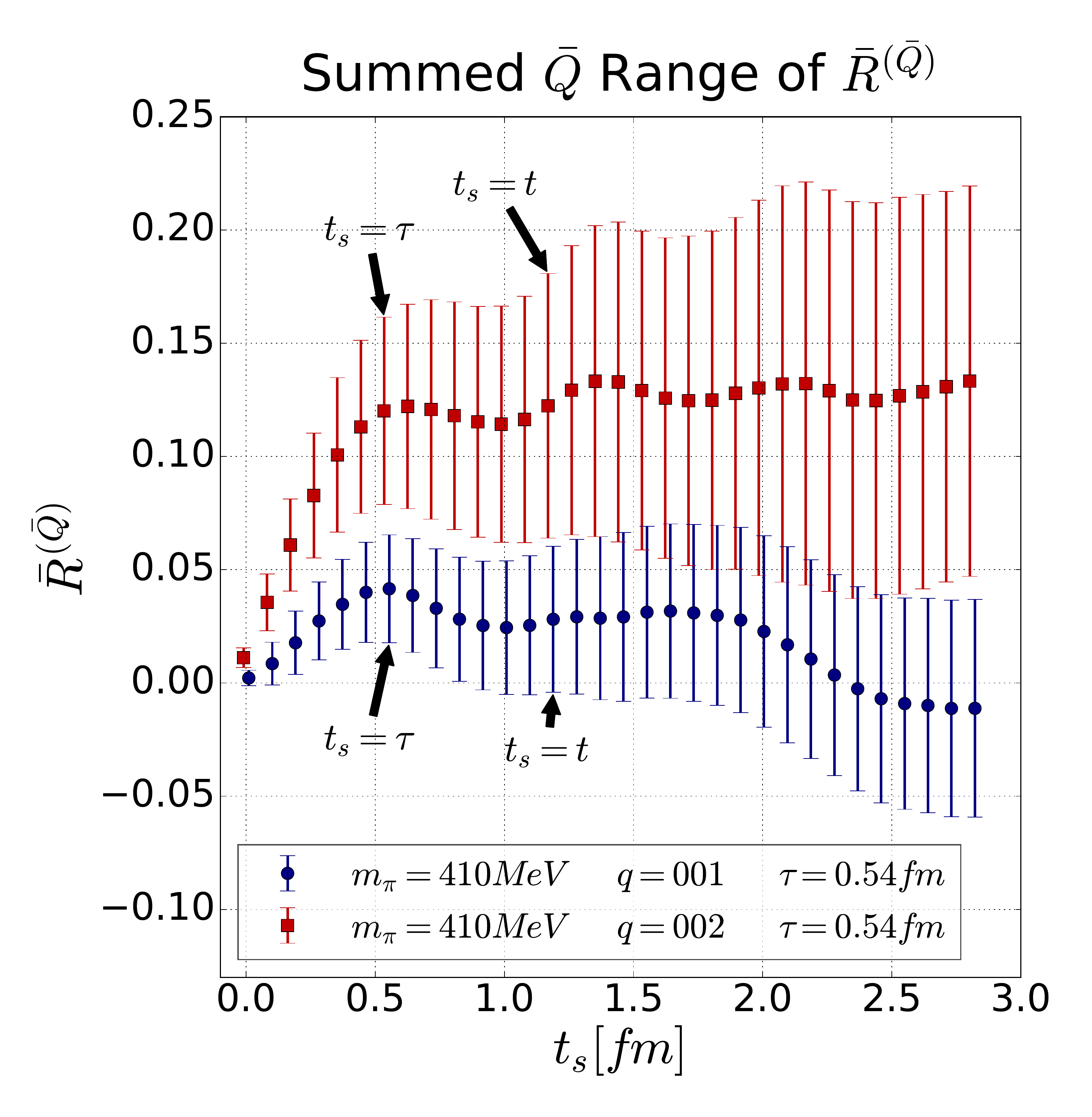}
  \caption{\label{fig:G3_imp_sum_mpi410}}
\end{subfigure}
\quad
\begin{subfigure}{.35\textwidth}
  \centering
  \includegraphics[trim={8mm 0cm 11mm 0cm},clip,width=\linewidth]{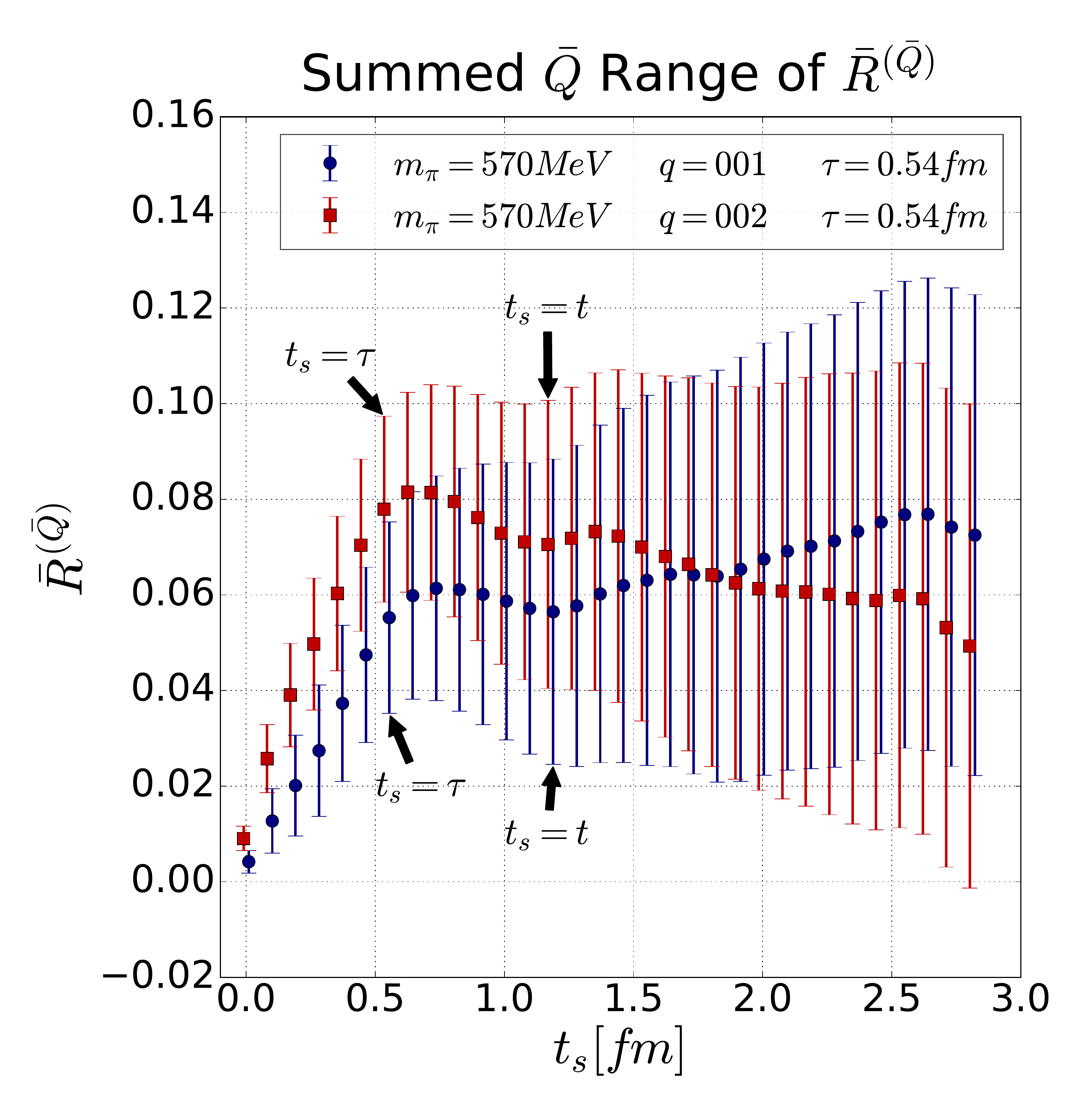}
  \caption{\label{fig:G3_imp_sum_mpi570}}
\end{subfigure}
\quad
\begin{subfigure}{.35\textwidth}
  \centering
  \includegraphics[trim={8mm 0cm 11mm 0cm},clip,width=\linewidth]{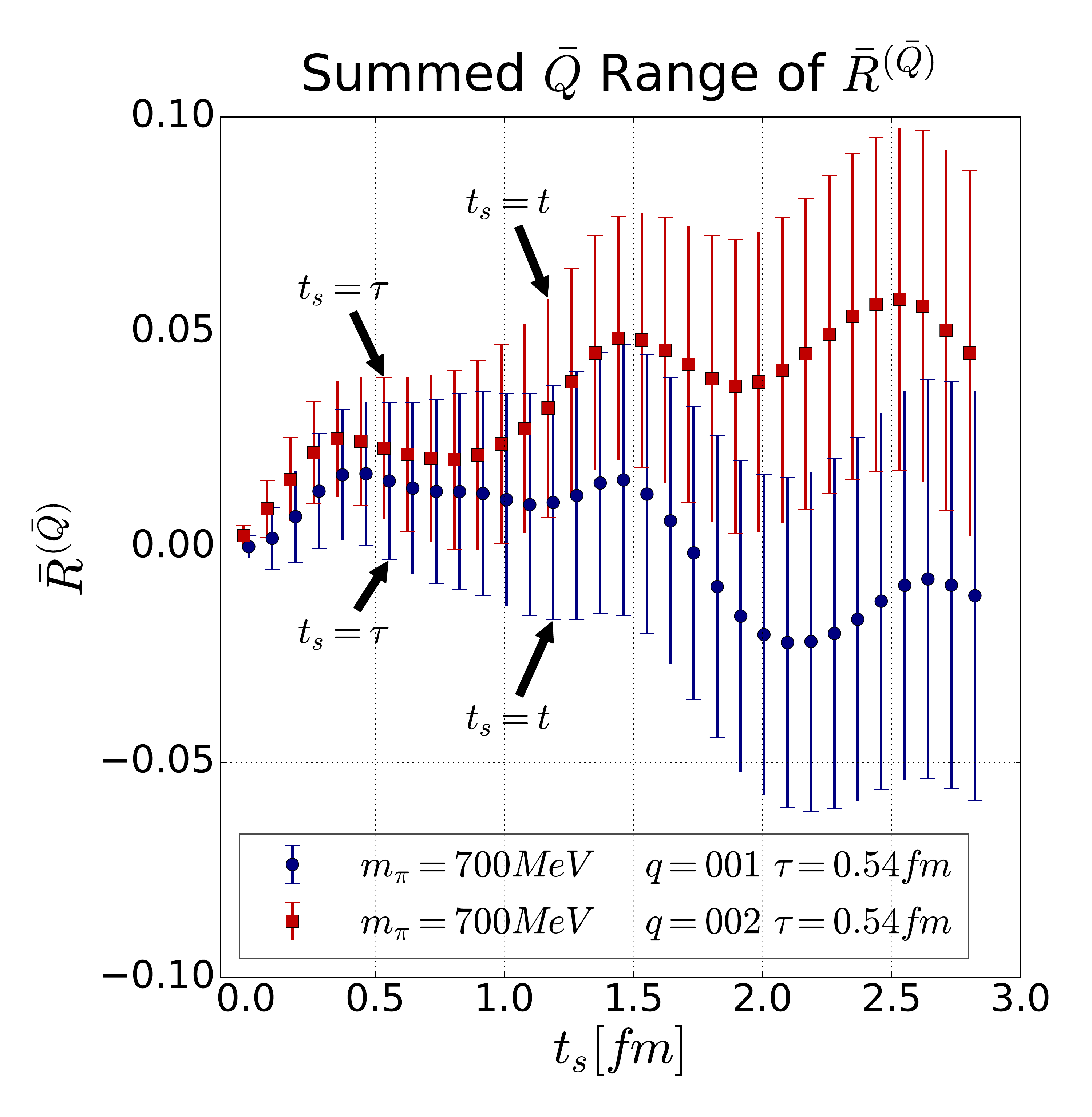}
  \caption{\label{fig:G3_imp_sum_mpi700}}
\end{subfigure}
\vspace*{\floatsep}
\begin{subfigure}{.35\textwidth}
  \centering
  \includegraphics[trim={8mm 0cm 11mm 0cm},clip,width=\linewidth]{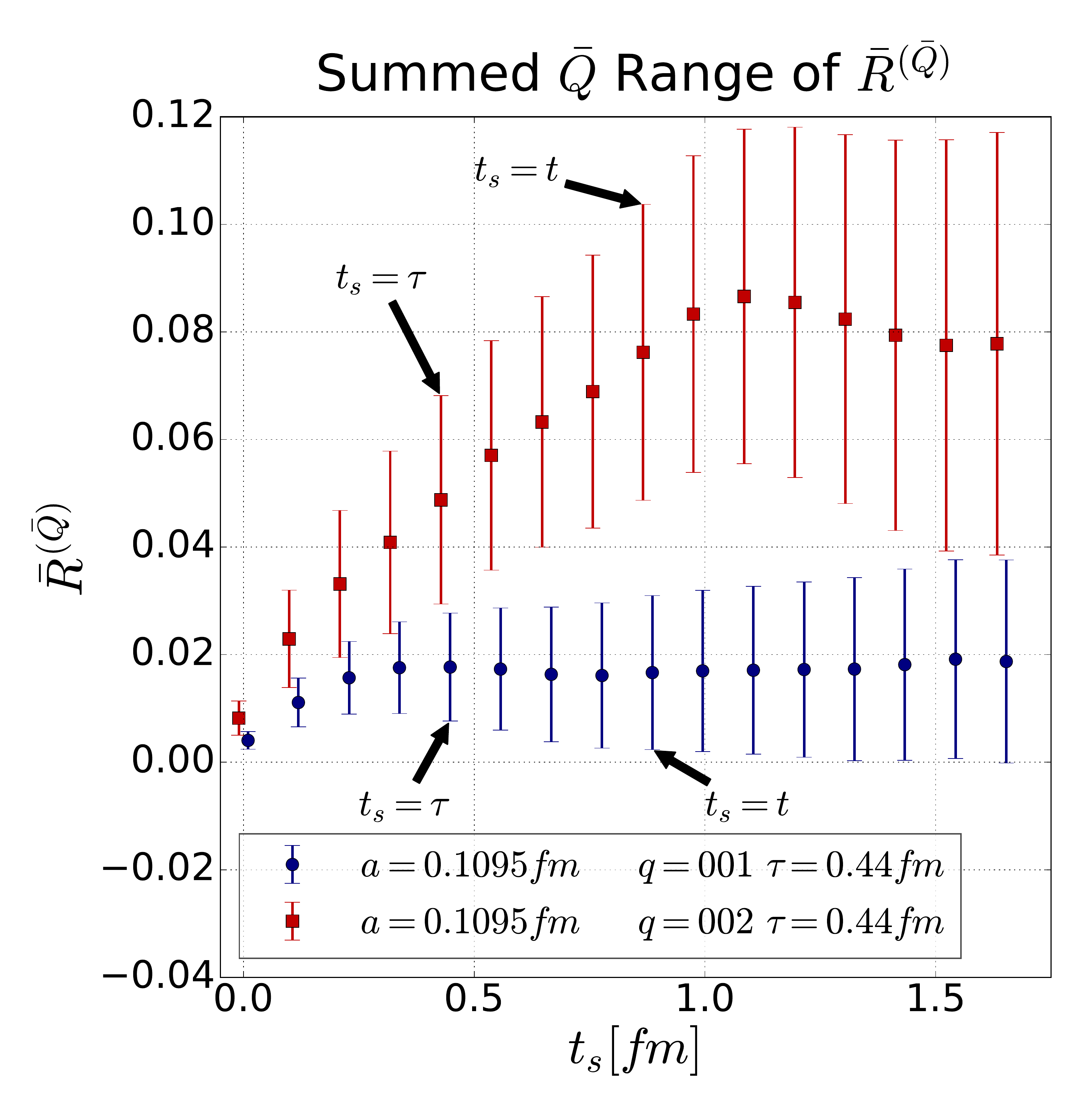}
  \caption{\label{fig:G3_imp_sum_L16}}
\end{subfigure}
\quad
\begin{subfigure}{.35\textwidth}
  \centering
  \includegraphics[trim={8mm 0cm 11mm 0cm},clip,width=\linewidth]{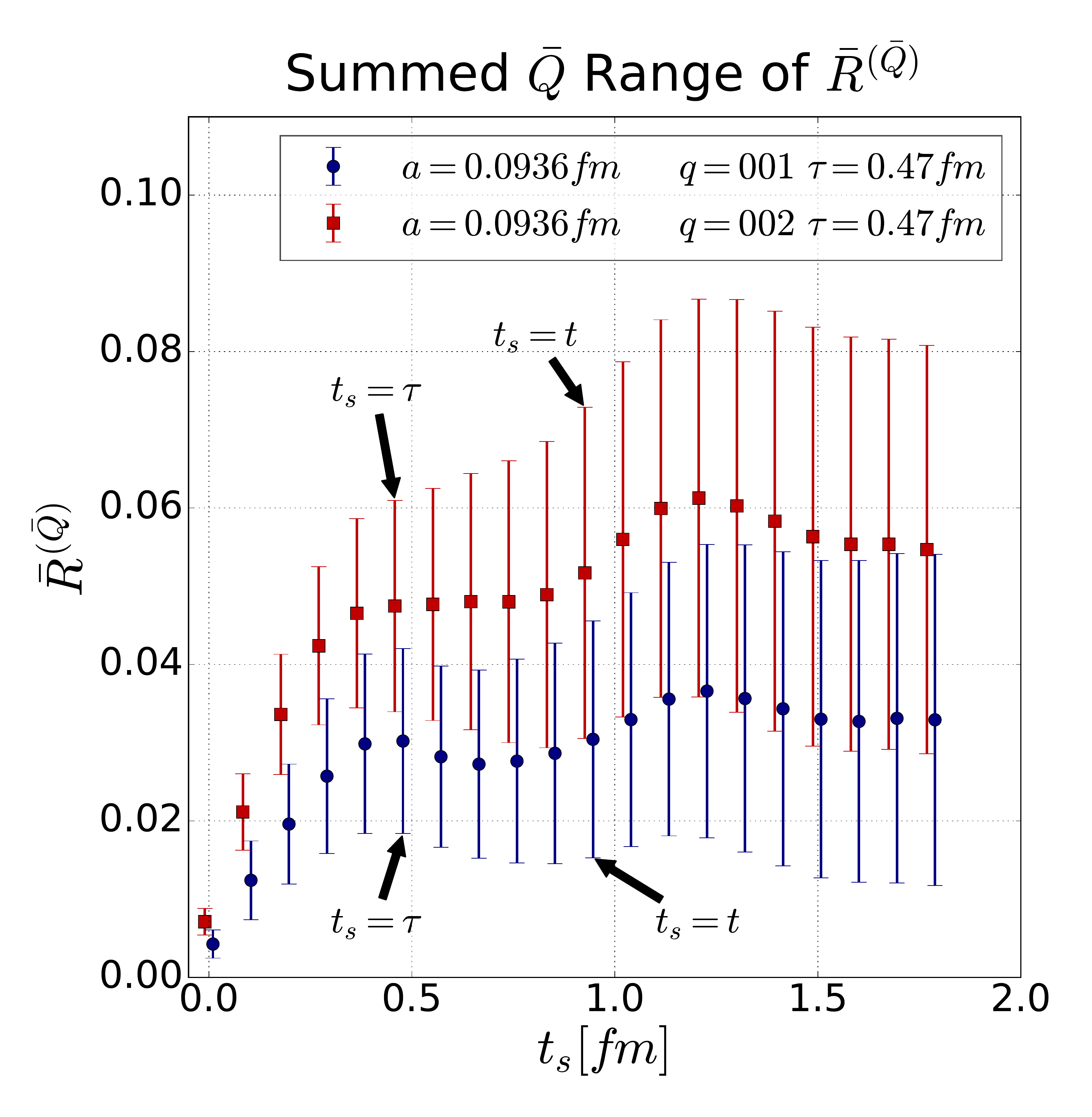}
  \caption{\label{fig:G3_imp_sum_L20}}
\end{subfigure}
\quad
\begin{subfigure}{.35\textwidth}
  \centering
  \includegraphics[trim={8mm 0cm 11mm 0cm},clip,width=\linewidth]{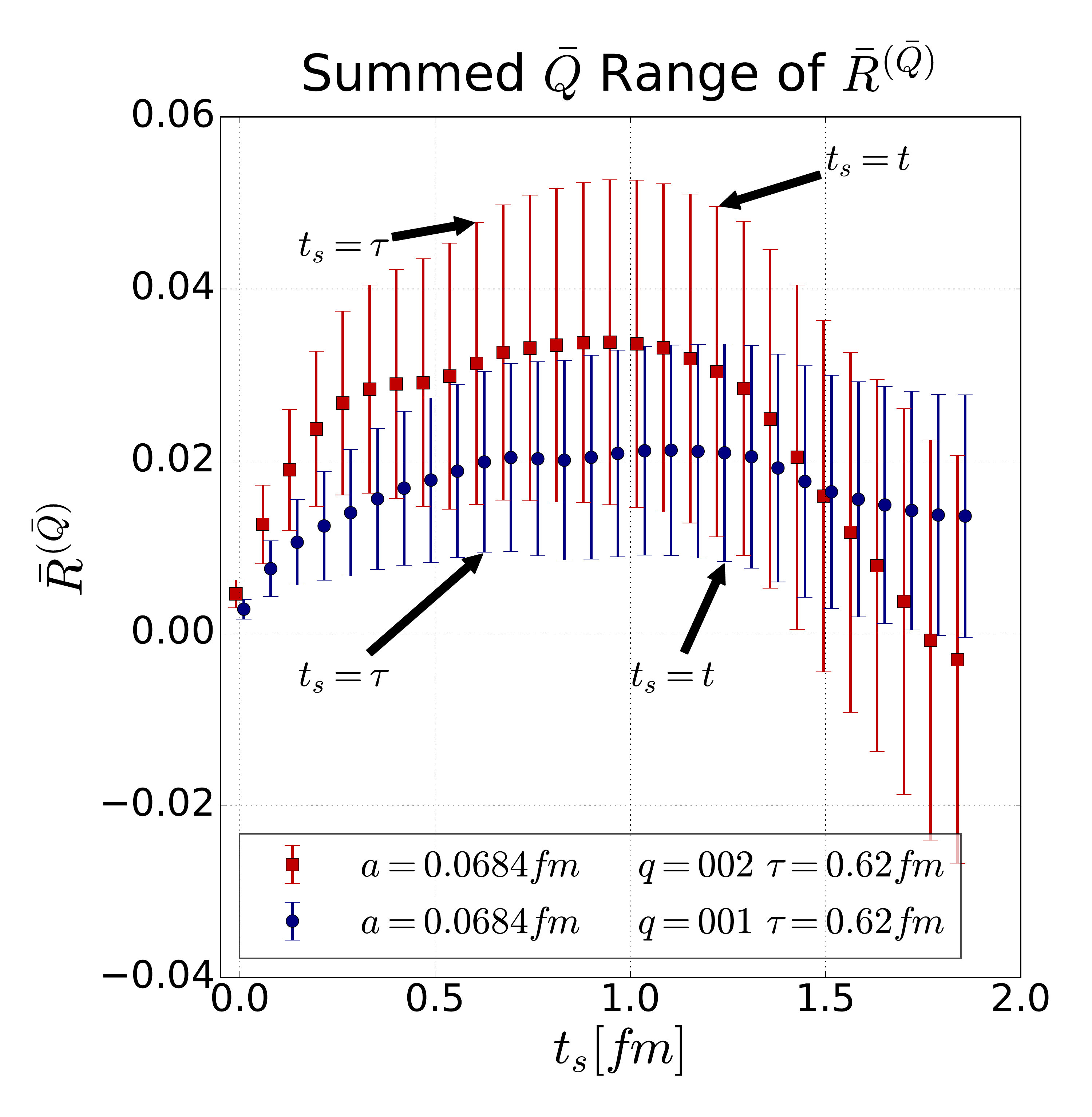}
  \caption{\label{fig:G3_imp_sum_L28}}
\end{subfigure}
\end{adjustwidth}
\caption{
  Plot of the ratio function \(\overbar{R}^{(\overbar{Q})}\) summed over
  \(\tau_{Q}\) (see fig.~\ref{fig:G3_imp}) from \(0\) to \(t_{s}\) and from \(T-t_{s}\)
  to \(T\), as a function of the summation window \(t_{s}\).
  We show the result for momentum
   \(\bm{q}=\frac{2\pi}{L}(0,0,2)\),
  \(\gamma_{\mu} = \gamma_{4}\), \(\Pi=\Pi_{+}i\gamma_{5}\gamma_{3}\) and
  the current insertion time \(\tau\) indicated in legend.
  The upper left, middle and right plots are the \(m_{\pi}=\{410,570,700\}\) MeV M-ensembles and
  the lower left, middle and right plots are the \(a=\{0.1095,0.0936,0.0684\}\) fm A-ensembles.
  The standard \(R^{(Q)}\) value for this quantity is obtained by taking the final \(t_{s}=\frac{T}{2}\) value.
  \label{fig:G3_imp_sum}}
\end{figure}

In fig.~\ref{fig:G3_imp_sum}, the results for the symmetrically summed topological charge ratio function \(\overbar{R}^{(\overbar{Q})}\)
are shown as a function of the sum range \(t_{s}\). In all cases, a plateau can be observed at
\(t_{s} = \tau\). This indicates that  all the exponential terms in eq.~\eqref{eq:G3_imp_fit_full} are suppressed
for \(t_{s}>\tau\). Coupled with the large statistical noise inherent in the
data, we fit the result with a constant value once the plateau has formed.
These fit ranges are displayed in tables~\ref{tab:F3_Q_ext_mpi_Rimp}
and~\ref{tab:F3_Q_ext_latspace_Rimp}, and
are used for the form factor analysis in sec.~\ref{sec:FF_Rat_imp}.

\begin{figure}
\begin{adjustwidth}{-0.12\textwidth}{-0.12\textwidth}
\centering
\begin{subfigure}{.35\textwidth}
  \centering
  \includegraphics[trim={8mm 0cm 11mm 0cm},clip,width=\linewidth]{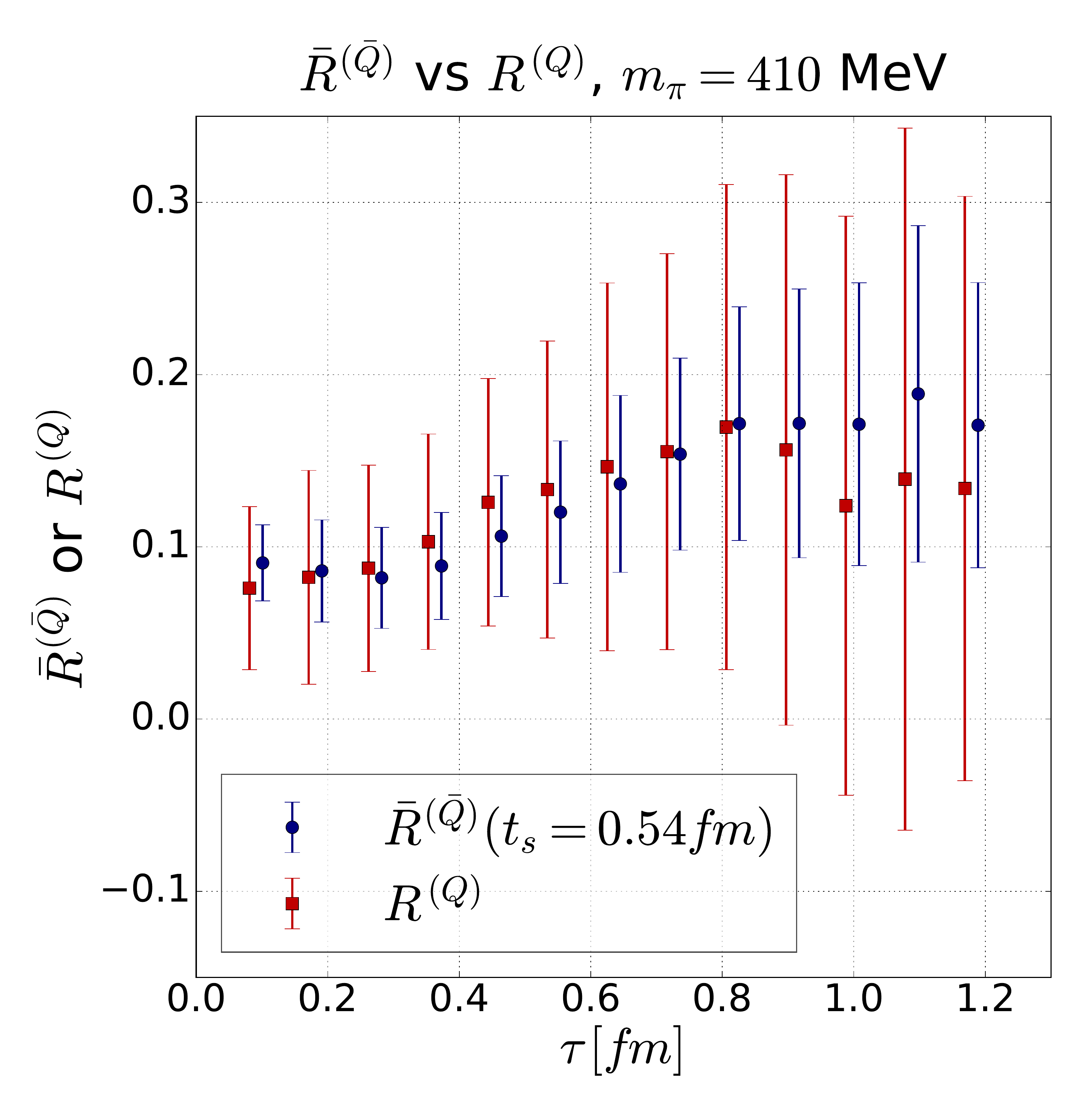}
  \caption{\label{fig:G3_imp_curr_mpi410}}
\end{subfigure}
\quad
\begin{subfigure}{.35\textwidth}
  \centering
  \includegraphics[trim={8mm 0cm 11mm 0cm},clip,width=\linewidth]{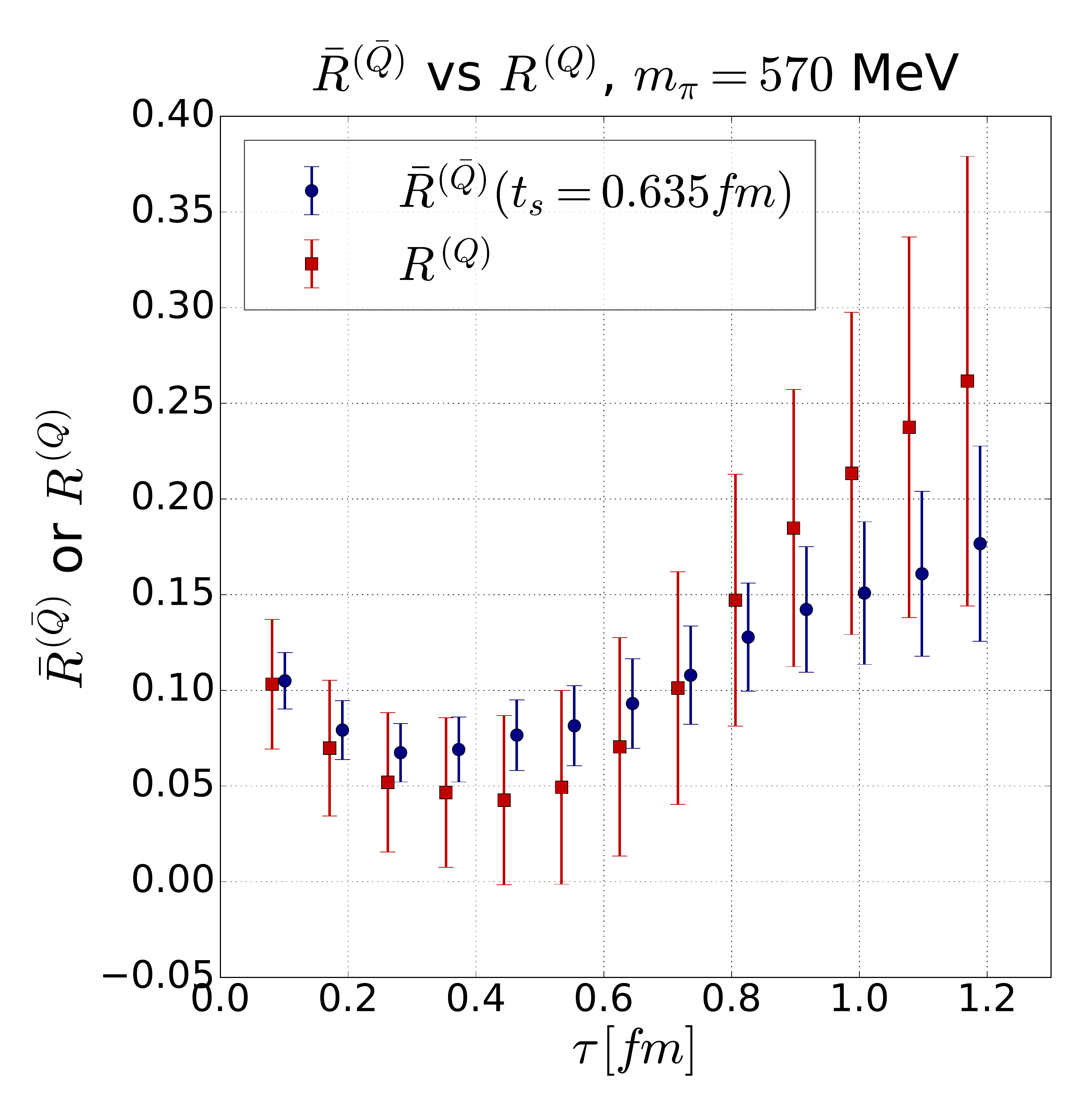}
  \caption{\label{fig:G3_imp_curr_mpi570}}
\end{subfigure}
\quad
\begin{subfigure}{.35\textwidth}
  \centering
  \includegraphics[trim={8mm 0cm 11mm 0cm},clip,width=\linewidth]{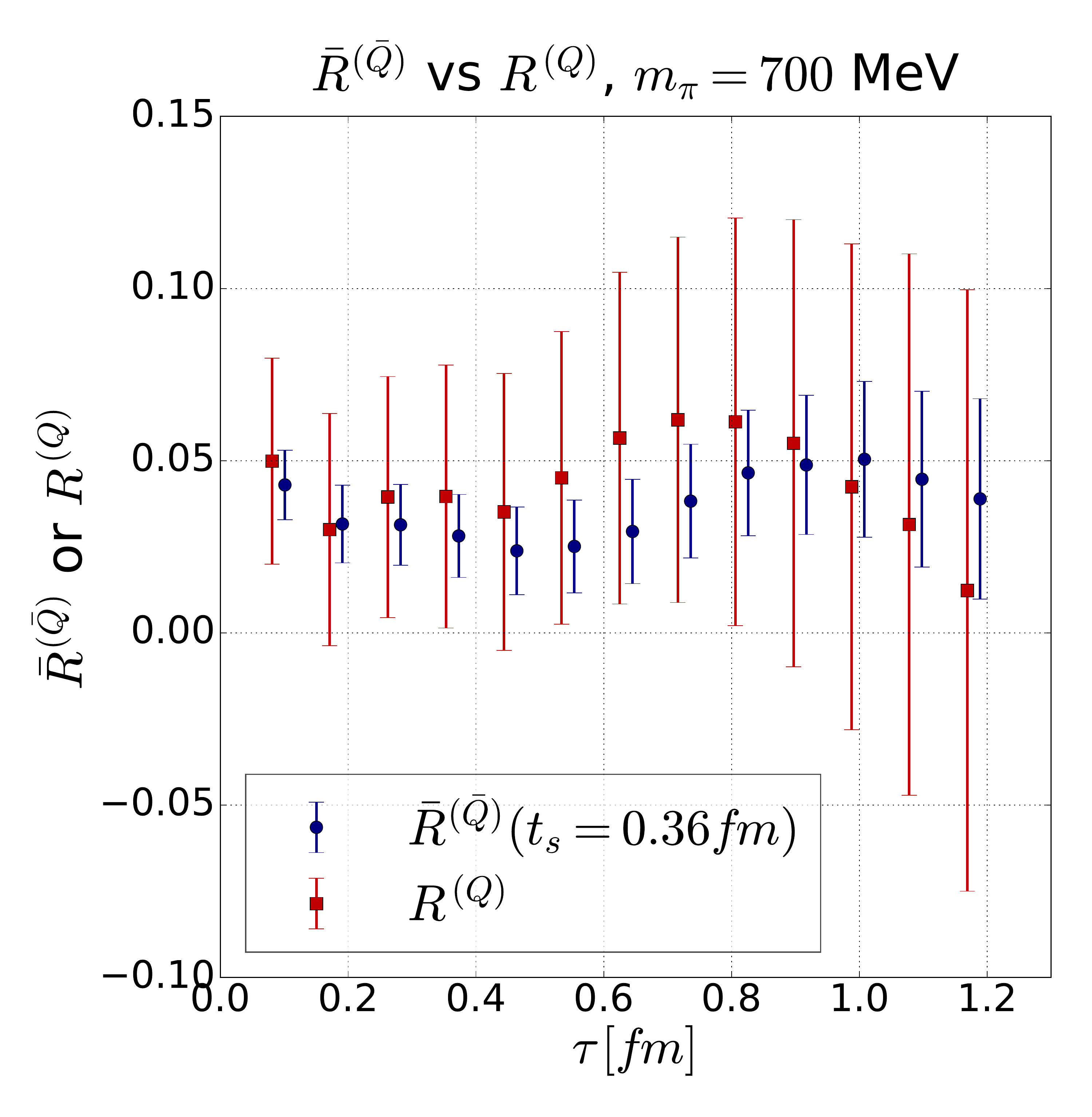}
  \caption{\label{fig:G3_imp_curr_mpi700}}
\end{subfigure}
\vspace*{\floatsep}
\begin{subfigure}{.35\textwidth}
  \centering
  \includegraphics[trim={8mm 0cm 11mm 0cm},clip,width=\linewidth]{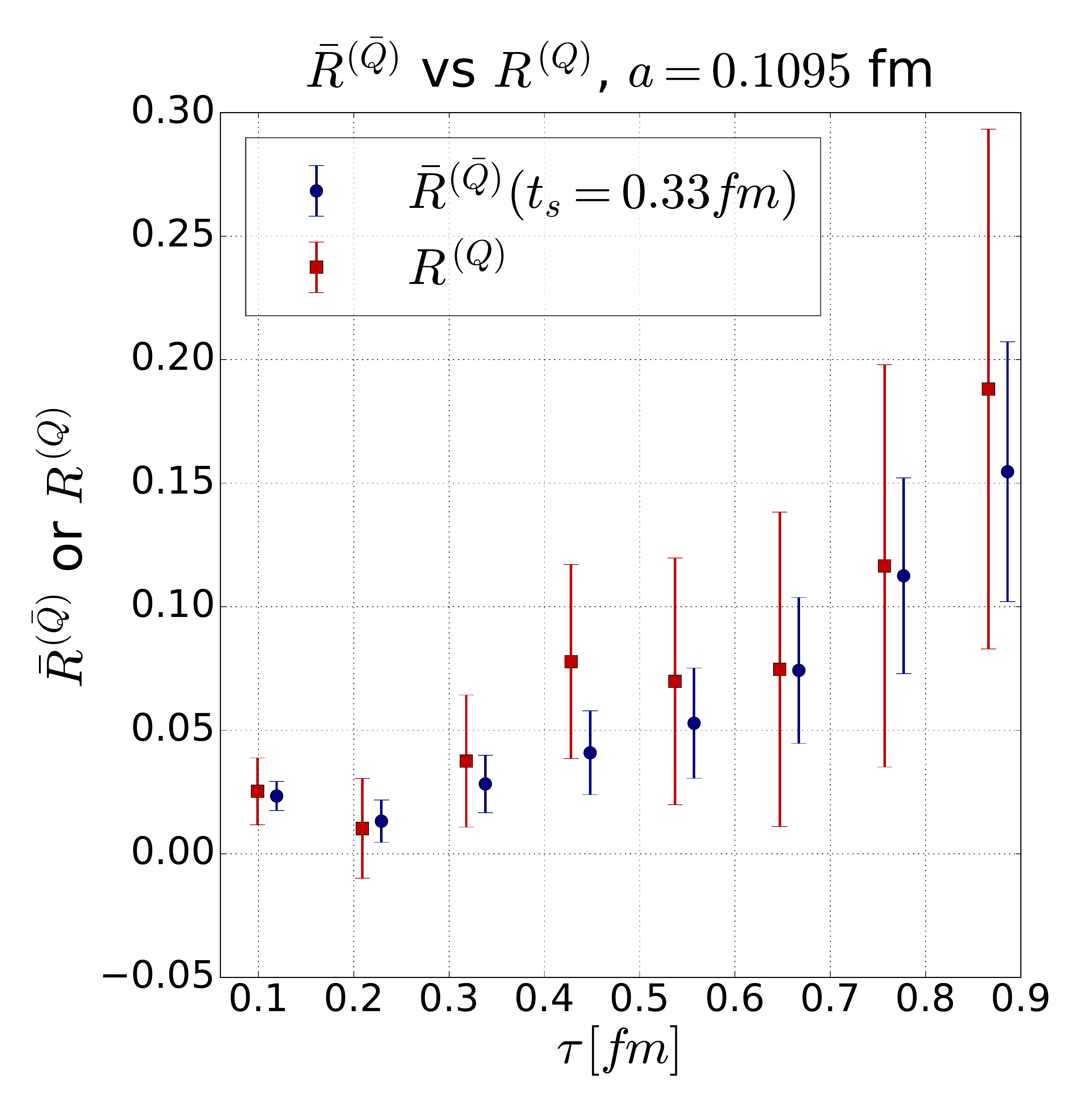}
  \caption{\label{fig:G3_imp_curr_L16}}
\end{subfigure}
\quad
\begin{subfigure}{.35\textwidth}
  \centering
  \includegraphics[trim={8mm 0cm 11mm 0cm},clip,width=\linewidth]{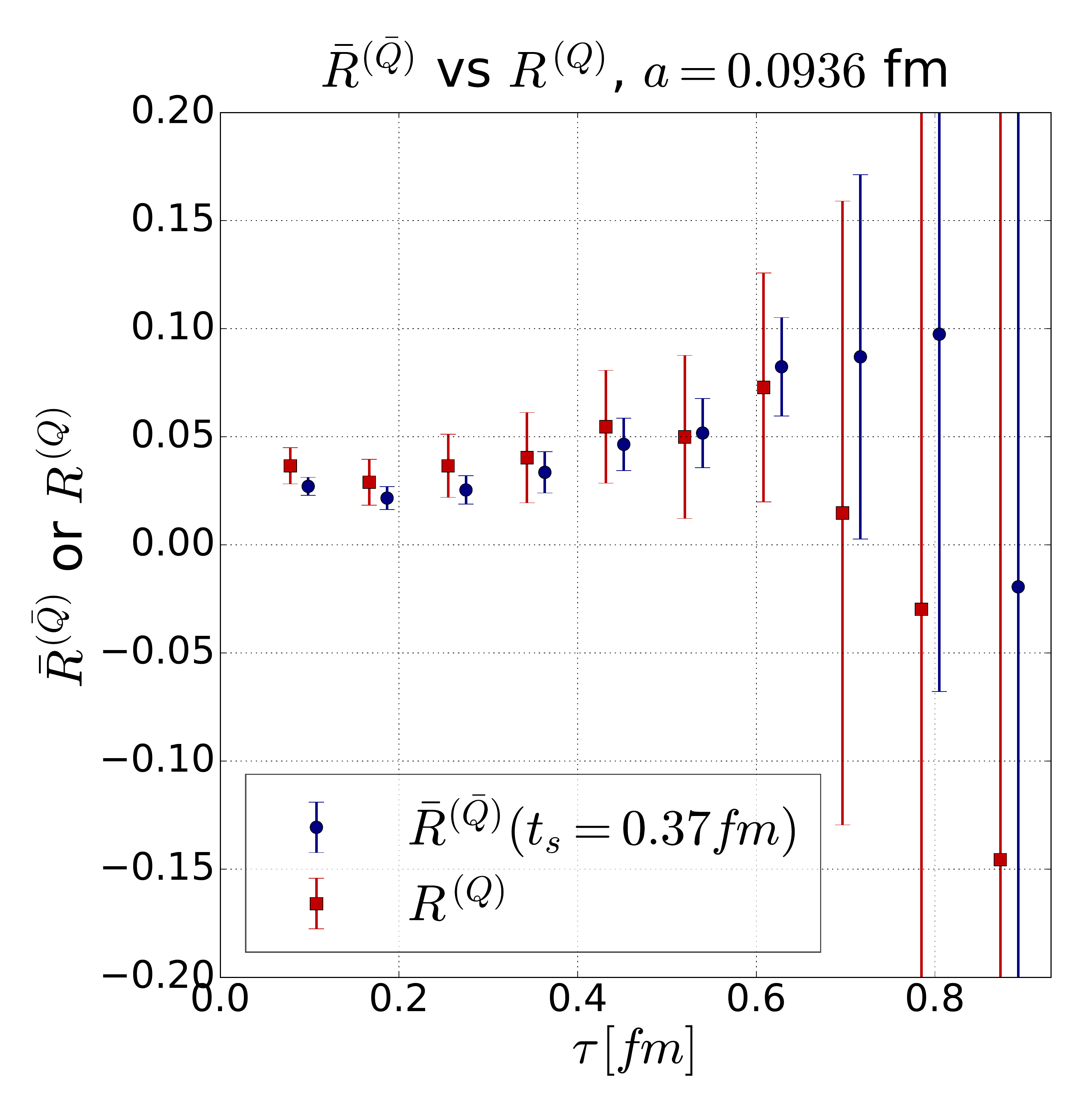}
  \caption{\label{fig:G3_imp_curr_L20}}
\end{subfigure}
\quad
\begin{subfigure}{.35\textwidth}
  \centering
  \includegraphics[trim={8mm 0cm 11mm 0cm},clip,width=\linewidth]{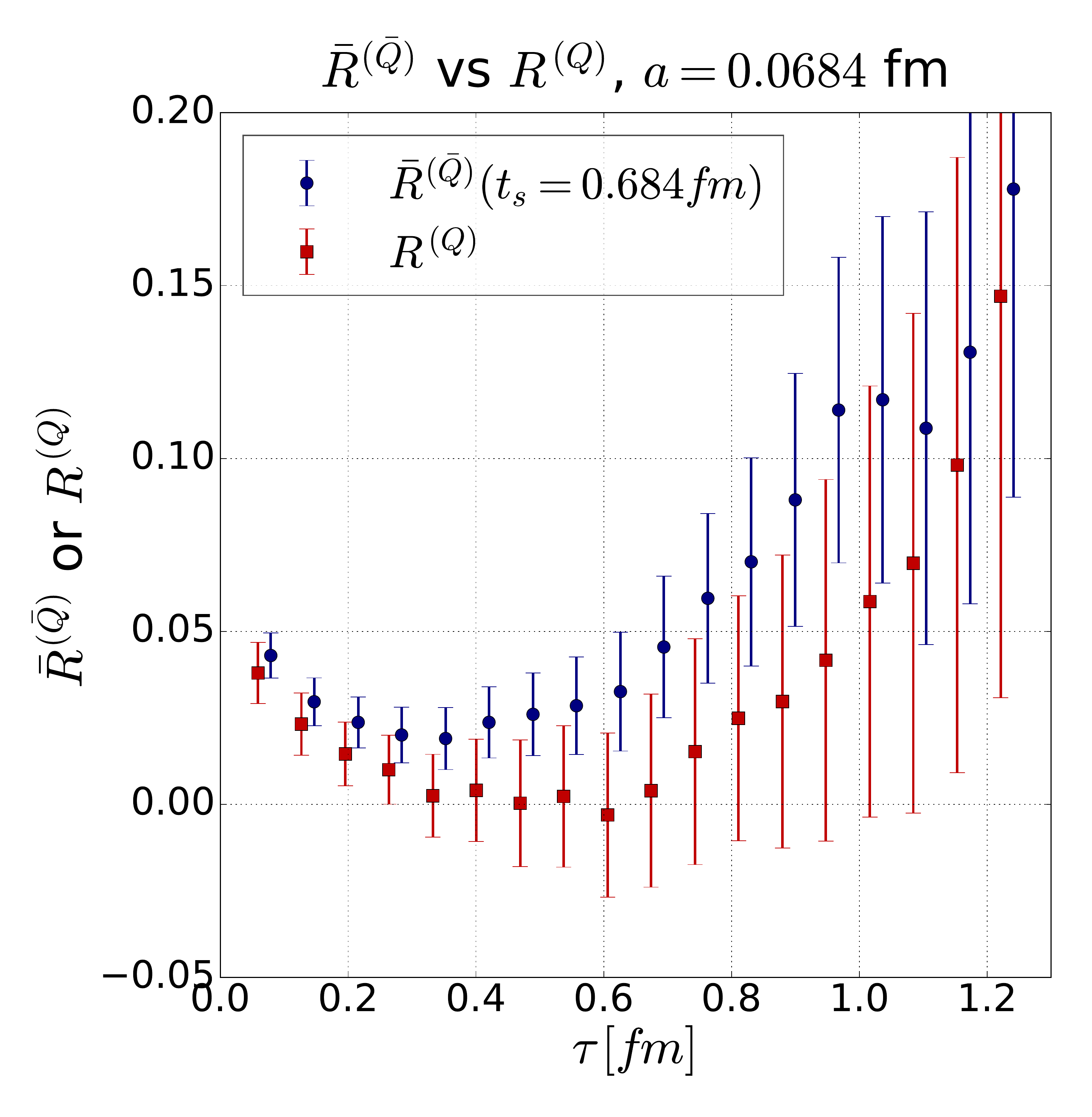}
  \caption{\label{fig:G3_imp_curr_L28}}
\end{subfigure}
\end{adjustwidth}
\caption{
  Comparison of improved \(\overbar{R}^{(\overbar{Q})}\) (blue) and unimproved
   \(R^{(Q)}\) (red) method for computing the ratio function as a fucntion of
   the vector current insertion time \(\tau\).
   We show the result for momentum
   \(\bm{q}=\frac{2\pi}{L}(0,0,2)\),
  \(\gamma_{\mu} = \gamma_{4}\), \(\Pi=\Pi_{+}i\gamma_{5}\gamma_{3}\).
  The upper left, middle and right plots are the \(m_{\pi}=\{410,570,700\}\) MeV M-ensembles and
  the lower left, middle and right plots are the \(a=\{0.1095,0.0936,0.0684\}\) fm A-ensembles.
  The \(t_{s}\) values in the legends were selected as the $t_{s}^{min}$ values from tabs.~\ref{tab:F3_Q_ext_mpi_Rimp},~\ref{tab:F3_Q_ext_latspace_Rimp}.
  \label{fig:G3_imp_curr}}
\end{figure}

Finally, fig.~\ref{fig:G3_imp_curr} displays a standard modified ratio function \(R_{3}^{(Q)}\)
plot over current insertion time \(\tau\),
where the improved ratio function (blue) is compared with the standard method (red). The improved
ratio function uses the ``min'' time from tables~\ref{tab:F3_Q_ext_mpi_Rimp} and~\ref{tab:F3_Q_ext_latspace_Rimp}.

\begin{figure}
\begin{adjustwidth}{-0.12\textwidth}{-0.12\textwidth}
\centering
\begin{subfigure}{.35\textwidth}
  \centering
  \includegraphics[trim={11mm 0cm 11mm 0cm},clip,width=\linewidth]{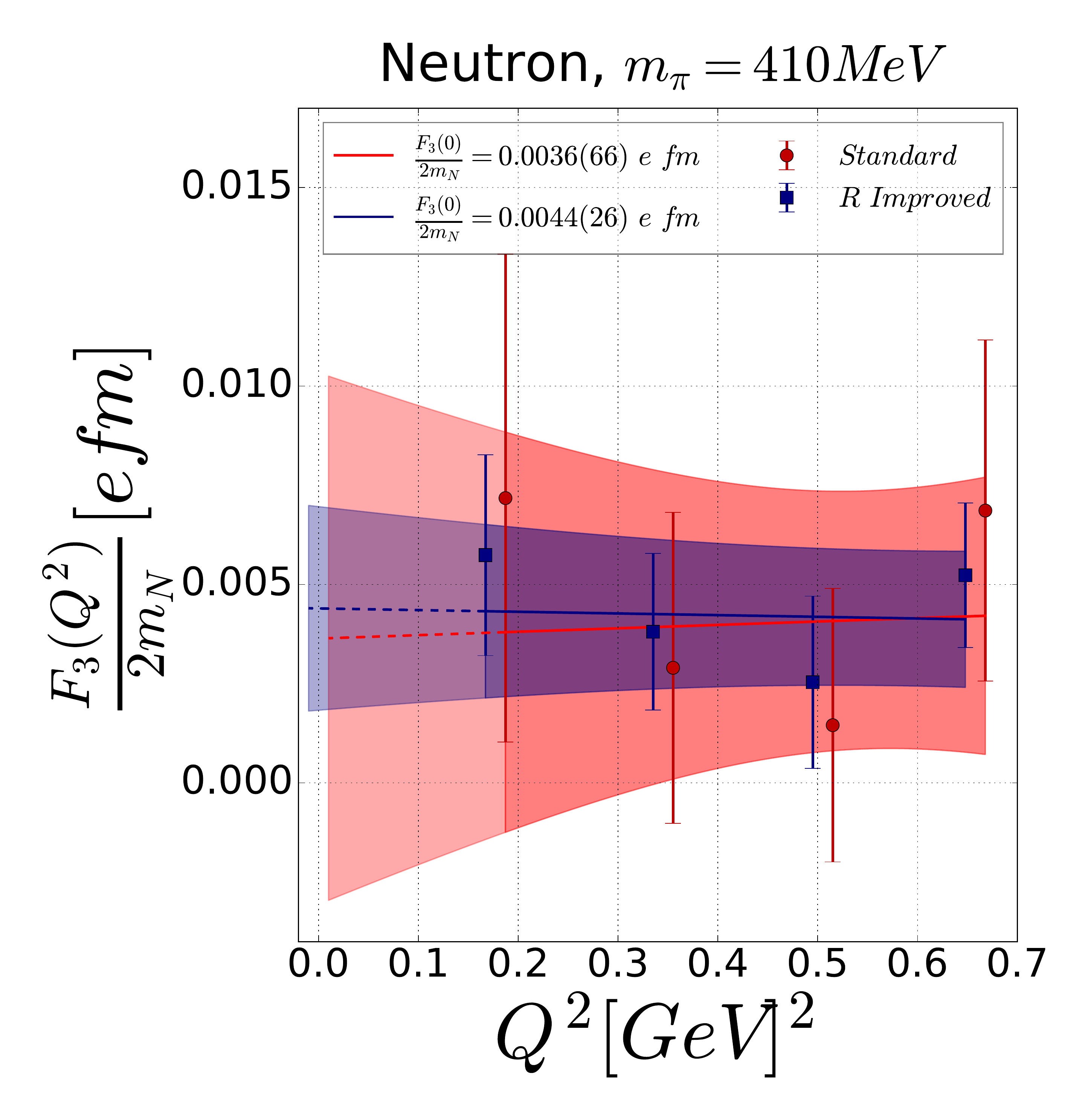}
  \caption{\label{fig:F3_Rimp_mpi410_proton}}
\end{subfigure}
\quad
\begin{subfigure}{.35\textwidth}
  \centering
  \includegraphics[trim={11mm 0cm 11mm 0cm},clip,width=\linewidth]{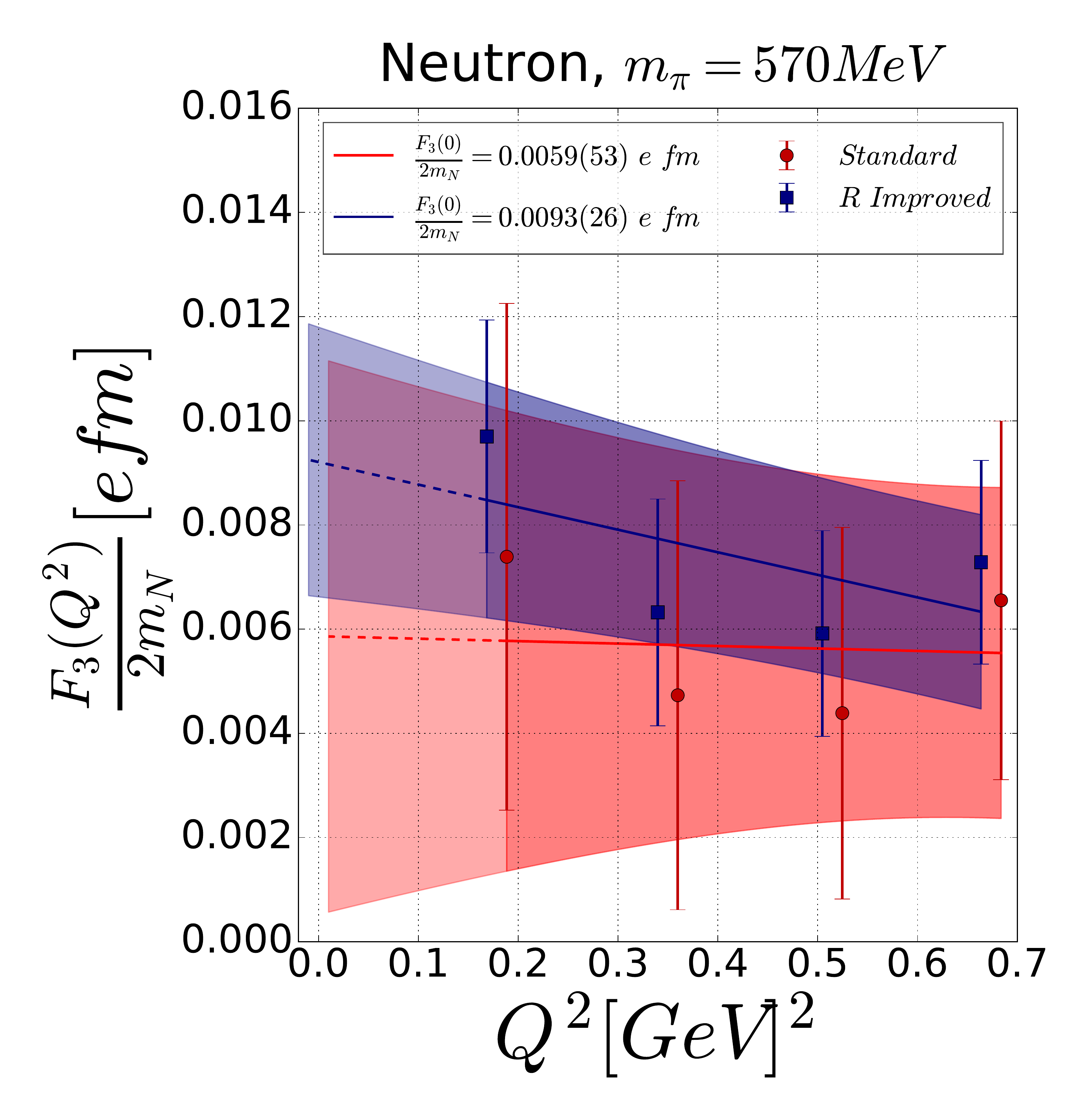}
  \caption{\label{fig:F3_Rimp_mpi570_proton}}
\end{subfigure}
\quad
\begin{subfigure}{.35\textwidth}
  \centering
  \includegraphics[trim={11mm 0cm 11mm 0cm},clip,width=\linewidth]{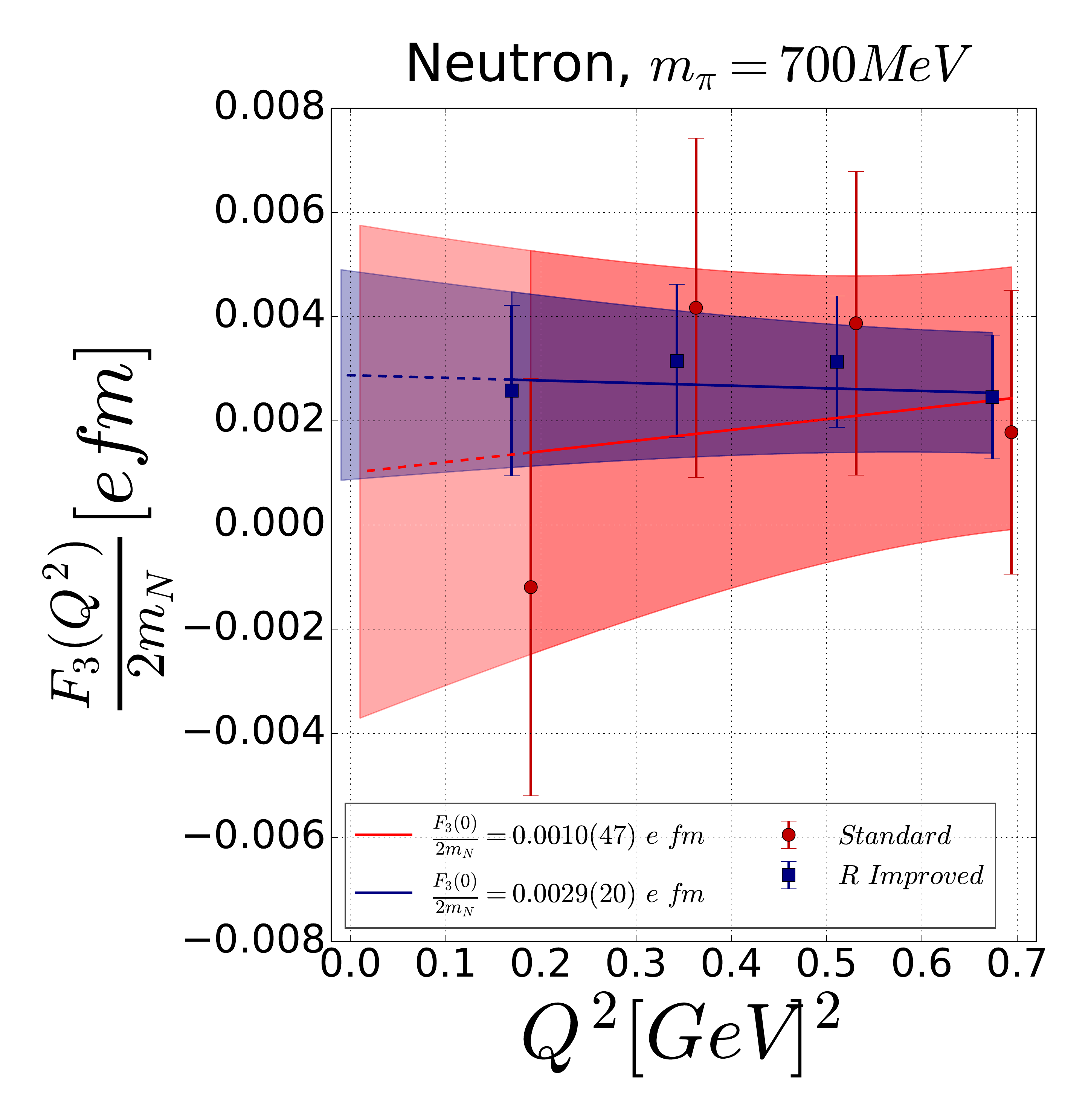}
  \caption{\label{fig:F3_Rimp_mpi700_proton}}
\end{subfigure}
\vspace*{\floatsep}
\begin{subfigure}{.35\textwidth}
  \centering
  \includegraphics[trim={11mm 0cm 11mm 0cm},clip,width=\linewidth]{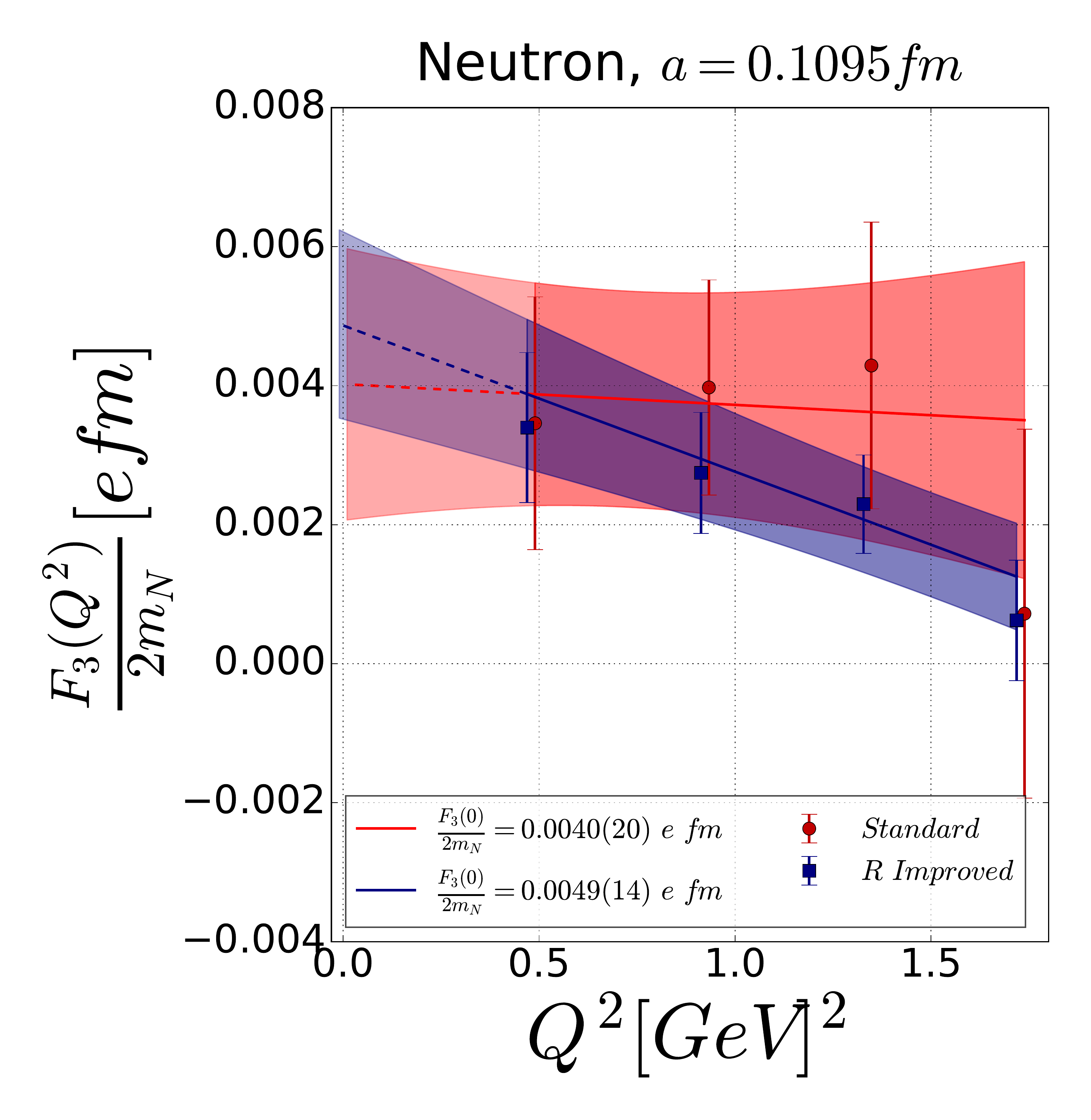}
  \caption{\label{fig:F3_Rimp_L16_proton}}
\end{subfigure}
\quad
\begin{subfigure}{.35\textwidth}
  \centering
  \includegraphics[trim={11mm 0cm 11mm 0cm},clip,width=\linewidth]{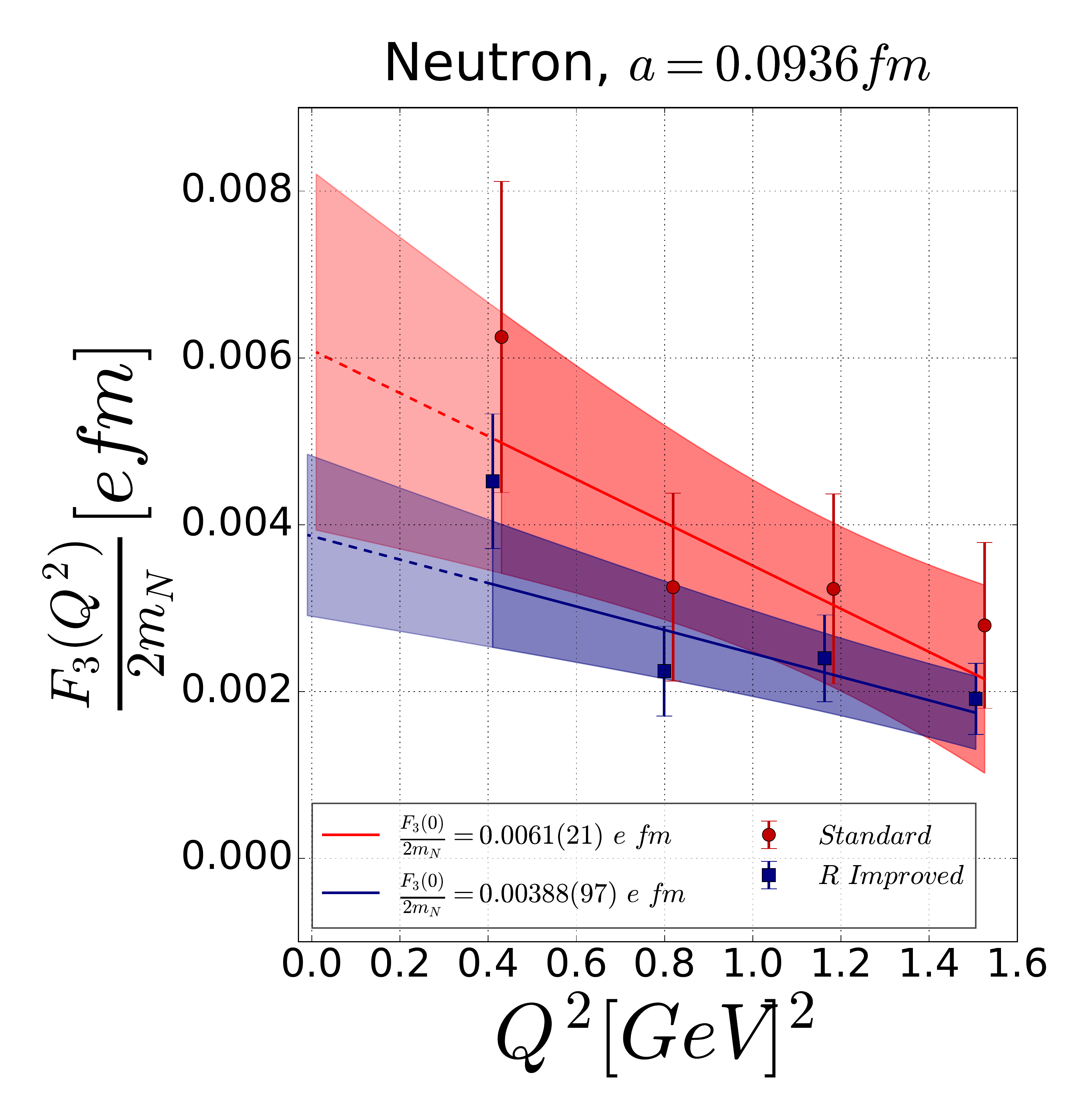}
  \caption{\label{fig:F3_Rimp_L20_proton}}
\end{subfigure}
\quad
\begin{subfigure}{.35\textwidth}
  \centering
  \includegraphics[trim={11mm 0cm 11mm 0cm},clip,width=\linewidth]{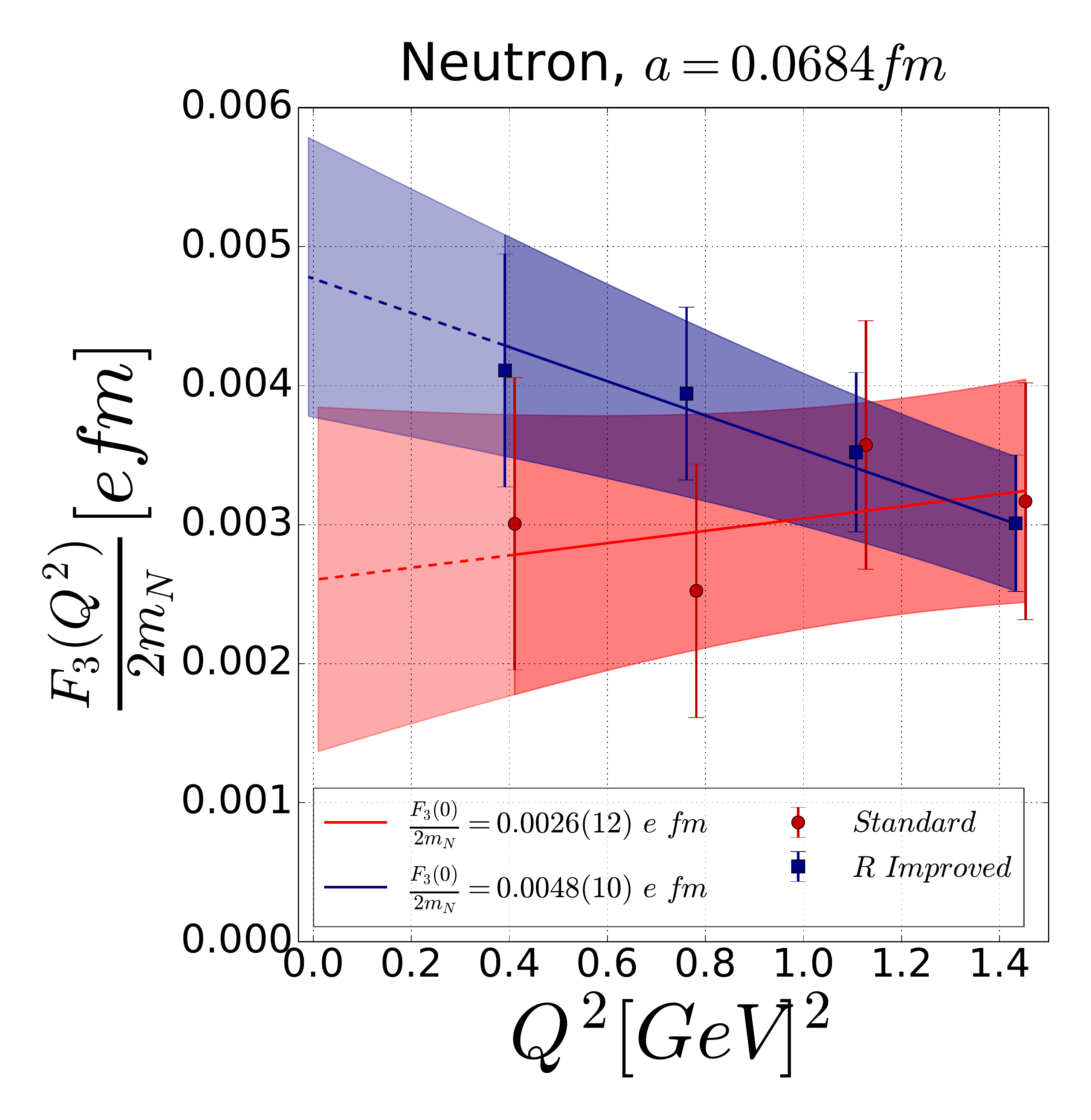}
  \caption{\label{fig:F3_Rimp_L28_proton}}
\end{subfigure}
\end{adjustwidth}
\caption{
  The neutron CP-odd form factor \(\frac{F_{3}(Q^{2})}{2M_{N}}\) results plotted
  against the transfer momentum \(Q^{2}\).
  The $m_{\pi}$ results (upper) were computed on $m_{\pi}=\{410,570,700\}$ MeV (left, middle and right) M-ensembles,
  and the lattice spacing results (lower) were computed at $a=\{0.1095,0.0936,0.0684\}$ fm (left, middle and right) A-ensembles.
  The form factors computed with the improved ratio functions (blue) is compared
  with the standard ratio functions (red). The bands are linear fits to the data,
  which are used to extrapolate to \(Q^{2}\rightarrow 0\) to determine the final EDM.
  Similar results are obtained for the proton.
  \label{fig:F3_Rimp_neutron}}
\end{figure}

In fig.~\ref{fig:F3_Rimp_neutron}, a comparison between
the improved ratio functions (blue) and the standard integrated topological charge (red) used in the
extraction of the neutron CP-violating form factor \(\frac{F_{3}(Q^{2})}{2M_{N}}\) is shown.
In all cases, a two-to-three times increase in the signal-to-noise is observed and all results are
statistically consistent\footnote{We have at most \(1.5~\sigma\) disagreement between the two methods at
\(Q^{2}\rightarrow 0\) for the \(a=0.0684\) fm ensemble.}.

\begin{table}
\caption{Fit ranges \([t_{s}^{min},\frac{T}{2}]\), over the symmetrically summed $\overbar{Q}$ time \(t_{s}\) and
resulting in the improved EDM determination \(\frac{F^{n}_{3}(Q^{2}\rightarrow 0)}{2M_{N}}\equiv \overbar{d}_{n}\), over the M-ensembles,
taken from fig.~\ref{fig:F3_Rimp_neutron}. The unimproved results \(d_{n}\) from tab.~\ref{tab:F3_Q_ext_mpi}
are included for comparison. The values determined at \(t_{s}^{min}\) differ by the fit results
at most by \(10\%\) of the error associated.
\label{tab:F3_Q_ext_mpi_Rimp}}
\centering
  \begin{tabular}{ r|c|c|c }
  ensemble &  \(m_{\pi}=410\) MeV &\(m_{\pi}=570\) MeV&\(m_{\pi}=700\) MeV \\
  \hline
  fit range & [6,32] & [7,32] & [4,32]  \\
  \hline
  fitr [fm] & [0.54,2.9] & [0.63,2.9] & [0.63,2.9]  \\
  \hline
  $d_{n}$ [$e$~fm] & -0.0045(26) & -0.0090(27) & -0.0027(20)\\
  \hline
  $\overbar{d}_{n}$ [$e$~fm] & -0.0035(66) & -0.0060(53) &-0.0009(47)\\
  \end{tabular}
\end{table}

\begin{table}
\caption{Fit ranges \([t_{s}^{min},\frac{T}{2}]\), over the symmetrically summed $\overbar{Q}$ time \(t_{s}\) and
resulting in the improved EDM determination \(\frac{F^{n}_{3}(Q^{2}\rightarrow 0)}{2M_{N}}\equiv \overbar{d}_{n}\), over the A-ensembles,
taken from fig.~\ref{fig:F3_Rimp_neutron}. The unimproved results \(d_{n}\) from tab.~\ref{tab:F3_Q_ext_latspace}
are included for comparison. The values determined at \(t_{s}^{min}\) differ by the fit results
at most by \(10\%\) of the error associated.
\label{tab:F3_Q_ext_latspace_Rimp}}
\centering
  \begin{tabular}{ r|c|c|c }
  ensemble &  \(a=0.1095\) fm  &\(a=0.0936\) fm &\(a=0.0684\) fm  \\
  \hline
  fit range & [3,16] & [4,20] & [10,28]  \\
  \hline
  fitr [fm] & [0.36,1.9] & [0.39,2.0] & [0.69,1.9]  \\
  \hline
  $d_{n}$ [$e$~fm] & -0.0048(13) & -.00393(97) & -0.0044(10) \\
  \hline
  $\overbar{d}_{n}$ [$e$~fm] & -0.0043(20) & -0.0063(20) & -0.0023(13) \\
  \end{tabular}
\end{table}

\subsection{Continuum Extrapolated Results with Improved Ratio Functions \label{sec:FF_Rat_imp}}
\begin{figure}
\begin{adjustwidth}{-0.1\textwidth}{-0.1\textwidth}
\centering
\begin{subfigure}{.55\textwidth}
  \centering
  \includegraphics[trim={11mm 0cm 0mm 0cm},clip,width=\linewidth]{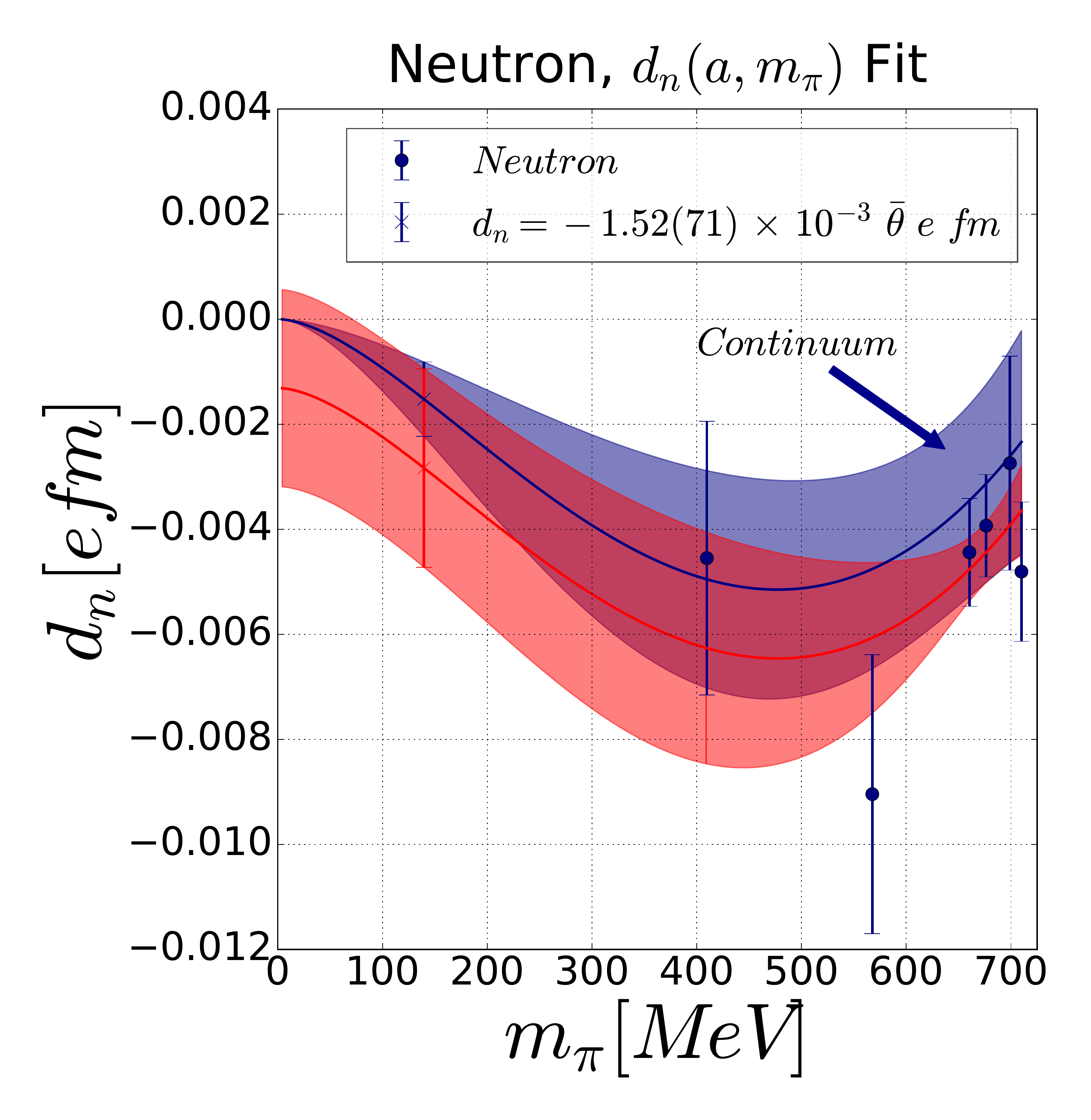}
  \caption{\label{fig:F3_Rimp_mpia_a9_extrap_neutron}}
\end{subfigure}
\qquad
\begin{subfigure}{.55\textwidth}
  \centering
  \includegraphics[trim={11mm 0cm 0mm 0cm},clip,width=\linewidth]{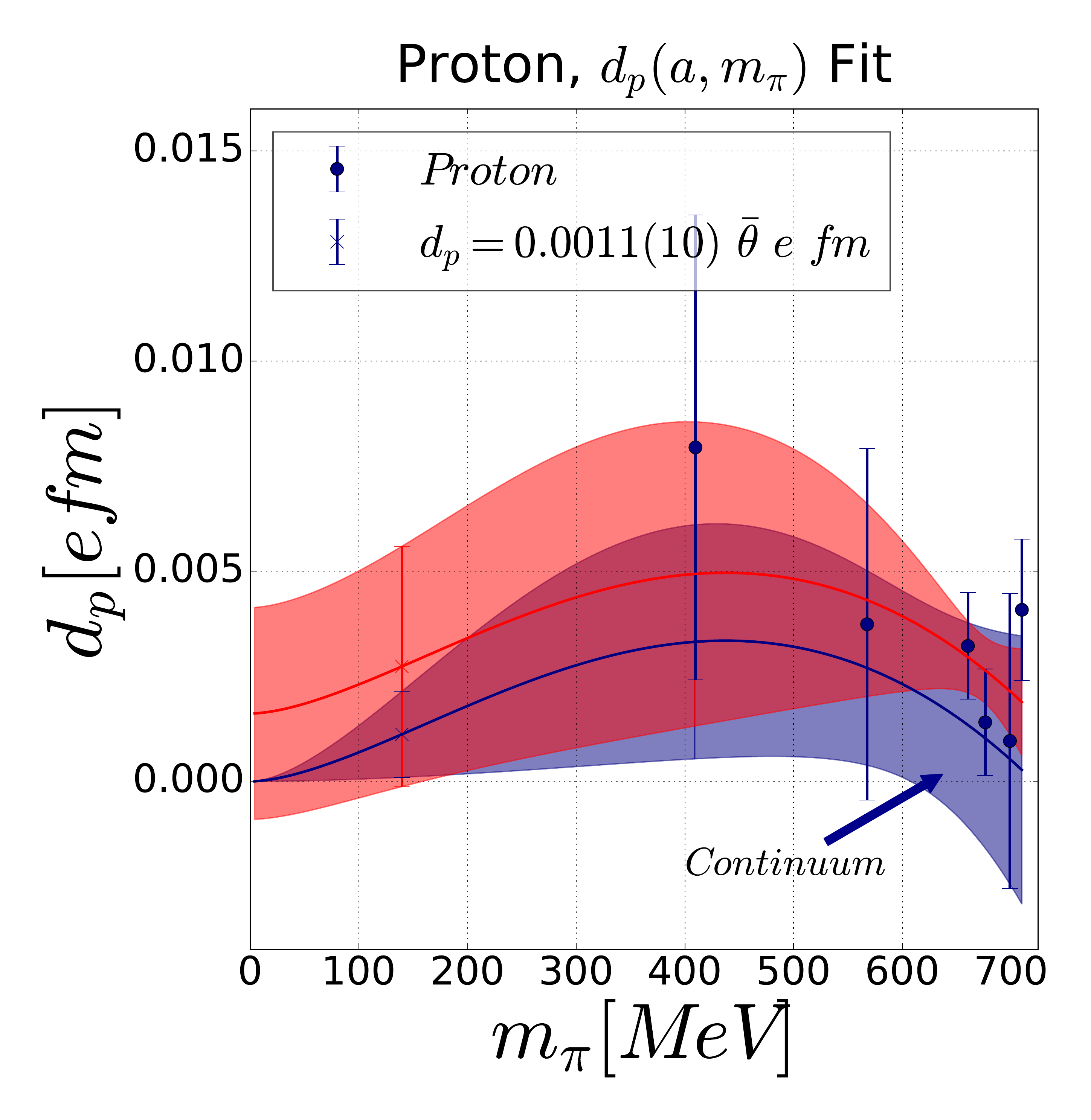}
  \caption{\label{fig:F3_Rimp_mpia_a9_extrap_proton}}
\end{subfigure}
\end{adjustwidth}
\caption{
  Determination of the EDM \(d_{p/n}\) for the neutron (left) and proton (right) for all 6 of our ensembles,
  plotted against their respective \(m_{\pi}\) values.
  The bands are the fits to all the ensembles using eq.~\eqref{eq:cont_fitfun},
  evaluated in the continuum \(a = 0\) (blue) and at \(a=0.0907\) fm (red) which coincides with the lattice spacing
  of the M-ensembles.
\label{fig:F3_Rimp_mpi_extrap}}
\end{figure}

\begin{figure}
\begin{adjustwidth}{-0.1\textwidth}{-0.1\textwidth}
\centering
\begin{subfigure}{.55\textwidth}
  \centering
  \includegraphics[trim={11mm 0cm 0mm 0cm},clip,width=\linewidth]{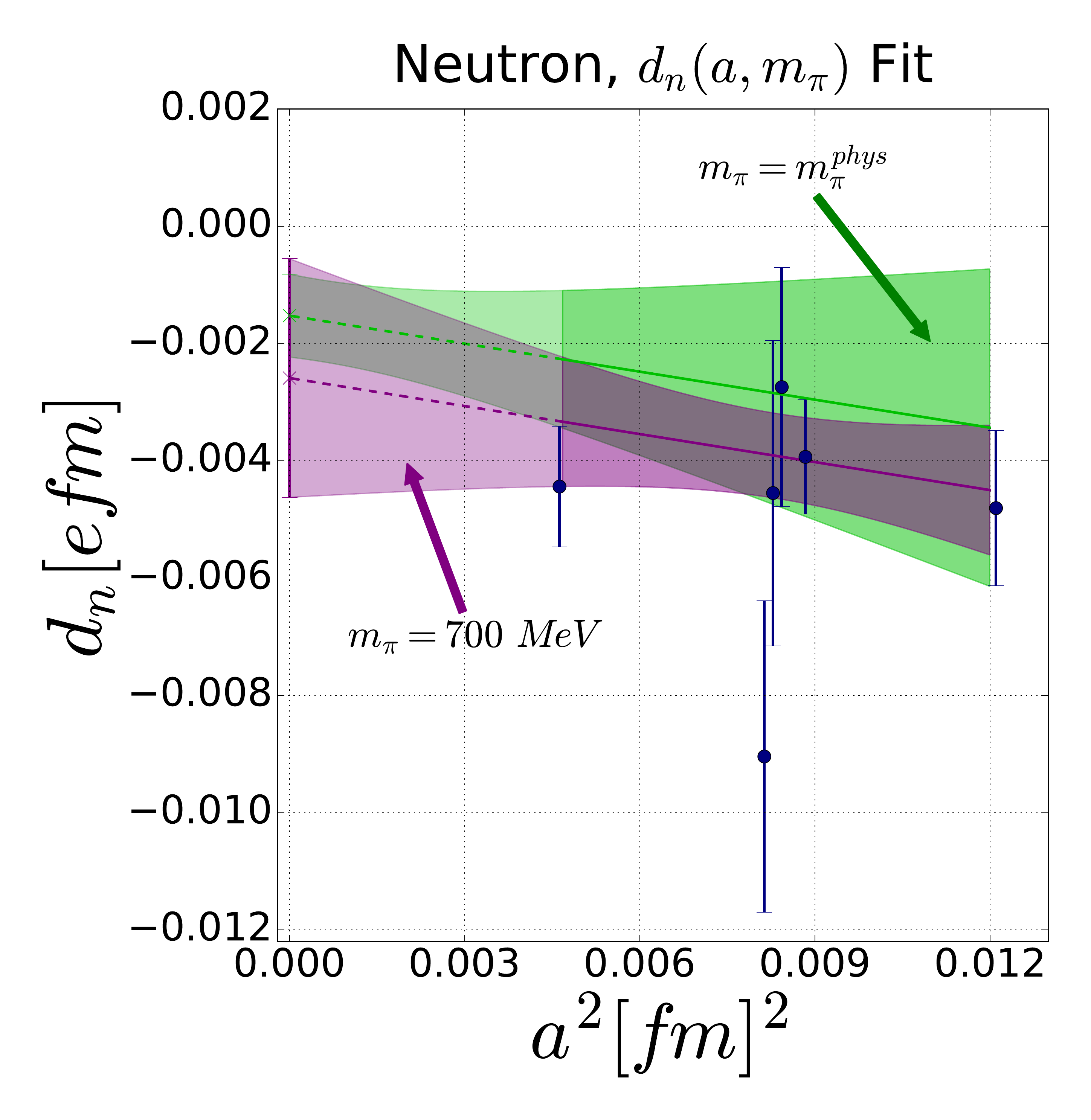}
  \caption{\label{fig:F3_mpia_latspace_extrap_neutron}}
\end{subfigure}
\qquad
\begin{subfigure}{.55\textwidth}
  \centering
  \includegraphics[trim={11mm 0cm 0mm 0cm},clip,width=\linewidth]{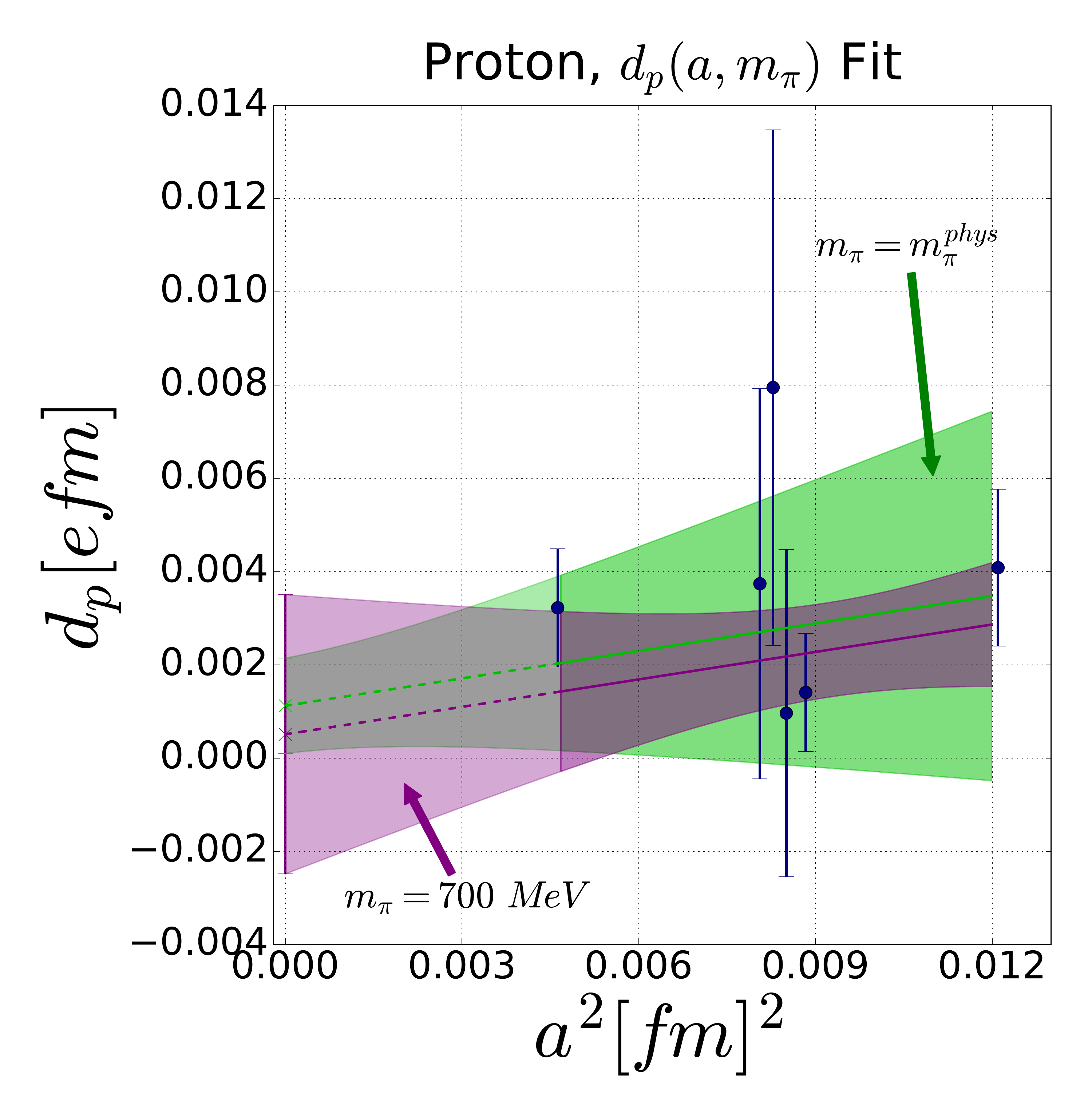}
  \caption{\label{fig:F3_mpia_latspace_extrap_proton}}
\end{subfigure}
\end{adjustwidth}
\caption{
  Determination of the EDM \(d_{p/n}\) for the neutron (left) and proton (right) for all 6 of our ensembles,
  plotted against their respective lattice spacing values.
  The bands are the fits to all the ensembles using eq.~\eqref{eq:cont_fitfun},
  evaluated at the physical point, \(m_{\pi}= m_{\pi}^{phys}\), (green) and in the
  chiral limit \(m_{\pi} = 700\)~MeV (purple).
\label{fig:F3_Rimp_a_extrap}}
\end{figure}

Armed with the improved results for the nucleon EDMs,
the next step entails the extrapolation to the physical pion mass
and the continuum limit.
From $\chi$PT we learn that the leading dependence of the nucleon EDMs
on the pion mass is given by \cite{Hockings:2005cn}
\begin{equation}\label{eq:chiral_fitfun}
  d_{p/n}(m_{\pi}) = C_1\ m_{\pi}^{2} + C_2\ m_{\pi}^{2}\log(\frac{m_{\pi}^{2}}{m_{N,phys}^{2}})\ ,
\end{equation}
where $C_1$ and $C_2$ are fit constants.
To account for the finite lattice spacing, we include an additional fit parameter, $C_3$,
\begin{equation}\label{eq:cont_fitfun}
  d_{p/n}(a,m_{\pi}) = C_1\ m_{\pi}^{2} + C_2\ m_{\pi}^{2}\log(\frac{m_{\pi}^{2}}{m_{N,phys}^{2}}) + C_3 a^{2}\,.
\end{equation}
The additional term ensures that the EDM only vanishes in the chiral limit
after taking the continuum limit.
We have performed a global fit with eq.~\eqref{eq:cont_fitfun}
taking into account our $6$ data points from ensembles A$_1$-A$_3$ and M$_1$-M$_3$.
In the four plots in figs.~\ref{fig:F3_Rimp_mpi_extrap},~\ref{fig:F3_Rimp_a_extrap}, we show the EDM results
 for the proton and neutron separately as function of the pion mass and lattice spacing.

Specifically, in Fig.~\ref{fig:F3_Rimp_mpi_extrap} we show the extraction of
 the neutron (left) and proton (right) EDM plotted against their
 \(m_{\pi}^{2}\) values (in MeV).
The blue band shows the extrapolation using the fit function in eq.~\eqref{eq:cont_fitfun},
 evaluated at $d_{p/n}(a=0,m_{\pi})$. This function evaluated at the physical pion mass is
 what we are interested in.
 In red we show the same extrapolation, where the fit is evaluated instead at
 $d_{p/n}(a=0.09\text{ fm},m_{\pi})$, to study the role of discretization errors. In particular, we observe
 an uncertainty of the EDMs at the physical pion mass that is roughly
 twice larger at $a=0.09$~fm.
 It is perhaps surprising that the uncertainty at the physical point
 reduces in the continuum limit. But the reason is clear.
 By fitting the nucleon EDMs to the fit function in eq.~\eqref{eq:cont_fitfun},
 the uncertainty on the fit parameters $C_1$ and $C_2$ is increased by the presence
 of the $C_3$ term. Now that the $a^2$ dependence is taken into account,
 we can perform an interpolation between the EDM in the chiral limit
 and the pion masses of our ensembles.
 In the continuum limit, $a=0$, the resulting nucleon
 EDM at the physical pion mass has now less uncertainty because
 $d_{n,p}(a=0,m_{\pi}=0)$ while $d_{n,p}(a>0,m_{\pi}=0)\neq 0$ and unconstrained.

The final continuum extrapolation values for the neutron and proton EDM are
\begin{eqnarray}\label{finalEDM}
  d_{n}(a=0,m_{\pi}=m_{\pi}^{phys})&=&-0.00152(71)\ \bar\theta\  e\text{~fm}, \nonumber\\
  d_{p}(a=0,m_{\pi}=m_{\pi}^{phys})&=&\phantom{-}0.0011(10)\ \bar\theta\  e\text{~fm},
\end{eqnarray}
and we include the determination for the fit parameters $C_1$, $C_2$, $C_3$
of eq.~\eqref{eq:cont_fitfun}, as well as the chi-squared per degree of freedom parameter,
$\chi^{2}_{PDF}$, in tab.~\ref{tab:EDM_fit_results}. The error on $\chi^{2}_{PDF}$ is determined
from the bootstrap samples distribution.
Since the correlators for the proton and the neutron EDM
are different, it is possible to obtain different relative uncertainties
in the two cases. It is not clear to us though, why we observe
a relative larger uncertainty for the proton than for the neutron.

In fig.~\ref{fig:F3_Rimp_a_extrap}  we show the dependence
of our EDM results on the lattice spacing $a$ for the neutron (left) and proton (right) EDM.
Overlaid on top, we have the evaluation of the fit function eq.~\eqref{eq:cont_fitfun}
at two different values of \(m_\pi\): $m_\pi = 700$ MeV (purple band) and $m_\pi = m_\pi^{phys}$ (green band).
The ensembles analyized in this work 
do not allow us to study mass-dependent discretization effects,
but we can still observe the impact of the chiral interpolation
on the continuum limit. The continuum extrapolation has less uncertainty,
thanks to the constraint that the EDM vanishes in the chiral limit.
Adding more ensembles to study mass dependence cutoff effects is certainly desirable,
but it does not change the main conclusion of this analysis.

We can extract a value of the CP-odd pion-nucleon LEC, $\bar{g}_{0}$,
which plays an important role in the EDMs of nuclei and diamagnetic atoms, by identifying
our result for fit parameter $C_2$ with the coefficient of the log term in eq.~\eqref{neutronEDM}
for the neutron EDM.
This gives the relation
\begin{equation}\label{C2}
\bar g_0 = - \frac{8\pi^2 f_\pi}{g_A} \frac{C_2 m_\pi^2}{e}\,,
\end{equation}
leading to the extraction
\begin{equation}
\bar g_0 = -12.8(6.4)\times 10^{-3}\,\bar \theta\,
\end{equation}
at the physical pion mass.
This result is in good agreement with the chiral perturbation
theory prediction in eq.~\eqref{eq:g0theta_eq}
and confirms the applicability of the fit function in eq.~\eqref{eq:cont_fitfun}.
A consistent result, with larger uncertainties
is obtained for the proton EDM (see tab.~\ref{tab:fit_edm}).


\begin{table}
\centering
\caption{Neutron and proton EDM fit parameters $C_{1}$, $C_{2}$, $C_{3}$ extracted from the combine fits to all
  6 ensembles using eq.~\eqref{eq:cont_fitfun}, as well as the resulting $\chi^{2}_{PDF}$.
  We also estimate \(\bar{g}_{0}\) using $C_2$ and eq.~\eqref{neutronEDM}.
\label{tab:EDM_fit_results}}
  \begin{tabular}{ r|c|c|c|c|c }
   &  $C_{1}$ $\left[\bar{\theta}\, e\text{~fm}^{3}\right]$ & $C_{2}$ $\left[\bar{\theta}\, e\text{~fm}^{3}\right]$ & $C_{3}$ $\left[\frac{\bar{\theta}\, e\text{~fm}}{\text{fm}^{2}}\right]$ & $\chi^{2}_{PDF}$
      &$\bar{g}_{0}^{\bar{\theta}}$ $\left[\bar{\theta}\right]$  \\
\hline
proton &   $-3.6(5.3) \times 10^{-4}$ &
            $-6.8(6.6) \times 10^{-4}$ &
            $0.20(31)$ &
            $2.0(1.4)$ &
            $-9.9(9.6) \times 10^{-3}$ \\
\hline
neutron  &  $3.1(3.2) \times 10^{-4}$ &
            $8.8(4.4) \times 10^{-4}$ &
            $-0.16(23)$ &
             $1.8(1.5)$ &
             $-12.8(6.4) \times 10^{-3}$ \\
  \end{tabular}
  \label{tab:fit_edm}
\end{table}

\subsection{Schiff Moment of the Proton and Neutron \label{sec:Schiff_res}}
Apart from the EDMs of the neutron and the proton, the nucleon electric dipole form factor (EDFF) contains additional information. The EDFF can be decomposed as
\begin{equation}
 \frac{F^{p/n}_{3}(Q^{2})}{2 M_N} = d_{p/n} - S_{p/n} Q^2 + H_{p/n}(Q^2)\,,
\end{equation}
where $d_{p/n}$ denotes the proton or neutron EDM, $S_{p/n}$ denotes the proton or neutron Schiff moments defined by ${S_{p/n} = (2 M_N)^{-1}(dF^{p/n}_{3}/dQ^2)|_{Q^2=0}}$, and $H_{p/n}$ are functions that capture the remaining $Q^2$ dependence. Chiral perturbation theory allows for a calculation of the Schiff Moments and the $H_{p/n}$ functions
from the analogous isovector and isoscalar quantities \cite{Hockings:2005cn}. At leading order in the chiral expansion the nucleon EDMs are given in eq.~\eqref{neutronEDM}. The leading-order Schiff moments are isovector and given by
\begin{eqnarray}
S_p=- S_n= -\frac{eg_A\bar{g}_0}{48 \pi^2 F_{\pi} m_{\pi}^2} =  1.7(3)\times 10^{-4}\,\tb\,e\,\mathrm{fm}^3\,,
\label{eq:S_radius}
\end{eqnarray}
where we have used eq.~\eqref{eq:g0theta_eq}. NLO corrections
have been calculated in ref.~\cite{Mereghetti:2010kp}
and reduce the leading-order result by roughly $50\%$
and provide a tiny contribution to $S_n + S_p = \mathcal O(10^{-5}\,\tb\,e\,\mathrm{fm}^3)$.
$\chi$PT thus predicts that the neutron and proton Schiff
moments are equal in magnitude but with opposite sign.
The leading-order $H_{p/n}$ are also isovector and given by
\begin{eqnarray}
H_p=- H_n=-\frac{eg_{A}\bar{g}_{0}}{30 \pi^2 F_{\pi} }
               \left[h_1^{(0)}\left(\frac{Q^{2}}{4m_{\pi}^{2}}\right) \right]\,,
\label{eq:H_radius}
\end{eqnarray}
where
\begin{equation}\label{eq:h0_fun}
h_1^{(0)}(x)=-\frac{15}{4}\left[
        \sqrt{1+\frac{1}{x}} \;
        \ln{\left(\frac{\sqrt{1+1/x}+1}{\sqrt{1+1/x}-1}\right)}
        -2\left(1+\frac{x}{3}\right)\right]\,.
\end{equation}

In the limit $Q^2 \ll m_\pi^2$ the nucleon EDFFs become
\begin{eqnarray}
 \frac{F^{p}_{3}(Q\ll m_\pi)}{2 M_N} &=& d_p + \frac{eg_A\bar{g}_0}{48 \pi^2 F_{\pi}}  \left(\frac{Q^2}{m_{\pi}^2}+ \dots \right) \,,\\
 \frac{F^{n}_{3}(Q\ll m_\pi)}{2 M_N}  &=&  d_n - \frac{eg_A\bar{g}_0}{48 \pi^2 F_{\pi}}  \left(\frac{Q^2}{m_{\pi}^2}+ \dots \right) \,,
\label{EDFFQ}
\end{eqnarray}
such that the Schiff moments provide the dominant $Q^2$ dependence of the EDFFs.  The nucleon EDMs and the LEC $\bar g_0$ are induced by the $\bar \theta$ term and scale as $d_{p/n} \sim \bar g_0 \sim \bar m_* \bar \theta \sim m_\pi^2 \bar \theta$. As such, the Schiff moments scale as $S_{p/n} \sim \bar g_0/m_\pi^2$ which is pion mass independent.
This statement is potentially confusing as we infer from eq.~\eqref{QCD2} that the $\bar \theta$ term decouples in the chiral limit and the whole nucleon EDFF should vanish. eq.~\eqref{EDFFQ}, however, requires $Q^2 \ll m_\pi^2$. In the opposite limit, we obtain
\begin{eqnarray}\label{eq:Qlargem}
 \frac{F^{p}_{3}(Q\gg m_\pi)}{2 M_N} &=& d_p - \frac{eg_A\bar{g}_0}{8 \pi^2 F_{\pi}}  \left(2 + \log \frac{m_\pi^2}{Q^2} \right)\,,\\
 \frac{F^{n}_{3}(Q\gg m_\pi)}{2 M_N}  &=& d_n + \frac{eg_A\bar{g}_0}{8 \pi^2 F_{\pi}}  \left(2 + \log \frac{m_\pi^2}{Q^2} \right)\,,
\end{eqnarray}
and the EDFFs vanish in the chiral limit as expected.

The goal is to extract $S_{p/n}$ from our lattice data as this allows for a direct comparison to the $\chi$PT prediction in eq.~\eqref{eq:S_radius} and the extraction in the previous section based on the pion mass dependence of the nucleon EDMs. To extract $S_{p/n}$, we first extrapolate our results to small $Q^2$ by fitting the EDFF to the function
\begin{equation}\label{eq:EDFF_fit}
  \frac{F^{p/n}_{3}(Q^{2},m_{\pi}^{2},a^{2})}{2M_{N}} = d_{p/n}(m_{\pi}^{2},a^{2}) - S_{p/n}(m_{\pi}^{2},a^{2})
  \left[
  Q^{2}-\frac{8m_{\pi}^{2}}{5}h_{1}^{(0)}\left(\frac{Q^{2}}{4m_{\pi}^{2}}\right)
  \right]\,.
\end{equation}
The effects of the $h_{1}^{(0)}$ function turns out to have minimal impact on the extraction of $d_{p/n}(m_{\pi}^{2},a^{2})$ and
$S_{p/n}(m_{\pi}^{2},a^{2})$, and we obtain similar results if we use the fit function
\begin{equation}
  \frac{F^{p/n}_{3}(Q^{2},m_{\pi}^{2},a^{2})}{2M_{N}} = d_{p/n}(m_{\pi}^{2},a^{2}) - S_{p/n}(m_{\pi}^{2},a^{2})Q^{2}\,.
\end{equation}
This shows that our results are not precise enough to isolate the more subtle $Q^2$ behavior.

\begin{figure}
\begin{adjustwidth}{-0.1\textwidth}{-0.1\textwidth}
\centering
\begin{subfigure}{.55\textwidth}
  \centering
  \includegraphics[trim={11mm 0cm 9mm 0cm},clip,width=\linewidth]{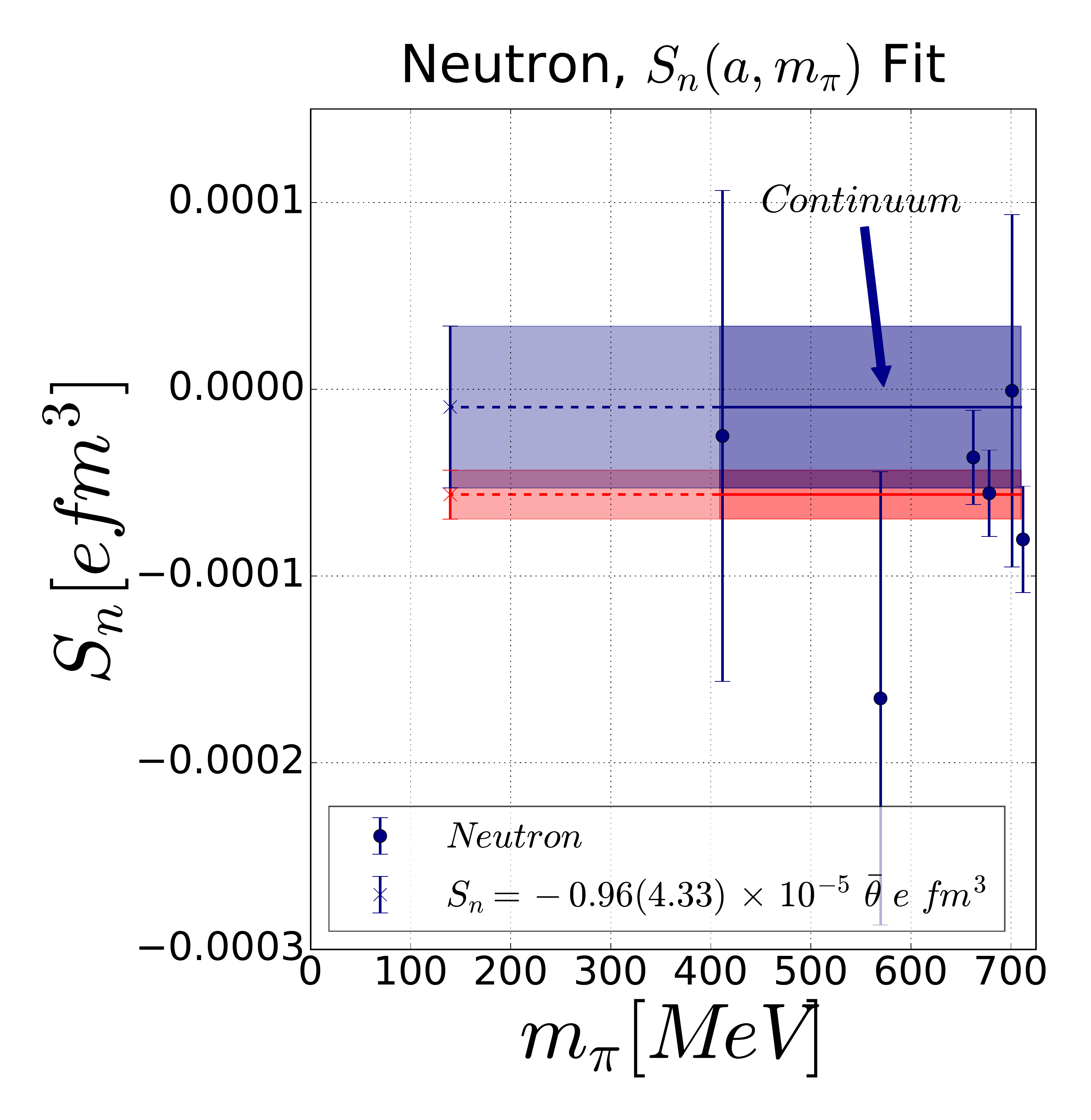}
  \caption{\label{fig:Schiff_mpi_neutron}}
\end{subfigure}
\quad
\begin{subfigure}{.55\textwidth}
  \centering
  \includegraphics[trim={11mm 0cm 9mm 0cm},clip,width=\linewidth]{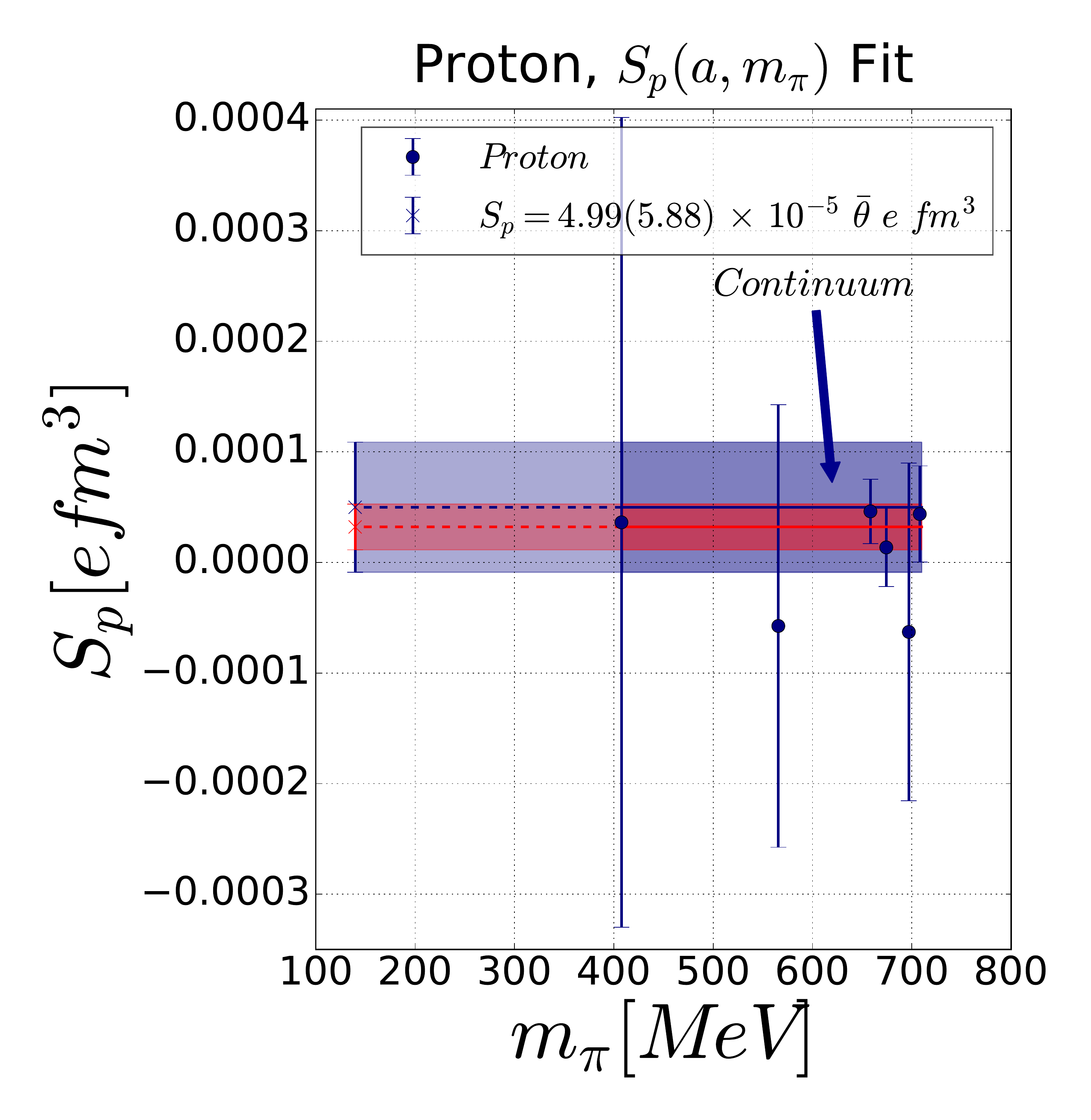}
  \caption{\label{fig:Schiff_mpi_proton}}
\end{subfigure}
\end{adjustwidth}
\caption{\label{fig:Schiff_mpi}
  Determination of the Schiff moment \(S_{p/n}\) for the neutron (left) and proton (right) for all 6 of our ensembles,
  plotted against their respective \(m_{\pi}\) values.
  The bands are the fits to all the ensembles using eq.~\eqref{eq:schiff_fit},
  evaluated in the continuum limit \(a=0\) (blue) and the lattice spacing corresponding to the
  M-ensembles (red).
  }
\end{figure}

\begin{figure}
\begin{adjustwidth}{-0.1\textwidth}{-0.1\textwidth}
\centering
\begin{subfigure}{.55\textwidth}
  \centering
  \includegraphics[trim={11mm 0cm 9mm 0cm},clip,width=\linewidth]{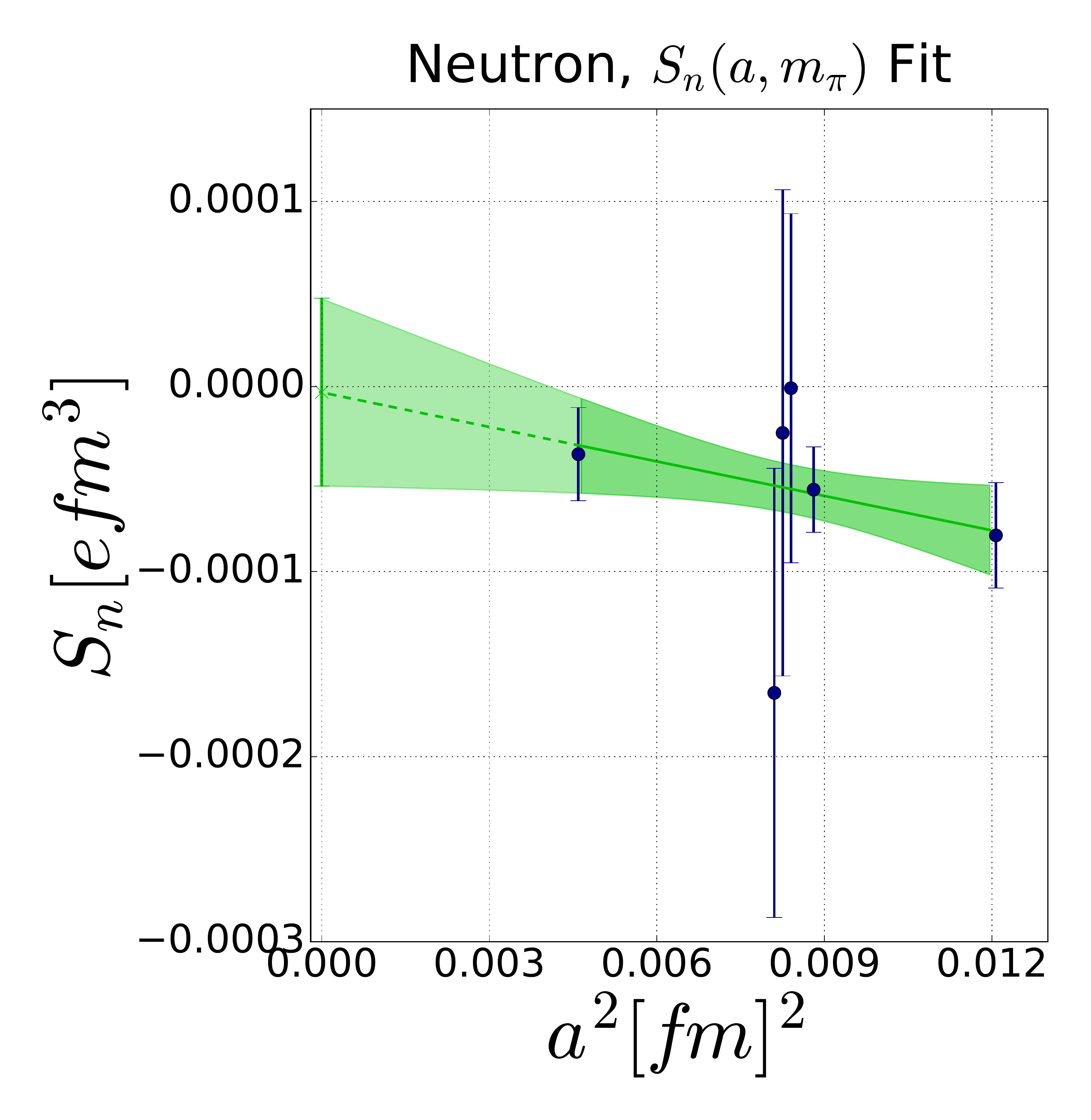}
  \caption{\label{fig:Schiff_a_neutron}}
\end{subfigure}
\quad
\begin{subfigure}{.55\textwidth}
  \centering
  \includegraphics[trim={11mm 0cm 9mm 0cm},clip,width=\linewidth]{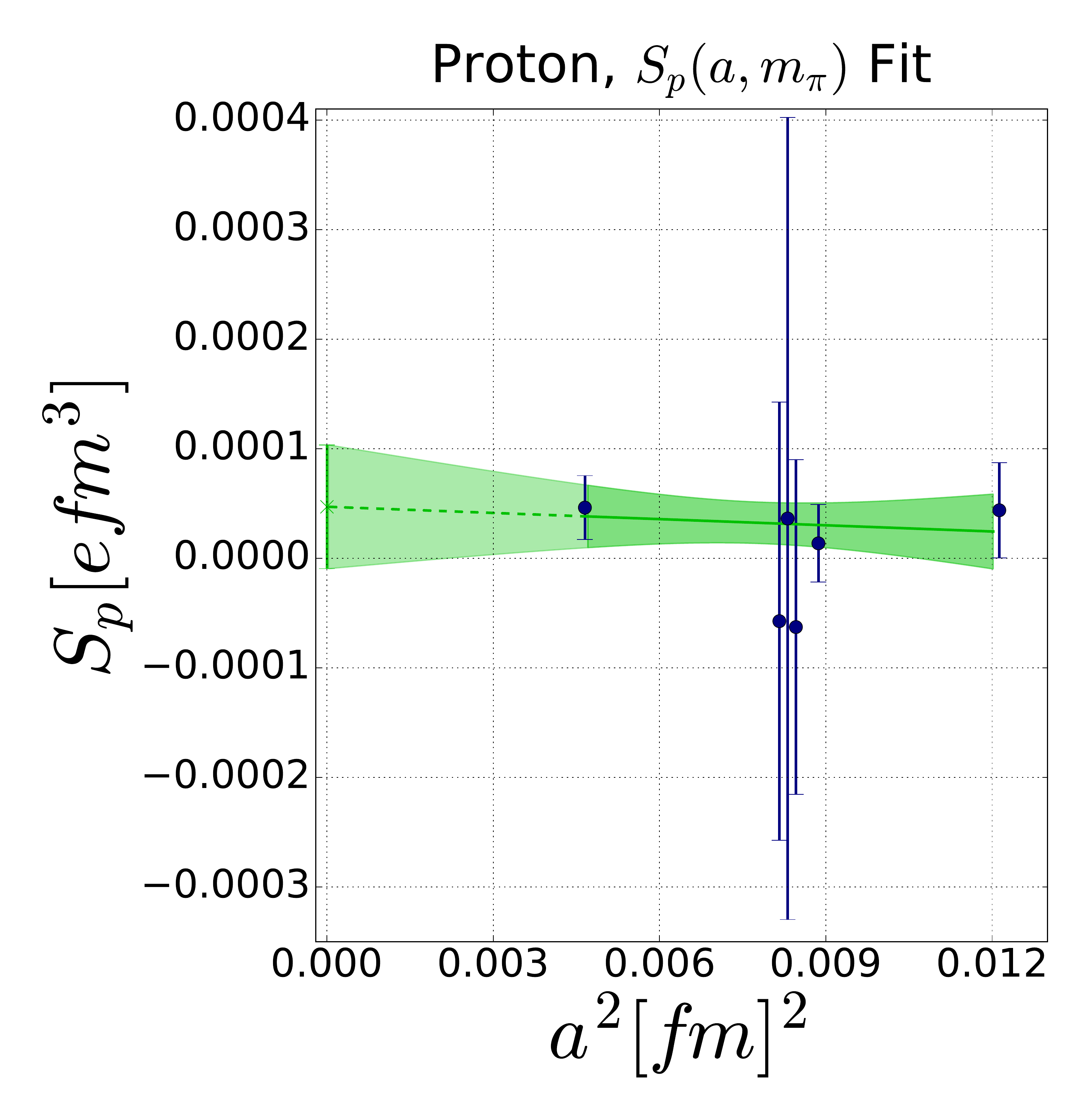}
  \caption{\label{fig:Schiff_a_proton}}
\end{subfigure}
\end{adjustwidth}
\caption{\label{fig:Schiff_a}
  Determination of the Schiff moment \(S_{p/n}\) for the neutron (left) and proton (right) for all 6 of our ensembles,
  plotted against their respective lattice spacing values.
  The green band is the lattice spacing dependence of the fit to all the ensembles using eq.~\eqref{eq:schiff_fit}.
  }
\end{figure}

Once we have obtained $S_{p/n}(m_{\pi}^{2},a^{2})$  we can extrapolate to the continuum limit and the physical pion mass. LO $\chi$PT predicts no dependence on the pion mass, and, having an $\mathcal O(a)$ improved lattice action, we add a quadratic dependence on the lattice spacing $a$
\begin{equation}\label{eq:schiff_fit}
  S_{p/n}(m_{\pi}^{2},a^{2}) = C_4 + C_5 a^{2},
\end{equation}
with $C_4$ and $C_5$ fit constants.
The results for the Schiff moments along with the
continuum extrapolation are shown in figs.~\ref{fig:Schiff_mpi},~\ref{fig:Schiff_a}.
In fig.~\ref{fig:Schiff_a} we show the fit results with the $a^2$ dependence. We observe
minimal discretization effects over the range \(a=\{0\rightarrow 0.12\}\) fm.
In fig.~\ref{fig:Schiff_mpi} we show the fit results as a function of the pion mass $m_\pi$.
At $a=0$ we perform a constant fit in the pion mass,
to obtain the proton and neutron Schiff moments at the physical point.
We do not extrapolate to the chiral limit because the $\chi$PT
prediction that $S_{p/n}$ are pion-mass independent
will break down at some point as inferred from eq.~\eqref{eq:Qlargem}.
We obtain for the Schiff moments at the physical point
\begin{eqnarray}
  S_p &=& \phantom{-}0.50(59) \times 10^{-4}\,\tb\,e\,\mathrm{fm}^3\,,\\
  S_n &=& -0.10(43) \times 10^{-4}\,\tb\,e\,\mathrm{fm}^3\,,
\end{eqnarray}
as well as the fit parameters \(C_4 = S_{p/n}\) and \(C_5\)
from performing this fit in tab.~\ref{tab:Schiff_fit_results}.
The uncertainties are significant and the magnitudes are somewhat below the LO $\chi$PT predictions in eq.~\eqref{eq:S_radius}, but in better agreement once $\chi$PT NLO corrections are included. There is some evidence for a dominantly isovector Schiff moment as predicted from $\chi$PT, but the uncertainties are too large to make strong statements. We perform a sanity check of our result by comparing the ChPT predictions for the fit coefficient $C_2$ and $C_4$. From eqs.~\eqref{C2} and \eqref{eq:S_radius}, we infer the LO ChPT prediction
\begin{equation}
\frac{C_2}{C_4} = -6\,.
\end{equation}
Our fit values for this ratio are given in tab.~\ref{tab:Schiff_fit_results}, and agree with this prediction within (large) statistical errors.


\begin{table}
\centering
\caption{Neutron and proton Schiff fit parameters $C_{4}$, $C_{5}$ extracted from the combine fits to all
  6 ensembles using eq.~\eqref{eq:schiff_fit}, as well as the resulting $\chi^{2}_{PDF}$.
  We additionally include the ratio $\frac{C_2}{C_4}$, where
$C_2$ is the second fit parameter result from tab.~\ref{tab:EDM_fit_results}.
\label{tab:Schiff_fit_results}}
  \begin{tabular}{ r|c|c|c|c }
   &  $C_{4}$ $\left[\bar{\theta}\, e\text{~fm}^{3}\right]$ & $C_{5}$ $\left[\frac{\bar{\theta}\, e\text{~fm}^{3}}{\text{fm}^{2}}\right]$ & $\chi^{2}_{PDF}$
   &$\frac{C_{2}}{C_{4}}\approx -6$\\
  \hline
proton &                        $0.50(59) \times 10^{-4}$ &                                     $-0.0022(73)$ &              $1.25(80)$ & $-20(200)$\\
\hline
neutron  &                       $-0.10(43) \times 10^{-4}$ &                                     $-0.0057(51)$ &              $1.37(97)$ & $70(970)$ \\
\end{tabular}
\end{table}


\section{Discussion \label{sect:discussion}}
In this section, we discuss the EDM and Schiff moment results for the neutron and proton.
The succeeding Section~\ref{sec:comp} compares our determination of the
EDM to previous lattice QCD EDM computations.
Then following in Section~\ref{sect:ram}, the phenomenological
ramifications of our results for the EDM and Schiff moments
are discussed.

\subsection{Comparison with other works\label{sec:comp}}

\begin{figure}
\centering
\begin{subfigure}{\textwidth}
  \centering
  \includegraphics[trim={11mm 0cm 11mm 0cm},clip,width=\linewidth]{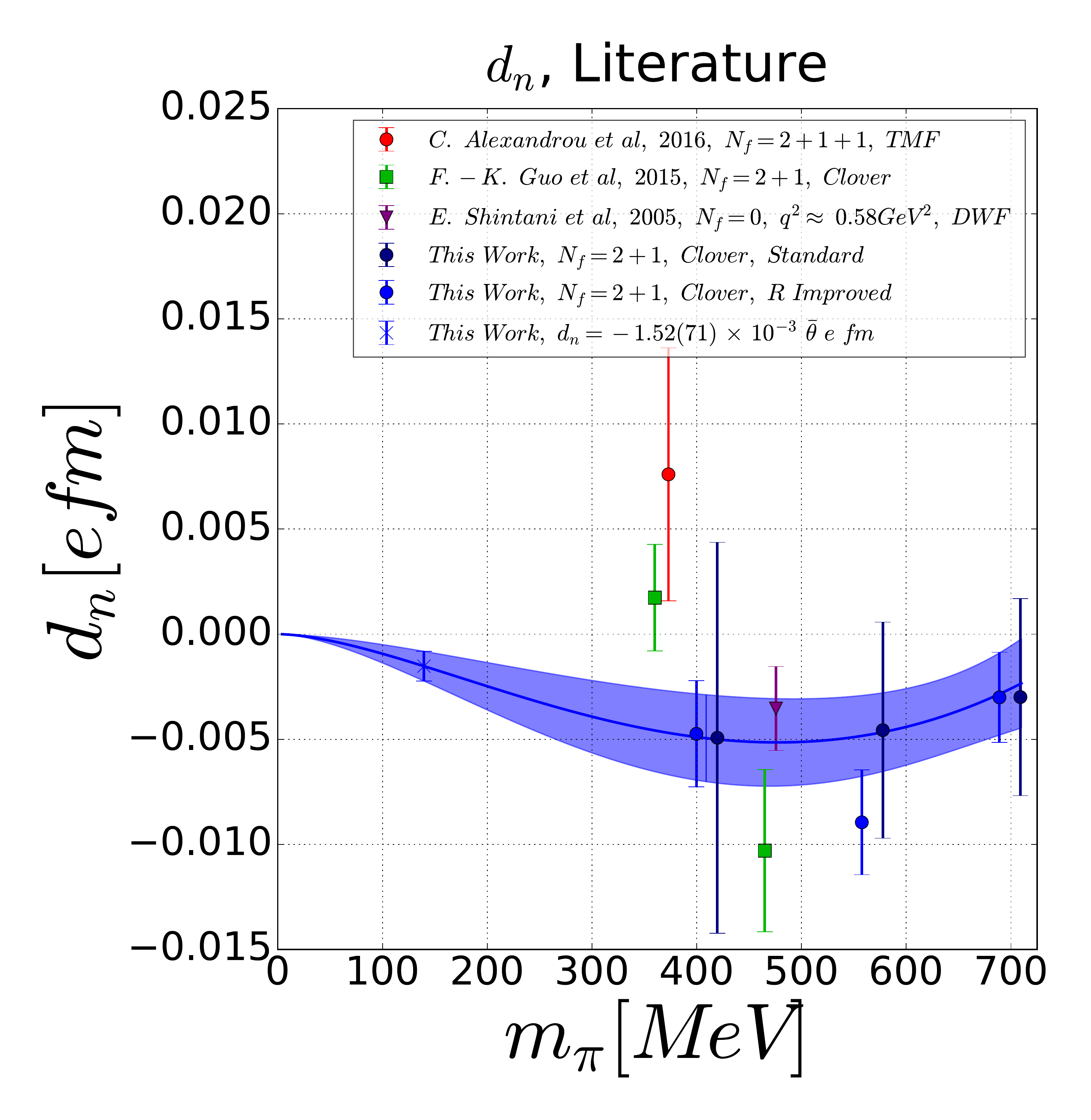}
\end{subfigure}
\caption{\label{fig:F3_comp_Rot}
  The results of \(d_{n}\) from this paper, improved in light blue and not-improved in dark blue,
  compared to other lattice QCD results~\cite{Alexandrou:2015spa,Guo:2015tla,Shintani:2005xg}.
  The light blue bands correspond to a chiral extrapolation, using the improved data (light blue),
  after having performed the continuum limit as described in sec.~\ref{sec:FF_Rat_imp}.
  To perform the rotation of the CP odd form factor \(F_3\) of other lattice calculations see the main text.
  We underline that our error determination for other works is purely
  illustrative since, not having at our disposal the raw data,
  we do not take into account correlations in the data.}
\end{figure}

In this section, we compare the results obtained
for the neutron EDM \(d_{n}\) with few lattice QCD results
from the literature. As noted in \cite{Abramczyk:2017oxr},
it is sometimes problematic to compare different EDM calculations,
as most results preceding this paper do not consider the rotation of the CP-odd form factor
\(F_{3}\) with \(F_{2}\) and \(\alpha_N\)
computed on the lattice.~\footnote{We note that no general consensus
has been reached about the need to perform this rotation of the form factors.}
The EDM \(d_{n}\) (rotated to \(F_{3}\)) is shown in Fig.~\ref{fig:F3_comp_Rot}.
Good agreement is seen from our results to the others \cite{Shintani:2005xg,Guo:2015tla}
at \(m_{\pi} \approx 475\) MeV,
but we see a slight tension between the results of \cite{Alexandrou:2015spa,Guo:2015tla}
and our results at \(m_\pi \approx 350\) MeV.
It must be stressed that the rotation requires knowledge
of the phase $\alpha_N$ and the unrotated form factors $F_2$ and $F_3$
which are not always easy to extract.
To rotate the ``\textit{C. Alexandrou et al, 2016}'' \cite{Alexandrou:2015spa} results,
an estimation of \(F_{2}\) was determined from \cite{Abdel-Rehim:2015owa}.
To rotate the ``\textit{F.-K. Guo et al, 2015}'' \cite{Guo:2015tla} results,
\(F_{2}\) was determined from \cite{Shanahan:2014uka} (at \(\bar{\theta} = 0\))
and \(\alpha_{N}\) and \(F_{3}\) estimated
via a linear+cubic fit in \(\bar{\theta}\)
performed by \cite{Abramczyk:2017oxr}.

In particular, the lattice results for \(F_{3}\)
not obtained in this work do not take into account correlations between
\(F_{2}\), \(F_{3}\) and \(\alpha_N\).
As such, Fig.~\ref{fig:F3_comp_Rot} is mainly shown for illustrative
purposes and the error estimates for results not obtained in this work
should be taken with a grain of salt.

\subsection{Impact on EDMs of light nuclei\label{sect:ram}}
Armed with a non-perturbative determination of the nucleon EDMs as a function of $\bar \theta$
we can revisit EDMs of systems with more than a single nucleon.
EDM experiments so far have mainly focused on neutral systems, but EDMs of charged particles can be probed if the particles are trapped in electromagnetic storage rings  \cite{Farley:2003wt}. This technique has lead to a direct limit on the EDM of the muon \cite{Bennett:2008dy}, and to plans to pursue EDM measurements of protons and light nuclei in dedicated storage rings. Such measurements are still far away but impressive progress has been reported in refs.~\cite{Eversmann:2015jnk, Guidoboni:2016bdn}. EDMs of light nuclei have been calculated using chiral EFT \cite{deVries:2011an, Bsaisou:2014zwa}
\begin{eqnarray} \label{eq:h2edm}
d_{{}^2\text{H}} &=&
0.94(1)(d_n + d_p) + \bigl [ 0.18(2) \,\bar g_1\bigr] \,e \,{\rm fm} \, ,\\
d_{{}^3\text{H}} &=& -0.03(1)d_n + 0.92(1) d_p  -\left[ 0.11(1) \bar g_0 - 0.14(2) \bar g_1\right] \,e  \,{\rm fm} \, \\
d_{{}^3\text{He}} &=& 0.90(1)d_n -0.03(1) d_p + \left[ 0.11(1) \bar g_0 + 0.14(2) \bar g_1\right] \,e  \,{\rm fm} \, .
\end{eqnarray}
in terms of the EDMs of nucleons and the CPV pion-nucleon coupling constants $\bar g_0$ and $\bar g_1$ associated to the interactions
\begin{equation}
\mathcal L_{\pi N}(\bar\theta) = \bar g_0\, \bar N \vec \pi \cdot \vec \tau N + \bar g_1\,\bar N \pi_3 N\,.
\end{equation}
Values of $\bar g_0$ and $\bar g_1$ can be obtained from chiral-symmetry arguments \cite{Bsaisou:2012rg,deVries:2015una} that link these LECs to the hadron spectrum
\begin{eqnarray}\label{g0theta}
\bar g_0 &=& -14.7(2.3)\times 10^{-3}\,\bar \theta\,,\nonumber\\
\bar g_1 &=&  \phantom{-1}3.4(2.4)\times 10^{-3}\,\bar \theta\,,
\end{eqnarray}
and the smallness of $\bar g_1/\bar g_0$ is due to approximate isospin symmetry.

In absence of direct lattice calculations of $d_n$ and $d_p$
we could only predict values for the combinations
\begin{eqnarray}
d_{{}^2\mathrm{H}}(\bar\theta)-d_n(\bar\theta) -d_p(\bar\theta) &=&\phantom{-}0.6 (4)\times 10^{-3}\,\tb\,e\,\mathrm{fm}\,,\nonumber\\
d_{{}^3\mathrm{H}}(\bar\theta)-0.9\,d_p(\bar\theta)&=& \phantom{-}2.1(5)\times 10^{-3}\,\tb\,e\,\mathrm{fm}\,,\nonumber\\
d_{{}^3\mathrm{He}}(\bar\theta)-0.9\,d_n(\bar\theta)&=&  -1.1(5)\times 10^{-3}\,\tb\,e\,\mathrm{fm}\,.
\end{eqnarray}
But with our lattice determination
of $d_n(\bar\theta)$ and $d_p(\bar\theta)$, we can now estimate
the EDMs of light ions directly in terms of $\bar \theta$
\begin{eqnarray}\label{pattern}
d_{{}^2\mathrm{H}}(\bar\theta) &=& \phantom{-}0.2(1.2)\times 10^{-3}\,\tb\,e\,\mathrm{fm}\, ,\nonumber\\
d_{{}^3\mathrm{H}}(\bar\theta)&=& \phantom{-}3.2(1.0)\times 10^{-3}\,\tb\,e\,\mathrm{fm}\, ,\nonumber\\
d_{{}^3\mathrm{He}}(\bar\theta)&=&  -2.5(0.8)\times 10^{-3}\,\tb\,e\,\mathrm{fm}\,.
\end{eqnarray}
Due the dependence on the isoscalar nucleon EDM, $d_n+d_p$, the deuteron EDM is still very uncertain. The tri-nucleon EDMs however are predicted more than three standard deviations from zero and with a fixed sign. The total uncertainty arises in roughly equal amounts from uncertainties in eq.~\eqref{g0theta} and in the determination of the nucleon EDMs in eq.~\eqref{finalEDM}.
If nonzero EDMs are measured these relations can be used to differentiate between the SM theta term and BSM sources of CP violation. They can also provide indirect evidence for the existence of a Peccei-Quinn mechanism, by finding EDM patterns in disagreement with eq.~\eqref{pattern} \cite{deVries:2018mgf}.


\section{Conclusion\label{sect:conclusion}}

In this paper we computed the proton and neutron EDM from dynamical
lattice QCD using various pion masses at different lattice spacings and volumes,
as enumerated in tabs.~\ref{tab:latpar},~\ref{tab:latpar2}.
We found our results have rather small (within our statistical uncertainties)
discretization effects, which greatly simplified our continuum limit extrapolations.
We found satisfactory agreement with existing results,
as discussed in \autoref{sect:discussion}.
With our measurements at multiple pion masses, we performed a
chiral~\emph{interpolation} to obtain, at the physical pion mass and in the continuum limit
\({d_{n}=-0.00152(71)\ \bar\theta\  e\text{~fm}}\) and
\({d_{p}=0.0011(10)\ \bar\theta\  e\text{~fm}}\).
The nonzero result for the neutron EDM 
confirms the existence of the strong CP problem at two standard deviations and
limits $\left|\bar \theta\right| < 1.98\times 10^{-10}$ at $90\%$ C.L. The dependence of the nucleon EDMs on the pion mass allowed us to extract of the CP-odd pion-nucleon coupling, $\bar g_0$. The resulting value is in good agreement with chiral perturbation theory.

Important to our analysis was the implementation of the
gradient flow on the topological charge. In fact we can perform the continuum
limit at fixed flow time with no need to calculate the normalization
of the topological charge.
As we have discussed and documented in \autoref{sec:twopt} and ensuing subsections,
the gradient flow also allows a more robust determination
of the integrated autocorrelation time
that must be taken into account when estimating statistical uncertainties.

Be that as it may, the extraction of a non-zero EDM is notoriously
difficult due to its poor signal-to-noise ratio.
To address this issue we have employed a novel technique,
first documented in~\cite{Dragos:2018uzd}, for reducing the noise
in our measured observables.  Instead of summing all space-time points
in the calculation of ratios relevant for the extraction of
our 3-point function and $\alpha_N$ term,
this method focuses on the space-time region where the signal is strongest.
We argued that the neglected space-time region gives exponentially suppressed contributions
to the correlation functions and this expectation has been confirmed by our numerical results.
On some ensembles, this method enabled us to increase the signal
to noise by a factor of \(\approx 2\).
This method was described in detail in \autoref{sec:R_imp} and \autoref{app:alpha_imp}.

We have also analysed the $Q^2$ dependence of our
form factors and performed an extraction of the Schiff moment with dynamical fermions.
Our results of \(S_{p} = 0.50(59)\times 10^{-4}\ \bar\theta\ e\)~fm\textsuperscript{3}
for the proton and \(S_{n} = -0.10(43)\times 10^{-4}\ \bar\theta\ e\)~fm\textsuperscript{3}
for the neutron that are in reasonable agreement 
 with chiral perturbation theory predictions.
Our estimates for this value can be improved
upon with more statistics and calculations on larger lattices
(and thus lower $Q^2$ points), which would allow for a more robust extraction.

Our calculation is a big step towards a precise determination of the nucleon EDM and Schiff moment.
Improvement of these results will most definitely
come from increased statistics, and more calculations
at different pion masses at several lattice spacings.
We comment here on the necessity to perform calculations at the physical pion mass.
In the chiral limit the EDM induced by the $\theta$ term vanishes (i.e. $d_{p/n}=0$ at $m_\pi=0$).
In our view, given the small value of the EDM induced by the $\theta$ term
and the additional standard reduction in signal-to-noise as the pion mass is lowered
for nucleon correlators, calculations of these quantities at the physical
pion mass have possibly less to gain than those at higher pion masses.
Because of this constraint, it could be advantageous to have results
at slightly heavier-than-physical pion masses and then
robustly \emph{interpolate} to the physical pion mass using $\chi$PT.
The subsequent errors of the interpolation are stable
and easily quantified precisely because one is doing
an interpolation and not an extrapolation.
We remark though that, for a chirally--breaking lattice action,
such as the non-perturbatively O($a$) improved
Wilson-clover fermion action we have adopted, the nucleon EDMs vanish in the chiral limit
only after performing the continuum limit.
This emphasize the importance of the continuum limit when using
a chirally-breaking action. In this respect the gradient flow
allows us to perform a safe study of discretization effects.

To summarize, the ideal scenario of a direct determination
at the physical point with statistical uncertainties under control,
can be circumvented by simply investing more time
in lattice QCD calculations at slightly heavier pion masses
(where the signal-to-noise is not as prohibitive).
It goes without saying that calculations of the EDM at heavier-than-physical
pion masses can potentially be more cost effective
than the physical-pion-mass calculations only if one has a robust
description of the lattice data with $\chi$PT.


\section*{\label{sect:thanks}Acknowledgments}
We would like to thank Mattia Bruno, Taku Izubuchi and
Sergey Syritsyn for valuable discussions.
This work was supported in part by Michigan State University through
computational resources, provided by the Institute for Cyber-Enabled Research, and in part through financial support from the Deutsche Forschungsgemeinschaft (Sino-German CRC 110).
The authors gratefully acknowledge the computing time granted through JARA-HPC on the supercomputer JURECA~\cite{jureca} at Forschungszentrum J\"ulich.

\appendix

\section{Alpha Improvement Derivation \label{app:alpha_imp}}
Staring with the general three-point correlation function
\begin{equation}\label{eq:start_alpha_imp}
\Delta_{2}^{(O)}(\mathbf{p}^{\, \prime},t;\mathbf{q},\tau;\Pi) =
a^{6}\sum_{\mathbf{x},\mathbf{y}}
e^{-i\mathbf{p}^{\, \prime}\cdot(\mathbf{x}-\mathbf{y})}
e^{i\mathbf{q}\cdot\mathbf{y}}
\text{Tr}\left\lbrace
\Pi
\braket{\mathcal{N}(\mathbf{x},t) O(\mathbf{y},\tau)
\overbar{\mathcal{N}}(\mathbf{0},0) }
\right\rbrace ,
\end{equation}
we handle the time ordering in the next two sections, by performing the spectral decomposition for \(t>\tau\) and \(\tau<t\). This is the general expression for an arbitrary operator \(O\) and spin projector \(\Pi\).
For the computation of the nucleon mixing angle \(\alpha_{N}\), we have \(O=\overbar{Q}\).
\subsection{Case \(t > \tau\)}
Starting with the specific time ordering \(t > \tau\) in eq.~\eqref{eq:start_alpha_imp}, we perform the standard spectral decomposition to produce the correlation function
\begin{equation}
  \begin{split}
  \Delta_{2}^{(O)}(\mathbf{p}^{\, \prime},t;\mathbf{q},\tau;\Pi) =&
  \sum_{\alpha,\beta,\gamma}
  \frac{1}{8E_{\alpha}E_{\beta}E_{\gamma}}
  e^{-E_{\alpha}(T-t)}
  e^{-E_{\beta}(t-\tau)}
  e^{-E_{\gamma}\tau} \\
  &\text{Tr}\lbrace \Pi
    \bra{\alpha} \mathcal{N} \ket{\beta}
    \bra{\beta}O\ket{\gamma}
    \bra{\gamma}\overbar{\mathcal{N}}\ket{\alpha}
  \rbrace,
  \end{split}
\end{equation}
where the sum over states \(\alpha, \beta, \gamma\) have been reduced to states
that only contain momenta \(\mathbf{p}_{\gamma} = \mathbf{q}\),
\(\mathbf{p}_{\beta} = \mathbf{p}-\mathbf{q}\) and \(\mathbf{p}_{\alpha} = \mathbf{p}\).
The two approximation one can apply to this equation are \(T \gg t\) and \(t \gg 0\),
which are related to the source-sink separation of the two-point correlation function
\begin{equation}
  \begin{split}
  \Delta_{2}^{(O)}(\mathbf{p}^{\, \prime},t;\mathbf{q},\tau;\Pi) =&
  \sum_{\beta,\gamma}
  \frac{1}{8E_{\alpha_{0}}E_{\beta}E_{\gamma}}
  e^{-E_{\alpha_{0}}(T-t)}
  e^{-E_{\beta}(t-\tau)}
  e^{-E_{\gamma}\tau} \\
  &\text{Tr}\lbrace \Pi
    \bra{\alpha_{0}} \mathcal{N} \ket{\beta}
    \bra{\beta}O\ket{\gamma}
    \bra{\gamma}\overbar{\mathcal{N}}\ket{\alpha_{0}}
  \rbrace,
  \end{split}
\end{equation}
where $\alpha_{0}$ is the lowest lying energy state that gives a non-zero contribution to $\Delta_{2}^{(O)}$
\begin{equation}
    \text{Tr}\lbrace \Pi
    \bra{\alpha_{0}} \mathcal{N} \ket{\beta}
    \bra{\beta}O\ket{\gamma}
    \bra{\gamma}\overbar{\mathcal{N}}\ket{\alpha_{0}}\rbrace \ne 0.
\end{equation}
\subsection{Case \(t < \tau \)}
This has the same form, with replacing \(\mathcal{N} \leftrightarrow O\)
\begin{equation}
  \begin{split}
  \Delta_{2}^{(O)}(\mathbf{p}^{\, \prime},t;\mathbf{q},\tau;\Pi) =&
  \sum_{\alpha,\beta,\gamma}
  \frac{1}{8E_{\alpha}E_{\beta}E_{\gamma}} \\
  &e^{-E_{\alpha}(T-\tau)}
  e^{-E_{\beta}(\tau-t)}
  e^{-E_{\gamma}t} \\
  &\text{Tr}\lbrace \Pi
    \bra{\alpha} O \ket{\beta}
    \bra{\beta}\mathcal{N}\ket{\gamma}
    \bra{\gamma}\overbar{\mathcal{N}}\ket{\alpha}
  \rbrace.
  \end{split}
\end{equation}
The two approximations \(T \gg t\) and \(t \gg 0\) are again applied
\begin{equation}
  \begin{split}
  \Delta_{2}^{(O)}(\mathbf{p}^{\, \prime},t;\mathbf{q},\tau;\Pi) =&
  \sum_{\alpha,\beta}
  \frac{1}{8E_{\alpha}E_{\beta}E_{\gamma_{0}}} \\
  &e^{-E_{\alpha}(T-\tau)}
  e^{-E_{\beta}(\tau-t)}
  e^{-E_{\gamma_{0}}t} \\
  &\text{Tr}\lbrace \Pi
    \bra{\alpha} O \ket{\beta}
    \bra{\beta}\mathcal{N}\ket{\gamma_{0}}
    \bra{\gamma_{0}}\overbar{\mathcal{N}}\ket{\alpha}
  \rbrace,
  \end{split}
\end{equation}
where this time, \(\gamma_{0}\) is the lowest lying state that gives a non-zero contribution to $\Delta_{2}^{(O)}$
\begin{equation}
  \text{Tr}\lbrace \Pi
  \bra{\alpha} O \ket{\beta}
  \bra{\beta}\mathcal{N}\ket{\gamma_{0}}
  \bra{\gamma_{0}}\overbar{\mathcal{N}}\ket{\alpha}
\rbrace \ne 0.
\end{equation}
\subsection{Total Form}
Over the total range \(\tau \in [0,T]\), the expression is
\begin{equation}
\Delta_{2}^{(O)}(\mathbf{p}^{\, \prime},t;\mathbf{q},\tau;\Pi) =
\left\{
	\begin{array}{ll}
    \sum_{\beta,\gamma}
    \frac{1}{8E_{\alpha_{0}}E_{\beta}E_{\gamma}}
    e^{-E_{\alpha_{0}}(T-t)}
    e^{-E_{\beta}(t-\tau)}
    e^{-E_{\gamma}\tau} & \\
    \text{Tr}\lbrace \Pi
      \bra{\alpha_{0}} \mathcal{N} \ket{\beta}
      \bra{\beta}O\ket{\gamma}
      \bra{\gamma}\overbar{\mathcal{N}}\ket{\alpha_{0}}
    \rbrace & t > \tau \\
    \vspace{1mm} \\
    \sum_{\alpha,\beta}
    \frac{1}{8E_{\alpha}E_{\beta}E_{\gamma_{0}}}
    e^{-E_{\alpha}(T-\tau)}
    e^{-E_{\beta}(\tau-t)}
    e^{-E_{\gamma_{0}}t} & \\
    \text{Tr}\lbrace \Pi
      \bra{\alpha} O \ket{\beta}
      \bra{\beta}\mathcal{N}\ket{\gamma_{0}}
      \bra{\gamma_{0}}\overbar{\mathcal{N}}\ket{\alpha}
    \rbrace & \tau > t
	\end{array}
\right.
\end{equation}

\subsection{Symmetric Partially Summed Current \(t_{s} > t\)}
This region is where the fit will take place, so we omit the derivation for  \(t_{s} < t\).
\begin{equation}
  \begin{split}
    &\overbar{G}_{2}^{(O)}(\mathbf{p}^{\, \prime},t;\mathbf{q},t_{s};\Pi) = \\
    &a\sum_{\tau/a=0}^{t/a} \Delta_{2}^{(O)}(\mathbf{p}^{\, \prime},t;\mathbf{q},\tau;\Pi) +
    a\sum_{\tau/a=t/a+1}^{t_{s}/a} \Delta_{2}^{(O)}(\mathbf{p}^{\, \prime},t;\mathbf{q},\tau;\Pi) +
    a\sum_{\tau/a=0}^{t_{s}/a} \Delta_{2}^{(O)}(\mathbf{p}^{\, \prime},t;\mathbf{q},T-\tau;\Pi) \\
    = &a\sum_{\tau/a=0}^{t/a}
    \sum_{\beta,\gamma}
    \frac{1}{8E_{\alpha_{0}}E_{\beta}E_{\gamma}}
    e^{-E_{\alpha_{0}}(T-t)}
    e^{-E_{\beta}(t-\tau)}
    e^{-E_{\gamma}\tau}
    \text{Tr}\lbrace \Pi
      \bra{\alpha_{0}} \mathcal{N} \ket{\beta}
      \bra{\beta}O\ket{\gamma}
      \bra{\gamma}\overbar{\mathcal{N}}\ket{\alpha_{0}}
    \rbrace + \\
    &a\sum_{\tau/a=t/a+1}^{t_{s}/a}
    \sum_{\alpha,\beta}
    \frac{1}{8E_{\alpha}E_{\beta}E_{\gamma_{0}}}
    e^{-E_{\alpha}(T-\tau)}
    e^{-E_{\beta}(\tau-t)}
    e^{-E_{\gamma_{0}}t}
    \text{Tr}\lbrace \Pi
      \bra{\alpha} O \ket{\beta}
      \bra{\beta}\mathcal{N}\ket{\gamma_{0}}
      \bra{\gamma_{0}}\overbar{\mathcal{N}}\ket{\alpha}
    \rbrace + \\
    &a\sum_{\tau/a=0}^{t_{s}/a}
    \sum_{\alpha,\beta}
    \frac{1}{8E_{\alpha}E_{\beta}E_{\gamma_{0}}}
    e^{-E_{\alpha}\tau}
    e^{-E_{\beta}(T-t-\tau)}
    e^{-E_{\gamma_{0}}t}
    \text{Tr}\lbrace \Pi
      \bra{\alpha} O \ket{\beta}
      \bra{\beta}\mathcal{N}\ket{\gamma_{0}}
      \bra{\gamma_{0}}\overbar{\mathcal{N}}\ket{\alpha}
    \rbrace,
  \end{split}
\end{equation}
noting the second sum is shifted to \(\tau \in [t+a,t_{s}]\) using the lattice spacing increment \(a\).

One thing to note here, is the terms \(\tau = 0\) and \(\tau = t\) are contact terms,
which need to be handled properly (operator product expansion, or gradient flow). Next we group
\(\tau\) terms in preparation for the \(t_{s}\) sum
\begin{equation}
  \begin{split}
    & \overbar{G}_{2}^{(O)}(\mathbf{p}^{\, \prime},t;\mathbf{q},t_{s};\Pi) = \\
    = &
    \sum_{\beta,\gamma}
    \left[
    a\sum_{\tau/a=0}^{t/a}
    e^{-(E_{\gamma}-E_{\beta})\tau}
    \right]
    \frac{1}{8E_{\alpha_{0}}E_{\beta}E_{\gamma}}
    e^{-E_{\alpha_{0}}(T-t)}
    e^{-E_{\beta}t}
    \text{Tr}\left\lbrace \Pi
      \bra{\alpha_{0}} \mathcal{N} \ket{\beta}
      \bra{\beta}O\ket{\gamma}
      \bra{\gamma}\overbar{\mathcal{N}}\ket{\alpha_{0}}
    \right\rbrace + \\
    &\sum_{\alpha,\beta}
    \left[
    a\sum_{\tau/a=t/a+1}^{t_{s}}
    e^{-(E_{\beta}-E_{\alpha})\tau}
    \right]
    \frac{1}{8E_{\alpha}E_{\beta}E_{\gamma_{0}}}
    e^{-E_{\alpha}T}
    e^{-(E_{\gamma_{0}}-E_{\beta})t}
    \text{Tr}\lbrace \Pi
      \bra{\alpha} O \ket{\beta}
      \bra{\beta}\mathcal{N}\ket{\gamma_{0}}
      \bra{\gamma_{0}}\overbar{\mathcal{N}}\ket{\alpha}
    \rbrace + \\
    &
    \sum_{\alpha,\beta}
    \left[
    a\sum_{\tau/a=0}^{t_{s}/a}
    e^{-(E_{\alpha}-E_{\beta})\tau}
    \right]
    \frac{1}{8E_{\alpha}E_{\beta}E_{\gamma_{0}}}
    e^{-E_{\beta}(T-t)}
    e^{-E_{\gamma_{0}}t}
    \text{Tr}\lbrace \Pi
      \bra{\alpha} O \ket{\beta}
      \bra{\beta}\mathcal{N}\ket{\gamma_{0}}
      \bra{\gamma_{0}}\overbar{\mathcal{N}}\ket{\alpha}
    \rbrace.
  \end{split}
\end{equation}

The sums can be computed by using:
\begin{equation}
  \sum_{\tau/a=T_{0}/a}^{T/a} ae^{-E\tau}=
  \sum_{\tau/a=T_{0}/a}^{T/a} ae^{-Ea(\tau/a)}=
  a\frac{e^{E-ET_{0}}- e^{-ET}}{e^{E}-1} =
  a\frac{e^{E(1-T_{0})}- e^{-ET}}{e^{E}-1},
\end{equation}
substituting in this expression, and including the terms where \(\alpha =\beta\) and \(\gamma = \beta\)
separately
\begin{equation}
  \begin{split}
    &\overbar{G}_{2}^{(O)}(\mathbf{p}^{\, \prime},t;\mathbf{q},t_{s};\Pi)= \\
    &\sum_{\beta}
    \frac{t}{8E_{\alpha_{0}}E_{\beta}^{2}}
    e^{-E_{\alpha_{0}}(T-t)}
    e^{-E_{\beta}t}
    \text{Tr}\left\lbrace \Pi
      \bra{\alpha_{0}} \mathcal{N} \ket{\beta}
      \bra{\beta}O\ket{\beta}
      \bra{\beta}\overbar{\mathcal{N}}\ket{\alpha_{0}}
    \right\rbrace + \\
    &
    \sum_{\alpha}
    \frac{2t_{s}-t-a}{8E_{\alpha}^{2}E_{\gamma_{0}}}
    e^{-E_{\alpha}(T-t)}
    e^{-E_{\gamma_{0}}t}
    \text{Tr}\lbrace \Pi
      \bra{\alpha} O \ket{\alpha}
      \bra{\alpha}\mathcal{N}\ket{\gamma_{0}}
      \bra{\gamma_{0}}\overbar{\mathcal{N}}\ket{\alpha}
    \rbrace + \\
    a&\sum_{\beta\ne\gamma}
    \frac{
    e^{(E_{\gamma}-E_{\beta})a}-
    e^{-(E_{\gamma}-E_{\beta})t}
    }{
    8[e^{(E_{\gamma}-E_{\beta})a}-1]
    E_{\alpha_{0}}E_{\beta}E_{\gamma}
    }
    e^{-E_{\alpha_{0}}(T-t)}
    e^{-E_{\beta}t}
    \text{Tr}\lbrace \Pi
      \bra{\alpha_{0}} \mathcal{N} \ket{\beta}
      \bra{\beta}O\ket{\gamma}
      \bra{\gamma}\overbar{\mathcal{N}}\ket{\alpha_{0}}
    \rbrace + \\
    a&\sum_{\alpha\ne\beta}
    \frac{
    e^{(E_{\beta}-E_{\alpha})t}-
    e^{-(E_{\beta}-E_{\alpha})t_{s}}
    }{
    8[e^{(E_{\beta}-E_{\alpha})a}-1]
    E_{\alpha}E_{\beta}E_{\gamma_{0}}
    }
    e^{-E_{\alpha}T}
    e^{-(E_{\gamma_{0}}-E_{\beta})t}
    \text{Tr}\lbrace \Pi
      \bra{\alpha} O \ket{\beta}
      \bra{\beta}\mathcal{N}\ket{\gamma_{0}}
      \bra{\gamma_{0}}\overbar{\mathcal{N}}\ket{\alpha}
    \rbrace + \\
    a&\sum_{\alpha\ne\beta}
    \frac{
    e^{(E_{\alpha}-E_{\beta})t}-
    e^{-(E_{\alpha}-E_{\beta})t_{s}}
    }{
    8[e^{(E_{\alpha}-E_{\beta})a}-1]
    E_{\alpha}E_{\beta}E_{\gamma_{0}}
    }
    e^{-E_{\beta}(T-t)}
    e^{-E_{\gamma_{0}}t}
    \text{Tr}\lbrace \Pi
      \bra{\alpha} O \ket{\beta}
      \bra{\beta}\mathcal{N}\ket{\gamma_{0}}
      \bra{\gamma_{0}}\overbar{\mathcal{N}}\ket{\alpha}
    \rbrace.
  \end{split}
\end{equation}

As the final 2 terms are only exponentially dependent on $t_{s}$, we write these terms
as exponentials of single energy indices
\begin{equation}
  \begin{split}
    &\overbar{G}_{2}^{(O)}(\mathbf{p}^{\, \prime},t;\mathbf{q},t_{s};\Pi)= \\
    &\sum_{\beta}
    \frac{t}{8E_{\alpha_{0}}E_{\beta}^{2}}
    e^{-E_{\alpha_{0}}(T-t)}
    e^{-E_{\beta}t}
    \text{Tr}\left\lbrace \Pi
      \bra{\alpha_{0}} \mathcal{N} \ket{\beta}
      \bra{\beta}O\ket{\beta}
      \bra{\beta}\overbar{\mathcal{N}}\ket{\alpha_{0}}
    \right\rbrace + \\
    &
    \sum_{\alpha}
    \frac{2t_{s}-t-a}{8E_{\alpha}^{2}E_{\gamma_{0}}}
    e^{-E_{\alpha}(T-t)}
    e^{-E_{\gamma_{0}}t}
    \text{Tr}\lbrace \Pi
      \bra{\alpha} O \ket{\alpha}
      \bra{\alpha}\mathcal{N}\ket{\gamma_{0}}
      \bra{\gamma_{0}}\overbar{\mathcal{N}}\ket{\alpha}
    \rbrace + \\
    a&\sum_{\beta\ne\gamma}
    \frac{
    e^{(E_{\gamma}-E_{\beta})a}-
    e^{-(E_{\gamma}-E_{\beta})t}
    }{8
    [e^{(E_{\gamma}-E_{\beta})a}-1]
    E_{\alpha_{0}}E_{\beta}E_{\gamma}
    }
    e^{-E_{\alpha_{0}}(T-t)}
    e^{-E_{\beta}t}
    \text{Tr}\lbrace \Pi
      \bra{\alpha_{0}} \mathcal{N} \ket{\beta}
      \bra{\beta}O\ket{\gamma}
      \bra{\gamma}\overbar{\mathcal{N}}\ket{\alpha_{0}}
    \rbrace + \\
    a&\sum_{\alpha\ne\beta}
    \frac{
    e^{-E_{\alpha}(T+t)}
    e^{-(E_{\gamma_{0}}-2E_{\beta})t}
    -
    e^{-E_{\alpha}(T-t_{s})}
    e^{-E_{\gamma_{0}}t}
    e^{-E_{\beta}(t_{s}-t)}
    }{8
    [e^{(E_{\beta}-E_{\alpha})a}-1]
    E_{\alpha}E_{\beta}E_{\gamma_{0}}
    }
    \text{Tr}\lbrace \Pi
      \bra{\alpha} O \ket{\beta}
      \bra{\beta}\mathcal{N}\ket{\gamma_{0}}
      \bra{\gamma_{0}}\overbar{\mathcal{N}}\ket{\alpha}
    \rbrace + \\
    a&\sum_{\alpha\ne\beta}
    \frac{
    e^{-E_{\beta}T}
    e^{-E_{\gamma_{0}}t}
    e^{E_{\alpha}t}
    -
    e^{-E_{\beta}(T-t-t_{s})}
    e^{-E_{\gamma_{0}}t}
    e^{-E_{\alpha}t_{s}}
    }{8
    [e^{(E_{\alpha}-E_{\beta})a}-1]
    E_{\alpha}E_{\beta}E_{\gamma_{0}}
    }
    \text{Tr}\lbrace \Pi
      \bra{\alpha} O \ket{\beta}
      \bra{\beta}\mathcal{N}\ket{\gamma_{0}}
      \bra{\gamma_{0}}\overbar{\mathcal{N}}\ket{\alpha}
    \rbrace,
  \end{split}
\end{equation}
and clumping like terms
\begin{equation}
  \begin{split}
    &\overbar{G}_{2}^{(O)}(\mathbf{p}^{\, \prime},t;\mathbf{q},t_{s};\Pi)= \\
    &\sum_{\beta}
    \frac{t}{8E_{\alpha_{0}}E_{\beta}^{2}}
    e^{-E_{\alpha_{0}}(T-t)}
    e^{-E_{\beta}t}
    \text{Tr}\left\lbrace \Pi
      \bra{\alpha_{0}} \mathcal{N} \ket{\beta}
      \bra{\beta}O\ket{\beta}
      \bra{\beta}\overbar{\mathcal{N}}\ket{\alpha_{0}}
    \right\rbrace + \\
    &
    \sum_{\alpha}
    \frac{2t_{s}-t-a}{8E_{\alpha}^{2}E_{\gamma_{0}}}
    e^{-E_{\alpha}(T-t)}
    e^{-E_{\gamma_{0}}t}
    \text{Tr}\lbrace \Pi
      \bra{\alpha} O \ket{\alpha}
      \bra{\alpha}\mathcal{N}\ket{\gamma_{0}}
      \bra{\gamma_{0}}\overbar{\mathcal{N}}\ket{\alpha}
    \rbrace + \\
    a&\sum_{\beta\ne\gamma}
    \frac{
    e^{(E_{\gamma}-E_{\beta})a}-
    e^{-(E_{\gamma}-E_{\beta})t}
    }{8
    [e^{(E_{\gamma}-E_{\beta})a}-1]
    E_{\alpha_{0}}E_{\beta}E_{\gamma}
    }
    e^{-E_{\alpha_{0}}(T-t)}
    e^{-E_{\beta}t}
    \text{Tr}\lbrace \Pi
      \bra{\alpha_{0}} \mathcal{N} \ket{\beta}
      \bra{\beta}O\ket{\gamma}
      \bra{\gamma}\overbar{\mathcal{N}}\ket{\alpha_{0}}
    \rbrace + \\
    a&\sum_{\alpha\ne\beta}
    \text{Tr}\lbrace \Pi
      \bra{\alpha} O \ket{\beta}
      \bra{\beta}\mathcal{N}\ket{\gamma_{0}}
      \bra{\gamma_{0}}\overbar{\mathcal{N}}\ket{\alpha}
    \rbrace
    e^{-E_{\gamma_{0}}t} \\
    &\hspace*{0.7cm}\Bigg[
    \frac{
    e^{-E_{\alpha}(T+t)}
    e^{2E_{\beta}t}
    -
    e^{-E_{\alpha}(T-t_{s})}
    e^{-E_{\beta}(t_{s}-t)}
    }{8
    [e^{(E_{\beta}-E_{\alpha})a}-1]
    E_{\alpha}E_{\beta}E_{\gamma_{0}}
    }
    +
    \frac{
    e^{-E_{\beta}T}
    e^{E_{\alpha}t}
    -
    e^{-E_{\beta}(T-t-t_{s})}
    e^{-E_{\alpha}t_{s}}
    }{8
    [e^{(E_{\alpha}-E_{\beta})a}-1]
    E_{\alpha}E_{\beta}E_{\gamma_{0}}
    }\Bigg].
  \end{split}
\end{equation}

\subsection{Explicit form for $O=\overbar{Q}$}
As \(\overbar{Q}\) is a parity violating operator, the nucleon states that propagate before and
after this operator must be opposite in parity. This removes the first and second terms as
$\bra{\beta}\overbar{Q}\ket{\beta} = 0$. As well as this, the terms with sums over two terms either
require (\(\alpha,\beta = \alpha_{+},\beta_{-}\)) or (\(\alpha,\beta = \alpha_{-},\beta_{+}\))
where the subscript \(\pm\) refers to the state having positive or negative parity.
The projector is selected to be \(\Pi=\gamma_{5}\Pi_{+} = \gamma_{5}\frac{I+\gamma_{4}}{2}\),
which results in only the trace term with (\(\alpha,\beta = \alpha_{+},\beta_{-}\)) being non-zero
\begin{equation}
  \begin{split}
    &\overbar{G}_{2}^{(\overbar{Q})}(\mathbf{p}^{\, \prime},t;\mathbf{q},t_{s};\gamma_{5}\Pi_{+})= \\
    &a\sum_{\beta\ne\gamma}
    \frac{
    e^{(E_{\gamma}-E_{\beta})a}-
    e^{-(E_{\gamma}-E_{\beta})t}
    }{8
    [e^{(E_{\gamma}-E_{\beta})a}-1]
    E_{\alpha_{0}}E_{\beta}E_{\gamma}
    }
    e^{-E_{\alpha_{0}}(T-t)}
    e^{-E_{\beta}t}
    \text{Tr}\lbrace \gamma_{5}\Pi_{+}
      \bra{\alpha_{0}} \mathcal{N} \ket{\beta}
      \bra{\beta}\overbar{Q}\ket{\gamma}
      \bra{\gamma}\overbar{\mathcal{N}}\ket{\alpha_{0}}
    \rbrace + \\
    &a\sum_{\alpha_{+},\beta_{-}}
    \text{Tr}\lbrace \gamma_{5}\Pi_{+}
      \bra{\alpha_{+}} \overbar{Q} \ket{\beta_{-}}
      \bra{\beta_{-}}\mathcal{N}\ket{\gamma_{0}}
      \bra{\gamma_{0}}\overbar{\mathcal{N}}\ket{\alpha_{+}}
    \rbrace
    e^{-E_{\gamma_{0}}t} \\
    &\hspace*{0.7cm}\Bigg[
    \frac{
    e^{-E_{\alpha_{+}}(T+t)}
    e^{2E_{\beta_{-}}t}
    -
    e^{-E_{\alpha_{+}}(T-t_{s})}
    e^{-E_{\beta_{-}}(t_{s}-t)}
    }{8
    [e^{(E_{\beta_{-}}-E_{\alpha_{+}})a}-1]
    E_{\alpha_{+}}E_{\beta_{-}}E_{\gamma_{0}}
    }
    +
    \frac{
    e^{-E_{\beta_{-}}T}
    e^{E_{\alpha_{+}}t}
    -
    e^{-E_{\beta_{-}}(T-t-t_{s})}
    e^{-E_{\alpha_{+}}t_{s}}
    }{8
    [e^{(E_{\alpha_{+}}-E_{\beta_{-}})a}-1]
    E_{\alpha_{+}}E_{\beta_{-}}E_{\gamma_{0}}
    }\Bigg].
  \end{split}
\end{equation}

The terms $e^{-E_{\alpha_{+}}(T+t)}$ and $e^{-E_{\alpha_{+}}(T-t_{s})}$ in the final sum are exponentially
suppressed as \({T\gg T/2 \geq t_{s}}\) and \({T\gg t}\)
\begin{equation}
  \begin{split}
    &\overbar{G}_{2}^{(\overbar{Q})}(\mathbf{p}^{\, \prime},t;\mathbf{q},t_{s};\gamma_{5}\Pi_{+})= \\
    &a\sum_{\beta\ne\gamma}
    \frac{
    e^{(E_{\gamma}-E_{\beta})a}-
    e^{-(E_{\gamma}-E_{\beta})t}
    }{8
    [e^{(E_{\gamma}-E_{\beta})a}-1]
    E_{\alpha_{0}}E_{\beta}E_{\gamma}
    }
    e^{-E_{\alpha_{0}}(T-t)}
    e^{-E_{\beta}t}
    \text{Tr}\lbrace \gamma_{5}\Pi_{+}
      \bra{\alpha_{0}} \mathcal{N} \ket{\beta}
      \bra{\beta}\overbar{Q}\ket{\gamma}
      \bra{\gamma}\overbar{\mathcal{N}}\ket{\alpha_{0}}
    \rbrace + \\
    &a\sum_{\alpha_{+},\beta_{-}}
    \frac{
    e^{-E_{\beta_{-}}T}
    e^{E_{\alpha_{+}}t}
    }{8
    [e^{(E_{\alpha_{+}}-E_{\beta_{-}})a}-1]
    E_{\alpha_{+}}E_{\beta_{-}}E_{\gamma_{0}}
    }
    e^{-E_{\gamma_{0}}t}
    \text{Tr}\lbrace \gamma_{5}\Pi_{+}
      \bra{\alpha_{+}} \overbar{Q} \ket{\beta_{-}}
      \bra{\beta_{-}}\mathcal{N}\ket{\gamma_{0}}
      \bra{\gamma_{0}}\overbar{\mathcal{N}}\ket{\alpha_{+}}
    \rbrace - \\
    &a\sum_{\alpha_{+},\beta_{-}}
    \frac{
    e^{-E_{\beta_{-}}(T-t-t_{s})}
    e^{-E_{\alpha_{+}}t_{s}}
    }{8
    [e^{(E_{\alpha_{+}}-E_{\beta_{-}})a}-1]
    E_{\alpha_{+}}E_{\beta_{-}}E_{\gamma_{0}}
    }
    e^{-E_{\gamma_{0}}t}
    \text{Tr}\lbrace \gamma_{5}\Pi_{+}
      \bra{\alpha_{+}} \overbar{Q} \ket{\beta_{-}}
      \bra{\beta_{-}}\mathcal{N}\ket{\gamma_{0}}
      \bra{\gamma_{0}}\overbar{\mathcal{N}}\ket{\alpha_{+}}
    \rbrace.
  \end{split}
\end{equation}

From this complicated expression, the \(t_{s}\) dependence only appears exponentially in the final term.
Therefore, we can fit the two-point correlation function with
\begin{equation}
fit(t_{s}) = A + Be^{-Et_{s}}.
\end{equation}

Due to the statistical noise of the data and high correlation in the data with respect to $t_{s}$,
we elected to neglect the excited state term by fitting a constant in the region where \(Be^{-Et_{s}} \ll A\).


\section{Ratio Function Fit Range Selection \label{app:fitr_select}}

In this appendix, we present the technique used for extracting the CP-odd form factor $F_3(Q^2)$
from the ratio function in eq.~\eqref{eq:Rat_fit}.
Since only constant ("one-state") fits are implemented for the ratio functions, careful consideration to
excited state effects is needed.

The method employed to account for fit range dependence in our error estimates, is to include multiple fit ranges
that satisfy some \(\chi^{2}\) per degree of freedom (\(\chi^{2}_{PDF}\)) criterium.
For this study, we only select fits that satisfy \(\chi^{2}_{PDF} \in [0.5,1]\).
Using the multiple fit range deturminations of \(R\) and \(R^{Q}\), we extend eq.~\eqref{eq:FF_system}
to include different fit ranges:

\begin{equation} \label{eq:FF_system_fits}
  \sum_{i=1}^{3}
  \mathcal{A}(Q^{2})_{Ai} F_{i,f(A)}(Q^{2}) =
  \left\{
	\begin{array}{ll}
    &R_{f(A)}(\bm{0},t,\bm{q}_{j},\Pi_{k},\gamma_{l}) \\
    &R^{(Q)}_{f(A)}(\bm{0},t,\bm{q}_{j},\Pi_{k},\gamma_{l},t_{f}) \\
	\end{array}
  \right. ,
\end{equation}
where the extra index \(f(A)\) refers to which fit range is used, which depends on the collective
index \(A=\{j,k,l\},\, A\in[1,\ldots,N_{A}]\).

Since each ratio function selected by index \(A\) has \(f(A)\) different ways to extract the quantity, the system is solved
for every combination of \(f(A)\, \forall\, A \in [1,\ldots,N_{A}]\). This results in \(\prod_{A} F(A)\) independent system of equations to solve,
where \(F(A)\) is the number of different fits accepted (using the \(\chi^{2}_{PDF}\) criterium) for index \(f(A) = 1,2,\ldots, F(A)\).

Once the form factors have been solved over different fit range combinations, the result we obtain is \(F_{i,f}(Q^{2})\),
where the (\(A\) missing) index \(f\) refers to which combined set of fit ranges were used. Since the extrapolation
to \(Q^{2}\rightarrow 0\) must be performed to compute the nucleon EDM, this must be performed for every \(f(Q^{2})\)
combination (analogous to \(f(A)\) above). So in addition to above, we increase the number of fits to
\(\prod_{Q^{2}}F(Q^{2})\), where \(F(Q^{2})\) is the number of fits computed for index \(f(Q^{2})= 1,2,\ldots, F(Q^{2})\).

Combining both these studies together, the resulting nucleon EDM has been computed using different fit ranges,
indexed by \(\frac{F^{p/n}_{3,f}(Q^{2}\rightarrow 0)}{2M_{N}}=d_{p/n,f}\). So to obtain a final result where
the statistical uncertainty from the gauge fields and the systematic errors arising from the fit ranges
can be combine into a single uncertainty, we extend the bootstrap samples which are already used to compute
the statistical uncertainty \(d_{p/n,B}\) where \(B\) runs over \([1,N_{b}] \otimes [1,N_{F}]\) where \(N_{f}\)
is the number of fits which each have \(N_{b}\) bootstrap samples.

\subsection{Computational Viability \label{app:fitr_comp}}

As one may notice, the above formulation is of order \(O(A!)\), assuming a fixed number of fit ranges selected.
A stochastic estimation of the fit range variation is highly recommended, which can be employed when solving the form factor eq.~\eqref{eq:FF_system_fits}, as well as when taking the form factor \(Q^{2}\rightarrow 0\) extrapolation.

At the form factor solving stage, this is employed by randomly selecting \(N_{\chi}\) different fit range that satisfy
the \(\chi^{2}_{PDF}\) criterium. The resulting number of systems of equations to be solved are \(N_{\chi}^{N_{A}}\).

For the form factor extrapolation in \(Q^{2} \rightarrow 0\), a random selection of \(N_{F}\) results
of index \(f(Q^{2})\) in  \(F_{N,f(Q^{2})}(Q^{2})\). The resulting number of fits to be performed
using this estimation is \(N_{F}^{N_{Q}}\) for \(N_{Q}\) number of transfer momentum \(Q^{2}\).

\subsection{Results computed in this paper \label{app:paper_fitr}}
The results computed in this paper use the fit criterium \(\chi^{2}_{PDF} \in [0.5,1]\), and excluded fits of length 2.
The cutoff for the number of fit ranges per ratio function is \(N_{\chi}=4\) and the cutoff for the
form factor extrapolation is \(N_{Q}=4\) as well.

\begin{figure}
\centering
\begin{subfigure}{.3\textwidth}
  \centering
  \includegraphics[trim={0cm 0cm 5mm 0cm},width=\linewidth]{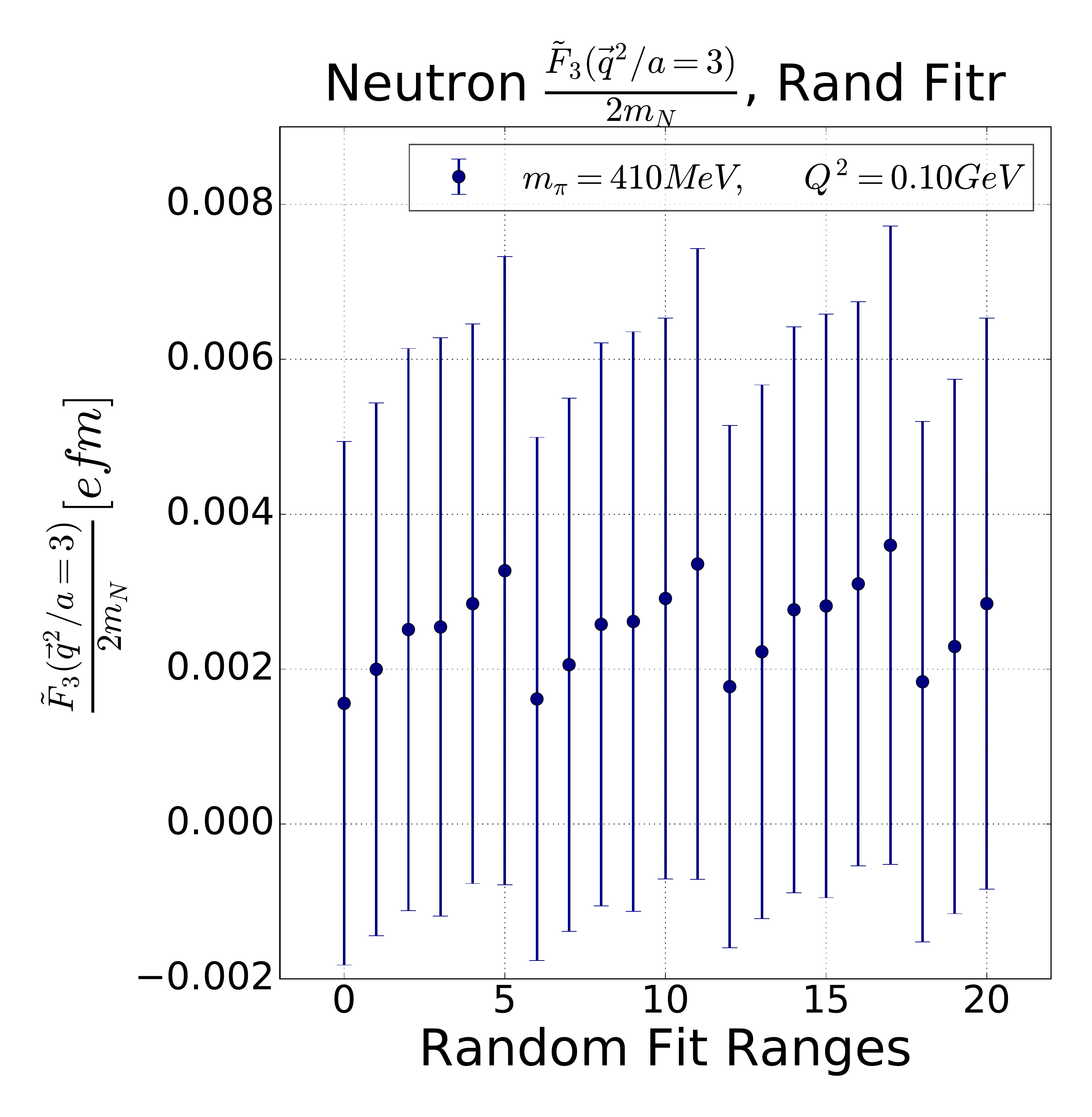}
  \caption{\label{fig:F3_chi_mpi410}}
\end{subfigure}
\quad
\begin{subfigure}{.3\textwidth}
  \centering
  \includegraphics[trim={0cm 0cm 5mm 0cm},width=\linewidth]{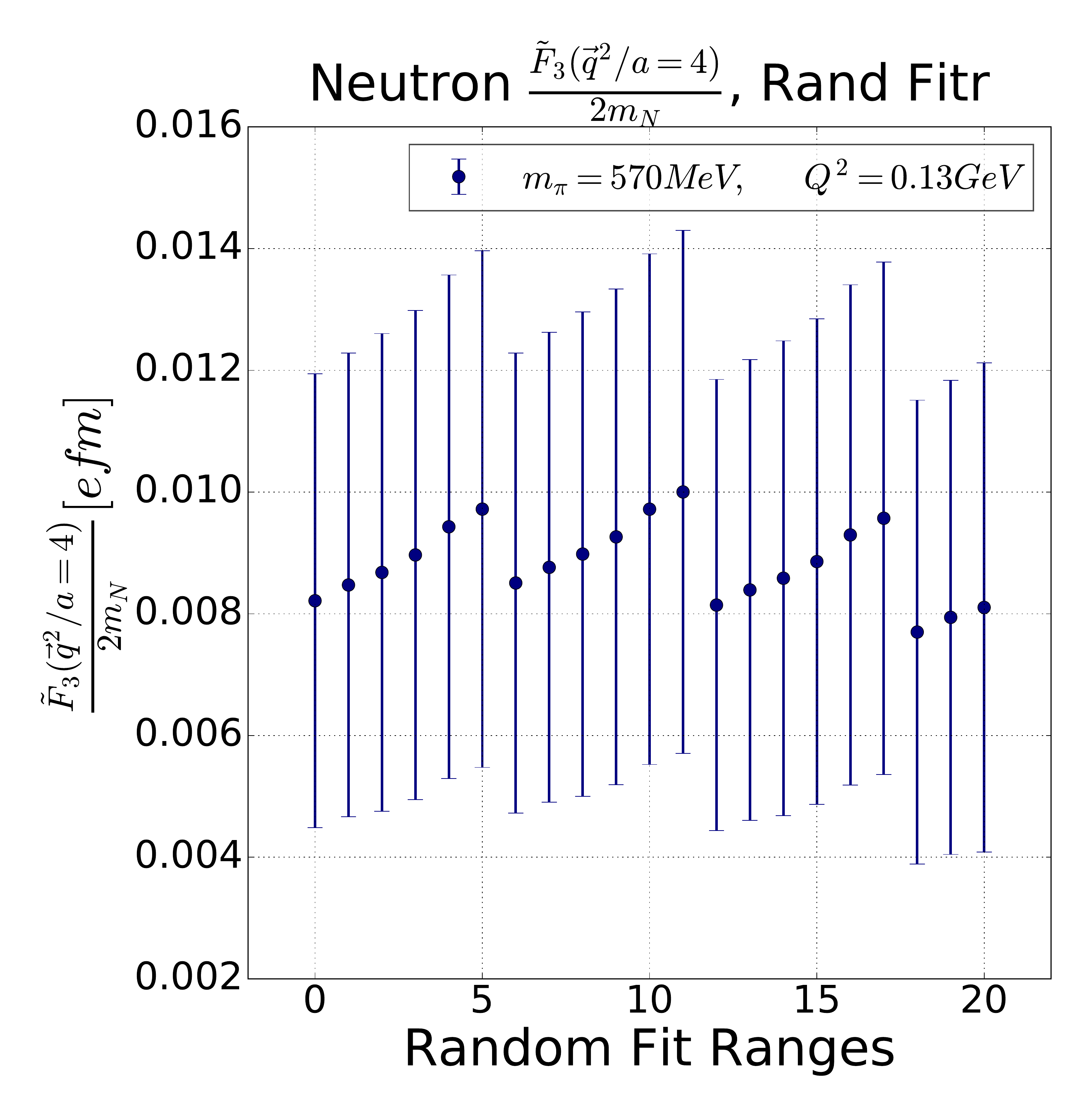}
  \caption{\label{fig:F3_chi_mpi570}}
\end{subfigure}\quad
\begin{subfigure}{.3\textwidth}
  \centering
  \includegraphics[trim={0cm 0cm 5mm 0cm},width=\linewidth]{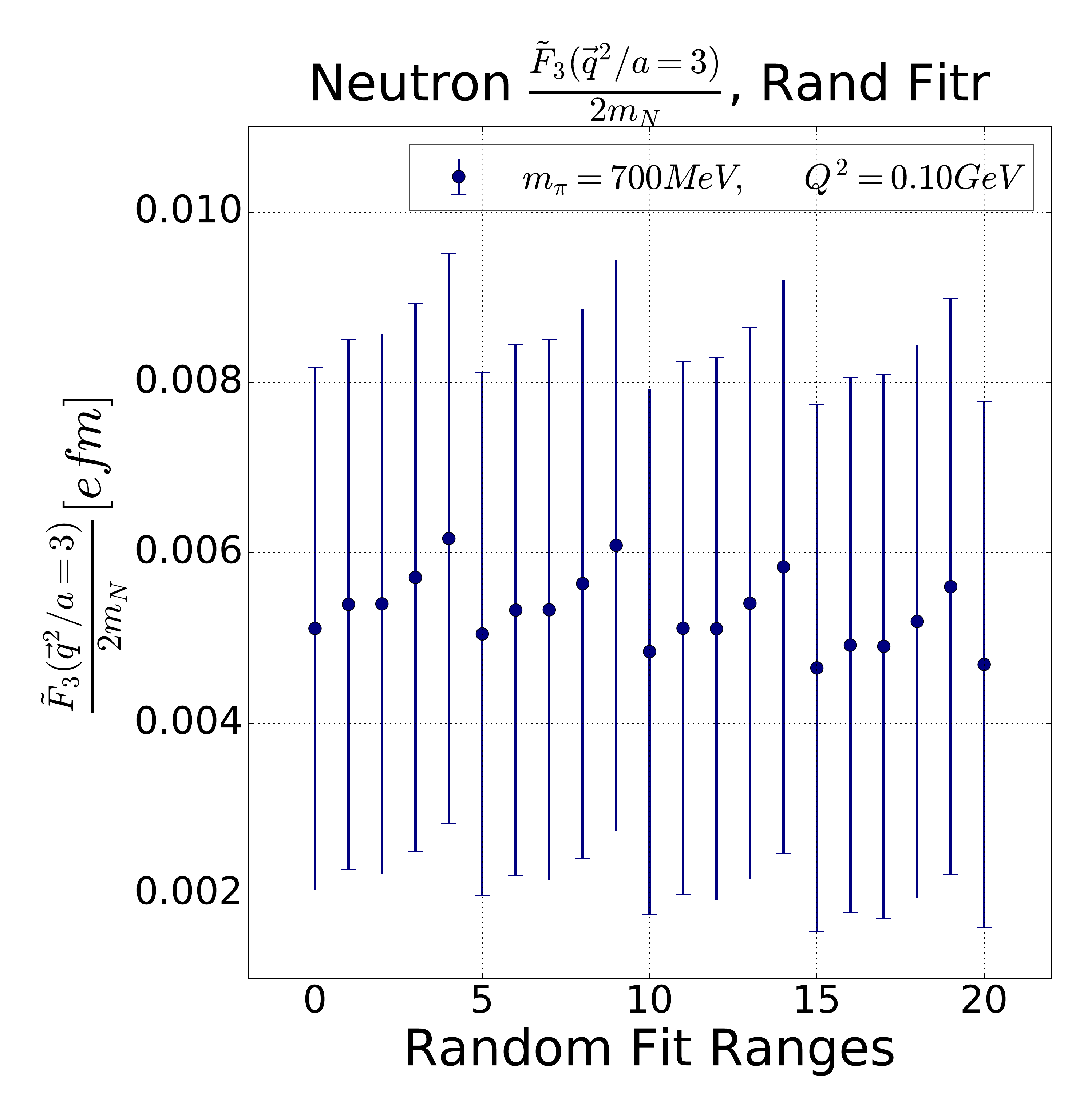}
  \caption{\label{fig:F3_chi_mpi700}}
\end{subfigure}
\caption{
  Comparison of different randomly selected fit ranges used in the solving of
  the CP-odd form factor \(\frac{F_{3}(Q^{2})}{2M_{N}}\)
  using the \(m_{\pi}=410,570,700\) MeV (left, middle and right) ensembles.
  Although lattice transfer momentum increment \((\bm{q}/a)^{2} = 3, 4\) were selected, all
  other momenta exhibited the same (lack of) behavior.}
  \label{fig:F3_chi}
\end{figure}

As for the the number of equations to solve, results at lattice \((\bm{q}/a)^{2} = 1, 4\) has \(1024\) equations,
\((\bm{q}/a)^{2} = 2\) has \(4096\) equations and \((\bm{q}/a)^{2} = 3\) has  \(16384\) equations. Multiplying these numbers by \(200\)
bootstrap samples, will give the individual number of system of equations solved. Once this is complete,
we avoid computing \(\sim 70\) trillion equations by performing the stochastic estimate which only requires \(256\)
equations to solve for. Although it may seem the values for \(N_{\chi}\) and \(N_{F}\) are insufficient in size, the results shown in
fig.~\ref{fig:F3_chi}
demonstrates minimal variation when analyzing each individual \(\frac{F^{p/n}_{3,f}(Q^{2})}{2M_{N}}\)
over different fit ranges \(f\).

\bibliographystyle{h-physrev3}
\bibliography{references}

\end{document}